\documentclass[conference,compsoc]{IEEEtran}
\ifCLASSOPTIONcompsoc
  \usepackage[nocompress]{cite}
\else
  \usepackage{cite}
\fi
\ifCLASSINFOpdf
\else
\fi

\usepackage{url}
\usepackage{graphicx}
\usepackage{cite}
\usepackage{threeparttable}

\usepackage{subfigure}
\usepackage{bm}
\usepackage{longtable} 
\usepackage{color}
\usepackage{booktabs}
\usepackage{float}
\usepackage{makecell}
\usepackage{amsmath,amsthm,amssymb}
\usepackage{multirow}
\usepackage{array}
\usepackage{stfloats}
\usepackage{algorithm}
\usepackage{algorithmic}

\usepackage{bbding} 
\usepackage{cancel}

\begin{document}
%
\title{SoK: Comparing Different Membership Inference Attacks with a Comprehensive Benchmark}

\IEEEoverridecommandlockouts

%
\author{\IEEEauthorblockN{Jun Niu\IEEEauthorrefmark{1},
Xiaoyan Zhu\thanks{$\star$ Xiaoyan Zhu and Yuqing Zhang are the corresponding authors.}\IEEEauthorrefmark{1},
Moxuan Zeng\IEEEauthorrefmark{2}, 
Ge Zhang\IEEEauthorrefmark{1}, Qingyang Zhao\IEEEauthorrefmark{1}, Chunhui Huang\IEEEauthorrefmark{2}, \\ Yangming Zhang\IEEEauthorrefmark{2}, Suyu An\IEEEauthorrefmark{2}, Yangzhong Wang\IEEEauthorrefmark{2}, Xinghui Yue\IEEEauthorrefmark{3}, Zhipeng He\IEEEauthorrefmark{4}, Weihao Guo\IEEEauthorrefmark{2},\\ Kuo Shen\IEEEauthorrefmark{1}, Peng Liu\IEEEauthorrefmark{5}, Yulong Shen\IEEEauthorrefmark{1}, Xiaohong Jiang\IEEEauthorrefmark{6}, Jianfeng Ma\IEEEauthorrefmark{1}, Yuqing Zhang\IEEEauthorrefmark{7}\IEEEauthorrefmark{1}}
\IEEEauthorblockA{\IEEEauthorrefmark{1}Xidian University, \{niujun,21151213588\}@stu.xidian.edu.cn},\{xyzhu, ylshen, jfma\}@mail.xidian.edu.cn, zhangg@nipc.org.cn, shenkuo\_xdu@163.com
\IEEEauthorblockA{\IEEEauthorrefmark{2}Hainan University, zengmoxuan@hainanu.edu.cn, \{huangch,zhangym,ansy,wangyz21,guowh\}@nipc.org.cn}
\IEEEauthorblockA{\IEEEauthorrefmark{3}Yanshan University, yxh0314@stumail.ysu.edu.cn}
\IEEEauthorblockA{\IEEEauthorrefmark{4}Xi'an University of Posts \& Telecommunications, hezhp@nipc.org.cn}
\IEEEauthorblockA{\IEEEauthorrefmark{5}Pennsylvania State University, pliu@ist.psu.edu}
\IEEEauthorblockA{\IEEEauthorrefmark{6}Future University of Hakodate, jiang@fun.ac.jp}
\IEEEauthorblockA{\IEEEauthorrefmark{7}University of Chinese Academy of Sciences, zhangyq@ucas.ac.cn}
}


\maketitle

\begin{abstract}
Membership inference (MI) attacks threaten user privacy through determining if a given data example has been used to train a target model. However, it has been increasingly recognized that the ``comparing different MI attacks'' methodology used in the existing works has serious limitations. Due to these limitations, we found (through the experiments in this work) that some comparison results reported in the literature are quite misleading. In this paper, 
we seek to develop a comprehensive benchmark for comparing 
different MI attacks, 
called \textbf{MIBench}, which consists not only the evaluation metrics, but also the evaluation scenarios. And we design the evaluation scenarios 
from {\bf four perspectives}: the distance distribution of data samples in the target dataset, the distance between data samples of the target dataset, the differential distance between two datasets (i.e., the target dataset and a generated dataset with only nonmembers), and the ratio of the samples that are made no inferences by an MI attack. The evaluation metrics consist of ten typical evaluation metrics. We have identified three principles for the proposed ``comparing different
MI attacks'' methodology, and we have designed and implemented 
the MIBench benchmark with 84 evaluation scenarios for each dataset. In total, we have used our benchmark to fairly and systematically 
compare 15 state-of-the-art MI attack algorithms across 588 evaluation scenarios, and these evaluation scenarios cover 7 widely used datasets and 7 representative types of models.  All codes and evaluations of MIBench are publicly available at \url{https://github.com/MIBench/MIBench.github.io/blob/main/README.md}.
\end{abstract}


%
\IEEEpeerreviewmaketitle


\section{Introduction}
\label{sebsec:Introduction}

Recently, machine learning (ML), especially deep learning (DL), has achieved tremendous progress in various domains such as image recognition~\cite{he2016deep}, speech recognition~\cite{hannun2014deep}, 

 \begin{figure}[htp]
\vspace{0.1pt}
\includegraphics[width=0.5\textwidth]{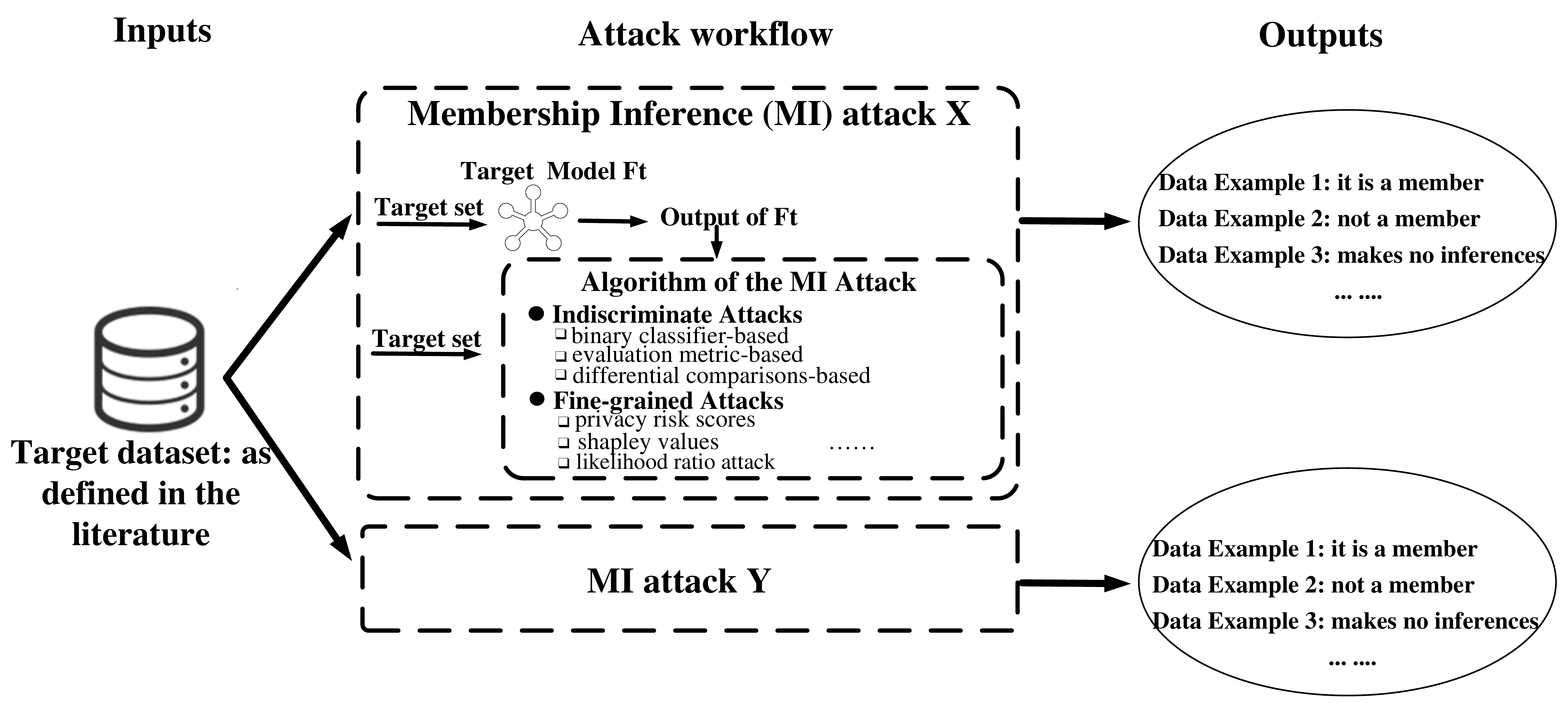}
\caption{\label{figs:Attack workflow_V8} A summary of MI attack workflows.}
\vspace{0.1pt}
\end{figure}

\noindent natural language processing~\cite{devlin2018bert} and medical analysis~\cite{choi2016doctor}. 
  However, many researches have demonstrated that ML models are vulnerable to various attacks, such as 
  property   
  inference attacks~\cite{
  song2020overlearning
  },  adversarial attacks~\cite{rosenberg2021adversarial}, model poisoning attacks~\cite{jere2020taxonomy} and membership inference attacks~\cite{shokri2017membership,salem2019ml,carlini2021extracting,jia2019memguard}.

  Among these attacks, Membership Inference (MI) attacks ~\cite{sablayrolles2019white,salem2019ml,shokri2017membership,yeom2018privacy,hui2021practical} determine whether a data example is inside the training dataset of the target model or not. The workflows of the existing MI attacks are summarized in Figure \ref{figs:Attack workflow_V8}: although all the existing MI attack algorithms have the same inputs, indiscriminate attack algorithms (e.g., \cite{shokri2017membership,song2019robust,liu2022ml,hisamoto2020membership,liew2020faceleaks,chen2021machine,he2020segmentations,nasr2019comprehensive,ha2022membership,gu2022cs,liu2021encodermi,webster2021person,li2022user,yang2021vulnerability,zhang2021membership,zhang2022inference,yeom2018privacy,salem2019ml,song2017machine,sablayrolles2019white,li2020membership,farokhi2020modelling,chang2021privacy,kaya2020effectiveness,carlini2018secret,salem2020updates,choquette2021label,chen2020gan,li2021membership,chen2020beyond,zhang2022evaluating,del2022leveraging,mahloujifar2022optimal,zhong2022understanding,li2022leaks,carlini2021extracting,he2021quantifying,bentley2020quantifying,leino2020stolen,rezaei2021difficulty,hui2021practical}) and fine-grained attack algorithms (e.g., \cite{jayaraman2020revisiting, long2020pragmatic, song2021systematic, ye2021enhanced, duddu2021shapr, carlini2022membership, watson2021importance}) could have different (kinds of) outputs. In particular, every attack algorithm has two inputs: (1) the {\bf target dataset}, which holds the set of given to-be-inferred data examples; (2) the outputs of the {\bf target model} against each given data example. 
For indiscriminate attack algorithms, their output is binary, telling whether a given data example is a member or not. In contrast, fine-grained attack algorithms firstly make (fine-grained) Yes/No decisions on making an inference or not.
When a No decision is made on data example $x$, the output 
will be ``no inference is made on $x$'' (e.g., Data Example 3 showed in Figure ~\ref{figs:Attack workflow_V8}). 

Due to the importance of MI attacks, researchers have proposed many MI attacks in recent years. However, it has been increasingly recognized that the ``comparing different MI attacks'' methodology used in the existing works has serious limitations. Due to these limitations, we found (through the experiments in this work)  
that some comparison results reported in the literature are quite misleading. 

To illustrate the serious limitations, let's summarize the main factors determining when an MI attack is more effective and when it is less effective. 
First of all, two primary factors have been recognized in all the existing works: the kind of data 
in the target dataset; and the type of the target model. In order to incorporate these two factors in 
evaluating attack effectiveness, all the existing works use multiple target datasets and multiple types (e.g., MLP, ResNet, DenseNet) of target models. 
Second, based on the experiment results which we will shortly present in Section \ref{sec:Using the Evaluation Framework}, we found the following factors. 
{\bf Factor 1.} It is found in our experiments (see Section \ref{sebsec:Effect of the Distance Distribution of Data Samples in the Target Dataset})   
that the effectiveness of 
an MI attack could be highly sensitive to the distance distribution of data samples in the target dataset. 
{\bf Factor 2.} It is found in our experiments, which we will shortly present in Section \ref{sebsec:Effect of Distance between members and nonmembers}, 
that the effectiveness of  
an MI attack could be highly sensitive to the distances 
between 
data samples of the target dataset.
The larger the distance between data samples of the target dataset, 
the higher the attacker's {\bf membership advantage} (MA). 
By ``membership advantage'', we mean the difference between an MI attack’s true and false positive rates (e.g., MA = TPR-FPR~\cite{yeom2018privacy}). 
In addition, we found that 
the existing state-of-the-art MI attacks can achieve high inference accuracy 
and low FPR (false positive rate) simultaneously when the distances between 
data samples of the target dataset are carefully controlled. 
{\bf Factor 3.} It is found in our experiments (see Section \ref{sebsec:Effect of Differential Distances between two datasets}) that the effectiveness of 
an MI attack could be very sensitive to the differential distance (i.e., Maximum Mean Discrepancy (MMD)~\cite{gretton2012kernel}) 
before and after a data example is moved from the target dataset 
to a generated dataset with only nonmembers.  
In addition, we found that in general Factor 3 has  
greater influence (on attack effectiveness) than Factor 2. 
{\bf Factor 4.} Since fine-grained attack algorithms make no inferences 
on certain data examples, it seems {\em unfair} to use the same number of 
data examples when comparing indiscriminate attack algorithms and 
fine-grained attack algorithms. 

These four factors clearly indicate that a 
``comparing different MI attacks'' methodology will not be convincing 
unless the following requirements are met. 
{\bf (R1)} A set of target datasets following 
different representative distance distributions should be used when 
comparing different MI attacks. Attack A could be more effective 
than Attack B when the target dataset follows one distribution 
but less effective under another distribution. 
{\bf (R2)} A set of target datasets having different 
distances between data samples should be used when comparing different MI attacks. 
{\bf (R3)} A set of target datasets having different 
differential distances should be used when comparing different MI attacks. 
{\bf (R4)} When comparing an indiscriminate attack algorithm and 
a fine-grained attack algorithm, the target datasets used for 
evaluating the first algorithm should hold a smaller number of data examples.  

The last 4 columns in Table \ref{tabs:The Comparations of 13 MI attacks} summarize whether these four requirements 
are satisfied in thirteen most representative existing works on MI attacks. Unfortunately, we find 
that (a) 8 out of the 15 existing works fail to meet any of the four requirements; 
(b) 7 out of the 15 existing works only manage to meet one of the four requirements. 
In particular, none of the 15 existing works meets requirement R1;  
none of the 15 existing works meets requirement R2; 
12 out of the 15 existing works don't meet requirement R3.  

In this work, we seek to develop a comprehensive benchmark for comparing 
different MI attacks. The primary design goal of our benchmark, called \textbf{MIBench}, is 
to meet {\em all} of these 4 requirements. In order to achieve this goal, our key insights are as follows. 
{\bf Insight A. } Although it is widely recognized in the existing works
that evaluation metrics (e.g., the ``membership advantage'' metric) play an essential 
role in comparing different MI attacks, we find that even if the employed metrics are 
appropriate, the comparison results can still suffer from {\em poor explanations}. 
For example, researchers have already noticed the following:  
(a1) one MI attack algorithm could suffer from fairly different evaluation 
results (i.e., metric measurements) against different target datasets: it is 
usually difficult to explain the lack of coherence cross datasets. 
(a2) MI attack A is more effective than attack B against target dataset D1, but attack B is 
more effective than attack A against target dataset D2. In such situations, it  
is usually very {\em difficult to draw general conclusions} on when attack A would be
more effective and when attack B would be more effective. 
(a3) Many MI attacks suffer from low precision against widely-used test datasets. 
For example, BlindMI-Diff~\cite{hui2021practical} suffers from a precision of 50.01\%. 
However, it is usually difficult to explain why higher precision is not achieved. 
{\bf Insight B. } We find that the above-mentioned four factors can  
be used to significantly improve the {\bf explainability} of a ``comparing different MI attacks''
methodology. For example, the four factors in many cases enable one to 
draw such general conclusions as ``attack A is more effective than attack B
when the following two conditions are simultaneously met: (c1) 
the distance distribution of data samples in a target dataset follows normal distribution; (c2) the    
distances between data samples of the target dataset are 6.000.''

Based on these insights, our benchmark consists not only the evaluation metrics, but also 
the evaluation scenarios. Based on the above-mentioned four factors, we design the evaluation scenarios 
from {\bf four perspectives}: the distance distribution of data samples in the target dataset, the distance between data samples of the target dataset, the differential distance between two datasets (i.e., the target dataset and a generated dataset with only nonmembers),  
and the ratio of the samples that are made no inferences by an MI attack. 
Moreover, each perspective corresponds to a control variable and so 
there are in total four {\bf control variables} in our evaluation scenarios. 
The evaluation metrics consist of ten evaluation metrics 
(i.e., accuracy, FPR, TPR, precision, f1-score, FNR, MA, AUC, TPR @ fixed (low) FPR, Threshold at maximum MA): they are 
widely used in the existing works on 
MI attacks (see Section~\ref{sebsec:Part II: Evaluation Metrics}) . 

\noindent \textbf{Contributions.} The main contribution of this work are as follows. 
First, we have identified three principles for the proposed ``comparing different
MI attacks'' methodology.  
Second, following these principles, we have designed and implemented 
the MIBench benchmark with 84 evaluation scenarios for each dataset. 
For each evaluation scenario, we have designed the particular values of the four 
above-mentioned control variables; and we have created the 
corresponding target datasets through customizing 
seven widely-used datasets in the research area of MI attacks. 
(It should be noticed that all the evaluation scenarios 
use the same set of evaluation metrics.) 
Third, we have used our benchmark to systematically 
compare 15 state-of-the-art MI attack algorithms.  
For this purpose, we have downloaded the code of the following 15 MI attacks from GitHub: NN\_attack \cite{shokri2017membership}, Loss-Threshold \cite{yeom2018privacy}, Label-only \cite{yeom2018privacy}, Top3-NN attack\cite{salem2019ml}, Top1-Threshold \cite{salem2019ml}, BlindMI-Diff-w \cite{hui2021practical}, BlindMI-Diff-w/o \cite{hui2021practical}, BlindMI-1CLASS \cite{hui2021practical}, Top2+True \cite{hui2021practical}, Privacy Risk Scores \cite{song2021systematic}, Shapley Values \cite{duddu2021shapr}, Positive Predictive Value \cite{jayaraman2020revisiting}, Calibrated Score \cite{watson2021importance}, 
Distillation-based \cite{ye2021enhanced}, 
Likelihood ratio attack \cite{carlini2022membership}.   
And we have conducted comparative evaluations of the 15 MI attacks 
using the above-mentioned evaluation scenarios. 
In total, we used 588 evaluation scenarios to fairly and systematically 
compare the existing MI attacks. These evaluation scenarios cover 7 datasets, which 
are widely used in the existing works on MI attacks, and 7 representative types of models.  
\section{Background}

\subsection{Membership Inference Attacks}
\label{subsec:subsec2_1}

A target ML model $\mathbb{F}_{T}$ is described as a ML model trained on a certain target training dataset $\mathcal{D}_{train}$. And the MI attacks aim to determine whether a data point $\emph{x}$ is in the target training dataset $\mathcal{D}_{train}$. More generally, given a data point $\emph{x}_{T}$, extra knowledge of the adversary $\mathcal{K}$, a targeted machine learning model $\mathbb{F}_{T}$, and the workflow of an MI attack (called the attack model $\mathbb{AM}$) is defined as the following. $\mathbb{AM} : \{\mathbb{F}_{T}(\emph{x}_\emph{T}), \mathcal{K}\} \Rightarrow \{0,1\}$, here the MI attack model $\mathbb{AM}$ is a binary classifier and 1 means the target data point $\emph{x}_{T}$ is a member of the $\mathcal{D}_{train}$ and 0 
is a non-member (inside ``T'' means a target model or data point).



\begin{table*}[]\tiny
    \caption{Categorizations of 15 membership inference attack algorithms in MIBench, according to ten perspectives, including \emph{treat models}, \emph{shadow model}, \emph{attack model}, \emph{extra data}, \emph{ground-truth}, \emph{the number of inferring}, \emph{evaluation metrics}, \emph{inputs}, \emph{outputs} and \emph{different kinds of evaluation requirements}.}
    \centering
    \begin{threeparttable}
    \setlength{\tabcolsep}{1.8mm}{
    \begin{tabular}{cccccccccccccccc}
        \toprule[1.1pt]
                 \multirow{2}{*}{\textbf{Attack } } &
            \multicolumn{3}{c}{\textbf{Threat Model}} & \multirow{2}{*}{\textbf{Shadow} } & \multirow{2}{*}{\textbf{Attack} } & \multirow{2}{*}{\textbf{Extra} } & \multirow{2}{*}{\textbf{Ground} } & \multirow{2}{*}{\textbf{\#} }& \multirow{2}{*}{\textbf{Evaluation} }& \multirow{2}{*}{\textbf{Inputs} } & \multirow{2}{*}{\textbf{Outputs} } & \multicolumn{4}{c}{\textbf{Evaluation Requirements (ERs)} }
            \\
            \cmidrule(r){2-4} \cmidrule(r){13-16} \textbf{Algorithm} & \textbf{blk.} & \textbf{gry.} & \textbf{whi.} & \textbf{Model} & \textbf{Model} & \textbf{Data} & \textbf{Truth} & \textbf{Inference} & \textbf{Metrics} & & & DisDistri. & DataSampleDist. & DifferDist. & RatNoInf. \\
         \midrule[1.1pt]

          NN\_attack\cite{shokri2017membership} &  \Checkmark & \XSolidBrush &  \XSolidBrush &  m &  m & \XSolidBrush & \Checkmark & w & 1+2 &  $\mathcal{D}_{t}$&  1 & \XSolidBrush & \XSolidBrush & \XSolidBrush&\XSolidBrush \\
        
        Loss-Threshold\cite{yeom2018privacy}  & \Checkmark & \XSolidBrush & \XSolidBrush & 1 & \XSolidBrush & \XSolidBrush & \Checkmark & w & 1+2+7  & $\mathcal{D}_{t}$ & 1 & \XSolidBrush & \XSolidBrush & \XSolidBrush & \XSolidBrush \\
        
          Label-only\cite{yeom2018privacy}  & \Checkmark & \XSolidBrush  & \XSolidBrush &  \XSolidBrush & \XSolidBrush & \XSolidBrush & \Checkmark & w & 1+2+7 & $\mathcal{D}_{t}$ & 1 & \XSolidBrush & \XSolidBrush &\XSolidBrush & \XSolidBrush \\  
         
         Top3-NN attack\cite{salem2019ml}  & \Checkmark & \XSolidBrush & \XSolidBrush & 1 & 1 & \XSolidBrush & \Checkmark & w & 1+2 & $\mathcal{D}_{t}$ & 1 & \XSolidBrush &\XSolidBrush & \XSolidBrush& \XSolidBrush\\ 

        Top1-Threshold\cite{salem2019ml} & \Checkmark & \XSolidBrush & \XSolidBrush &  \XSolidBrush & \XSolidBrush & \Checkmark & \XSolidBrush & w & 1+2+8 & $\mathcal{D}_{t}$ & 1 & \XSolidBrush & \XSolidBrush & \XSolidBrush & \XSolidBrush \\ 

         BlindMI-Diff-w\cite{hui2021practical}  & \Checkmark & \Checkmark & \Checkmark &  \XSolidBrush & \XSolidBrush & \Checkmark & \Checkmark & w & 1+2+3 &  $\mathcal{D}_{t}$ &  1 & \XSolidBrush & \XSolidBrush & \Checkmark & \XSolidBrush \\

         BlindMI-Diff-w/o\cite{hui2021practical}   & \Checkmark & \Checkmark & \Checkmark &  \XSolidBrush & \XSolidBrush & \Checkmark & \Checkmark & w & 1+2+3 &  $\mathcal{D}_{t}$ &  1 & \XSolidBrush & \XSolidBrush & \Checkmark & \XSolidBrush \\

        BlindMI-1CLASS\cite{hui2021practical}  & \Checkmark & \Checkmark & \Checkmark &  \XSolidBrush & \XSolidBrush & \Checkmark & \Checkmark & w & 1+2+3 &  $\mathcal{D}_{t}$ &  1 & \XSolidBrush & \XSolidBrush & \Checkmark & \XSolidBrush \\

          Top2+True\cite{hui2021practical} & \Checkmark & \XSolidBrush & \XSolidBrush & m & m &  \XSolidBrush &  \Checkmark & w & 1+2+3 & $\mathcal{D}_{t}$ & 1 &\XSolidBrush & \XSolidBrush & \XSolidBrush &\XSolidBrush \\

          Privacy Risk Scores\cite{song2021systematic} & \Checkmark & \XSolidBrush & \XSolidBrush & m &  \XSolidBrush &  \XSolidBrush & \Checkmark & p &  1+2+4 & $\mathcal{D}_{t}$ &  1+2 & \XSolidBrush & \XSolidBrush & \XSolidBrush & \Checkmark \\ 

         Shapley Values\cite{duddu2021shapr} &  \Checkmark &  \XSolidBrush & \XSolidBrush & m &  \XSolidBrush &  \XSolidBrush & \Checkmark & p &  1+2+4+5 & $\mathcal{D}_{t}$ & 1+2 & \XSolidBrush & \XSolidBrush & \XSolidBrush& \Checkmark \\
         
          Positive Predictive Value\cite{jayaraman2020revisiting} &  \Checkmark & \XSolidBrush &  \XSolidBrush & m & \XSolidBrush &  \XSolidBrush &  \Checkmark & p &  1+2+6+7+9 &  $\mathcal{D}_{t}$ &  1+2 & \XSolidBrush & \XSolidBrush & \XSolidBrush &\Checkmark \\

          Calibrated Score\cite{watson2021importance} &  \Checkmark &  \XSolidBrush & \XSolidBrush & m &  \XSolidBrush &  \XSolidBrush & \Checkmark & w & 1+2+4+8+9 & $\mathcal{D}_{t}$ &  1 & \XSolidBrush & \XSolidBrush & \XSolidBrush & \XSolidBrush \\  

          Distillation-based Thre.\cite{ye2021enhanced} &  \Checkmark &  \XSolidBrush &  \XSolidBrush & m &  \XSolidBrush &  \XSolidBrush &  \Checkmark &  w & 1+2+8+9 &  $\mathcal{D}_{t}$ &  1 & \XSolidBrush & \XSolidBrush & \XSolidBrush & \XSolidBrush \\

          Likelihood ratio attack\cite{carlini2022membership} &  \Checkmark &  \XSolidBrush & \XSolidBrush &  m &  \XSolidBrush &  \XSolidBrush & \Checkmark &  w & 8+9+10 & $\mathcal{D}_{t}$ &  1 & \XSolidBrush & \XSolidBrush & \XSolidBrush & \Checkmark \\
         
         \bottomrule[1.1pt]
    \end{tabular}}
     \begin{tablenotes}
    \scriptsize
    \item a) \textbf{Threat Model:} blk.$\rightarrow$ Black-box; gry.$\rightarrow$ Gray-box; whi.$\rightarrow$ White-box;
    \item b) \textbf{\# Inference:} w $\rightarrow$ the whole data in the target dataset; p $\rightarrow$ the part of data in the target dataset;
   \item c) \textbf{Shadow Model:} 1 $\rightarrow$ one shadow model; m $\rightarrow$ multiple shadow models;
   \item d) \textbf{Attack Model:} 1 $\rightarrow$ one attack model; m $\rightarrow$ multiple attack models;
   \item e) \textbf{Evaluation Metrics:} \textbf{1 $\rightarrow$ precision}, namely the ratio of true positives to the sum of true positive and false positives;
\item \hspace{1em} \textbf{2 $\rightarrow$ recall}, namely the ratio of true positives to the sum of true positives and false negatives;
\item \hspace{1em} \textbf{3 $\rightarrow$ f1-score}, namely the harmonic mean of precision and recall;
\item \hspace{1em} \textbf{4 $\rightarrow$ accuracy}, namely the percentage of data records with correct membership predictions by the attack model;
\item \hspace{1em} \textbf{5 $\rightarrow$ average membership privacy risk score}, namely the average over the membership privacy risk scores assigned to training data records; 
\item \hspace{1em} \textbf{6 $\rightarrow$ positive predictive value (PPV)}, namely the ratio of true members predicted among all the positive membership predictions; 
\item \hspace{1em} \textbf{7 $\rightarrow$ membership advantage (MA)}, namely the difference between the true positive rate and the false positive rate (e.g., MA = TPR - FPR);
\item \hspace{1em} \textbf{8 $\rightarrow$ the area under ROC curve (AUC)}, namely the area under the Receiver Operating Characteristic (ROC) curve;
\item \hspace{1em} \textbf{9 $\rightarrow$ false positive rate (FPR)}, namely the ratio of nonmember samples are erroneously predicted as members;
\item \hspace{1em} \textbf{10 $\rightarrow$ a TPR at low FPR}, namely the true-positive rate at low (e.g., $\leq$ 0.1\%) false-positive rates;
\item f) \textbf{Outputs:} 1 $\rightarrow$ predict the target data is (or is not) a member; 2 $\rightarrow$ make no inferences about the target data;
\item g) \textbf{Evaluation Requirements:} DisDistri. $\rightarrow$ the distance distribution of data samples in the target dataset; DataSampleDist. $\rightarrow$ distance between data samples of \ \ \ \ \\the target dataset; DifferDist. $\rightarrow$ the differential distance before and after moving a data sample; RatNoInf. $\rightarrow$ the ratio of the samples that are made
no inferences by an MI attack;
\item h) \textbf{$\mathcal{D}_{t}$: }means a target data or dataset; ``\Checkmark '': means the adversary needs (considers) the knowledge (ER); ``\XSolidBrush'': indicates the knowledge (ER) is not necessary (be considered).

    \end{tablenotes}
    \end{threeparttable}
    \label{tabs:The Comparations of 13 MI attacks}
    \vspace{3pt}
\end{table*}

\subsection{Threat Model}
\label{subsec:subsec2_2}

\noindent \textbf{\emph{Black-box attack.}} The attackers only know the outputs of the target model (e.g., confidence score vectors) and mainly utilize them for training a binary classifier to determine the membership of a data sample.
 \textbf{\emph{Gray-box attack.}} The attackers not only control the certain target models' inputs and outputs, but know the distributions of training data trained the target model. 
\textbf{\emph{{White-box attack.}}} The attackers know all information of the target model and training data. 

\section{Motivation and Overview}
\label{sec: Motivation and Overview}

\subsection{Motivation}
\label{sec: Motivation}

The proposed benchmark is motivated by the following observations. {\bf Observation 1. }We find that even if the
employed metrics are appropriate, the comparison results can
still suffer from poor explanations. For example, researchers
have already noticed the following: (a1) One MI attack algorithm
could suffer from fairly different evaluation results (i.e.,
metric measurements) against different target datasets: it is
usually difficult to explain the lack of coherence cross datasets.
In many cases, although the different target datasets have different 
data semantics (e.g., CIFAR100 holds 100 kind of data, while 
CH-MNIST holds 8 kind of data), strong correlation between 
the data semantics and the evaluation results is hard to identify. (a2) MI attack A is more effective than attack B against target
dataset D1, but attack B is more effective than attack A against
target dataset D2. In such situations, it is usually very difficult
to draw general conclusions on when attack A would be more
effective and when attack B would be more effective.
For example, the NN\_attack\cite{shokri2017membership} (e.g., MA=34.51\%) is more effective than Label-only \cite{yeom2018privacy} attack (e.g., MA=32.77\%) against CIFAR10, but the Label-only \cite{yeom2018privacy} attack (e.g., MA=70.6\%) is more effective than NN\_attack\cite{shokri2017membership} (e.g., MA=58.54\%) against CIFAR100. 
(a3) Many MI attacks suffer from low precision (e.g., NN\_attack\cite{shokri2017membership} (51.7\%), Label-only \cite{yeom2018privacy} (50.5\%) and BlindMI-Diff~\cite{hui2021practical}) (50.01\%)) against widely-used
test datasets. However, it is usually difficult to explain why higher precision is not achieved.

{\bf Observation 2. }We find that although all the existing works pay adequate attention to two primary factors determining when an MI attack is more effective, i.e.,  the semantics of data examples
in the target dataset and the type of the target model, 
the existing works pay inadequate attention to three additional main factors. (Factor 1) We find that the effectiveness of an MI attack could be highly sensitive to the distance distribution of data samples in the target dataset. (Factor 2) We find that the effectiveness of  
an MI attack could be highly sensitive to the distance between data samples of the target dataset. (Factor 3) We find that the effectiveness of 
an MI attack could be very sensitive to the differential distance (i.e., MMD~\cite{gretton2012kernel}) 
before and after a data example is moved from the target dataset 
to a generated dataset with only nonmembers. For example, (b1) without incorporating Factor 1, the comparasion evaluation results in \cite{shokri2017membership,yeom2018privacy,salem2019ml,hui2021practical,song2021systematic,duddu2021shapr,jayaraman2020revisiting,watson2021importance,ye2021enhanced,carlini2022membership} are difficult to 
provide insightful explannations on low precision (e.g., NN\_attack\cite{shokri2017membership} (51.7\%), Label-only \cite{yeom2018privacy} (50.5\%) and BlindMI-Diff~\cite{hui2021practical} (50.01\%)) or high FPR~\cite{shokri2017membership,rezaei2021difficulty}. 
(b2) Without incorporating Factor 2, the comparasion evaluation results 
in \cite{shokri2017membership,yeom2018privacy,salem2019ml,hui2021practical,song2021systematic,duddu2021shapr,jayaraman2020revisiting,watson2021importance,ye2021enhanced,carlini2022membership} are difficult to explain 
why the higher precision is not achieved. (b3) Without incorporating Factor 3, the comparasion evaluation results 
in \cite{shokri2017membership,yeom2018privacy,salem2019ml,song2021systematic,duddu2021shapr,jayaraman2020revisiting,watson2021importance,ye2021enhanced,carlini2022membership} are difficult to explain 
why the low FPR and higher accuracy are not achieved simultaneously.

Based on above key observations, we are strongly motivated to 
incorporate the above-mentioned factors into a new benchmark 
and use the benchmark to fairly and systematically 
compare different MI attacks. In addition, 
we are strongly motivated to develop a new 
``comparing different MI attacks'' methodology which 
 has {\bf significantly improved ability to explain} the comparative evaluation results. 

\subsection{Overview}
\label{subsec:Overview}

To alleviate the dilemma of the existing MI attacks, 
in this work, we seek to develop a comprehensive benchmark for comparing 
different MI attacks. The primary design goal of our benchmark, called \textbf{MIBench} (Section \ref{sec Evaluation Framework}), is 
to meet {\em all} of above-mentioned 4 requirements (see Section \ref{sebsec:Introduction}). Our benchmark consists not only the evaluation metrics(Section \ref{sebsec:Part II: Evaluation Metrics}), but also 
the evaluation scenarios (Section \ref{sebsec:Part I: Evaluation Scenarios}). Based on the above-mentioned four factors (see Section \ref{sebsec:Introduction}), we design the evaluation scenarios 
from {\bf four perspectives}: the distance distribution of data samples in the target dataset, the distance between data samples of the target dataset, the differential distance between two datasets (i.e., the target dataset and a generated dataset with only nonmembers),  
and the ratio of the samples that are made no inferences by an MI attack. 
Moreover, each perspective corresponds to a control variable and so 
there are in total four {\bf control variables} in our evaluation scenarios. 
The evaluation metrics consist of ten widely used evaluation metrics 
(i.e., accuracy, FPR, TPR, precision, f1-score, FNR, MA, AUC \cite{carlini2022membership}, TPR @ fixed (low) FPR \cite{carlini2022membership}, Threshold at maximum MA) (see Section~\ref{sebsec:Part II: Evaluation Metrics}) .

Based on the evaluation framework (Section \ref{sec Evaluation Framework}), 
we also present how to use the evaluation framework (Section \ref{sec:Using the Evaluation Framework}) to fairly and effectively evaluate the real performance of an MI attack. 
There are four steps before using the evaluation framework. \textbf{Step 1: Construct the distance distribution of data samples in the target dataset (CV1) (Section \ref{sebsec:Effect of the Distance Distribution of Data Samples in the Target Dataset}).} For a target dataset, we first input all data samples (e.g., {\em n}) in the target dataset to the target model and get the corresponding output probabilities, and we classify these data into two categories according to their output probabilities (such as the data samples with high output probability and the data samples with low output probability), and the number of the two categories is the same (e.g., {\em $\frac{n}{2}$}). We extract the same number of data samples (e.g., $\eta$) from these two kinds of target data samples, and calculate the MMD~\cite{gretton2012kernel} of these two categories target data samples (e.g., $2\eta$). This calculation iterates until the MMDs are calculated for all samples in the target dataset, and we calculate a total of $\frac{n}{2\eta}$ groups of  MMDs. According to the calculated $\frac{n}{2\eta}$ groups of MMDs of the target dataset, we rank these $\frac{n}{2\eta}$ groups MMDs and construct the distance distributions of data samples in the target dataset obey normal, uniform and bernoulli distributions, respectively. \textbf{Step 2: Calculate the distance between data samples of the target dataset (CV2) (Section \ref{sebsec:Effect of Distance between members and nonmembers}).} For target datsests whose distance distribution of data samples obeying different distance distributions, we calculate the distance between data samples of the target dataset by the MMD method. \textbf{Step 3: Calculate the differential distance between two datasets (CV3) (Section \ref{sebsec:Effect of Differential Distances between two datasets}).} Next, we generate a dataset with nonmembers via transforming existing samples into new samples or directly using
the test data~\cite{hui2021practical}, and then 
we compute the differential distance between two datasets (i.e., the target dataset and a generated dataset with only nonmembers) before and after a data example (i.e., a data example with high output probability and a data example with low output probability) is moved from the target dataset to the nonmember dataset. 
  \textbf{Step 4: Set the ratios of the samples that are
made no inferences by an MI attack (CV4) (Section \ref{sebsec:Effect of Ignored Ratios}).} Finally, 
we set different ratios of the samples that are
made no inferences by an MI attack for image datasets (i.e., 20\%, 40\%, 45\% and 49\%) and text dagtasets (i.e., 2\%, 4\%, 10\% and 12\%) as the prior Privacy Risk Scores-based ~\cite{song2021systematic} and the Shapley Values-based MI attacks~\cite{duddu2021shapr}. 

\begin{figure*}[htp]
\vspace{0.1pt}
\begin{center}
\subfigure[CIFAR100]{
\begin{minipage}{0.20\textwidth}
\label{Fig.sub.1.1}
\includegraphics[height=0.8\textwidth,width=1\textwidth]{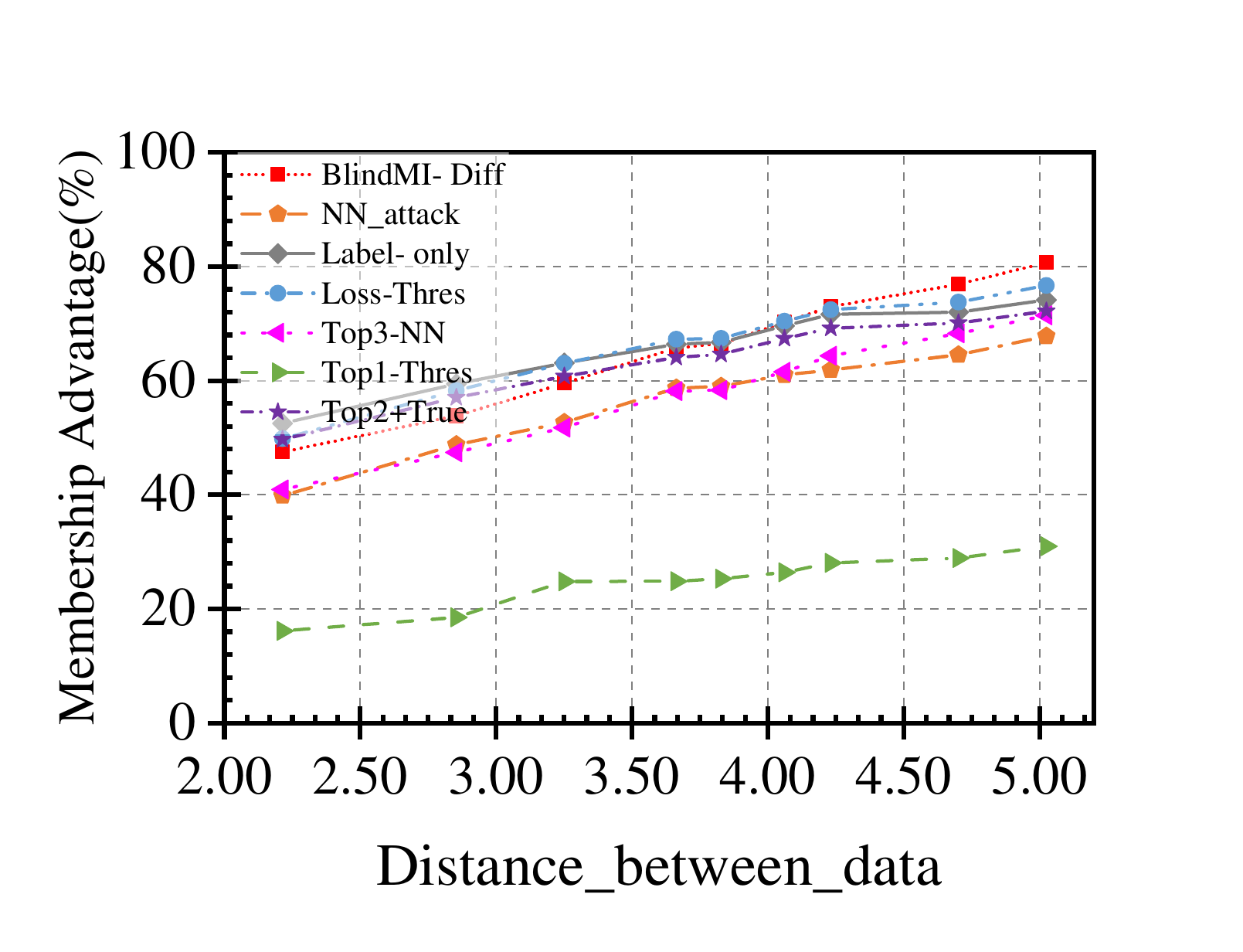}
\end{minipage}}
\subfigure[CIFAR10]{
\begin{minipage}{0.20\textwidth}
\label{Fig.sub.1.2}
\includegraphics[height=0.8\textwidth,width=1\textwidth]{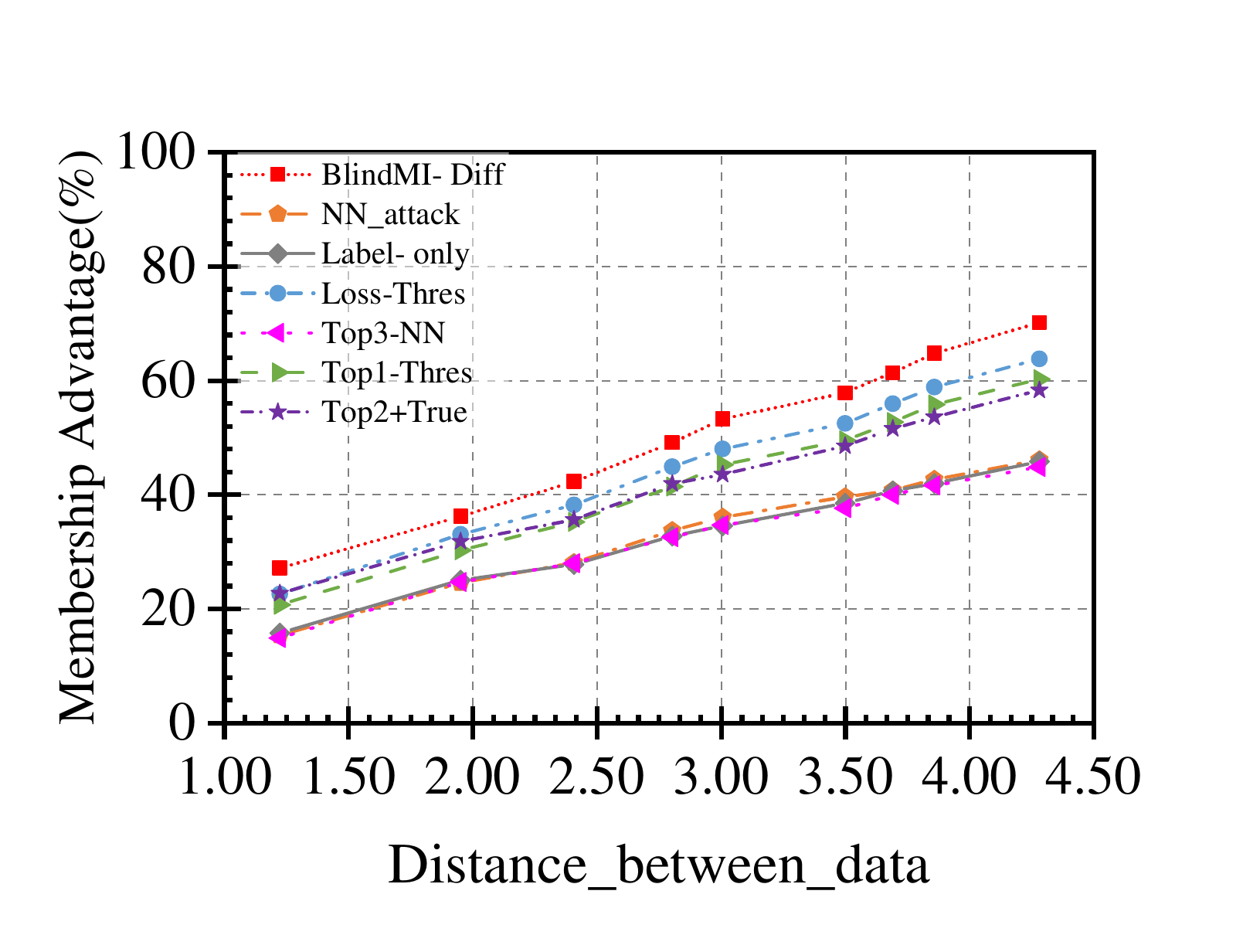}
\end{minipage}}
\subfigure[CH\_MNIST]{
\begin{minipage}{0.20\textwidth}
\label{Fig.sub.1.3}
\includegraphics[height=0.8\textwidth,width=1\textwidth]{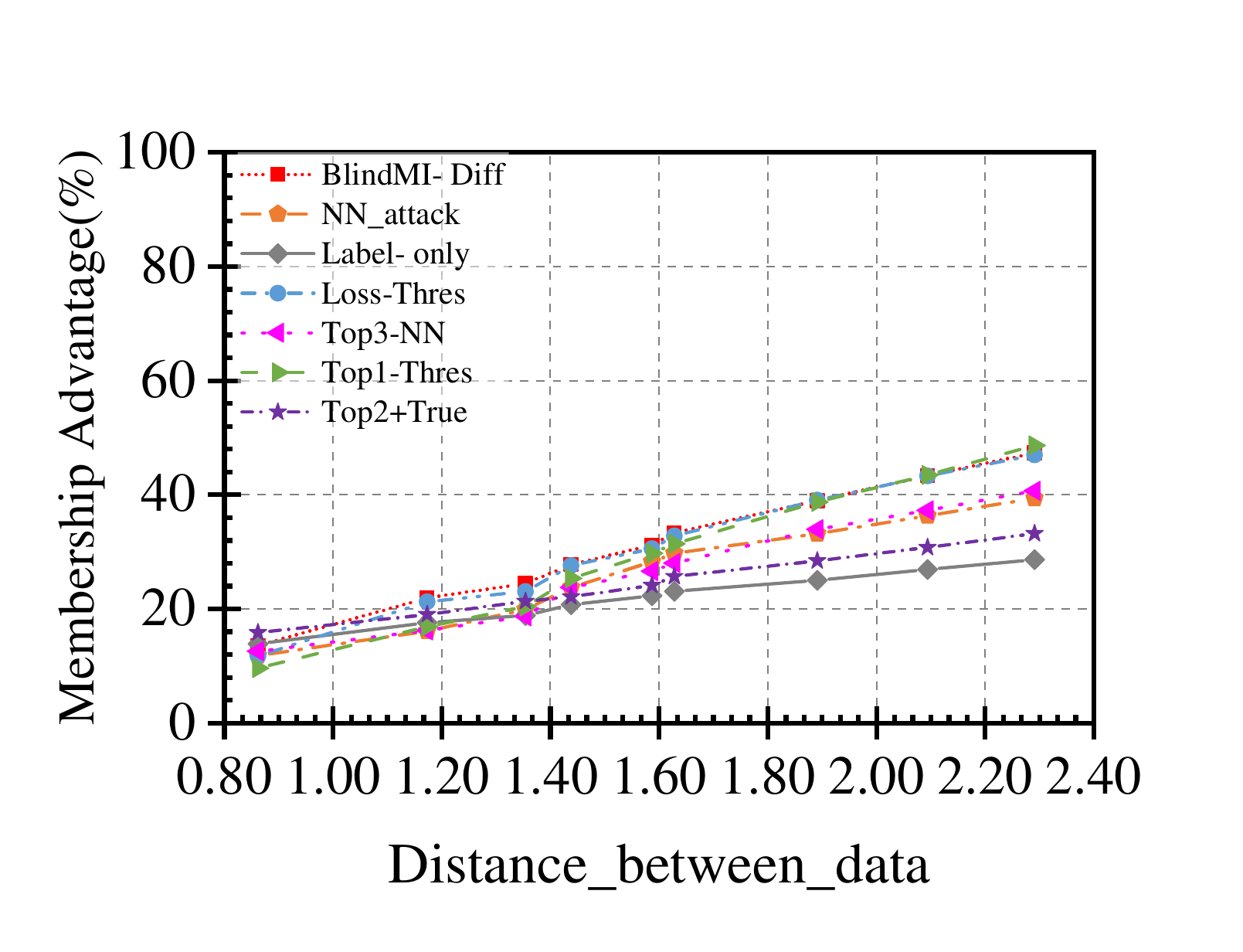}
\end{minipage}}
\subfigure[ImageNet]{
\begin{minipage}{0.20\textwidth}
\label{Fig.sub.1.4}
\includegraphics[height=0.8\textwidth,width=1\textwidth]{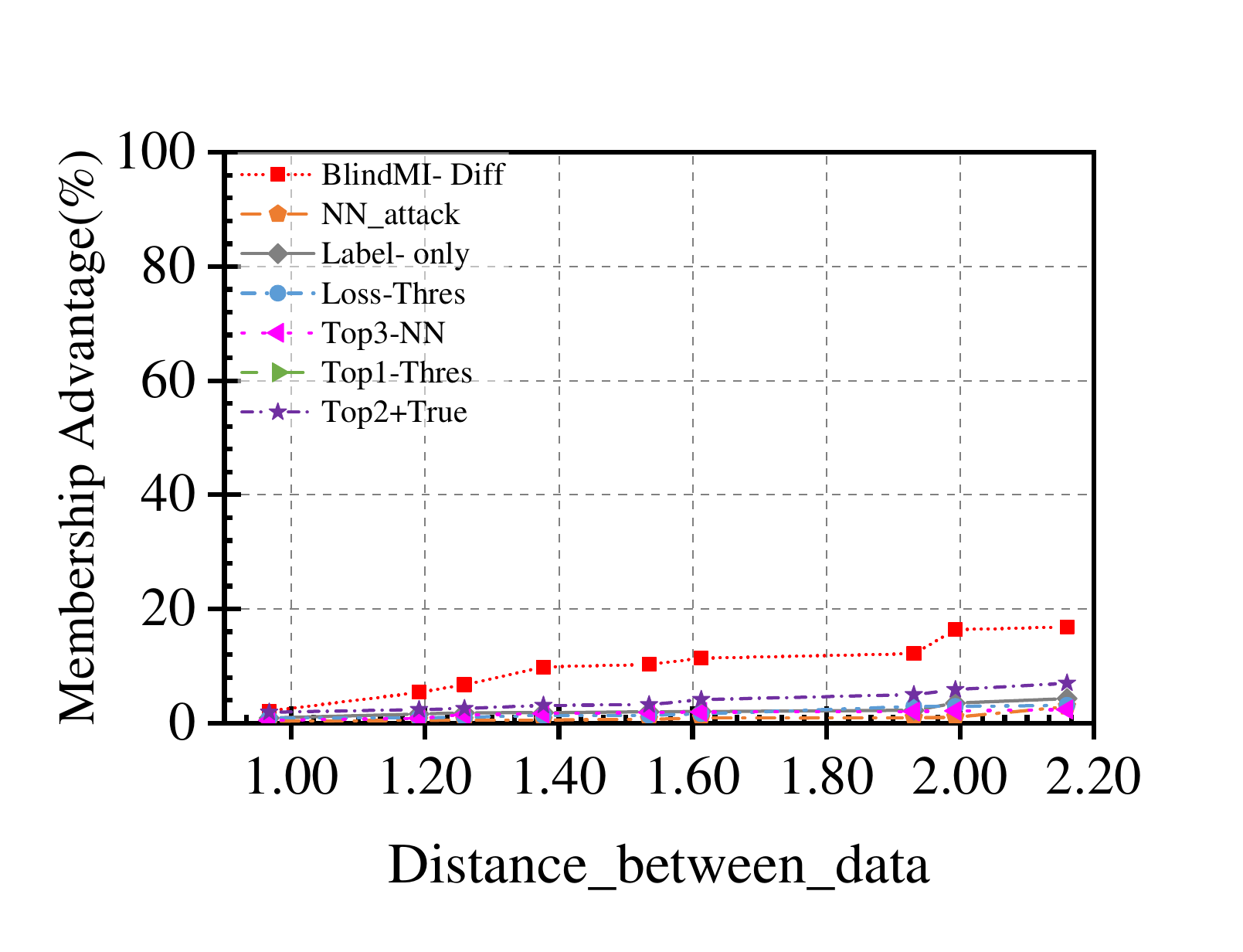}
\end{minipage}}
\end{center}
\caption{\label{figs:Fig22} The effect of the distances between
data samples of the target dataset on the Membership Advantage.}
\vspace{0.1pt}
\end{figure*}

\begin{figure*}[]
\vspace{0.1pt}
\begin{center}
\subfigure[CIFAR100]{
\begin{minipage}{0.20\textwidth}
\label{Fig.sub.2.1}
\includegraphics[height=0.8\textwidth,width=1\textwidth]{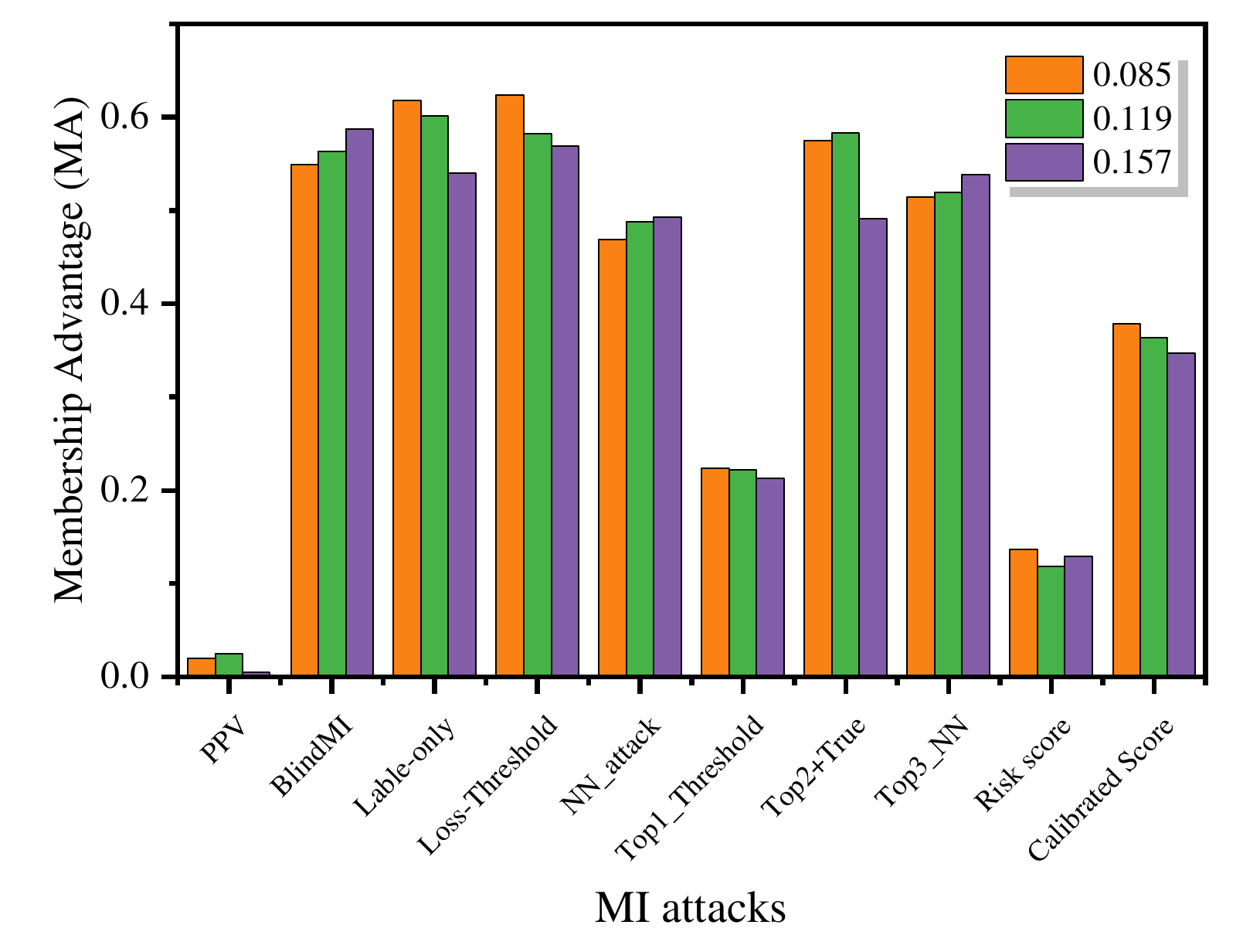}
\end{minipage}}
\subfigure[CIFAR10]{
\begin{minipage}{0.20\textwidth}
\label{Fig.sub.2.2}
\includegraphics[height=0.8\textwidth,width=1\textwidth]{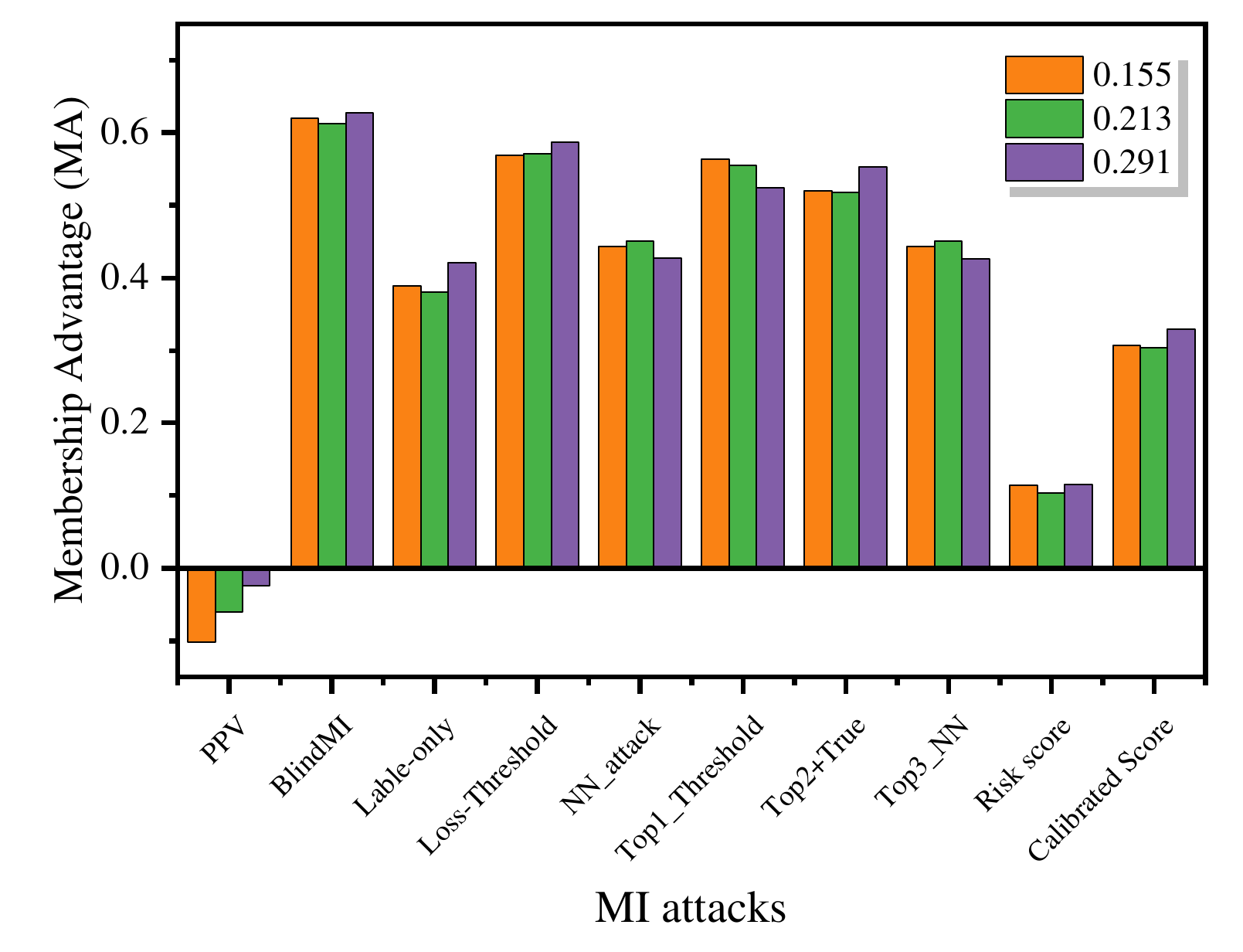}
\end{minipage}}
\subfigure[CH\_MNIST]{
\begin{minipage}{0.20\textwidth}
\label{Fig.sub.2.3}
\includegraphics[height=0.8\textwidth,width=1\textwidth]{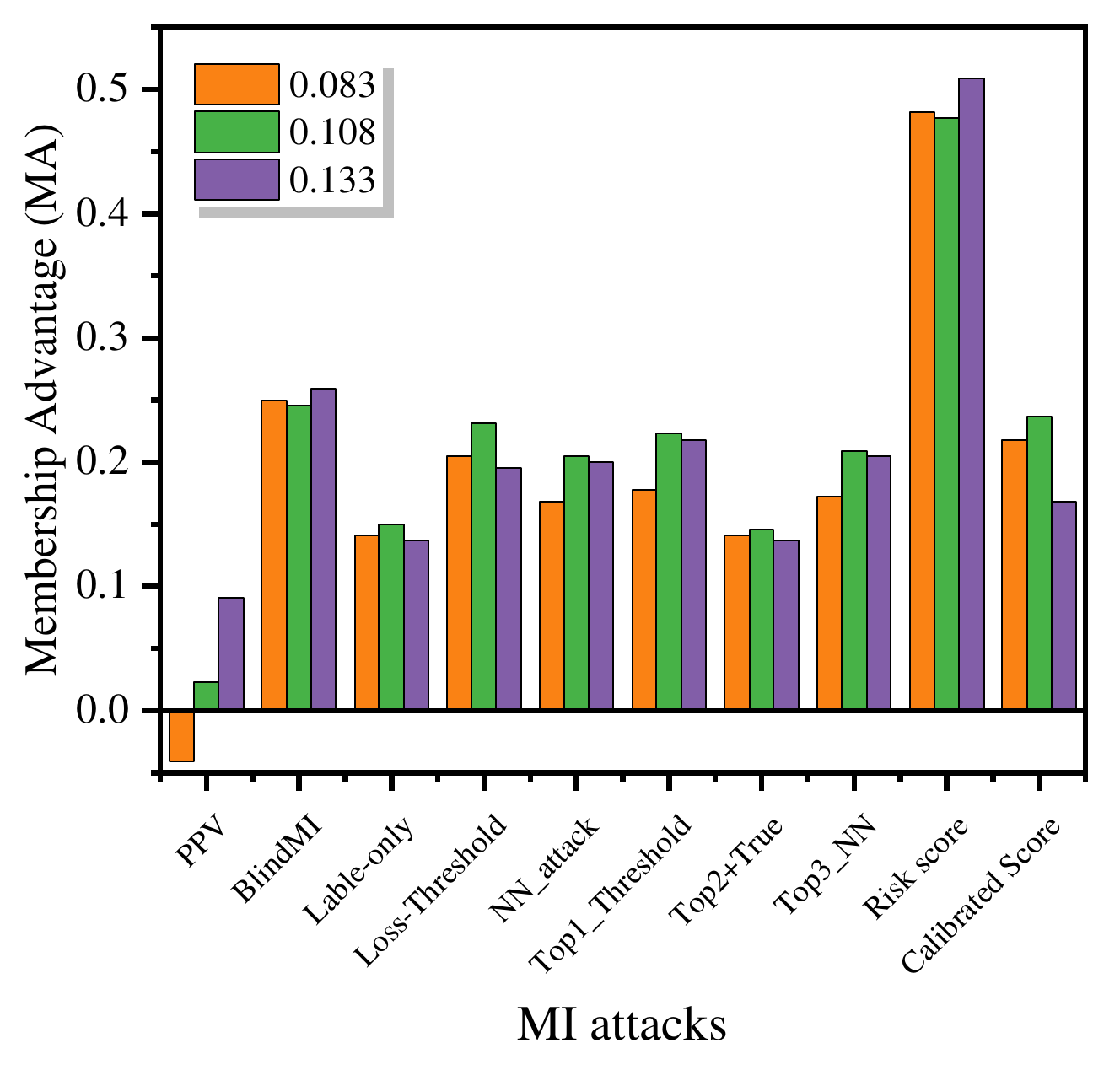}
\end{minipage}}
\subfigure[ImageNet]{
\begin{minipage}{0.20\textwidth}
\label{Fig.sub.2.4}
\includegraphics[height=0.8\textwidth,width=1\textwidth]{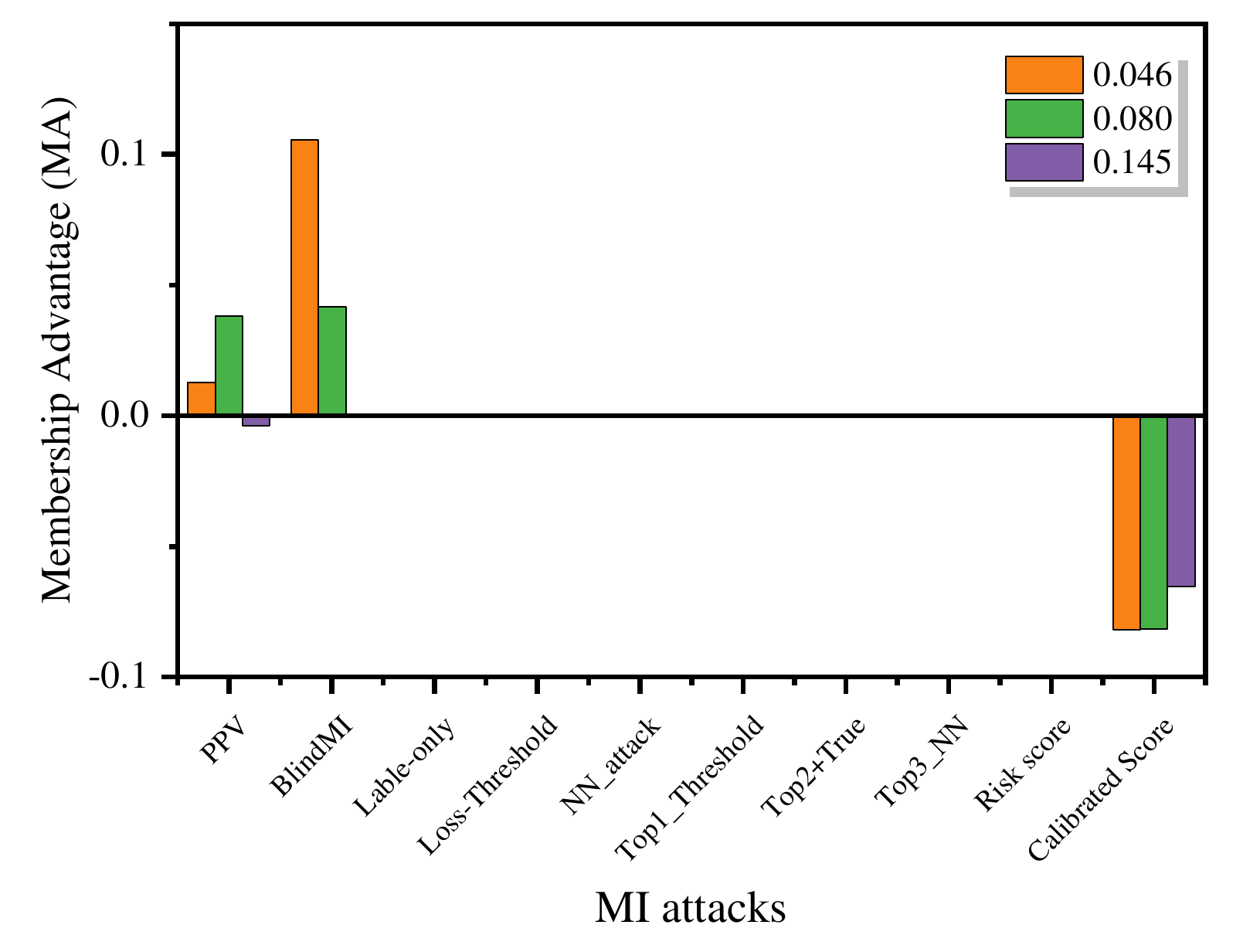}
\end{minipage}}
\end{center}
\caption{\label{figs:Fig33} The effect of differential distance between two datasets on the Membership Advantage.}
\vspace{0.1pt}
\end{figure*}

After achieving these four steps, we use the four calculated CVs to build the evaluation scenarios (ESs), and each ES consists of four CVs, and we 
study a group level evaluation, where the concept of group is that a group consists of multiple ESs, only one CV changes in a group and the other three CVs remain unchanged and the values of other three CVs must correspond to the same 
(see Section \ref{sec:Using the Evaluation Framework}), and we make comprehensive evaluations of the existing MI attacks through the 10 typical evaluation metrics (Section~\ref{sebsec:Part II: Evaluation Metrics}). In this work, we have identified three principles for the proposed ``comparing different MI attacks'' methodology, and we have designed and implemented 
the MIBench benchmark with 84 ESs for each dataset. In total, we have used our benchmark to fairly and systematically 
compare 15 state-of-the-art MI attack algorithms (shown in Table \ref{tabs:The Comparations of 13 MI attacks})
across 588 ESs, and these ESs cover 7 widely used datasets and 7 representative types of models.

\section{Principles for better Explaining the attack Comparative Evaluation Results}
\label{sec:Principles for better Explaining the attack Comparison Evaluation Results}

\subsection{Principle 1: The Two Primary Factors are Necessary but Insufficient}
\label{sebsec:Principle 1: the two primary factors are necessary but insufficient}

\noindent The two primary factors, i.e., the kind of data
in the target dataset and the type of the target model, are
necessary but insufficient to let a ``comparing different MI attacks'' methodology have desirable ability to explain the comparative evaluation results.

In order to incorporate these two factors in evaluating attack effectiveness, all the existing works use multiple target datasets and multiple types (e.g., MLP, ResNet, DenseNet) of target models. However, 
we find that even if the
employed metrics are appropriate, the comparison results can
still suffer from poor explanations. For example, researchers
have already noticed the following: (a1) One MI attack algorithm
could suffer from fairly different evaluation results (i.e.,
metric measurements) against different target datasets: it is
usually difficult to explain the lack of coherence cross datasets.
In many cases, although the different target datasets have different 
data semantics (e.g., CIFAR100 holds 100 kind of data, while 
CH-MNIST holds 8 kind of data), strong correlation between 
the data semantics and the evaluation results is hard to identify. (a2) MI attack A is more effective than attack B against target
dataset D1, but attack B is more effective than attack A against
target dataset D2. In such situations, it is usually very difficult
to draw general conclusions on when attack A would be more
effective and when attack B would be more effective.
For example, the Top2+True \cite{hui2021practical} attack (e.g., MA=43.71\%) is more effective than the NN\_attack\cite{shokri2017membership} (e.g., MA=34.51\%) against CIFAR10, whereas the NN\_attack\cite{shokri2017membership} (e.g., MA=0.008\%) is more effective than Top2+True \cite{hui2021practical} attack (e.g., MA=0.003\%) against ImageNet. (a3) Many MI attacks suffer from  high FPR (see Figure~\ref{figs:fig3}) and low precision (see Figure~\ref{figs:fig4}) against widely-used
test datasets (e.g., NN\_attack\cite{shokri2017membership} (51.7\%), Label-only \cite{yeom2018privacy} (50.5\%) and BlindMI-Diff~\cite{hui2021practical} (50.01\%)). However, it is usually difficult to explain why higher precision is not achieved. 



\begin{table*}[]\tiny
    \caption{The Evaluation Results of the target datasets obeying different distance distributions of data samples in the target dataset, according to ten evaluation metrics such as accuracy, precision, recall, f1-score, FNR, FPR, MA, AUC, TPR @ 0.01\% FPR (T@0.01\%F) and the Threshold at maximum MA (Thres@max MA).}
    \centering
    \begin{threeparttable}
    \setlength{\tabcolsep}{0.7mm}{
    \begin{tabular}{c|cccc|cccccccccccccccc}
        \hline
        \toprule[1.1pt]
            \multirow{3}{*}{\textbf{Dataset}} & \multirow{3}{*}{\textbf{DisDistr.}} & \multirow{3}{*}{\textbf{DisBD.}} & \multirow{3}{*}{\textbf{DiffDis.}} & \multirow{3}{*}{\textbf{RatNoI.}} &  \multirow{3}{*}{{\textbf{Metric}}}&
         \multicolumn{15}{c}{\textbf{Membership Inference Attack Algorithms}} \\
            \cmidrule(r){7-21} & & & & & &
            \textbf{BlindMI} & \textbf{BlindMI} & \textbf{BlindMI} & \textbf{NN} & \textbf{Label} & \textbf{Loss} & \textbf{Top3-NN} & \textbf{Top1-Thres} & \textbf{Top2+True} & \textbf{Risk Scores} & \textbf{PPV} & \textbf{Calibrated} & \textbf{Shapley} & \textbf{Distillation} & \textbf{LiRA} \\ 
            
             & & & & & & \textbf{-w} & \textbf{-w/o} & \textbf{-1CLASS} & \textbf{attack} & \textbf{-only} & \textbf{-Thres} & \textbf{attack} & \textbf{attack} & \textbf{attack} & \textbf{attack} & \textbf{attack} & \textbf{Score} & \textbf{Values} & \textbf{-based Thre.} & \textbf{attack} \\

         \midrule[1.1pt]
        \multirow{30}{*}{CIFAR100} & \multirow{10}{*}{normal} & \multirow{10}{*}{2.893} & \multirow{10}{*}{0.085} & \multirow{10}{*}{20\%} & accuracy & \textbf{73.23\%} & 74.50\% & \textbf{68.88\%} & \textbf{73.06\%} & \textbf{81.13\%} & 98.50\% & \textbf{75.94\%} & 76.13\% & 96.38\% & 89.40\% & 51.56\% & 74.44\% & 52.75\% & 54.50\% & 53.54\% \\
      
        &  &  &  & & precision & \textbf{81.59\%} & 82.82\% & \textbf{61.99\%} & \textbf{65.57\%} & \textbf{72.94\%} & 97.95\% & \textbf{70.38\%} & 89.49\% & 98.85\% & 83.43\% & 57.14\% & 67.16\% & 43.48\% & 54.20\% & 52.59\% \\

       &  &  &  & & recall & \textbf{73.00\%} & 70.22\% & \textbf{93.46\%} & \textbf{94.23\%} & \textbf{97.44\%} & 98.96\% & \textbf{91.59\%} & 69.94\% & 94.02\% & 98.31\% & 12.50\% & 95.63\% & 56.82\% & 72.44\% & 64.86\% \\

        &  &  &  & & f1-score & \textbf{75.35\%} & 76.00\% & \textbf{74.54\%} & \textbf{77.33\%} & \textbf{83.42\%} & 98.45\% & \textbf{79.60\%} & 78.51\% & 96.38\% & 90.26\% & 20.51\% & 78.91\% & 49.26\% & 62.00\% & 58.09\%\\

        &  &  &  & & FNR & \textbf{27.00\%} & 29.78\% & \textbf{6.54\%} & \textbf{5.77\%} & \textbf{2.56\%} & 1.04\% & \textbf{8.41\%} & 30.06\% & 5.98\% & 1.69\% & 87.50\% & 4.38\% & 43.18\%& 27.56\% & 35.14\% \\

        &  &  &  & & FPR & \textbf{18.13\%} & 19.71\% & \textbf{54.51\%} & \textbf{47.07\%} & \textbf{34.39\%} & 1.93\% & \textbf{40.51\%} & 13.62\% & 1.15\% & 19.52\% & 9.38\% & 46.75\% & 50.00\% & 64.36\% & 57.63\% \\

        &  &  &  & & MA & 54.88\% & 50.51\% & 38.95\% & 47.16\% & 63.05\% & \textbf{97.03\%} & 51.07\% & 56.32\% & 92.87\% & 78.79\% & 3.13\% & 48.88\% & 6.82\% & 8.08\% & 7.24\% \\

        &  &  &  & & AUC & \textbf{0.790} & 0.747 & \textbf{0.695} & \textbf{0.846} & \textbf{0.810} & 0.025& 0.840 & 0.492 & 0.960 & 0.225 & 0.012 & 0.152 & 0.100 & 0.538 & 0.456 \\

        &  &  &  & & T@0.01\%F & \textbf{0.014\%} & 0.013\% & \textbf{0.011\%} & \textbf{0.26\%} & \textbf{0.00\%} & 0.00\% & \textbf{0.12\%} & 0.00\% & 0.12\% & 0.00\% & 0.63\% & 0.00\% & 0.00\% & 0.00\% & 0.74\% \\

        &  &  &  & & Thres@max MA & \textbf{-} & - & \textbf{0.900} & \textbf{-} & - & 1.985 & - & 0.623 & - & -2.98$e^{-3}$ & 0.640 & 1$e^{-6}$ & -7.22$e^{-7}$& 0.031 & 4.609 \\ \cline{2-21}

        & \multirow{10}{*}{uniform} & \multirow{10}{*}{2.893} & \multirow{10}{*}{0.085} & \multirow{10}{*}{20\%} & accuracy & \textbf{71.19\%} & 73.31\% & 69.56\% & 74.25\% & 79.88\% & \textbf{98.13\%} & 76.56\% & 76.81\% & 97.25\% & 89.42\% & 53.38\% & 72.25\% & 51.91\% & 53.94\% & 45.56\% \\
      
        &  &  &  & & precision & \textbf{81.44\%} & 79.49\% & \textbf{62.94\%} & \textbf{67.03\%} & \textbf{71.91\%} & 97.47\% & \textbf{70.77\%} & 83.42\% & 98.60\% & 84.05\% & 57.85\% & 65.61\% & 43.13\% & 53.38\% & 44.98\% \\

       &  &  &  & & recall & \textbf{68.88\%} & 70.32\% & \textbf{93.29\%} & \textbf{94.18\%} & \textbf{97.22\%} & 98.72\% & \textbf{91.48\%} & 73.30\% & 95.93\% & 97.30\% & 24.88\% & 93.50\% & 66.35\% & 71.11\% & 57.29\% \\

        &  &  &  & & f1-score & \textbf{73.86\%} & 74.63\% & \textbf{75.17\%} & \textbf{78.32\%} & \textbf{82.67\%} & 98.09\% & \textbf{79.81\%} & 78.03\% & 97.25\% & 90.20\% & 34.79\% & 77.11\% & 52.27\% & 60.98\% & 50.39\% \\

        &  &  &  & & FNR & \textbf{31.13\%} & 29.68\% & \textbf{6.71\%} & \textbf{5.82\%} & \textbf{2.78\%} & 1.28\% & \textbf{8.52\%} & 26.70\% & 4.07\% & 2.70\% & 75.13\% & 6.50\% & 33.65\% & 28.89\% & 42.71\%\\

        &  &  &  & & FPR & \textbf{17.00\%} & 22.91\% & \textbf{53.58\%} & \textbf{45.19\%} & \textbf{37.04\%} & 2.44\% & \textbf{38.73\%} & 18.69\% & 1.39\% & 18.47\% & 18.13\% & 49.00\% & 57.59\% & 63.67\% & 65.39\% \\

        &  &  &  & & MA & \textbf{51.88\%} & 47.41\% & \textbf{39.71\%} & \textbf{48.99\%} & \textbf{60.18\%} & 96.28\% & \textbf{52.75\%} & 54.61\% & 94.53\% & 78.84\% & 6.75\% & 44.50\% & 8.75\% & 7.44\% & -8.10\% \\

        &  &  &  & & AUC & \textbf{0.772} & 0.734 & \textbf{0.699} & \textbf{0.840} & \textbf{0.820} & 0.023 & 0.833 & 0.336 & 0.980 & 0.223 & 0.020 & 0.132 & 0.090 & 0.534 & 0.541 \\

        &  &  &  & & T@0.01\%F & \textbf{0.014\%} & 0.013\% & \textbf{0.010\%} & \textbf{4.43\%} & \textbf{0.00\%} & 0.00\% & \textbf{0.13\%} & 0.00\% & 0.10\% & 0.00\% & 0.88\% & 0.00\% & 0.00\% & 0.00\%& 1.53\%\\

        &  &  &  & & Thres@max MA & - & - & \textbf{0.900} & \textbf{0.220} & - & 1.984 & - & 0.717 & - & -3.02$e^{-3}$ & 0.560 & 1$e^{-6}$ & -5.49$e^{-6}$ & 0.014 & 4.538 \\ \cline{2-21}

        & \multirow{10}{*}{bernoulli} & \multirow{10}{*}{2.893} & \multirow{10}{*}{0.085} & \multirow{10}{*}{20\%} & accuracy & \textbf{72.78\%} & 75.50\% & \textbf{69.81\%} & \textbf{74.56\%} & \textbf{80.94\%} & 80.38\% & \textbf{75.94\%} & 75.19\% & 78.50\% & 89.68\% & 50.81\% & 71.69\% & 54.61\% & 51.00\% &49.88\%  \\
      
        &  &  &  & & precision & \textbf{81.29\%} & 78.37\% & \textbf{63.54\%} & \textbf{67.75\%} & \textbf{73.37\%} & 96.25\% & \textbf{69.74\%} & 89.12\% & 70.47\% & 83.96\% & 53.51\% & 64.87\% & 43.92\% & 50.75\% & 50.33\%\\

       &  &  &  & & recall & \textbf{70.88\%} & 74.11\% & \textbf{93.00\%} & \textbf{93.75\%} & \textbf{97.12\%} & 73.06\% & \textbf{91.63\%} & 69.70\% & 98.13\% & 98.11\% & 12.38\% & 94.63\% & 58.48\% & 67.37\% & 56.19\%  \\

        &  &  &  & & f1-score & \textbf{74.90\%} & 76.18\% & \textbf{75.49\%} & \textbf{78.66\%} & \textbf{83.59\%} & 83.06\% & \textbf{79.20\%} & 78.22\% & 82.03\% & 90.48\% & 20.10\% & 76.97\% & 50.17\% & 57.89\% &53.10\% \\

        &  &  &  & & FNR & \textbf{29.13\%} & 25.89\% & \textbf{7.00\%} & \textbf{6.25\%} & \textbf{2.88\%} & 26.94\% & \textbf{8.37\%} & 30.30\% & 1.88\% & 1.89\% & 87.63\% & 5.38\% & 41.52\% & 32.63\% & 43.81\%\\

        &  &  &  & & FPR & \textbf{17.13\%} & 22.94\% & \textbf{53.37\%} & \textbf{44.63\%} & \textbf{35.25\%} & 5.49\% & \textbf{39.75\%} & 15.08\% & 41.13\% & 18.75\% & 10.75\% & 51.25\% & 47.88\% & 65.37\% & 56.57\% \\

        &  &  &  & & MA & \textbf{53.758\%} & 51.17\% & \textbf{39.62\%} & \textbf{49.12\%} & \textbf{61.87\%} & 67.56\% & \textbf{51.88\%} & 54.62\% & 57.00\% & 79.36\% & 1.63\% & 43.38\% & 10.60\% & 2.00\% & -0.38\% \\

        &  &  &  & & AUC & \textbf{0.782} & 0.755 & \textbf{0.698} & \textbf{0.850} & \textbf{0.810} & 0.032 & 0.840 & 0.106 & 0.880 & 0.234 & 0.012 & 0.132 & 0.107 & 0.505 & 0.514 \\

        &  &  &  & & T@0.01\%F & \textbf{0.018\%} & 0.015\% & \textbf{0.01\%} & \textbf{2.88\%} & \textbf{0.00\%} & 0.00\% & \textbf{0.13\%} & 0.00\% & 0.10\% & 0.00\% & 0.50\% & 0.00\% & 0.00\% & 0.00\% & 0.74\% \\

        &  &  &  & & Thres@max MA & - & - & \textbf{0.900} & - & \textbf{0.370} & 1.385 & \textbf{66.10\%} & 0.550 & - & -3.10$e^{-3}$ & 0.640 & 0.132 & -8.23$e^{-11}$& 0.012 & 4.538 \\

        \bottomrule[1.1pt]
    \end{tabular}}
    \begin{tablenotes}
    \scriptsize
    \item \emph{DisDistr.}: the distance distribution of data samples in the target dataset; \emph{DisBD.}: the distance between data samples of the target dataset; 
   \item \emph{DiffDis.}: the differential distance between two datasets; \emph{RatNoI.}: the ratio of the samples that are made no inferences by an MI attack.  
    \end{tablenotes}
    \end{threeparttable}
    \label{tabs:The Evaluation Results of the target datasets obeying different distance distributions of data samples in the target dataset.}
    \vspace{0.1pt}
\end{table*}

\subsection{Principle 2: Distance Distribution of Data Samples in the Target Dataset Matters}
\label{Principle 2: Data Distribution}

Incorporating the following factor could substantially
enhance the ability of a ``comparing different MI attacks''
methodology to explain the comparative evaluation results: (factor 1) the effectiveness of an MI attack could be highly sensitive
to the distance distribution of data samples in the target dataset.

From Table~\ref{tabs:The Evaluation Results of the target datasets obeying different distance distributions of data samples in the target dataset.}, 
we discover that the evaluation results of an MI attack are different when distance distributions of data samples in the target dataset obey different distance distributions (e.g., the MAs and the TPR @ 0.01\% FPR of the Positive Predictive Value (PPV) attack~\cite{jayaraman2020revisiting} are 3.13\%, 6.75\%, 1.63\% and 0.63\%, 0.88\%, 0.50\% when the distance distribution of data samples in the target dataset obeys normal, uniform, and bernoulli distributions, respectively.)

\subsection{Principle 3: Distance between Data Samples of the Target Dataset Matters}
\label{Principle 3: Distance between Members and Nonmembers}

Incorporating the following factor could substantially
enhance the ability of a ``comparing different MI attacks" methodology to explain the comparative evaluation results: (factor 2) the effectiveness of an
MI attack could be highly sensitive to the distances between data samples of the target dataset.

From Figure~\ref{figs:Fig22}, 
 we discover that 
 the evaluation results of an MI attack are different when target datasets have different distances between data samples (e.g., the MAs of the NN attack~\cite{shokri2017membership} are 28.89\%, 32.98\% and 39.39\% when the distance between data samples of the target dataset are 2.510, 3.813, 4.025, respectively). 


\subsection{Principle 4: Differential Distance between Two Datasets Matters}
\label{Principle 4: Differential Distance between Two Sets}

Incorporating the following factor could substantially
enhance the ability of a ``comparing different MI attacks''
methodology to explain the comparative evaluation results: (factor 3) the
effectiveness of an MI attack could be highly sensitive to the differential
distance (i.e., MMD~\cite{gretton2012kernel} ) before and after a data example is moved from the target dataset
to a generated dataset with only nonmembers.

From Figure~\ref{figs:Fig33}, we find that the evaluation results of an MI attack are different when target datasets have different differential distances between two datasets (e.g., the MAs of the Calibrated Score attack~\cite{watson2021importance} are 46.73\%, 39.82\% and 43.45\% when the differential distances between two datasets are 0.085, 0.119, 0.157, respectively).

\section{Evaluation Framework}
\label{sec Evaluation Framework} 

In this section, 
we first introduce the experimental setup of our benchmark. Then, we propose the composition of the evaluation framework. 

\subsection{Experimental Setup}
\label{sebsec:Experimental Setup}

\noindent \textbf {Implementation.} In this paper, we utilize the Python 3.7 to achieve our evaluations. And nonmembers utilized in our experiments is the same as~\cite{hui2021practical}.

\noindent \textbf {Models.} We adopt the same model architectures and hyperparameters as prior BlindMI-Diff attack~\cite{hui2021practical}, which are defined in table~\ref{tabs:tab6}. We also adopt a multilayer perceptron (MLP) model that has at most seven dense layers (e.g., 8192, 4096, 2048, 1024, 512, 256 and 128), and an additional Softmax layer. The standard CNN architectures and hyperparameters are the same as the prior black MI attack~\cite{shokri2017membership}. We also utilize the three kinds of popular DNNs (e.g., VGG, ResNet and DenseNet), which are standard architectures with pre-trained parameters from ImageNet.

\noindent  \textbf {Target Model.} Given a dataset, we randomly select a target model from the target model column of Table~\ref{tabs:tab6}, and train the target model with the specified hyperparameters. 

\noindent \textbf{Shadow Model.} For 
black-box attacks, we randomly select a shadow model architecture from the shadow model column of Table~\ref{tabs:tab6}. For the gray/white-box settings, we select the same architectures as the given target model as our shadow models' architectures.

\noindent \textbf {Datasets.} We utilize seven datasets (as shown in Table~\ref{tabs:tab7}) to perform the comprehensive evaluations of the existing MI attacks. We select half members and half nonmembers in each target dataset, and the details are shown in the Appendix~\ref{subsec:Datasets Description}.

\begin{table*}[]\tiny
    \caption{The description of different datasets used in our evaluation.}
    \centering
    \begin{threeparttable}
    \setlength{\tabcolsep}{3.5mm}{
    \begin{tabular}{ccccccccccc}
        \toprule[1.1pt]
                 \textbf{Dataset} &
            \textbf{Description}& \textbf{$\#$ of class} & \textbf{Resolution} &
            \textbf{\# Epochs(target model)} & $\mathcal{TS}_{t}$ & $\mathcal{TS}_{s}$ & $\mathcal{D}_{t1}$& $\mathcal{D}_{t2}$& $\mathcal{D}_{t3}$& $\mathcal{D}_{t4}$
            \\
         \midrule[1.1pt]
        
          CH\_MNIST &  histological images& 8 & 64$\times$64 &  (pre-trained + 15) or 150 & 2,500 & 2,500 & 5,000 & 800 & 800 & 800\\
        
        CIFAR10 & image classification& 10 & 32$\times$32 &  (pre-trained + 30) or 150 & 10,000 & 10,000 & 20,000 & 2,000 & 2,000 & 2,000 \\
        
         CIFAR100 & image classification & 100  & 32$\times$32 &  (pre-trained + 30) or 150 & 10,000 & 10,000 & 20,000 & 2,000 & 2,000 & 2,000\\
         
         ImageNet & image classification & 200 & 64$\times$64 &  (pre-trained + 30) or 150 & 10,000 & 10,000 & 20,000 & 2,000 & 2,000 & 2,000\\

        Location30 & text classification & 30 & 64$\times$64 &  (pre-trained + 30) or 150 & 1,000 & 1,000 & 2,000 & 500 & 500 & 500\\

         Purchase100 & text classification & 100 & 64$\times$64 &  (pre-trained + 30) or 150 & 19,732 & 19,732 & 39,464 & 2,000 & 2,000 & 2,000\\

         Texas100 & text classification & 100 & 64$\times$64 &  (pre-trained + 30) or 150 & 10,000 & 10,000 & 20,000 & 1,560 & 1,560 & 1,560\\
         \bottomrule[1.1pt]
    \end{tabular}}
     \begin{tablenotes}
    \scriptsize
    \item $\mathcal{TS}_{t}$ and $\mathcal{TS}_{s}$ are the training datasets of the target and shadow models, respectively. $\mathcal{D}_{t1}$ is the target dataset that doesn't obey any distribution, 
\item $\mathcal{D}_{t2}$-$\mathcal{D}_{t4}$ are the target datasets obeying normal, uniform and bernoulli distributions, respectively. 
    \end{tablenotes}
    \end{threeparttable}
    \label{tabs:tab7}
    \vspace{3pt}
\end{table*}

\begin{table}[]\tiny
    \caption{The Description of settings on target and shadow models' architectures and hyperparameters.}
    \centering
    \begin{threeparttable}
    \setlength{\tabcolsep}{2mm}{
    \begin{tabular}{cccccc}
        \toprule[1.1pt]
            \multirow{2}{*}{\textbf{ModArch.}} & \multirow{2}{*}{\textbf{\# of layers}}& \multicolumn{2}{c}{\textbf{target model}}  & \multicolumn{2}{c}{\textbf{shadow model}} \\ 
            \cmidrule(r){3-4} \cmidrule(r){5-6}&  & 
                \textbf{MaxEpoch.} & \textbf{LNRate.} &
             \textbf{MaxEpoch.} & \textbf{LNRate.}\\
         \midrule[1.1pt] 
       MLP  & [3-7] dense & $e_{m}$ & 5$e^{-5}$ & [0.3-2]$e_{m}$ & 1$e^{-4}$/$e^{-5}$   \\
       StandDNN & 2 & $e_{m}$ & 5$e^{-5}$ & 0.5$e_{m}$ & 1$e^{-4}$   \\
       VGG16  & 16 & $e_{p}^{*}$+$e_{m}^{**}$ & 5$e^{-5}$ & $e_{p}$+0.6$e_{m}$ & 5$e^{-5}$   \\
       VGG19  & 19 & $e_{p}$+$e_{m}$ & 5$e^{-5}$ & $e_{p}$+1.5$e_{m}$ & 5$e^{-5}$  \\
       ResNet50 & 50 & $e_{p}$+$e_{m}$ & 5$e^{-5}$ & $e_{p}$+0.2$e_{m}$ & 5$e^{-5}$ \\
       ResNet101 & 101 & $e_{p}$+$e_{m}$ & 5$e^{-5}$ & $e_{p}$+0.3$e_{m}$& 1$e^{-4}$  \\
        DenseNet121 & 121 & $e_{p}$+$e_{m}$ & 5$e^{-5}$ & $e_{p}$+$e_{m}$ & 1$e^{-4}$ \\
        \bottomrule[1.1pt]
    \end{tabular}}
    \begin{tablenotes}
    \scriptsize
    \item  $e_{m}$ is the maximum epochs of target model for each dataset in table~\ref{tabs:tab7}; 
\item $e_{p}$ is the epoch of a pre-trained weight on the ImageNet dataset.
    \end{tablenotes}
    \end{threeparttable}
    \label{tabs:tab6}
    \vspace{0.1pt}
\end{table}

\subsection{The Composition of the Evaluation Framework}
\label{sebsec:The Composition of the Evaluation Framework}

The primary design goal of our benchmark, called \textbf{MIBench}, is 
to meet {\em all} of above-mentioned 4 requirements (see Section \ref{sebsec:Introduction}). Our benchmark consists not only the evaluation metrics (Section \ref{sebsec:Part II: Evaluation Metrics}), but also the evaluation scenarios(Section \ref{sebsec:Part I: Evaluation Scenarios}). 
We design the evaluation scenarios 
from the {\bf four perspectives} mentioned in Section \ref{sebsec:Introduction}, 
and the ratio of the samples that are made no inferences by an MI attack. 
Moreover, each perspective corresponds to a control variable and so 
there are in total four {\bf control variables} in our evaluation scenarios. 
The evaluation metric module consists of ten typical evaluation metrics. 

\subsubsection{Part I: Evaluation Scenarios}
\label{sebsec:Part I: Evaluation Scenarios}

Since fine-grained attacks
make no inferences on certain data examples, it seems unfair
to use the same number of data examples when comparing
indiscriminate attack algorithms and fine-grained attack algorithms (shown in Figure ~\ref{figs: The effect of the ratio of the samples
that are made no inferences by an MI attack}).

The inaccurate and unfair evaluation results shown in Figure~\ref{figs: The effect of the ratio of the samples
that are made no inferences by an MI attack}, 
Principle 2 (Section~\ref{Principle 2: Data Distribution}), 
Principle 3 (Section~\ref{Principle 3: Distance between Members and Nonmembers}) and 
Principle 4 (Section \ref{Principle 4: Differential Distance between Two Sets}) indicate that a good evaluation framework should have at least 4 control variables, consisting of
the distance distribution of data samples in the target dataset (CV1), the distance between data samples of the target dataset (CV2), the differential distance between two
datasets (CV3), and the ratio of the samples that are made
no inferences by an MI attack (CV4).

The four control variables adopted in our evaluation scenarios (as shown in Figure~\ref{figs:Evaluation workflow_V5}) are as follows: \textbf{Control Variable 1:} \{C100\_N, C100\_U, C100\_B; C10\_N, C10\_U, C10\_B; CH\_N, CH\_U, CH\_B; I\_N, I\_U, I\_B, L30\_N, L30\_U, L30\_B; P100\_N, P100\_U, P100\_B; T100\_N, T100\_U, T100\_B\}; \textbf{Control Variable 2:} \{1.225, 1.24, \dots, 6.250\}; \textbf{Control Variable 3:} \{0.001, 0.002, \dots, 0.500\}; \textbf{Control Variable 4:} \{2\%, 4\%, 10\%, 12\%, 20\%, 40\%, 45\%, 49\%\}. Based on the four control variables, we build 84 evaluation scenarios for each dataset, and the 84 evaluation scenarios of the CIFAR100 are shown in Table~\ref{tabs:tab166}. 

\subsubsection{Part II: Evaluation Metrics}
\label{sebsec:Part II: Evaluation Metrics}

\begin{figure*}[htp]
\vspace{0.1pt}
\includegraphics[width=1\textwidth]{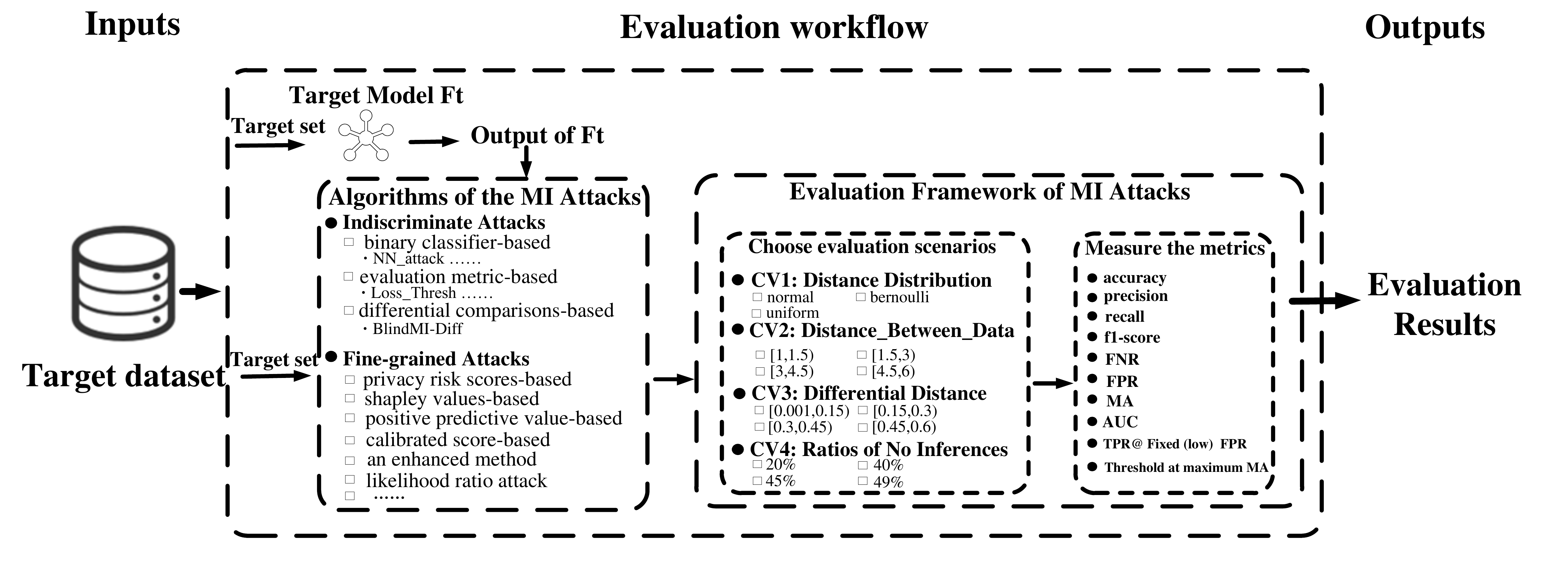}
\caption{\label{figs:Evaluation workflow_V5} The Evaluation Workflow of the Membership Inference Attacks.}
\vspace{0.1pt}
\end{figure*}

We mainly use attacker-side accuracy, precision, recall, f1-score, false positive rate (FPR), false negative rate (FNR), membership advantage (MA), the
Area Under the Curve (AUC) of attack Receiver
Operating Characteristic (ROC) curve \cite{carlini2022membership}, TPR @ fixed (low) FPR \cite{carlini2022membership}, threshold at maximum MA, as our evaluation metrics (as shown in Figure~\ref{figs:Evaluation workflow_V5}). Specifically, {\em accuracy} is the percentage of data samples with correct membership predictions by MI attacks; {\em precision} represents the ratio of real-true members predicted among all the positive membership predictions made by an adversary; {\em recall} demonstrates the ratio of true members predicted by an adversary among all the real-true members; {\em f1-score} is the harmonic mean of precision and recall; {\em FPR} is the ratio of nonmembers are erroneously predicted as members; {\em FNR} is the difference of the 1 and recall (e.g., FNR=1-recall); {\em MA} is the difference between the true positive rate and the false positive rate (e.g., MA = TPR - FPR~\cite{yeom2018privacy}); {\em AUC} is computed as the
Area Under the Curve of attack ROC in logarithmic scale; {\em TPR @ fixed (low) FPR} is an attack’s true-positive rate at (fixed) low false-positive rates in logarithmic scale; and the {\em threshold at maximum MA} is a threshold to achieve maximum MA.

\section{Using MIBench to Compare Different MI Attacks}
\label{sec:Using the Evaluation Framework}

The high-level idea for using the proposed benchmark is the controlled variable approach, where only one control variable (CV) is allowed to change at a time and the other three CVs are left unchanged, and then the evaluation results of the different MI attacks in
this case are tested. Our evaluation aims to answer the following Research Questions (RQs). \textbf{\emph{RQ1:}} Whether target datasets obeying different distance distributions of data samples have notable effects on the (effectiveness) ranking of state-of-the-art MI attacks when other three CVs keep the same values? \textbf{\emph{RQ2:}} Whether target datasets that have different distances between data samples have notable effects on the (effectiveness) ranking? \textbf{\emph{RQ3:}} Whether target datasets that have different differential distances between two datasets have notable effects on the (effectiveness) ranking? \textbf{\emph{RQ4:}} Whether target datasets that have different ratios of the
samples that are made no inferences by an MI attack have notable effects on the (effectiveness) ranking?

In our experiments, the target dataset sizes 
are shown in Table~\ref{tabs:tab7}. We first divide the target dataset into small batches of 20 data samples as the prior BlindMI-Diff attack~\cite{hui2021practical}, then we calculate the MMD of the output probabilities through target models by Equation~\ref{equs:equ1}. Next, we construct and calculate 4 CVs as described in Section~\ref{sebsec:Part I: Evaluation Scenarios}. Finally, we use our benchmark to fairly and systematically compare 15 state-of-the-art MI attack algorithms (see Table \ref{tabs:The Comparations of 13 MI attacks}) across 588 evaluation scenarios (84 evaluation scenarios for each dataset (see Table~\ref{tabs:tab166})), and study how every CV affects the evaluation results. 
Each attack is performed ten times with a new target and shadow model with different training datasets, model architectures and hyperparameters each time, and we obtain the average values of 10 evaluation metrics 
together with the standard error of the mean among the thirteen attacks. We perform experiments on CIFAR100, CIFAR10, CH\_MNIST and ImageNet with seven model architectures described in Section \ref{sebsec:Experimental Setup}, and for Location30, Purchase100 and Texas100, we use fully connected neural networks with 4 hidden layers, and the numbers of neurons for hidden layers are 1024, 512, 256, and 128, respectively. We use Tanh as the activation function on Purchase100 and Texas100 datasets, and ReLU on the Location30. For all MI attacks, we evaluate them with metrics in Section \ref{sebsec:Part II: Evaluation Metrics}, and for the fine-grained MI attacks based on thresholds (e.g., Shapley Values\cite{duddu2021shapr} and Risk Scores\cite{song2021systematic}), we report different thresholds for different classes at maximum MA (e.g., 100 thresholds on different classes on CIFAR100) and other MI attacks that use a single
threshold for all classes (e.g., Loss-Threshold\cite{yeom2018privacy}, Top1-Threshold\cite{salem2019ml}, PPV\cite{jayaraman2020revisiting}, Calibrated Score\cite{watson2021importance}, Distillation-based Thre.\cite{ye2021enhanced} and LiRA\cite{carlini2022membership}), we report the thresholds at maximum MA. All evaluation results of MIBench are shown in the \url{https://github.com/MIBench/MIBench.github.io/blob/main/README.md}.

Because of the (simplified) 84 evaluation scenarios, for each dataset, we select three distances among all distances between data samples under different distance distributions and three distances among all differential distance between two datasets, recectively. And the distances are the mean distance of the distance between data samples (or the differential distance between two datasets) and the two distances equally divided on the the left and right sides, respectively. For example, the three distances between data samples (differential distances between two datasets) selected on the CIFAR100 are 2.893, 3.813, 4.325 (0.085, 0.119, 0.157), respectively. (see Table \ref{tabs:The Selected Distances} in the Appendix C).

\begin{table*}[htp]\tiny
    \caption{Evaluation Scenarios of CIFAR100. }
    \centering
    \begin{threeparttable}
    \setlength{\tabcolsep}{1.5mm}{
    \begin{tabular}{ccccc|ccccc|ccccc|ccccc}
        \toprule[1.1pt]
            \textbf{ESN.} & \textbf{DisDistr.} &  \textbf{DisBD.} & \textbf{DiffDis.} & \textbf{RatNoI.} & \textbf{ESN.} & \textbf{DisDistr.} &  \textbf{DisBD.} & \textbf{DiffDis.} & \textbf{RatNoI.} &  \textbf{ESN.} & \textbf{DisDistr.} &  \textbf{DisBD.} & \textbf{DiffDis.} & \textbf{RatNoI.} & \textbf{ESN.} & \textbf{DisDistr.} &  \textbf{DisBD.} & \textbf{DiffDis.} & \textbf{RatNoI.} 
            \\ 
         \midrule[0.8pt] \midrule[0.8pt]
        ES01 & Normal & 2.893 & 0.085 & 20\% & 
        ES22 & Normal & 4.325 & 0.085 & 40\% &
        ES43 & Uniform & 3.813 & 0.119 & 45\% &
        ES64 & Bernoulli & 2.893 & 0.157 & 49\%  \\
       ES02 & Normal & 2.893 & 0.085 & 40\% &
        ES23 & Normal & 4.325 & 0.085 & 45\% & 
       ES44 & Uniform & 3.813 & 0.119 & 49\% &
       ES65 & Bernoulli & 3.813 & 0.085 & 20\% \\
       ES03 & Normal & 2.893 & 0.085 & 45\% &
       ES24 & Normal & 4.325 & 0.119 & 40\% & 
      ES45 & Uniform & 3.813 & 0.157 & 20\% & 
      ES66 & Bernoulli & 3.813 & 0.085 & 40\% \\ 
       ES04 & Normal & 2.893 & 0.119 & 40\% & 
       ES25 & Normal & 4.325 & 0.119 & 45\% &
      ES46 & Uniform & 3.813 & 0.157 & 40\% &
      ES67 & Bernoulli & 3.813 & 0.085 & 45\% \\
       ES05 & Normal & 2.893 & 0.119 & 45\% &
       ES26 & Normal & 4.325 & 0.119 & 49\% & 
        ES47 & Uniform & 3.813 & 0.157 & 45\%  &
        ES68 & Bernoulli & 3.813 & 0.085 & 49\% \\ 
       ES06 & Normal & 2.893 & 0.119 & 49\% &
       ES27 & Normal & 4.325 & 0.157 & 45\% &
       ES48 & Uniform & 3.813 & 0.157 & 49\%  & 
       ES69 & Bernoulli & 3.813 & 0.119 & 20\% \\
       ES07 & Normal & 2.893 & 0.157 & 45\% & 
       ES28 & Normal & 4.325 & 0.157 & 49\% &
        ES49 & Uniform & 4.325 & 0.085 & 20\% &
        ES70 & Bernoulli & 3.813 & 0.119 & 40\% \\

       ES08 & Normal & 2.893 & 0.157 & 49\% &
        ES29 & Uniform & 2.893 & 0.085 & 20\% &
       ES50 & Uniform & 4.325 & 0.085 & 40\% &
       ES71 & Bernoulli & 3.813 & 0.119 & 45\% \\
       ES09 & Normal & 3.813 & 0.085 & 20\% &
        ES30 & Uniform & 2.893 & 0.085 & 40\% &
      ES51 & Uniform & 4.325 & 0.085 & 45\% & 
      ES72 & Bernoulli & 3.813 & 0.119 & 49\% \\
       ES10 & Normal & 3.813 & 0.085 & 40\% &
       ES31 & Uniform & 2.893 & 0.085 & 45\% & 
       ES52 & Uniform & 4.325 & 0.119 & 40\% & 
       ES73 & Bernoulli & 3.813 & 0.157 & 20\% \\
       ES11 & Normal & 3.813 & 0.085 & 45\% & 
       ES32 & Uniform & 2.893 & 0.119 & 40\% & 
        ES53 & Uniform & 4.325 & 0.119 & 45\% &
        ES74 & Bernoulli & 3.813 & 0.157 & 40\% \\
       ES12 & Normal & 3.813 & 0.085 & 49\% &
        ES33 & Uniform & 2.893 & 0.119 & 45\% & 
       ES54 & Uniform & 4.325 & 0.119 & 49\% &
       ES75 & Bernoulli & 3.813 & 0.157 & 45\% \\
       ES13 & Normal & 3.813 & 0.119 & 20\% &
        ES34 & Uniform & 2.893 & 0.119 & 49\% &
      ES55 & Uniform & 4.325 & 0.157 & 45\% & 
      ES76 & Bernoulli & 3.813 & 0.157 & 49\% \\
       ES14 & Normal & 3.813 & 0.119 & 40\% &
        ES35 & Uniform & 2.893 & 0.157 & 45\% &
       ES56 & Uniform & 4.325 & 0.157 & 49\% &
       ES77 & Bernoulli & 4.325 & 0.085 & 20\% \\
       ES15 & Normal & 3.813 & 0.119 & 45\% & 
       ES36 & Uniform & 2.893 & 0.157 & 49\% & 
       ES57 & Bernoulli & 2.893 & 0.085 & 20\% &
       ES78 & Bernoulli & 4.325 & 0.085 & 40\% \\
       
       ES16 & Normal & 3.813 & 0.119 & 49\% &
       ES37 & Uniform & 3.813 & 0.085 & 20\% & 
       ES58 & Bernoulli & 2.893 & 0.085 & 40\% &
       ES79 & Bernoulli & 4.325 & 0.085 & 45\% \\
       ES17 & Normal & 3.813 & 0.157 & 20\% & 
       ES38 & Uniform & 3.813 & 0.085 & 40\% & 
       ES59 & Bernoulli & 2.893 & 0.085 & 45\% &
        ES80 & Bernoulli & 4.325 & 0.119 & 40\% \\
       ES18 & Normal & 3.813 & 0.157 & 40\% &
       ES39 & Uniform & 3.813 & 0.085 & 45\% & 
       ES60 & Bernoulli & 2.893 & 0.119 & 40\% &
       ES81 & Bernoulli & 4.325 & 0.119 & 45\% \\
       
       ES19 & Normal & 3.813 & 0.157 & 45\% & 
       ES40 & Uniform & 3.813 & 0.085 & 49\% & 
       ES61 & Bernoulli & 2.893 & 0.119 & 45\% &
       ES82 & Bernoulli & 4.325 & 0.119 & 49\% \\
       ES20 & Normal & 3.813 & 0.157 & 49\% & 
       ES41 & Uniform & 3.813 & 0.119 & 20\% & 
       ES62 & Bernoulli & 2.893 & 0.119 & 49\% &
        ES83 & Bernoulli & 4.325 & 0.157 & 45\% \\ 
       ES21 & Normal & 4.325 & 0.085 & 20\% &
       ES42 & Uniform & 3.813 & 0.119 & 40\% &
       ES63 & Bernoulli & 2.893 & 0.157 & 45\%  &
       ES84 & Bernoulli & 4.325 & 0.157 & 49\% \\ 
        \bottomrule[1.1pt]
    \end{tabular}}
    \begin{tablenotes}
    \scriptsize
    \item  \emph{ESN.}: the number of the Evaluation Scenarios; \emph{DisDistr.}: the distance distribution of data samples in the target dataset; \emph{DisBD.}: the distance between data samples 
    \item of the target dataset; \emph{DiffDis.}: the differential distance between two datasets; \emph{RatNoI.}: the ratio of the samples that are made no inferences by an MI attack.\ \ \ \ \ \   
    \end{tablenotes}
    \end{threeparttable}
   \label{tabs:tab166}
    \vspace{0.1pt}
\end{table*}

\begin{figure}[]
\vspace{0.1pt}
\begin{center}
\subfigure[CIFAR100]{
\begin{minipage}{0.2\textwidth}
\label{Fig.sub.3.1}
\includegraphics[height=0.75\textwidth,width=1\textwidth]{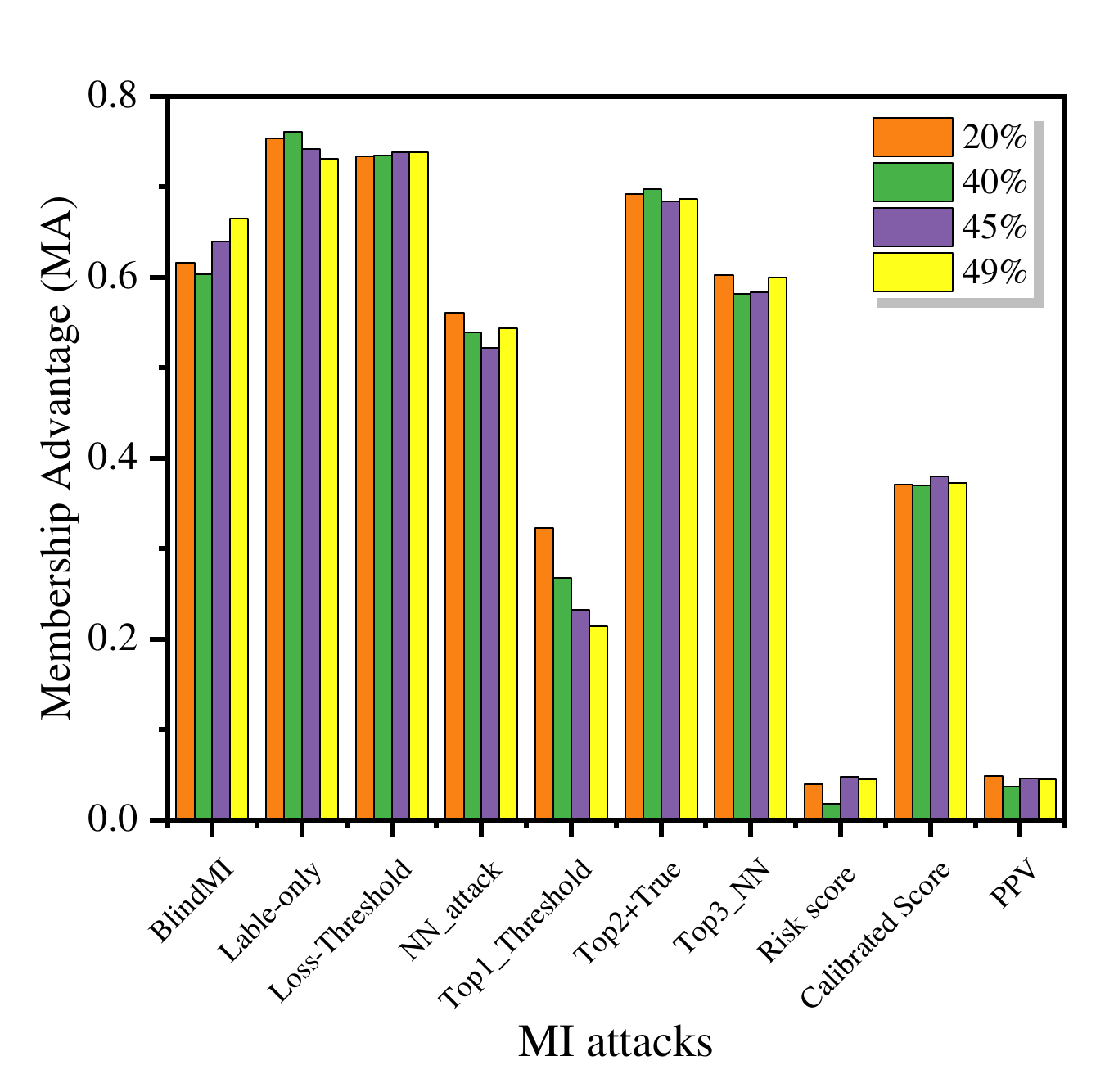}
\end{minipage}}
\subfigure[CIFAR10]{
\begin{minipage}{0.2\textwidth}
\label{Fig.sub.3.2}
\includegraphics[height=0.75\textwidth,width=1\textwidth]{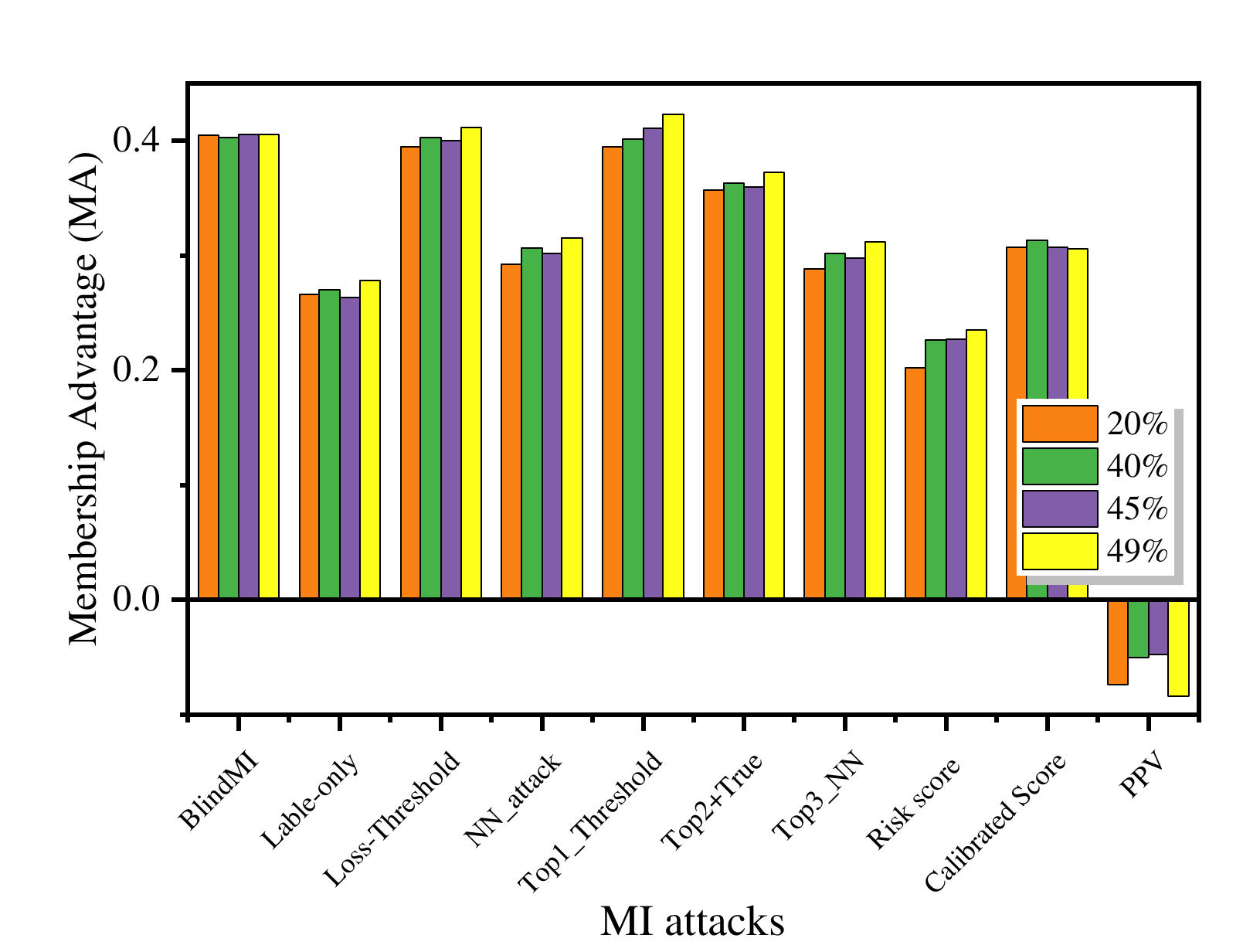}
\end{minipage}}
\subfigure[CH\_MNIST]{
\begin{minipage}{0.2\textwidth}
\label{Fig.sub.3.3}
\includegraphics[height=0.75\textwidth,width=1\textwidth]{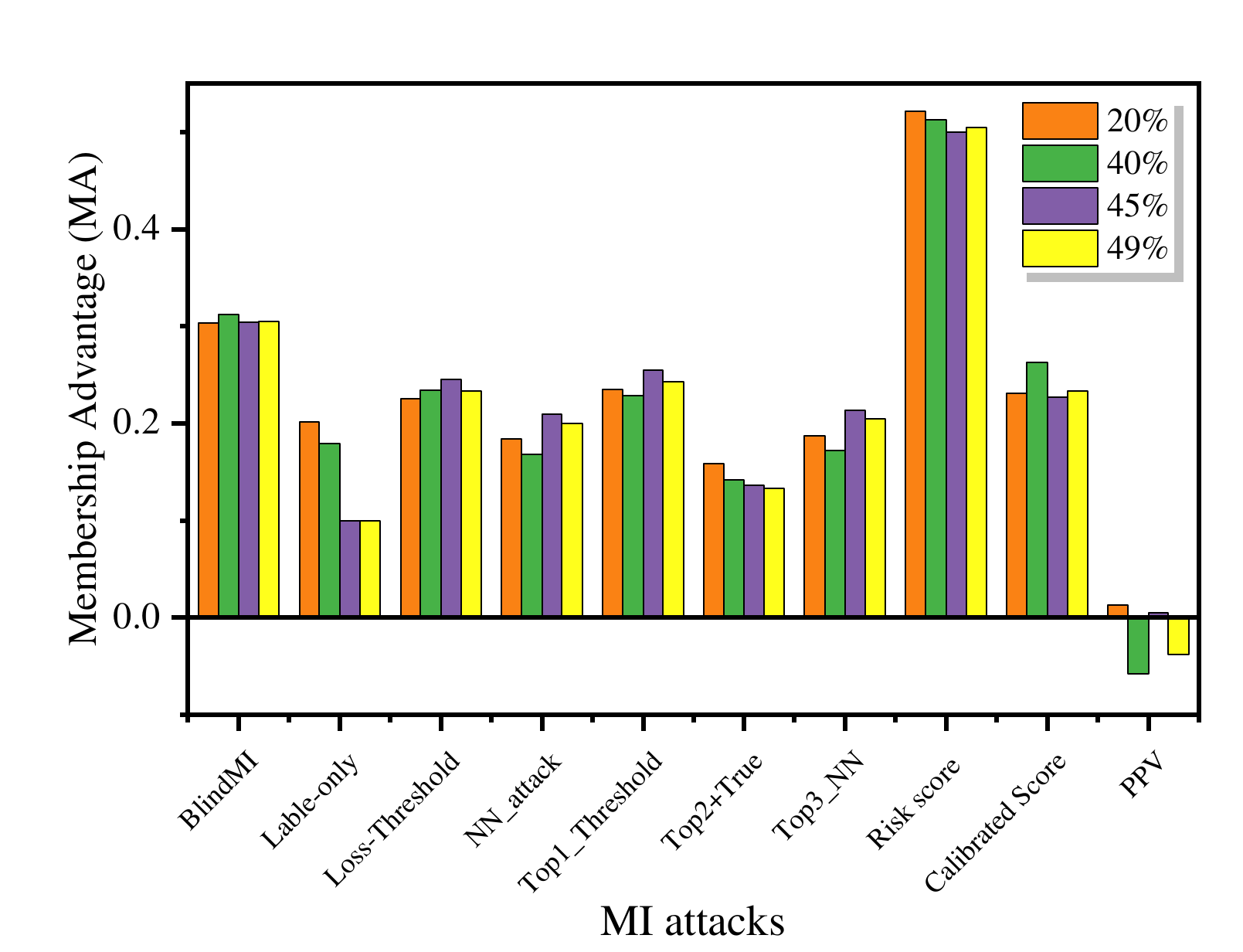}
\end{minipage}}
\subfigure[ImageNet]{
\begin{minipage}{0.2\textwidth}
\label{Fig.sub.3.4}
\includegraphics[height=0.75\textwidth,width=1\textwidth]{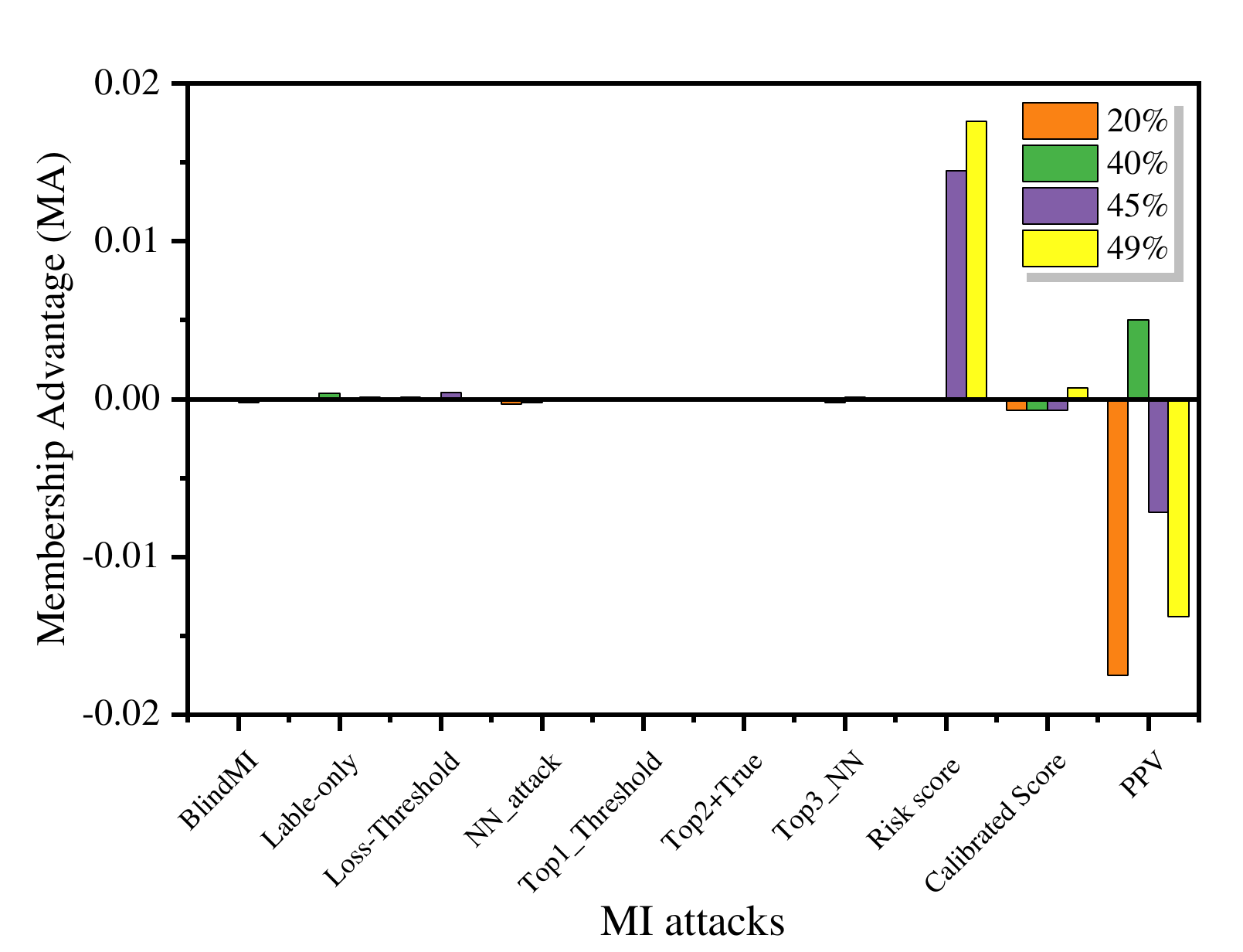}
\end{minipage}}
\end{center}
\caption{\label{figs: The effect of the ratio of the samples
that are made no inferences by an MI attack} The effect of the ratio of the samples that are made no inferences by an MI attack.}
\vspace{0.1pt}
\end{figure}

\subsection{RQ1: Effect of the Distance Distribution of Data Samples in the Target Dataset}
\label{sebsec:Effect of the Distance Distribution of Data Samples in the Target Dataset}

\noindent \textbf {Methodology.} For a target dataset, we first input all data samples (e.g., {\em n}) in the target dataset to the target model and get the corresponding output probabilities, and we classify these data into two categories according to their output probabilities (such as the data samples with high output probability and the data samples with low output probability), and the number of the two categories is the same (e.g., {\em $\frac{n}{2}$}). We extract the same number of data samples (e.g., $\eta$) from these two kinds of target data samples, and utilize Equation~\ref{equs:equ1} to calculate the MMD~\cite{gretton2012kernel} of these two categories target data samples' output probabilities through target models  (e.g., $2\eta$). 

\begin{equation}\label{equs:equ1}
  \textbf {MMD}[\mathcal{P}_{p},\mathcal{P}_{q}] = \left \| \frac{1}{n_p} \sum_{i=1}^{n_p}\phi(\mathcal{P}_{i}) - \frac{1}{n_q}\sum_{j=1}^{n_q}\phi(\mathcal{P}_{j}) \right\|_v
\end{equation}

where $\mathcal{P}_{i} \in \mathcal{P}_{p}$ and $\mathcal{P}_{j} \in \mathcal{P}_{q}$, $p$ and $q$ are data samples with high output probability and data samples with low output probability of the target dataset, respectively. And $n_p$ and $n_q$ are the size of $\mathcal{P}_{p}$ and $\mathcal{P}_{q}$, respectively. $\phi(.)$ is a feature space map from the probability of the target model to the Reproducing Kernel Hilbert Space (RKHS)\cite{borgwardt2006integrating}, namely $\emph{k} \mapsto \nu$. And most commonly used is a Gaussian kernel function, namely $\emph{k}(\mathcal{P}_{p},\mathcal{P}_{q}) = \left \langle \phi(\mathcal{P}_{p}),\phi(\mathcal{P}_{q})\right \rangle = \emph{exp}(-\left \| \mathcal{P}_{p} - \mathcal{P}_{q}\right \|/(2\sigma^2))$.

This calculation of MMD is performed until the MMDs are calculated for all samples in the target dataset, and we calculate a total of $\frac{n}{2\eta}$ groups of  MMDs. Next, according to the calculated $\frac{n}{2\eta}$ groups of MMDs of the target dataset, we first sort these $\frac{n}{2\eta}$ groups MMDs in ascending order, 

\begin{figure}[htp]
\vspace{0.1pt}
\includegraphics[width=0.5\textwidth]{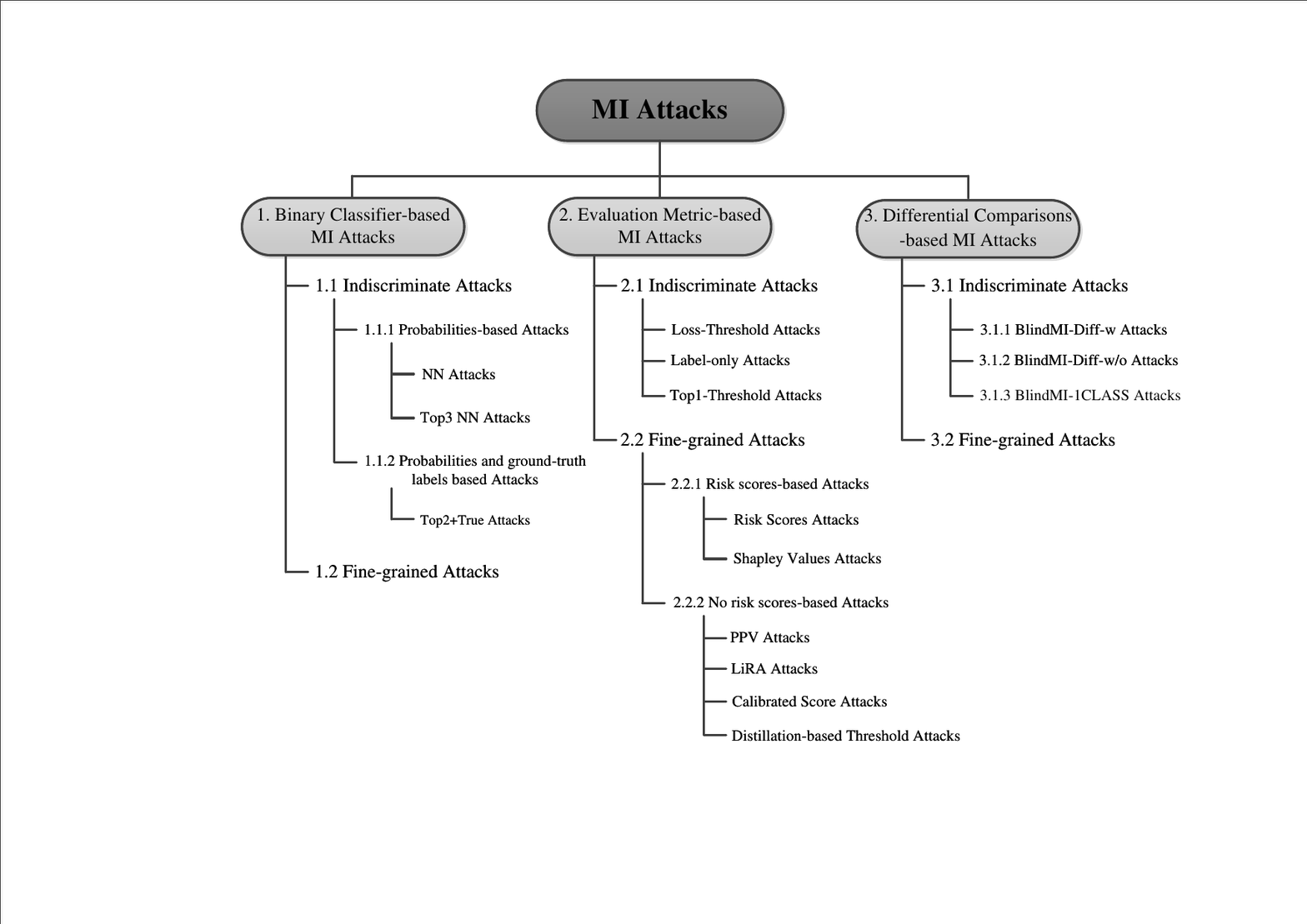}
\caption{\label{figs:Classification Tree_MI attacks} The Classification of the Membership Inference Attacks.}
\vspace{0.1pt}
\end{figure}

\noindent namely $\mathcal{MMD} = \{\mathcal{MMD}_{1}, \dots, \mathcal{MMD}_{n/2\eta}\}$. Then, we construct the distance distributions of data samples in the target dataset as normal, uniform and bernoulli distributions, respectively. (a) \textbf {Normal Distribution.} We select some MMDs from the calculated $\frac{n}{2\eta}$ groups of MMDs of the target dataset to construct a set of MMDs, $\{\mathcal{MMD}_{1}, \dots, \mathcal{MMD}_{h}\}$ ($1\leq h \leq n/2\eta $), which obeys a normal distribution, that is, $\{\mathcal{MMD}_{1}, \dots, \mathcal{MMD}_{h}\} \sim N(\mu, \sigma^2)\nonumber$, where $\mu$ and $\sigma$ are the mean and variance of the normal distribution, respectively. (b) \textbf {Uniform Distribution.} We assume two parameters $\emph{a}$ and $\emph{b}$, and the $\mathcal{MMD}_{1} \leq \emph{a} \leq \emph{b} \leq \mathcal{MMD}_{n/2\eta}$. We select $\mathcal{MMD}_{i}$ from the calculated $\frac{n}{2\eta}$ groups of MMDs of the target dataset, and $a \leq \mathcal{MMD}_{i} \leq b$ and divide the selected MMDs, $\{\mathcal{MMD}_{1}, \dots, \mathcal{MMD}_{s}\}$ ($1\leq s \leq n/2\eta $), in the distance intervals [a,b] equally into several parts (e.g., $\gamma$) and the number of MMD distance groups in each MMD distance part is equal (e.g., $s/\gamma$). Therefore, the selected MMDs obeys a uniform distribution of parameters $s/\gamma$ on [a,b]. (c) \textbf {Bernoulli Distribution.} We select a distance threshold (e.g., $\epsilon$) in the calculated $\frac{n}{2\eta}$ groups of MMDs of the target dataset as the split point of the Bernoulli distribution, and if the selected $\mathcal{MMD}_{i}$ is smaller than the selected distance threshold $\epsilon$, we regard the incident as a success, otherwise a failure. Specially, we select some MMDs, $\{\mathcal{MMD}_{1}, \dots, \mathcal{MMD}_{r}\}$ ($1\leq r \leq n/2\eta $) from the calculated $\frac{n}{2\eta}$ groups of MMDs of the target dataset, and construct these selected MMDs (e.g., $\{\mathcal{MMD}_{1}, \dots, \mathcal{MMD}_{r}\}$) obey Bernoulli distribution with a parameter p, where the ratio of MMDs greater than the distance threshold $\epsilon$ is p and the ratio of MMDs less than $\epsilon$ is (1-p).

In our experiments, for a target dataset on each dataset (see $\mathcal{D}_{t1}$ in Table \ref{tabs:tab7}), we first input all data samples (e.g., 20,000 on CIFAR100) in the target dataset to the target model and get the corresponding output probabilities, and we classify these data into two categories according to their output probabilities (such as the data samples with high output probability and the data samples with low output probability), and the number of the two categories is the same (e.g., {10,000}). We extract the same number of data samples (e.g., 10), as the prior BlindMI-Diff attack \cite{hui2021practical}, from these two kinds of target data samples, and combine them into small batches of 20
data samples and utilize Equation~\ref{equs:equ1} to calculate the MMD~\cite{gretton2012kernel} of them. And we in total get 1000 MMDs, and original distance distribution of the 1000 MMDs on CIFAR100 is shown as Figure \ref{CIFAR100_Normal_V4}, and from the Figure \ref{CIFAR100_Normal_V4} we find that the original distance distribution of the 1000 MMDs on CIFAR100 is approximately normal distribution. When the original distance is within a certain range, the original distance distribution of the MMDs on CIFAR100 dataset is approximately uniform distribution (see Figure \ref{CIFAR100_Uniform_V4}). We select different distance thresholds of the 1000 MMDs on target CIFAR100 dataset, and find that the distance of the 1000 MMDs on CIFAR100 dataset is Bernoulli distribution shown in Table \ref{tabs:The Parameter and thresholds}. We find the similar phenomenon in other six target datasets, and the results are shown in the Appendix B.

 Because of the fair evaluation of the attack results under these three distributions, we chose an equal number of samples to calculate the MMDs (see $\mathcal{D}_{t2}$-$\mathcal{D}_{t4}$ in Table \ref{tabs:tab7}), and we select equal number of MMDs (e.g., 100 MMDs on CIFAR100) to construct the distance distribution of data samples in target datasets as normal, uniform and bernoulli distributions, respectively (see Figure \ref{figs:Fig3333} in Appendix B).

\noindent \textbf {Results.} We verify the evaluation results 
of 15 state-of-the-art MI attacks when only the  distance distribution of data samples in the target dataset (CV1) is allowed to change and other three CVs 
(see Section \ref{sebsec:Part I: Evaluation Scenarios}) are left unchanged. From Table~\ref{tabs:The Evaluation Results of the target datasets obeying different distance distributions of data samples in the target dataset.} and Figure \ref{figs:ROC}, we discover that the evaluation results of an MI attack are different when target datasets obey different distance distributions. 
We observe that the ranking results of a particular MI attack (i.e. Loss-Threshold \cite{yeom2018privacy} (Class 2.1 see Figure \ref{figs:Classification Tree_MI attacks})) remain {\bf unchanged} across almost all the ESs in a particular group (i.e., when only the CV1 changed and other three CVs remained the same, the Loss threshold based attack \cite{yeom2018privacy} is 85.17\% of the top three most effective attacks). This indicates that the corresponding tuning variable probably does not affect the MI attack. To confirm this conclusion, we look into the algorithm of the MI attack and find the following: the Loss threshold-based attack \cite{yeom2018privacy} required a shadow model and computed the average cross-entropy
loss of all training samples in the shadow model to distinguish members. Therefore, when only the CV1 changes and other three CVs remain the same, it does not affect the average cross-entropy
loss of the Loss threshold based attack \cite{yeom2018privacy}, and thus has little effect on evaluation results. 

We find that the Top1-Threshold based attack \cite{salem2019ml} (Class 2.1 see Figure \ref{figs:Classification Tree_MI attacks})) is most affected when only the CV1 changes, other three CVs remain the same, for example, for one particular ES01 on Purchase100 could result in the following ranking:
Class 2.1 (except for the Label-only, Loss Threshold attacks)\,\textgreater\,Class 3.1.2\,\textgreater\,Class 2.2.1 (except for the Shapley values attack)\,\textgreater\,Class 1.1.2\,\textgreater\,Class 1.1.1 (except for the Top3\_NN attack)\,\textgreater\,Class 2.2.2 (except for the Calibrated Score, LiRA and PPV attacks)\,\textgreater\,Class 3.1.1\,\textgreater\,Class 3.1.3,
and for another particular ES29 on Purchase100 could result in the following ranking: 
Class 3.1.2\,\textgreater\,Class 1.1.2\,\textgreater\,Class 2.2.1 (except for the Shapley values attack)\,\textgreater\,Class 1.1.1 (except for the Top3\_NN attack)\,\textgreater\,Class 2.1 (except for the Loss Threshold attack)\,\textgreater\,Class 2.2.2 (except for the LiRA and PPV attacks)\,\textgreater\,Class 3.1.1\,\textgreater\,Class 3.1.3.
The reason is that the Top1-Threshold based attack \cite{salem2019ml} needs some extra nonmembers to get the nonmember threshold to infer members, and the evaluation results of Top1\_Threshold attacks depends on the choice of the nonmembers, and when only CV1 changes and the nonmembers remain unchanged, may cause the selected nonmembers to be unable to effectively distinguish between members and nonmembers of different distance distributions of data samples in the target dataset.

We also observe that the ranking order between two particular MI attacks (i.e. Top3-NN attack\cite{salem2019ml} and NN\_attack\cite{shokri2017membership} (Class 1.1.1)) remain {\bf unchanged} across almost all the ESs in a particular group. This indicates that the ranking order is independent of the control variable. In addition, this indicates that the ranking order is to certain extent determined by the three non-tuning factors. We look into the values of the three non-tuning factors, and we find that the ranking order between Top3-NN attacks and NN\_attacks is Top3\_NN\,\textgreater\,NN\_attack when only the CV1 changes and other three CVs remain the same, while the ranking order of them is NN\_attack\,\textgreater\,Top3\_NN when only the CV2 changes and the other three variables remain the same. The reason is that Top3-NN attacks and NN\_attacks are mainly based on the output probability of the target model, so it hardly affects the ranking order of them when only the CV1 changes, which does not effect the output probability of the target model; and changing only the CV2 will change the distance between the data samples with high output probability and data samples with low output probability of the target dataset to some extent, this will affect the attack decision, and finally make the ranking order of the two attacks change. Meanwhile, 
from Figure~\ref{figs:fig7}, 
We find that the 
FNR is almost the same when the target datasets obeying different distance distributions of data samples in the target dataset.

[Observation] \emph{The distance distributions of data samples in the target dataset have little effect on the 
FNR.}

\begin{table*}[htp]\tiny
    \caption{The MI attacks that rank opposite. }
    \centering
    \begin{threeparttable}
    \setlength{\tabcolsep}{0.1mm}{
    \begin{tabular}{cccccccccc|cccccccccc}
        \toprule[1.1pt]
            \textbf{ARN.} & \textbf{Attack A} &  \textbf{Attack B} & \textbf{EL.} & \textbf{RAR.} & \textbf{OAR.} & \textbf{dataset} &  \textbf{RES.} & \textbf{MR.} & \textbf{SMR.} & \textbf{ARN.} & \textbf{Attack A} &  \textbf{Attack B} & \textbf{EL.} & \textbf{RAR.} & \textbf{OAR.} & \textbf{dataset} &  \textbf{RES.} & \textbf{MR.} & \textbf{SMR.} \\
         \midrule[0.8pt] \midrule[0.8pt]
       \multirow{2}{*}{AR01}  & \multirow{2}{*}{Risk score} & \multirow{2}{*}{Loss-Threshold} & \multirow{2}{*}{1+2} & \multirow{2}{*}{A \textgreater B }& 
        \multirow{2}{*}{A \textless B} & \multirow{2}{*}{2+3+6}
    & ES03:C100\_N+2.893+0.085+45\% &\multirow{2}{*}{CV4} & \multirow{2}{*}{40\%} & \multirow{2}{*}{AR43}  & \multirow{2}{*}{BlindMI-1CLASS} & \multirow{2}{*}{Top3\_NN} & \multirow{2}{*}{3} & \multirow{2}{*}{A \textgreater B }& 
        \multirow{2}{*}{A \textless B} & \multirow{2}{*}{1+4+6}
    & ES63:C10\_B+1.908+0.291+45\%  &\multirow{2}{*}{CV3} & \multirow{2}{*}{40\%} \\

          &  &  & & & &
         & ES01:C100\_N+2.893+0.085+20\% & & &&  &  & & & &
         & ES59:C10\_B+1.908+0.155+45\% & & \\

         \multirow{2}{*}{AR02}  & \multirow{2}{*}{Loss-Threshold} & \multirow{2}{*}{Label-Only} & \multirow{2}{*}{3} & \multirow{2}{*}{A \textgreater B }& 
        \multirow{2}{*}{A \textless B} & \multirow{2}{*}{2}
    & ES01:C100\_N+2.893+0.085+20\%  &\multirow{2}{*}{CV2} & \multirow{2}{*}{40\%} & \multirow{2}{*}{AR44}  & \multirow{2}{*}{Risk score} & \multirow{2}{*}{PPV} & \multirow{2}{*}{5} & \multirow{2}{*}{A \textgreater B }& 
        \multirow{2}{*}{A \textless B} & \multirow{2}{*}{1}
    & ES68:C10\_B+2.501+0.155+49\%  &\multirow{2}{*}{CV4} & \multirow{2}{*}{40\%} \\

          &  &  & & & &
         & ES21:C100\_N+4.325+0.085+20\% & & &&  &  & & & &
         & ES65:C10\_B+2.501+0.155+20\% & & \\
        
       \multirow{2}{*}{AR03}  & \multirow{2}{*}{Top2+True} & \multirow{2}{*}{Risk score} & \multirow{2}{*}{-} & \multirow{2}{*}{A \textgreater B }& 
        \multirow{2}{*}{A \textless B} & \multirow{2}{*}{2+3+6}
    & ES09:CH\_M\_N+1.355+0.083+20\%  &\multirow{2}{*}{CV4} & \multirow{2}{*}{40\%} & \multirow{2}{*}{AR45}  & \multirow{2}{*}{LiRA} & \multirow{2}{*}{Risk score} & \multirow{2}{*}{1} & \multirow{2}{*}{A \textgreater B }& 
        \multirow{2}{*}{A \textless B} & \multirow{2}{*}{1}
    & ES13:C10\_N+2.501+0.213+20\%  &\multirow{2}{*}{CV4} & \multirow{2}{*}{40\%} \\

          &  &  & & & &
         & ES12:CH\_M\_N+1.355+0.083+49\% & & &&  &  & & & &
         & ES16:C10\_N+2.501+0.213+49\% & & \\
       
        \multirow{2}{*}{AR04}  & \multirow{2}{*}{Label-Only} & \multirow{2}{*}{Top2+True} & \multirow{2}{*}{3} & \multirow{2}{*}{A \textgreater B }& 
        \multirow{2}{*}{A \textless B} & \multirow{2}{*}{2+3}
    & ES07:CH\_M\_N+0.954+0.133+45\%  &\multirow{2}{*}{CV2} & \multirow{2}{*}{40\%} & \multirow{2}{*}{AR46}  & \multirow{2}{*}{Calibrated Score} & \multirow{2}{*}{BlindMI-w} & \multirow{2}{*}{-} & \multirow{2}{*}{A \textgreater B }& 
        \multirow{2}{*}{A \textless B} & \multirow{2}{*}{3+4+5+6+7}
    & ES02:T100\_N+0.530+0.038+4\%  &\multirow{2}{*}{CV3} & \multirow{2}{*}{40\%} \\

          &  &  & & & &
         & ES27:CH\_M\_N+1.720+0.133+45\% & & &&  &  & & & &
         & ES04:T100\_N+0.530+0.073+4\% & & \\
       
       \multirow{2}{*}{AR05}  & \multirow{2}{*}{Top1\_Threshold} & \multirow{2}{*}{Top2+True} & \multirow{2}{*}{3} & \multirow{2}{*}{A \textgreater B }& 
        \multirow{2}{*}{A \textless B} & \multirow{2}{*}{1+2+3+6+7}
    & ES52:C10\_U+3.472+0.213+40\%  &\multirow{2}{*}{CV2} & \multirow{2}{*}{40\%} & \multirow{2}{*}{AR47}  & \multirow{2}{*}{Calibrated Score} & \multirow{2}{*}{Shapley values} & \multirow{2}{*}{-} & \multirow{2}{*}{A \textgreater B }& 
        \multirow{2}{*}{A \textless B} & \multirow{2}{*}{3}
    & ES45:CH\_M\_U+1.355+0.133+20\%  &\multirow{2}{*}{CV4} & \multirow{2}{*}{40\%} \\

          &  &  & & & &
         & ES32:C10\_U+1.908+0.213+40\% & & &&  &  & & & &
         & ES48:CH\_M\_U+1.355+0.133+49\% & & \\
      
        \multirow{2}{*}{AR06}  & \multirow{2}{*}{BlindMI-w} & \multirow{2}{*}{Top2+True} & \multirow{2}{*}{3} & \multirow{2}{*}{A \textgreater B }& 
        \multirow{2}{*}{A \textless B} & \multirow{2}{*}{1+2+3+4}
    & ES46:C10\_U+2.501+0.291+40\%  &\multirow{2}{*}{CV4} & \multirow{2}{*}{40\%} & \multirow{2}{*}{AR48}  & \multirow{2}{*}{Calibrated Score} & \multirow{2}{*}{Distillation-based} & \multirow{2}{*}{1} & \multirow{2}{*}{A \textgreater B }& 
        \multirow{2}{*}{A \textless B} & \multirow{2}{*}{3}
    & ES83:CH\_M\_B+1.720+0.133+45\%  &\multirow{2}{*}{CV2} & \multirow{2}{*}{40\%} \\

          &  &  & & & &
         & ES38:C10\_U+2.501+0.155+40\% & & &&  &  & & & &
         & ES63:CH\_M\_B+0.954+0.133+45\% & & \\
       
       \multirow{2}{*}{AR07}  & \multirow{2}{*}{Top3\_NN} & \multirow{2}{*}{Top2+True} & \multirow{2}{*}{3} & \multirow{2}{*}{A \textgreater B }& 
        \multirow{2}{*}{A \textless B} & \multirow{2}{*}{1+2+3+5}
    & ES56:CH\_M\_U+1.720+0.133+49\%  &\multirow{2}{*}{CV2} & \multirow{2}{*}{40\%} & \multirow{2}{*}{AR49}  & \multirow{2}{*}{BlindMI-w} & \multirow{2}{*}{Distillation-based} & \multirow{2}{*}{-} & \multirow{2}{*}{A \textgreater B }& 
        \multirow{2}{*}{A \textless B} & \multirow{2}{*}{3+4}
    & ES75:ImaN\_B+1.130+0.145+45\%  &\multirow{2}{*}{CV3} & \multirow{2}{*}{40\%} \\

          &  &  & & & &
         & ES36:CH\_M\_U+0.954+0.133+49\% & & &&  &  & & & &
         & ES67:ImaN\_B+1.130+0.046+45\% & & \\
      
       \multirow{2}{*}{AR08}  & \multirow{2}{*}{BlindMI-without} & \multirow{2}{*}{Top2+True} & \multirow{2}{*}{3} & \multirow{2}{*}{A \textgreater B }& 
        \multirow{2}{*}{A \textless B} & \multirow{2}{*}{1+2+3}
    & ES07:C100\_N+2.893+0.157+45\%  &\multirow{2}{*}{CV3} & \multirow{2}{*}{40\%} & \multirow{2}{*}{AR50}  & \multirow{2}{*}{Top3\_NN} & \multirow{2}{*}{Shapley values} & \multirow{2}{*}{-} & \multirow{2}{*}{A \textgreater B }& 
        \multirow{2}{*}{A \textless B} & \multirow{2}{*}{3+4+5+7}
    & ES49:CH\_M\_U+1.720+0.083+20\%  &\multirow{2}{*}{CV2} & \multirow{2}{*}{40\%} \\

          &  &  & & & &
         & ES03:C100\_N+2.893+0.085+45\% & & &&  &  & & & &
         & ES29:CH\_M\_U+0.954+0.083+20\% & & \\

        \multirow{2}{*}{AR09}  & \multirow{2}{*}{NN\_attack
} & \multirow{2}{*}{Top2+True} & \multirow{2}{*}{3} & \multirow{2}{*}{A \textgreater B }& 
        \multirow{2}{*}{A \textless B} & \multirow{2}{*}{1+2+3+4}
    & ES26:CH\_M\_N+1.720+0.108+49\%  &\multirow{2}{*}{CV2} & \multirow{2}{*}{40\%} & \multirow{2}{*}{AR51}  & \multirow{2}{*}{Top3\_NN} & \multirow{2}{*}{Distillation-based} & \multirow{2}{*}{-} & \multirow{2}{*}{A \textgreater B }& 
        \multirow{2}{*}{A \textless B} & \multirow{2}{*}{3+4+5+7}
    & ES49:L30\_U+0.801+0.041+4\%  &\multirow{2}{*}{CV2} & \multirow{2}{*}{40\%} \\

          &  &  & & & &
         & ES06:CH\_M\_N+0.954+0.108+49\% & & &&  &  & & & &
         & ES29:L30\_U+0.570+0.041+4\% & & \\
        
        \multirow{2}{*}{AR10}  & \multirow{2}{*}{Top1\_Threshold} & \multirow{2}{*}{Label-only} & \multirow{2}{*}{3} & \multirow{2}{*}{A \textgreater B }& 
        \multirow{2}{*}{A \textless B} & \multirow{2}{*}{2+6}
    & ES02:P100\_N+0.550+0.087+4\%  &\multirow{2}{*}{CV2} & \multirow{2}{*}{40\%} & \multirow{2}{*}{AR52}  & \multirow{2}{*}{BlindMI-without} & \multirow{2}{*}{Risk score} & \multirow{2}{*}{-} & \multirow{2}{*}{A \textgreater B }& 
        \multirow{2}{*}{A \textless B} & \multirow{2}{*}{6+7}
    & ES07:T100\_N+0.530+0.107+10\%  &\multirow{2}{*}{CV3} & \multirow{2}{*}{40\%} \\

          &  &  & & & &
         & ES10:P100\_N+0.625+0.087+4\% & & &&  &  & & & &
         & ES03:T100\_N+0.530+0.038+10\% & & \\
       
        \multirow{2}{*}{AR11}  & \multirow{2}{*}{BlindMI-w} & \multirow{2}{*}{Label-only} & \multirow{2}{*}{3} & \multirow{2}{*}{A \textgreater B }& 
        \multirow{2}{*}{A \textless B} & \multirow{2}{*}{2+3+4}
    & ES07:C100\_N+2.893+0.157+45\%  &\multirow{2}{*}{CV3} & \multirow{2}{*}{40\%} & \multirow{2}{*}{AR53}  & \multirow{2}{*}{BlindMI-without} & \multirow{2}{*}{Top1\_Threshold} & \multirow{2}{*}{3} & \multirow{2}{*}{A \textgreater B }& 
        \multirow{2}{*}{A \textless B} & \multirow{2}{*}{6+7}
    & ES08:P100\_N+0.550+0.156+12\%  &\multirow{2}{*}{CV3} & \multirow{2}{*}{40\%} \\

          &  &  & & & &
         & ES03:C100\_N+2.893+0.085+45\% & & &&  &  & & & &
         & ES06:P100\_N+0.550+0.110+12\% & & \\
        
       \multirow{2}{*}{AR12}  & \multirow{2}{*}{Label-only} & \multirow{2}{*}{Top3\_NN} & \multirow{2}{*}{3} & \multirow{2}{*}{A \textgreater B }& 
        \multirow{2}{*}{A \textless B} & \multirow{2}{*}{1+2+3+5+6}
    & ES01:T100\_N+0.530+0.038+2\%  &\multirow{2}{*}{CV2} & \multirow{2}{*}{40\%} & \multirow{2}{*}{AR54}  & \multirow{2}{*}{Risk score} & \multirow{2}{*}{Top1\_Threshold} & \multirow{2}{*}{-} & \multirow{2}{*}{A \textgreater B }& 
        \multirow{2}{*}{A \textless B} & \multirow{2}{*}{3+6}
    & ES23:CH\_M\_N+1.720+0.083+45\%  &\multirow{2}{*}{CV4} & \multirow{2}{*}{40\%} \\

          &  &  & & & &
         & ES21:T100\_N+0.734+0.038+2\% & & &&  &  & & & &
         & ES21:CH\_M\_N+1.720+0.083+20\% & & \\
      
        \multirow{2}{*}{AR13}  & \multirow{2}{*}{BlindMI-without} & \multirow{2}{*}{Label-only} & \multirow{2}{*}{3} & \multirow{2}{*}{A \textgreater B }& 
        \multirow{2}{*}{A \textless B} & \multirow{2}{*}{1+2+3}
    & ES07:C100\_N+2.893+0.157+45\%  &\multirow{2}{*}{CV3} & \multirow{2}{*}{40\%} & \multirow{2}{*}{AR55}  & \multirow{2}{*}{NN\_attack} & \multirow{2}{*}{Top1\_Threshold} & \multirow{2}{*}{3} & \multirow{2}{*}{A \textgreater B }& 
        \multirow{2}{*}{A \textless B} & \multirow{2}{*}{6}
    & ES20:P100\_N+0.625+0.156+12\%  &\multirow{2}{*}{CV2} & \multirow{2}{*}{40\%} \\

          &  &  & & & &
         & ES03:C100\_N+2.893+0.085+45\% & & &&  &  & & & &
         & ES28:P100\_N+0.729+0.156+12\% & & \\
       
       \multirow{2}{*}{AR14}  & \multirow{2}{*}{Label-only} & \multirow{2}{*}{NN\_attack} & \multirow{2}{*}{3} & \multirow{2}{*}{A \textgreater B }& 
        \multirow{2}{*}{A \textless B} & \multirow{2}{*}{1+2+3+4+5+7}
    & ES01:CH\_M\_N+0.954+0.083+20\%  &\multirow{2}{*}{CV2} & \multirow{2}{*}{40\%} & \multirow{2}{*}{AR56}  & \multirow{2}{*}{Top1\_Threshold } & \multirow{2}{*}{Distillation-based} & \multirow{2}{*}{-} & \multirow{2}{*}{A \textgreater B }& 
        \multirow{2}{*}{A \textless B} & \multirow{2}{*}{6}
    & ES21:P100\_N+0.729+0.087+2\%  &\multirow{2}{*}{CV2} & \multirow{2}{*}{40\%} \\

          &  &  & & & &
         & ES21:CH\_M\_N+1.720+0.083+20\% & & &&  &  & & & &
         & ES09:P100\_N+0.625+0.087+2\% & & \\
       
        \multirow{2}{*}{AR15}  & \multirow{2}{*}{BlindMI-w} & \multirow{2}{*}{Top1-Threshold} & \multirow{2}{*}{3} & \multirow{2}{*}{A \textgreater B }& 
        \multirow{2}{*}{A \textless B} & \multirow{2}{*}{1+2+5}
    & ES63:C100\_B+2.893+0.157+45\%  &\multirow{2}{*}{CV3} & \multirow{2}{*}{40\%} & \multirow{2}{*}{AR57}  & \multirow{2}{*}{Top1\_Threshold } & \multirow{2}{*}{Shapley values} & \multirow{2}{*}{-} & \multirow{2}{*}{A \textgreater B }& 
        \multirow{2}{*}{A \textless B} & \multirow{2}{*}{6}
    & ES14:P100\_N+0.625+0.110+4\%  &\multirow{2}{*}{CV4} & \multirow{2}{*}{40\%} \\

          &  &  & & & &
         & ES59:C100\_B+2.893+0.085+45\% & & &&  &  & & & &
         & ES16:P100\_N+0.625+0.110+12\% & & \\
       
        \multirow{2}{*}{AR16}  & \multirow{2}{*}{BlindMI-w} & \multirow{2}{*}{Top3-NN} & \multirow{2}{*}{3} & \multirow{2}{*}{A \textgreater B }& 
        \multirow{2}{*}{A \textless B} & \multirow{2}{*}{1+2+3+4+6}
    & ES35:C100\_U+2.893+0.157+45\%  &\multirow{2}{*}{CV3} & \multirow{2}{*}{40\%} & \multirow{2}{*}{AR58}  & \multirow{2}{*}{Label-only} & \multirow{2}{*}{Shapley values} & \multirow{2}{*}{-} & \multirow{2}{*}{A \textgreater B }& 
        \multirow{2}{*}{A \textless B} & \multirow{2}{*}{3+4+6+7}
    & ES41:CH\_M\_U+1.355+0.108+20\% &\multirow{2}{*}{CV4} & \multirow{2}{*}{40\%} \\

          &  &  & & & &
         & ES31:C100\_U+2.893+0.085+45\% & & &&  &  & & & &
         & ES44:CH\_M\_U+1.355+0.108+49\% & & \\
        
        \multirow{2}{*}{AR17}  & \multirow{2}{*}{BlindMI-w} & \multirow{2}{*}{BlindMI-without} & \multirow{2}{*}{3} & \multirow{2}{*}{A \textgreater B }& 
        \multirow{2}{*}{A \textless B} & \multirow{2}{*}{1+2+3+4+5}
    & ES07:C100\_N+2.893+0.157+45\%  &\multirow{2}{*}{CV3} & \multirow{2}{*}{40\%} & \multirow{2}{*}{AR59}  & \multirow{2}{*}{Calibrated Score} & \multirow{2}{*}{PPV} & \multirow{2}{*}{5} & \multirow{2}{*}{A \textgreater B }& 
        \multirow{2}{*}{A \textless B} & \multirow{2}{*}{5+6+7}
    & ES50:L30\_U+0.801+0.041+8\%  &\multirow{2}{*}{CV1} & \multirow{2}{*}{40\%} \\

          &  &  & & & &
         & ES03:C100\_N+2.893+0.085+45\% & & &&  &  & & & &
         & ES78:L30\_B+0.801+0.041+8\% & & \\
        
        \multirow{2}{*}{AR18}  & \multirow{2}{*}{BlindMI-w} & \multirow{2}{*}{NN\_attack} & \multirow{2}{*}{3} & \multirow{2}{*}{A \textgreater B }& 
        \multirow{2}{*}{A \textless B} & \multirow{2}{*}{2+3+4}
    & ES35:C100\_U+2.893+0.157+45\%  &\multirow{2}{*}{CV3} & \multirow{2}{*}{40\%} & \multirow{2}{*}{AR60}  & \multirow{2}{*}{LiRA} & \multirow{2}{*}{Calibrated Score} & \multirow{2}{*}{1} & \multirow{2}{*}{A \textgreater B }& 
        \multirow{2}{*}{A \textless B} & \multirow{2}{*}{5+6+7}
    & ES24:L30\_N+0.801+0.076+8\%  &\multirow{2}{*}{CV2} & \multirow{2}{*}{40\%} \\

          &  &  & & & &
         & ES31:C100\_U+2.893+0.085+45\% & & &&  &  & & & &
         & ES04:L30\_N+0.570+0.076+8\% & & \\

   \multirow{2}{*}{AR19}  & \multirow{2}{*}{BlindMI-without} & \multirow{2}{*}{Top3\_NN} & \multirow{2}{*}{3} & \multirow{2}{*}{A \textgreater B }& 
        \multirow{2}{*}{A \textless B} & \multirow{2}{*}{1+2+3+4}
    & ES63:C100\_B+2.893+0.157+45\%  &\multirow{2}{*}{CV3} & \multirow{2}{*}{40\%} & \multirow{2}{*}{AR61}  & \multirow{2}{*}{BlindMI-w} & \multirow{2}{*}{PPV} & \multirow{2}{*}{-} & \multirow{2}{*}{A \textgreater B }& 
        \multirow{2}{*}{A \textless B} & \multirow{2}{*}{4+5+6+7}
    & ES27:L30\_N+0.801+0.094+12\%  &\multirow{2}{*}{CV3} & \multirow{2}{*}{40\%} \\

          &  &  & & & &
         & ES59:C100\_B+2.893+0.085+45\% & & &&  &  & & & &
         & ES23:L30\_N+0.801+0.041+12\% & & \\

         \multirow{2}{*}{AR20}  & \multirow{2}{*}{BlindMI-without} & \multirow{2}{*}{Calibrated Score} & \multirow{2}{*}{-} & \multirow{2}{*}{A \textgreater B }& 
        \multirow{2}{*}{A \textless B} & \multirow{2}{*}{1+2+3}
    & ES36:C100\_U+2.893+0.157+49\%  &\multirow{2}{*}{CV3} & \multirow{2}{*}{40\%} & \multirow{2}{*}{AR62}  & \multirow{2}{*}{BlindMI-w} & \multirow{2}{*}{LiRA} & \multirow{2}{*}{-} & \multirow{2}{*}{A \textgreater B }& 
        \multirow{2}{*}{A \textless B} & \multirow{2}{*}{4+5+6+7}
    & ES45:L30\_U+0.724+0.094+4\%  &\multirow{2}{*}{CV3} & \multirow{2}{*}{40\% }\\

          &  &  & & & &
         & ES34:C100\_U+2.893+0.119+49\% & & &&  &  & & & &
         & ES37:L30\_U+0.724+0.041+4\% & & \\

         \multirow{2}{*}{AR21}  & \multirow{2}{*}{BlindMI-without} & \multirow{2}{*}{NN\_attack} & \multirow{2}{*}{3} & \multirow{2}{*}{A \textgreater B }& 
        \multirow{2}{*}{A \textless B} & \multirow{2}{*}{1+2+3+4}
    & ES35:C100\_U+2.893+0.157+45\%  &\multirow{2}{*}{CV3} & \multirow{2}{*}{40\%} & \multirow{2}{*}{AR63}  & \multirow{2}{*}{BlindMI-1CLASS} & \multirow{2}{*}{PPV} & \multirow{2}{*}{-} & \multirow{2}{*}{A \textgreater B }& 
        \multirow{2}{*}{A \textless B} & \multirow{2}{*}{1+4+6+7}
    & ES74:P100\_B+0.625+0.156+4\%  &\multirow{2}{*}{CV3} & \multirow{2}{*}{40\%} \\

          &  &  & & & &
         & ES31:C100\_U+2.893+0.085+45\% & & &&  &  & & & &
         & ES70:P100\_B+0.625+0.110+4\%& & \\

         \multirow{2}{*}{AR22}  & \multirow{2}{*}{Calibrated Score} & \multirow{2}{*}{NN\_attack} & \multirow{2}{*}{1} & \multirow{2}{*}{A \textgreater B }& 
        \multirow{2}{*}{A \textless B} & \multirow{2}{*}{1+2+3+4}
    & ES01:C100\_N+2.893+0.085+20\%  &\multirow{2}{*}{CV2} & \multirow{2}{*}{40\%} & \multirow{2}{*}{AR64}  & \multirow{2}{*}{BlindMI-1CLASS} & \multirow{2}{*}{LiRA} & \multirow{2}{*}{-} & \multirow{2}{*}{A \textgreater B }& 
        \multirow{2}{*}{A \textless B} & \multirow{2}{*}{3+4+5+6}
    & ES20:L30\_N+0.724+0.094+16\%  &\multirow{2}{*}{CV3} & \multirow{2}{*}{40\%} \\

          &  &  & & & &
         & ES21:C100\_N+4.325+0.085+20\% & & &&  &  & & & &
         & ES16:L30\_N+0.724+0.076+16\% & & \\

         \multirow{2}{*}{AR23}  & \multirow{2}{*}{Calibrated Score} & \multirow{2}{*}{BlindMI-1CLASS} & \multirow{2}{*}{-} & \multirow{2}{*}{A \textgreater B }& 
        \multirow{2}{*}{A \textless B} & \multirow{2}{*}{1+2+3+6+7}
    & ES11:C100\_N+3.813+0.085+45\%  &\multirow{2}{*}{CV3} & \multirow{2}{*}{40\%} & \multirow{2}{*}{AR65}  & \multirow{2}{*}{Top3\_NN} & \multirow{2}{*}{PPV} & \multirow{2}{*}{-} & \multirow{2}{*}{A \textgreater B }& 
        \multirow{2}{*}{A \textless B} & \multirow{2}{*}{4+6}
    & ES49:P100\_U+0.729+0.087+2\%  &\multirow{2}{*}{CV2} & \multirow{2}{*}{40\%} \\

          &  &  & & & &
         & ES19:C100\_N+3.813+0.157+45\% & & &&  &  & & & &
         & ES29:P100\_U+0.550+0.087+2\% & & \\

         \multirow{2}{*}{AR24}  & \multirow{2}{*}{LiRA} & \multirow{2}{*}{Distillation-based} & \multirow{2}{*}{1} & \multirow{2}{*}{A \textgreater B }& 
        \multirow{2}{*}{A \textless B} & \multirow{2}{*}{1+2+3+4}
    & ES23:C100\_N+4.325+0.085+45\%  &\multirow{2}{*}{CV2} & \multirow{2}{*}{40\%} & \multirow{2}{*}{AR66}  & \multirow{2}{*}{Top3\_NN} & \multirow{2}{*}{LiRA} & \multirow{2}{*}{-} & \multirow{2}{*}{A \textgreater B }& 
        \multirow{2}{*}{A \textless B} & \multirow{2}{*}{4+6}
    & ES15:P100\_N+0.625+0.110+10\%  &\multirow{2}{*}{CV2} & \multirow{2}{*}{40\%} \\

          &  &  & & & &
         & ES03:C100\_N+2.893+0.085+45\% & & &&  &  & & & &
         & ES25:P100\_N+0.729+0.110+10\% & & \\

         \multirow{2}{*}{AR25}  & \multirow{2}{*}{Distillation-based} & \multirow{2}{*}{Shapley values} & \multirow{2}{*}{-} & \multirow{2}{*}{A \textgreater B }& 
        \multirow{2}{*}{A \textless B} & \multirow{2}{*}{1+2+3+5+6}
    & ES01:C100\_N+2.893+0.085+20\%  &\multirow{2}{*}{CV4} & \multirow{2}{*}{40\%} & \multirow{2}{*}{AR67}  & \multirow{2}{*}{Shapley values} & \multirow{2}{*}{BlindMI-w} & \multirow{2}{*}{-} & \multirow{2}{*}{A \textgreater B }& 
        \multirow{2}{*}{A \textless B} & \multirow{2}{*}{4}
    & ES40:ImaN\_U+1.130+0.046+49\%  &\multirow{2}{*}{CV4} & \multirow{2}{*}{40\%} \\

          &  &  & & & &
         & ES03:C100\_N+2.893+0.085+45\% & & &&  &  & & & &
         & ES37:ImaN\_U+1.130+0.046+20\% & & \\

         \multirow{2}{*}{AR26}  & \multirow{2}{*}{PPV} & \multirow{2}{*}{Distillation-based} & \multirow{2}{*}{1} & \multirow{2}{*}{A \textgreater B }& 
        \multirow{2}{*}{A \textless B} & \multirow{2}{*}{1+2+3+4}
    & ES21:C100\_N+4.325+0.085+20\%  &\multirow{2}{*}{CV2} & \multirow{2}{*}{40\%} & \multirow{2}{*}{AR68}  & \multirow{2}{*}{Shapley values} & \multirow{2}{*}{NN\_attack} & \multirow{2}{*}{-} & \multirow{2}{*}{A \textgreater B }& 
        \multirow{2}{*}{A \textless B} & \multirow{2}{*}{3+4+5+7}
    & ES44:ImaN\_U+1.130+0.080+49\%  &\multirow{2}{*}{CV4} & \multirow{2}{*}{40\%} \\

          &  &  & & & &
         & ES01:C100\_N+2.893+0.085+20\% & & &&  &  & & & &
         & ES41:ImaN\_U+1.130+0.080+20\% & & \\

         \multirow{2}{*}{AR27}  & \multirow{2}{*}{LiRA} & \multirow{2}{*}{ Shapley values} & \multirow{2}{*}{-} & \multirow{2}{*}{A \textgreater B }& 
        \multirow{2}{*}{A \textless B} & \multirow{2}{*}{2+3+5}
    & ES01:C100\_N+2.893+0.085+20\%  &\multirow{2}{*}{CV4} & \multirow{2}{*}{40\%} & \multirow{2}{*}{AR69}  & \multirow{2}{*}{Shapley values} & \multirow{2}{*}{PPV} & \multirow{2}{*}{-} & \multirow{2}{*}{A \textgreater B }& 
        \multirow{2}{*}{A \textless B} & \multirow{2}{*}{3+4}
    & ES56:CH\_M\_U+1.720+0.133+49\%  &\multirow{2}{*}{CV4} & \multirow{2}{*}{40\%} \\

          &  &  & & & &
         & ES03:C100\_N+2.893+0.085+45\% & & &&  &  & & & &
         & ES55:CH\_M\_U+1.720+0.133+45\% & & \\

         \multirow{2}{*}{AR28}  & \multirow{2}{*}{LiRA} & \multirow{2}{*}{PPV} & \multirow{2}{*}{1} & \multirow{2}{*}{A \textgreater B }& 
        \multirow{2}{*}{A \textless B} & \multirow{2}{*}{1+2+3+4+5+6}
    & ES01:C100\_N+2.893+0.085+20\%  &\multirow{2}{*}{CV2} & \multirow{2}{*}{40\%} & \multirow{2}{*}{AR70}  & \multirow{2}{*}{Distillation-based} & \multirow{2}{*}{NN\_attack} & \multirow{2}{*}{6} & \multirow{2}{*}{A \textgreater B }& 
        \multirow{2}{*}{A \textless B} & \multirow{2}{*}{3+4+5+7}
    & ES60:CH\_M\_B+0.954+0.108+40\%  &\multirow{2}{*}{CV2} & \multirow{2}{*}{40\%} \\

          &  &  & & & &
         & ES21:C100\_N+4.325+0.085+20\% & & &&  &  & & & &
         & ES80:CH\_M\_B+1.720+0.108+40\% & & \\

         \multirow{2}{*}{AR29}  & \multirow{2}{*}{Top1\_Threshold} & \multirow{2}{*}{Loss-Threshold} & \multirow{2}{*}{3} & \multirow{2}{*}{A \textgreater B }& 
        \multirow{2}{*}{A \textless B} & \multirow{2}{*}{1+3+6}
    & ES77:C10\_B+3.472+0.155+20\%  &\multirow{2}{*}{CV2} & \multirow{2}{*}{40\%} & \multirow{2}{*}{AR71}  & \multirow{2}{*}{Distillation-based} & \multirow{2}{*}{BlindMI-1CLASS} & \multirow{2}{*}{-} & \multirow{2}{*}{A \textgreater B }& 
        \multirow{2}{*}{A \textless B} & \multirow{2}{*}{3+4}
    & ES03:CH\_M\_N+0.954+0.083+45\%  &\multirow{2}{*}{CV3} & \multirow{2}{*}{40\%} \\

          &  &  & & & &
         & ES57:C10\_B+1.908+0.155+20\% & & &&  &  & & & &
         & ES07:CH\_M\_N+0.954+0.133+45\% & & \\

         \multirow{2}{*}{AR30}  & \multirow{2}{*}{BlindMI-w} & \multirow{2}{*}{Loss-Threshold} & \multirow{2}{*}{3} & \multirow{2}{*}{A \textgreater B }& 
        \multirow{2}{*}{A \textless B} & \multirow{2}{*}{1+5}
    & ES28:C10\_N+3.472+0.291+49\%  &\multirow{2}{*}{CV3} & \multirow{2}{*}{40\%} & \multirow{2}{*}{AR72}  & \multirow{2}{*}{Distillation-based} & \multirow{2}{*}{ BlindMI-without} & \multirow{2}{*}{-} & \multirow{2}{*}{A \textgreater B }& 
        \multirow{2}{*}{A \textless B} & \multirow{2}{*}{3+4}
    & ES23:ImaN\_N+1.388+0.046+45\%  &\multirow{2}{*}{CV3} & \multirow{2}{*}{40\%} \\

          &  &  & & & &
         & ES26:C10\_N+3.472+0.213+49\% & & &&  &  & & & &
         & ES25:ImaN\_N+1.388+0.080+45\% & & \\

         \multirow{2}{*}{AR31}  & \multirow{2}{*}{Loss-Threshold} & \multirow{2}{*}{Top3\_NN} & \multirow{2}{*}{3} & \multirow{2}{*}{A \textgreater B }& 
        \multirow{2}{*}{A \textless B} & \multirow{2}{*}{1}
    & ES79:C10\_B+3.472+0.155+45\%  &\multirow{2}{*}{CV2} & \multirow{2}{*}{40\%} & \multirow{2}{*}{AR73}  & \multirow{2}{*}{LiRA} & \multirow{2}{*}{NN\_attack} & \multirow{2}{*}{1} & \multirow{2}{*}{A \textgreater B }& 
        \multirow{2}{*}{A \textless B} & \multirow{2}{*}{4}
    & ES29:ImaN\_U+0.934+0.046+20\%  &\multirow{2}{*}{CV2} & \multirow{2}{*}{40\%} \\

          &  &  & & & &
         & ES59:C10\_B+1.908+0.155+45\% & & &&  &  & & & &
         & ES49:ImaN\_U+1.388+0.046+20\% & & \\

         \multirow{2}{*}{AR32}  & \multirow{2}{*}{Top1\_Threshold} & \multirow{2}{*}{BlindMI-1CLASS} & \multirow{2}{*}{3} & \multirow{2}{*}{A \textgreater B }& 
        \multirow{2}{*}{A \textless B} & \multirow{2}{*}{1}
    & ES09:C10\_N+2.501+0.155+20\%  &\multirow{2}{*}{CV3} & \multirow{2}{*}{40\%} & \multirow{2}{*}{AR74}  & \multirow{2}{*}{PPV} & \multirow{2}{*}{NN\_attack} & \multirow{2}{*}{7} & \multirow{2}{*}{A \textgreater B }& 
        \multirow{2}{*}{A \textless B} & \multirow{2}{*}{4}
    & ES78:ImaN\_B+1.388+0.046+40\%  &\multirow{2}{*}{CV2} & \multirow{2}{*}{40\%} \\

          &  &  & & & &
         & ES17:C10\_N+2.501+0.291+20\% & & &&  &  & & & &
         & ES66:ImaN\_B+1.130+0.046+40\% & & \\

         \multirow{2}{*}{AR33}  & \multirow{2}{*}{Top3-NN} & \multirow{2}{*}{Top1\_Threshold} & \multirow{2}{*}{4} & \multirow{2}{*}{A \textgreater B }& 
        \multirow{2}{*}{A \textless B} & \multirow{2}{*}{1}
    & ES59:C10\_B+1.908+0.155+45\%  &\multirow{2}{*}{CV2} & \multirow{2}{*}{40\%} & \multirow{2}{*}{AR75}  & \multirow{2}{*}{LiRA} & \multirow{2}{*}{BlindMI-without} & \multirow{2}{*}{-} & \multirow{2}{*}{A \textgreater B }& 
        \multirow{2}{*}{A \textless B} & \multirow{2}{*}{4+5}
    & ES37:L30\_U+0.724+0.041+4\% &\multirow{2}{*}{CV3} & \multirow{2}{*}{40\%} \\

          &  &  & & & &
         & ES79:C10\_B+3.472+0.155+45\% & & &&  &  & & & &
         & ES45:L30\_U+0.724+0.094+4\% & & \\

         \multirow{2}{*}{AR34}  & \multirow{2}{*}{BlindMI-w} & \multirow{2}{*}{BlindMI-1CLASS} & \multirow{2}{*}{3} & \multirow{2}{*}{A \textgreater B }& 
        \multirow{2}{*}{A \textless B} & \multirow{2}{*}{1+4+6+7}
    & ES38:P100\_U+0.625+0.087+4\%  &\multirow{2}{*}{CV3} & \multirow{2}{*}{40\%} & \multirow{2}{*}{AR76}  & \multirow{2}{*}{PPV} & \multirow{2}{*}{BlindMI-without} & \multirow{2}{*}{-} & \multirow{2}{*}{A \textgreater B }& 
        \multirow{2}{*}{A \textless B} & \multirow{2}{*}{3+4+5}
    & ES40:ImaN\_U+1.130+0.046+49\%  &\multirow{2}{*}{CV3} & \multirow{2}{*}{40\%} \\

          &  &  & & & &
         & ES46:P100\_U+0.625+0.156+4\% & & &&  &  & & & &
         & ES48:ImaN\_U+1.130+0.145+49\% & & \\

         \multirow{2}{*}{AR35}  & \multirow{2}{*}{BlindMI-without} & \multirow{2}{*}{BlindMI-1CLASS} & \multirow{2}{*}{3} & \multirow{2}{*}{A \textgreater B }& 
        \multirow{2}{*}{A \textless B} & \multirow{2}{*}{1}
    & ES31:C10\_U+1.908+0.155+45\%  &\multirow{2}{*}{CV3} & \multirow{2}{*}{40\%} & \multirow{2}{*}{AR77}  & \multirow{2}{*}{Top2+True} & \multirow{2}{*}{Distillation-based} & \multirow{2}{*}{1} & \multirow{2}{*}{A \textgreater B }& 
        \multirow{2}{*}{A \textless B} & \multirow{2}{*}{3+5}
    & ES50:L30\_U+0.801+0.041+8\%  &\multirow{2}{*}{CV2} & \multirow{2}{*}{40\%} \\

          &  &  & & & &
         & ES35:C10\_U+1.908+0.291+45\% & & &&  &  & & & &
         & ES30:L30\_U+0.570+0.041+8\% & & \\

         \multirow{2}{*}{AR36}  & \multirow{2}{*}{Top2+True} & \multirow{2}{*}{Calibrated Score} & \multirow{2}{*}{-} & \multirow{2}{*}{A \textgreater B }& 
        \multirow{2}{*}{A \textless B} & \multirow{2}{*}{1+3+4}
    & ES24:ImaN\_N+1.388+0.080+40\%  &\multirow{2}{*}{CV2} & \multirow{2}{*}{40\%} & \multirow{2}{*}{AR78}  & \multirow{2}{*}{Distillation-based} & \multirow{2}{*}{Label-only} & \multirow{2}{*}{-} & \multirow{2}{*}{A \textgreater B }& 
        \multirow{2}{*}{A \textless B} & \multirow{2}{*}{3+5+7}
    & ES21:T100\_N+0.734+0.038+2\% &\multirow{2}{*}{CV2} & \multirow{2}{*}{40\%} \\

          &  &  & & & &
         & ES04:ImaN\_N+0.934+0.080+40\% & & &&  &  & & & &
         & ES01:T100\_N+0.530+0.038+2\% & & \\

         \multirow{2}{*}{AR37}  & \multirow{2}{*}{BlindMI-1CLASS} & \multirow{2}{*}{Top2+True} & \multirow{2}{*}{3} & \multirow{2}{*}{A \textgreater B }& 
        \multirow{2}{*}{A \textless B} & \multirow{2}{*}{1}
    & ES07:C10\_N+1.908+0.291+45\% &\multirow{2}{*}{CV3} & \multirow{2}{*}{40\%} & \multirow{2}{*}{AR79}  & \multirow{2}{*}{Shapley values} & \multirow{2}{*}{BlindMI-1CLASS} & \multirow{2}{*}{-} & \multirow{2}{*}{A \textgreater B }& 
        \multirow{2}{*}{A \textless B} & \multirow{2}{*}{3+5}
    & ES39:L30\_U+0.724+0.041+12\% &\multirow{2}{*}{CV4} & \multirow{2}{*}{40\%} \\

          &  &  & & & &
         & ES03:C10\_N+1.908+0.155+45\% & & &&  &  & & & &
         & ES37:L30\_U+0.724+0.041+4\%  & & \\

         \multirow{2}{*}{AR38}  & \multirow{2}{*}{BlindMI-1CLASS} & \multirow{2}{*}{NN\_attack} & \multirow{2}{*}{-} & \multirow{2}{*}{A \textgreater B }& 
        \multirow{2}{*}{A \textless B} & \multirow{2}{*}{1+3+4}
    & ES35:C10\_U+1.908+0.291+45\%  &\multirow{2}{*}{CV3} & \multirow{2}{*}{40\%} & \multirow{2}{*}{AR80}  & \multirow{2}{*}{LiRA} & \multirow{2}{*}{Loss-Threshold} & \multirow{2}{*}{1} & \multirow{2}{*}{A \textgreater B }& 
        \multirow{2}{*}{A \textless B} & \multirow{2}{*}{5}
    & ES33:L30\_U+0.570+0.076+12\%  &\multirow{2}{*}{CV2} & \multirow{2}{*}{40\%} \\

          &  &  & & & &
         & ES31:C10\_U+1.908+0.155+45\% & & &&  &  & & & &
         & ES53:L30\_U+0.801+0.076+12\%  & & \\

         \multirow{2}{*}{AR39}  & \multirow{2}{*}{Top3\_NN} & \multirow{2}{*}{NN\_attack} & \multirow{2}{*}{3+4} & \multirow{2}{*}{A \textgreater B }& 
        \multirow{2}{*}{A \textless B} & \multirow{2}{*}{1+3+4+5+7}
    & ES50:C10\_U+3.472+0.155+40\%  &\multirow{2}{*}{CV2} & \multirow{2}{*}{40\%} & \multirow{2}{*}{AR81}  & \multirow{2}{*}{LiRA} & \multirow{2}{*}{Top1\_Threshold} & \multirow{2}{*}{-} & \multirow{2}{*}{A \textgreater B }& 
        \multirow{2}{*}{A \textless B} & \multirow{2}{*}{5}
    & ES22:L30\_N+0.801+0.041+8\% &\multirow{2}{*}{CV2} & \multirow{2}{*}{40\% }\\

          &  &  & & & &
         & ES30:C10\_U+1.908+0.155+40\% & & &&  &  & & & &
         & ES02:L30\_N+0.570+0.041+8\% & & \\

         \multirow{2}{*}{AR40}  & \multirow{2}{*}{Calibrated Score} & \multirow{2}{*}{Label-only} & \multirow{2}{*}{-} & \multirow{2}{*}{A \textgreater B }& 
        \multirow{2}{*}{A \textless B} & \multirow{2}{*}{1+3}
    & ES01:C10\_N+1.908+0.155+20\%  &\multirow{2}{*}{CV1} & \multirow{2}{*}{40\%} & \multirow{2}{*}{AR82}  & \multirow{2}{*}{Top1\_Threshold } & \multirow{2}{*}{PPV} & \multirow{2}{*}{-} & \multirow{2}{*}{A \textgreater B }& 
        \multirow{2}{*}{A \textless B} & \multirow{2}{*}{5}
    & ES53:L30\_U+0.801+0.076+12\% &\multirow{2}{*}{CV2} & \multirow{2}{*}{40\%} \\

          &  &  & & & &
         & ES57:C10\_B+1.908+0.155+20\% & & &&  &  & & & &
         & ES43:L30\_U+0.724+0.076+12\% & & \\

         \multirow{2}{*}{AR41}  & \multirow{2}{*}{Calibrated Score} & \multirow{2}{*}{Top3\_NN} & \multirow{2}{*}{-} & \multirow{2}{*}{A \textgreater B }& 
        \multirow{2}{*}{A \textless B} & \multirow{2}{*}{1+3+6}
    & ES01:C10\_N+1.908+0.155+20\%  &\multirow{2}{*}{CV2} & \multirow{2}{*}{40\%} & \multirow{2}{*}{AR83}  & \multirow{2}{*}{BlindMI-without} & \multirow{2}{*}{Shapley values} & \multirow{2}{*}{-} & \multirow{2}{*}{A \textgreater B }& 
        \multirow{2}{*}{A \textless B} & \multirow{2}{*}{3}
    & ES32:CH\_M\_U+0.954+0.108+40\%  &\multirow{2}{*}{CV4} & \multirow{2}{*}{40\%} \\

          &  &  & & & &
         & ES21:C10\_N+3.472+0.155+20\% & & &&  &  & & & &
         & ES34:CH\_M\_U+0.954+0.108+49\% & & \\

         \multirow{2}{*}{AR42}  & \multirow{2}{*}{BlindMI-1CLASS } & \multirow{2}{*}{Label-only} & \multirow{2}{*}{-} & \multirow{2}{*}{A \textgreater B }& 
        \multirow{2}{*}{A \textless B} & \multirow{2}{*}{1+3}
    & ES74:C10\_B+2.501+0.291+40\%  &\multirow{2}{*}{CV3} & \multirow{2}{*}{40\%} & \multirow{2}{*}{}  & \multirow{2}{*}{} & \multirow{2}{*}{} & \multirow{2}{*}{} & \multirow{2}{*}{ }& 
        \multirow{2}{*}{} & \multirow{2}{*}{}
    &   &\multirow{2}{*}{} & \multirow{2}{*}{} \\

          &  &  & & & &
         & ES66:C10\_B+2.501+0.155+40\% & & &&  &  & & & &
         &  & & \\

        \bottomrule[1.1pt]
    \end{tabular}}
    \begin{tablenotes}
    \scriptsize
    \item  \emph{ARN.}: the number of the attack ranking; \emph{EL.}: Existing literature, where 1 $\rightarrow$ the LiRA attacks \cite{carlini2022membership}; 2 $\rightarrow$ Risk score-based attacks \cite{song2021systematic}; 3 $\rightarrow$ BlindMI-Diff attacks \cite{hui2021practical}; 
    \item4 $\rightarrow$ the Top3-NN attacks \cite{salem2019ml}; 5 $\rightarrow$ the Calibrated Score-based attacks \cite{watson2021importance}; 6 $\rightarrow$ the Distillation-based Threshold attacks \cite{ye2021enhanced}; 7 $\rightarrow$ the PPV attacks \cite{jayaraman2020revisiting}; - $\rightarrow$ the 
    \item ranking of these two attacks does not existing in the existing literature; \emph{RAR.}: replicate the attack ranking of the existing literature; \emph{dataset}: 1 $\rightarrow$ CIFAR10 (C10); 2 $\rightarrow$ 
    \item CIFAR100 (C100); 3 $\rightarrow$ CH\_MNIST (CH\_M); 4 $\rightarrow$ ImageNet (ImaN); 5 $\rightarrow$ Location30 (L30); 6 $\rightarrow$ Purchase100 (P100); 7 $\rightarrow$ Texas100 (T100); \emph{RES.}: Evaluation 
    \item Scenarios of replicated attacks (upper) and our experiments (lower); \emph{MR.}: the main reason for the opposite rank of the attacks, where CV1 $\rightarrow$ the distance distribution
    \item of data samples in the target datase, CV2 $\rightarrow$ the distance between data samples of the
target dataset, CV3 $\rightarrow$ the differential distance between two
datasets, CV4 $\rightarrow$
\item the ratio of the samples that are made
no inferences by an MI attack; \emph{SMR.}: the summary of the main reason for the opposite rank of the attacks, where CV1, CV2,
\item CV3 and CV4 accounted for 2.41\%, 40.96\%, 37.35\% and 19.28\% respectively.
    \end{tablenotes}
    \end{threeparttable}
   \label{tabs:The MI attacks that rank opposite}
    \vspace{0.1pt}
\end{table*}

\subsection{RQ2: Effect of Distance between Data Samples of the Target Dataset}
\label{sebsec:Effect of Distance between members and nonmembers}

\noindent \textbf {Methodology.} For target datsests obeying different distance distributions, we first use the Equation~\ref{equs:equ1} to calculate the distance between data samples (e.g., data samples with high output probability and data samples with low output probability classified in Section \ref{sebsec:Effect of the Distance Distribution of Data Samples in the Target Dataset}) of the target dataset. Then, 
we verify the evaluation results 
of 15 state-of-the-art MI attacks when only the distance between data samples (e.g., data samples with high output probability and data samples with low output probability) (CV2) of the target dataset is allowed to change and other three CVs 
(see Section \ref{sebsec:Part I: Evaluation Scenarios}) are left unchanged.

\noindent \textbf {Results.} From Figure \ref{figs:Fig22}, 
we discover that the MA of the MI attacks increases with the increase of the distance between data samples of the target dataset. 
Meanwhile, Table~\ref{tabs:tab9} also shows the similar results that the distance between data samples on ImageNet, CH\_MNIST, CIFAR10 and CIFAR100 increases in turn, and the MA increases gradually. We observe that the ranking results of a particular MI attack (i.e. Top2+True attack\cite{hui2021practical} (Class 1.1.2 see Figure \ref{figs:Classification Tree_MI attacks})) remain {\bf unchanged} across almost all the ESs in a particular group (i.e., when only the CV2 changed and other three CVs remained the same, the Top2+True attack\cite{hui2021practical} is 70.83\% of the top three to top five most effective attacks). This indicates that the corresponding tuning variable probably does not affect the MI attack. To confirm this conclusion, we look into the algorithm of the MI attack and find the following: Top2+True attacks\cite{hui2021practical} mainly combined the top two output probabilities of the target model and the ground-truth labels. Therefore, when only the CV2 changes and other three CVs remain the same, it does not affect the ground-truth labels of the target dataset, and thus has little effect on evaluation results of the Top2+True attacks\cite{hui2021practical}. 

We find that the Top3\_NN attack\cite{salem2019ml} (Class 1.1.1 see Figure \ref{figs:Classification Tree_MI attacks}) is most affected when only CV2 changes, other three CVs remain the same, for example, for one particular evaluation scenario ES14 on CH\_MNIST could result in the following ranking: 
Class 2.1 (except for the Label-only and Loss-Threshold attacks)\,\textgreater\,Class 2.2.1 (except for the Shapley values attack)\,\textgreater\,Class 1.1.1 (except for the NN attack )\,\textgreater\,Class 3.1.1\,\textgreater\,Class 3.1.2\,\textgreater\,Class 2.2.2 (except for the Distillation-based, PPV, LiRA attacks)\,\textgreater\,Class 1.1.2\,\textgreater\,Class 3.1.3,
and for another particular evaluation scenario ES24 on CH\_MNIST could result in the following ranking: 
Class 2.1 (except for the Label-only and Loss-Threshold attacks)\,\textgreater\,Class 2.2.1 (except for the Shapley values)\,\textgreater\, Class 3.1.1 \textgreater\,Class 3.1.2\,\textgreater\,Class 2.2.2 (except for the Distillation-based, PPV, LiRA attacks)\,\textgreater\,Class 1.1.1 (except for the NN attack)\,\textgreater\,Class 1.1.2 \textgreater\,Class 3.1.3. The reason is that the Top3\_NN attack\cite{salem2019ml} mainly is based on the Top three output probability of the target model, and when only the CV2 changes, the difference between members and nonmembers could not be captured effectively by the Top3\_NN attack, which will affect evaluation results.


We also observe that the ranking order between two particular MI attacks (i.e. LiRA attack\cite{carlini2022membership} and PPV attack \cite{jayaraman2020revisiting} (Class 2.2.2 see Figure \ref{figs:Classification Tree_MI attacks})) remain {\bf unchanged} across almost all the ESs in a particular group. This indicates that the ranking order is independent of the control variable. In addition, this indicates that the ranking order is to certain extent determined by the three non-tuning factors. We look into the values of the three non-tuning factors, and we find that the ranking order between LiRA attacks and PPV attacks is PPV\,\textgreater\,LiRA when only the CV2 changes and other three CVs remain the same, while the ranking order of them is LiRA\,\textgreater\,PPV when only CV4 changes and other three variables remain the same. The reason is that PPV attacks mainly considered the prior distribution of the data, and they classified a target data example as a member when its per-instance loss was larger than a decision threshold, and the LiRA attacks combined per-example difficulty scores with a principled and well-calibrated Gaussian likelihood estimate, and they utilized a true-positive rate at low FPR
metric to help determine the value of the threshold. Therefore, it hardly affects the ranking order of them when only the CV2 changes, which does not effect the prior distribution of the training data of the target model and does not effect the differences of the members and nonmembers; and changing only the CV4 will change the ratio of the samples of the target dataset, and the removed samples that are made no inferences by an MI attack may have high difficulty scores, this will increase the evaluation results of the LiRA attacks, and finally make the ranking order of the two attacks change. Moreover, from Figure~\ref{figs:fig5}, 
we find that 
the FNR is almost the same with the increase in sample distances.

\begin{figure*}[]
\vspace{0.1pt} 
\begin{center}
\subfigure[CIFAR100]{
\begin{minipage}{0.18\textwidth}
\label{Fig.sub.4.1}
\includegraphics[height=0.65\textwidth,width=1\textwidth]{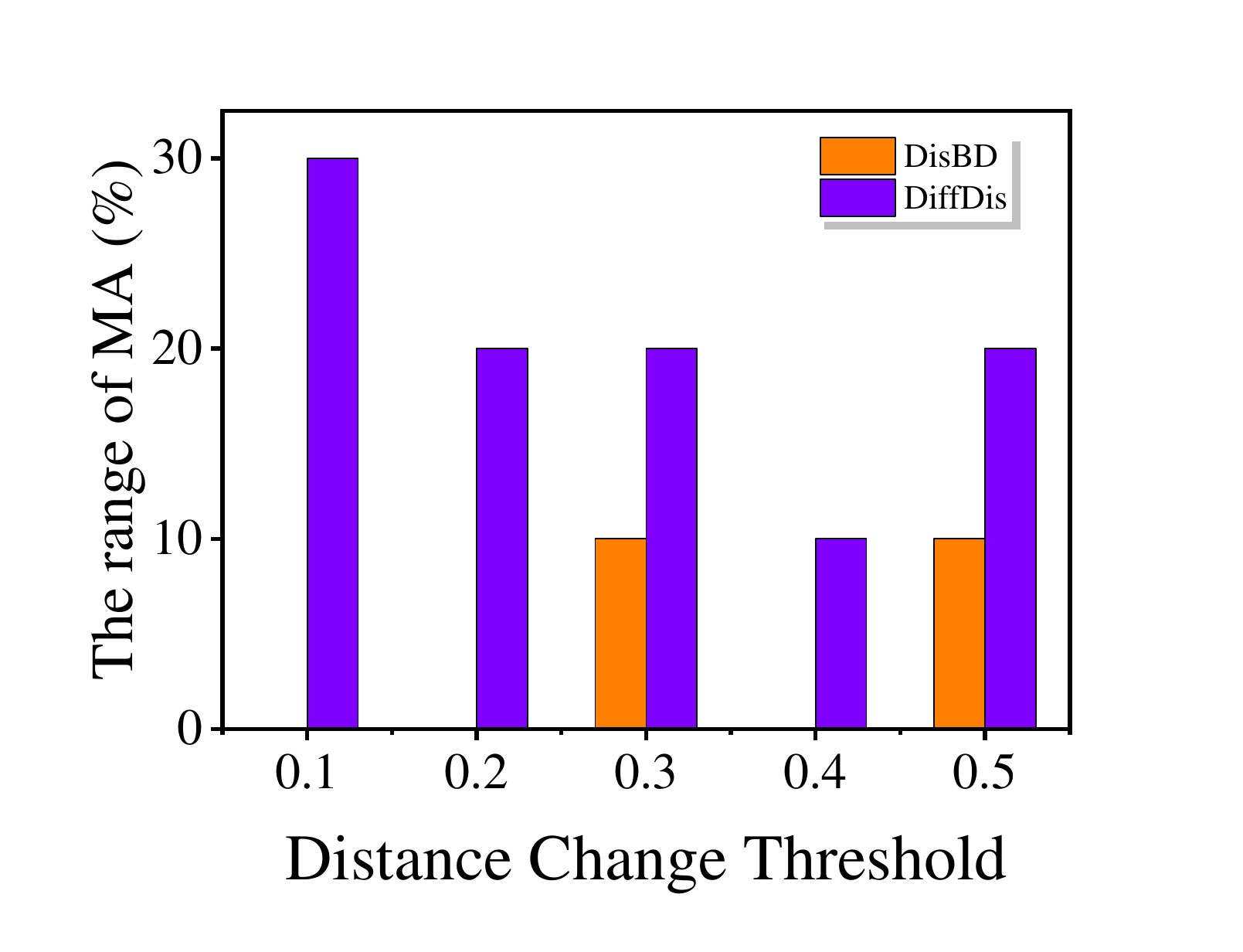}
\end{minipage}}
\subfigure[CIFAR10]{
\begin{minipage}{0.18\textwidth}
\label{Fig.sub.4.2}
\includegraphics[height=0.65\textwidth,width=1\textwidth]{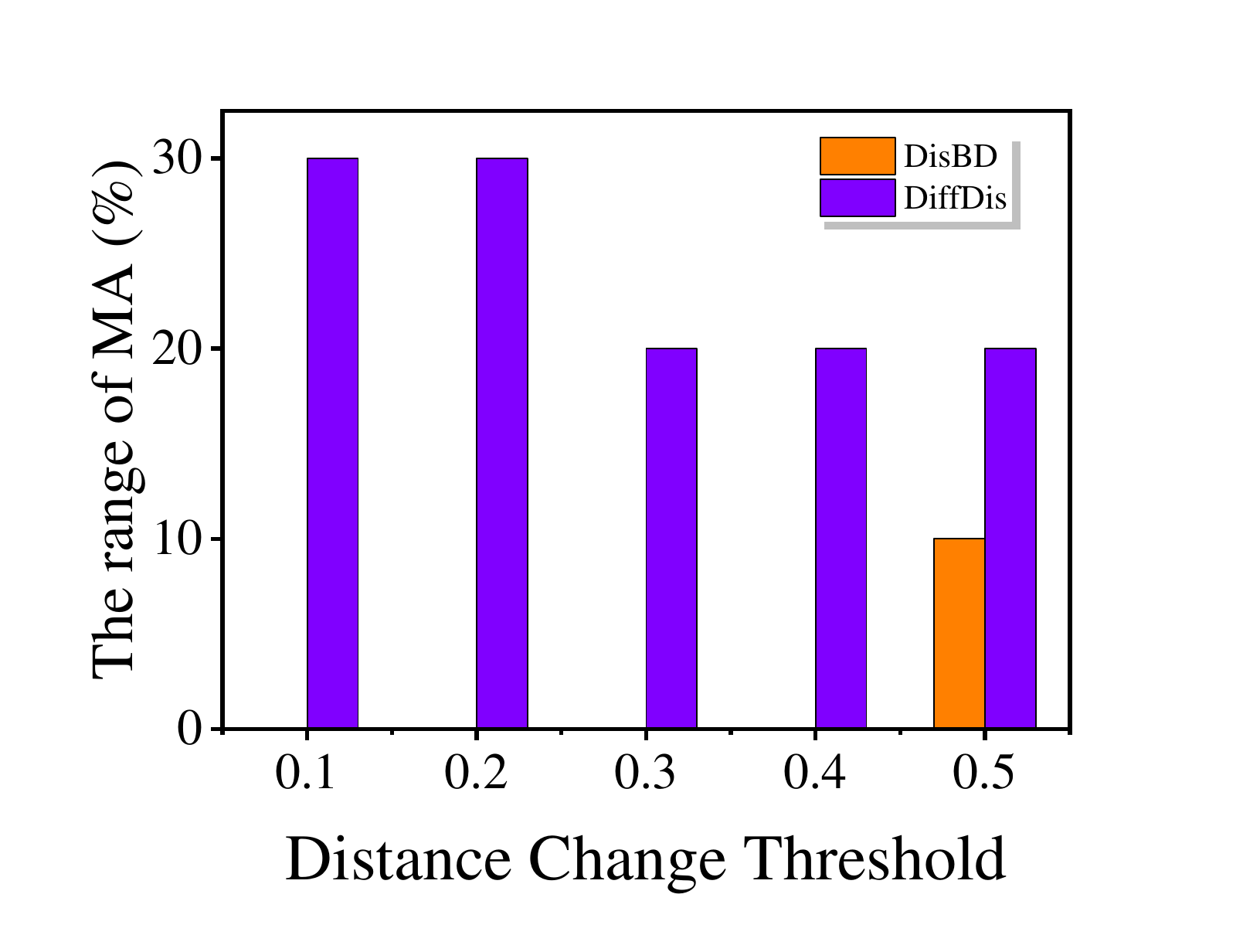}
\end{minipage}}
\subfigure[CH\_MNIST]{
\begin{minipage}{0.18\textwidth}
\label{Fig.sub.4.3}
\includegraphics[height=0.65\textwidth,width=1\textwidth]{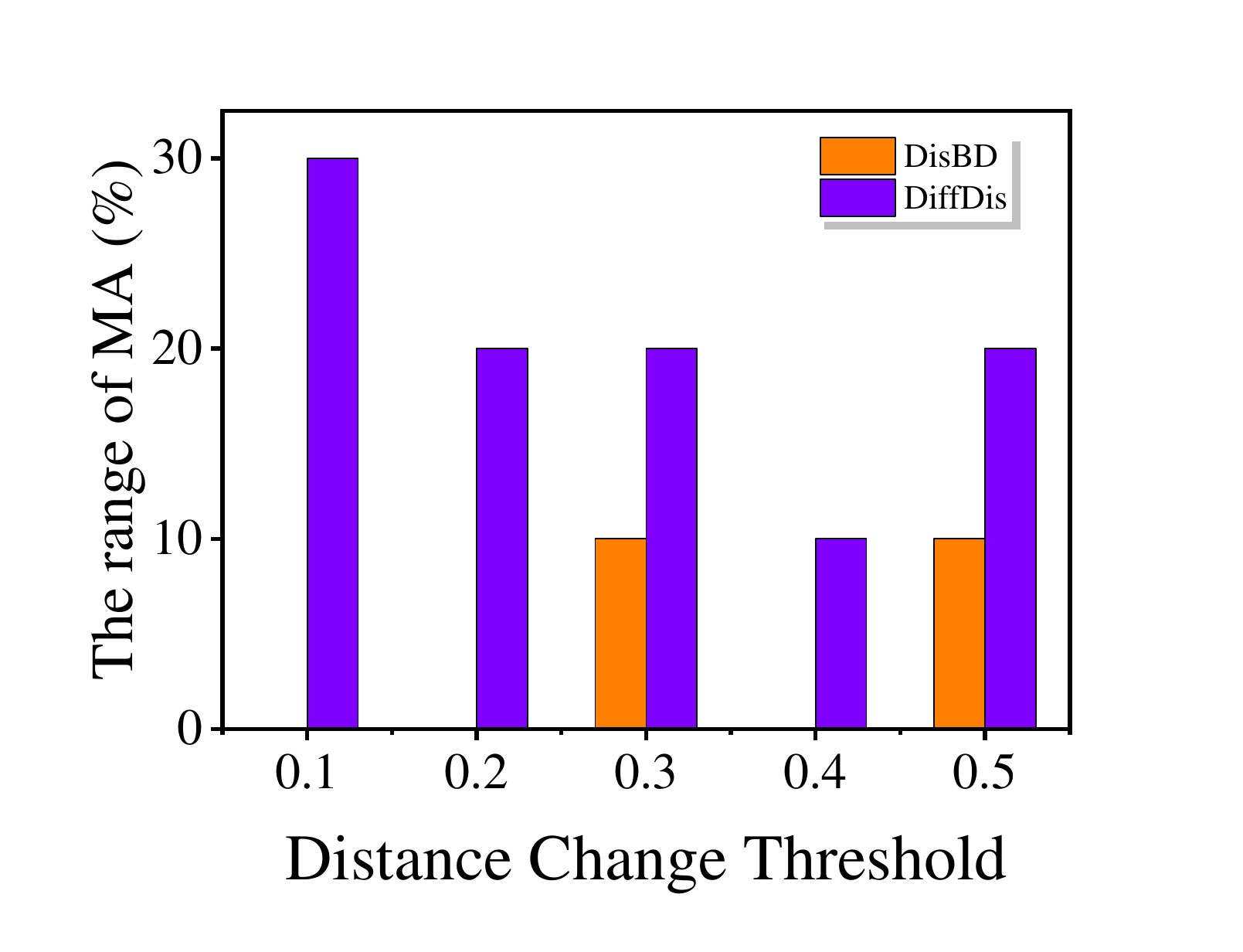}
\end{minipage}}
\subfigure[ImageNet]{
\begin{minipage}{0.18\textwidth}
\label{Fig.sub.4.4}
\includegraphics[height=0.65\textwidth,width=1\textwidth]{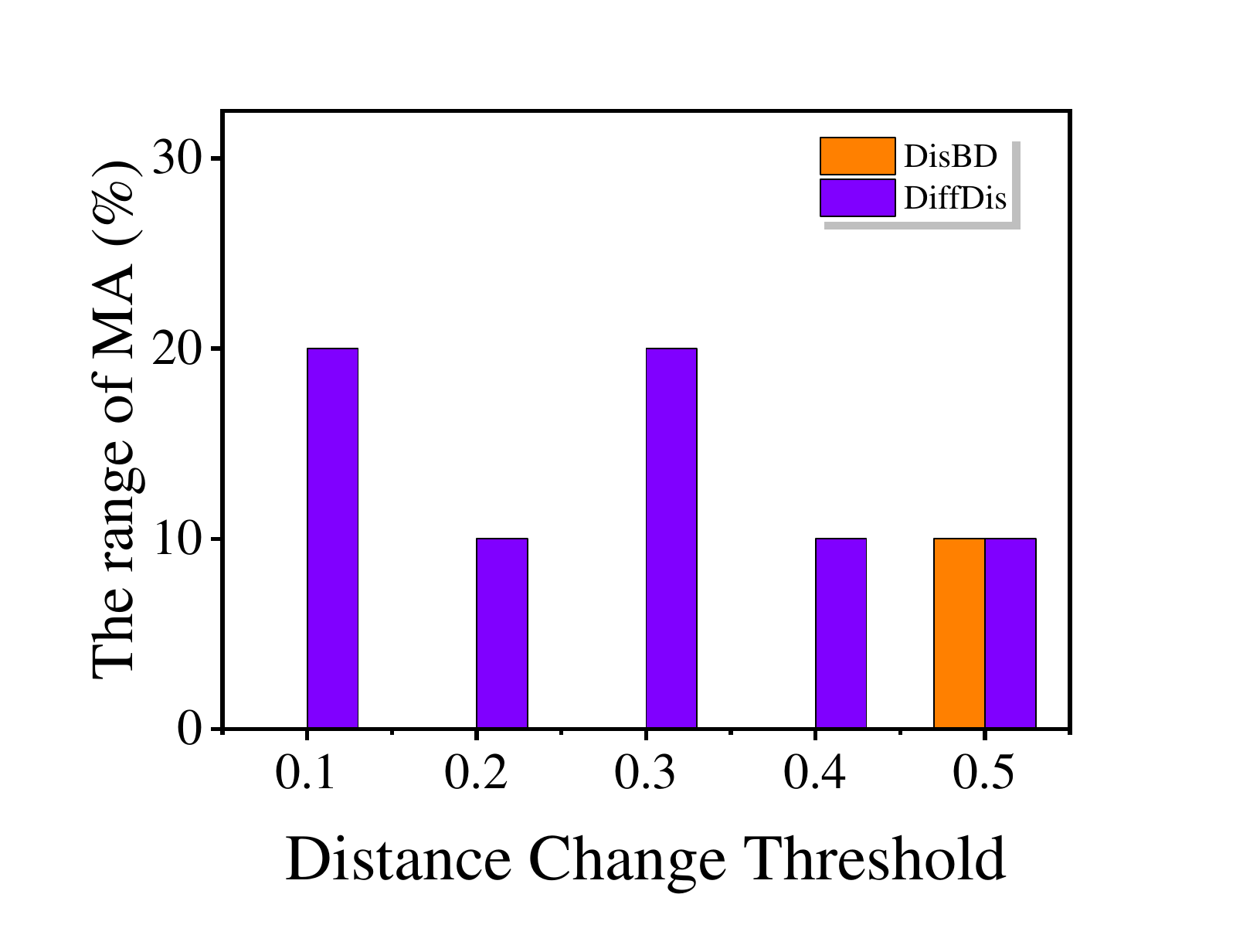}
\end{minipage}}
\subfigure[Purchase100]{
\begin{minipage}{0.18\textwidth}
\label{Fig.sub.4.5}
\includegraphics[height=0.65\textwidth,width=1\textwidth]{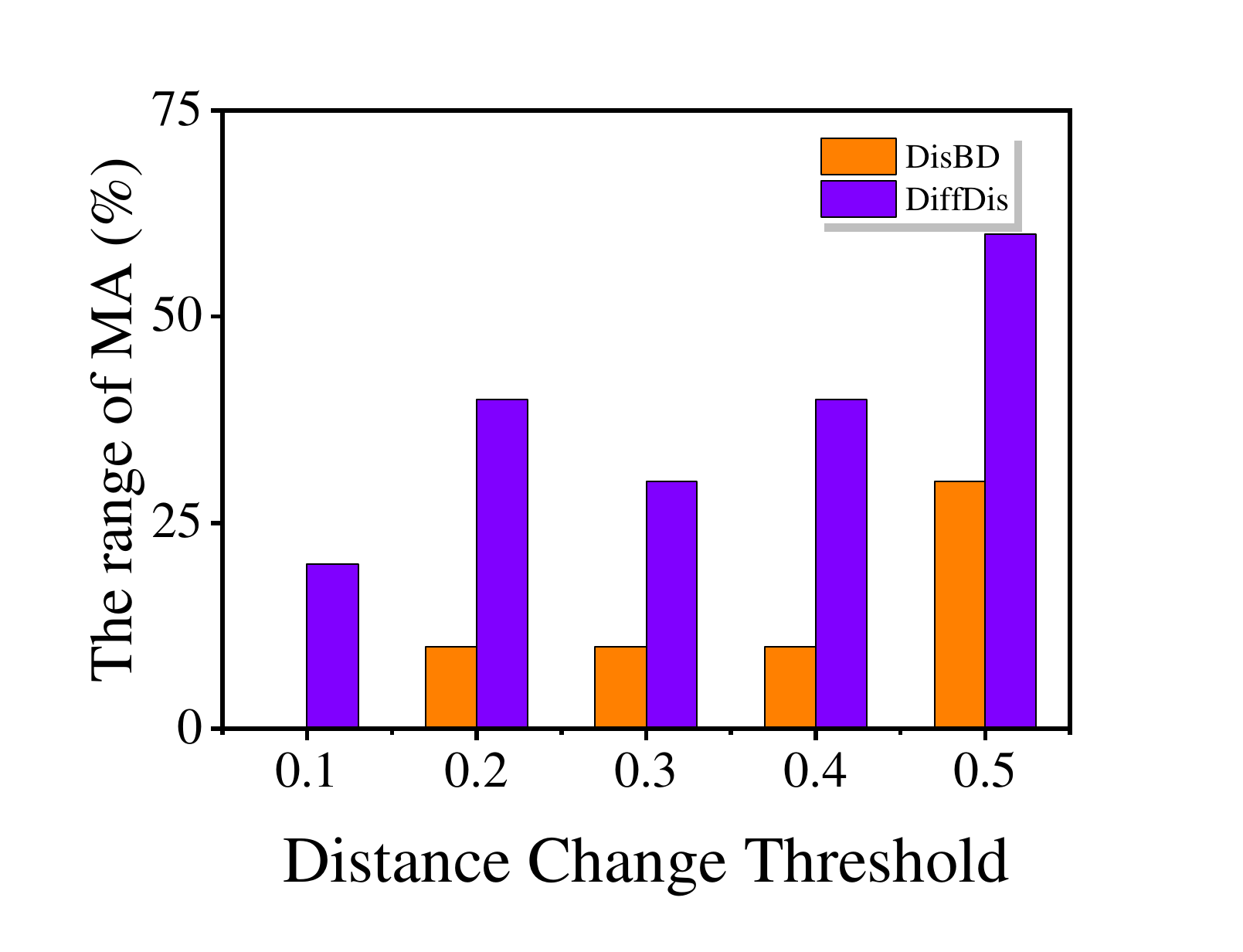}
\end{minipage}}
\end{center}
\caption{\label{figs:Comparisions DisMN and DiffDis on MA} The comparisions of the effect of the distance between data samples of the target dataset (DisBD) and the differential difference between two datasets (DiffDis) on MA.}
\vspace{0.1pt}
\end{figure*}

[Observation] \emph{The lager the distance between data samples 
of the target dataset, the higher the MA. 
}

[Observation] \emph{The distance between data samples of the target dataset have little effect on the 
FNR.}

\subsection{RQ3: Effect of Differential Distances between Two Datasets}
\label{sebsec:Effect of Differential Distances between two datasets}

\noindent \textbf {Methodology.} After constructing distance distributions of data samples and calculating the distance between data samples of the target dataset, we first generate a dataset with nonmembers via transforming existing samples into new samples or directly using the test data \cite{hui2021practical}, and then 
we compute the differential distance between two datasets before and after a data sample with high output probability (a data sample with low output probability) (classified in Section \ref{sebsec:Effect of the Distance Distribution of Data Samples in the Target Dataset}) is moved from the target dataset to a nonmember dataset. And we get the average differential distance of data samples with high output probability and the average differential distance of data samples with low output probability between two datasets. Finally, 
we verify the evaluation results 
of 15 state-of-the-art MI attacks when only the differential distance between two datasets (e.g., the target dataset and the nonmember dataset) (CV3) is allowed to change and other three CVs 
(see Section \ref{sebsec:Part I: Evaluation Scenarios}) are left unchanged. Moreover, we design a distance interval threshold $\delta$ and compare the evaluation results when the distance between data samples of the target dataset and the differential distance between two datasets 
are changed to the same distance threshold (e.g., $\delta$), respectively.

\noindent \textbf {Results.} From Figure~\ref{figs:Fig33} 
we find that the MA increases with the differential distance between two datasets (e.g., the target dataset and the nonmember dataset) increases. 


[Observation] \emph{The more the differential difference between two datasets, the higher the attacker-side MA.}

From Figure \ref{figs:Comparisions DisMN and DiffDis on MA}, we find that the effect of the distance between data samples of the target dataset on MA almost is 0, whereas that of the differential difference between two datasets is up to 30\%, when the distance between data samples of the target dataset and the differential difference between two datasets are changed to the same distance threshold (e.g., 0.1). 
Moreover, 
from Table \ref{tabs:The average distance change on the Attacker-side Membership Advantage.}, we find that 
the change of average distances between data samples of the target dataset (e.g., 0.113) on CIFAR10 are larger than that of the average differential differences between two datasets (e.g., 0.082) when the MA is changed to the same threshold (e.g., 10\%). 
From Figure~\ref{figs:fig6}, 
we find that the FNR is almost zero when the differential distance between two datasets changes.


[Observation] \emph{The effect of the differential distance between two datasets on MA is larger than that of the distance between data samples of the target dataset.}

[Observation] \emph{The differential distance between two datasets have little effect on the 
FNR.}

We observe that the ranking results of a particular MI attack (i.e. Label-only\cite{yeom2018privacy} (Class 2.1 see Figure \ref{figs:Classification Tree_MI attacks})) remain {\bf unchanged} across almost all the ESs in a particular group (i.e., when only the CV3 changed and other three CVs remained the same, the Label-only\cite{yeom2018privacy} is 70.88\% of the top three to top seven most effective attacks). This indicates that the corresponding tuning variable probably does not affect the MI attack. To confirm this conclusion, we look into the algorithm of the MI attack and find the following: Label-only\cite{yeom2018privacy} mainly classified a sample as a member if the predicted class was
the same as its ground-truth label. Therefore, when only the CV3 changes and other three CVs remain the same, it does not affect the ground-truth labels of the target dataset, and thus has little effect on the Label-only\cite{yeom2018privacy} attack.

We find that the BlindMI-Diff-w\cite{hui2021practical} (Class 3.1.1 see Figure \ref{figs:Classification Tree_MI attacks}) is most affected when only the CV3 changes, other three CVs remain the same, for example, for one particular evaluation scenario ES03 on ImageNet could result in the following ranking: Class 2.1 (except for the Label-only and Top1\_Threshold attacks)\,\textgreater\,Class 2.2.1\,\textgreater\,Class 3.1.1\,\textgreater\,Class 1.1.1 (except for the NN attack)\,\textgreater\,Class 2.2.2  (except for the Calibrated Score, LiRA, PPV attacks)\,\textgreater\,Class 3.1.3\,\textgreater\,Class 3.1.2\,\textgreater\,Class 1.1.2, 
 and for another particular evaluation scenario ES05 on ImageNet could result in the following ranking: Class 2.1 (except for the Label-only and Top1\_Threshold attacks)\,\textgreater\,Class 2.2.1\,\textgreater\,Class 1.1.1 (except for the Top3 NN attack)\,\textgreater\,Class 2.2.2 (except for the Calibrated Score, LiRA attack)\,\textgreater\,Class 3.1.3\,\textgreater\,Class 3.1.2\,\textgreater\,Class 3.1.1\,\textgreater\,Class 1.1.2. 
The reason is that the BlindMI-Diff-w attack \cite{hui2021practical} mainly utilized the differences between
two datasets before and after a sample was moved, and when only the CV3 changes, the differential distance between two datasets will change, which will affect evaluation results.

We also observe that the ranking order between two particular MI attacks (i.e. Label-only (Class 2.1 see Figure \ref{figs:Classification Tree_MI attacks}) and NN\_attack (Class 1.1.1 see Figure \ref{figs:Classification Tree_MI attacks})) remain {\bf unchanged} across almost all the ESs in a particular group. This indicates that the ranking order is independent of the control variable. In addition, this indicates that the ranking order is to certain extent determined by the three non-tuning factors. We look into the values of the three non-tuning factors, and we find that the ranking order between Label-only and NN\_attacks is NN\_attack\,\textgreater\,Label-only when only the CV3 changes and other three CVs remain the same, while the ranking order of them is Label-only\,\textgreater\,NN\_attack when only the CV1 changes and other three variables remain the same. The reason is that it hardly affects the ranking order of them when only the CV3 changes, which does not effect the ground-truth labels and output probabilities of the target dataset; and changing only the CV1 may cause NN\_attacks to be unable to effectively distinguish between members and nonmembers of different distance distributions of data samples in the target dataset, and finally make the ranking order of the two attacks change.

\subsection{RQ4: Effect of the Ratios of the samples that are made
no inferences by an MI attack}
\label{sebsec:Effect of Ignored Ratios}

\noindent \textbf {Methodology.} For each distance distribution of data samples, the distance between data samples and the differential distance between two datasets of the target dataset, 
we first set different ratios of the samples that are made no inferences by an MI attack for image (e.g., 20\%, 40\%, 45\% and 49\%) and text datasets (e.g., 2\%, 4\%, 10\% and 12\%) as the prior Privacy Risk Scores-based ~\cite{song2021systematic} and the Shapley Values-based MI attacks~\cite{duddu2021shapr}. 
Then, 
we verify the evaluation results 
of 15 state-of-the-art MI attacks when only the ratio of the samples that are made
no inferences by an MI attack (CV4) is allowed to change and other three CVs 
(see Section \ref{sebsec:Part I: Evaluation Scenarios}) are left unchanged.

\noindent \textbf {Results.} From Figure ~\ref{figs: The effect of the ratio of the samples
that are made no inferences by an MI attack}, we find that the evaluations are different when the ratios are different. We observe that the ranking results of a particular MI attack (i.e. NN\_attack\cite{shokri2017membership} (Class 1.1.1 see Figure \ref{figs:Classification Tree_MI attacks})) remain {\bf unchanged} across almost all the ESs in a particular group (i.e., when only the CV4 changed and other three CVs remained the same, the NN\_attack\cite{shokri2017membership} is 88.95\% of the top five to top nine most effective attacks). This indicates that the corresponding tuning variable probably does not affect the MI attack. To confirm this conclusion, we look into the algorithm of the MI attack and find the following: NN\_attack\cite{shokri2017membership} was mainly based on the output probability of the target
model. Therefore, when only the CV4 changes and other three CVs remain the same, it does not affect the output probability of data samples in the target dataset, and thus has little effect on the NN\_attack\cite{shokri2017membership}.

We find that the Shapley Values\cite{duddu2021shapr} attack (Class 2.2.1 see Figure \ref{figs:Classification Tree_MI attacks}) is most affected when only the CV4 changes, other three CVs remain the same, for example, for one particular ES49 on ImageNet could result in the following ranking: Class 2.1 (except for the Label-only and Top1\_Threshold)\,\textgreater\,Class 3.1.1\,\textgreater\,Class 1.1.1 (except for the Top3\_NN attack)\,\textgreater\,Class 2.2.1 (except for the Risk score attack)\,\textgreater\,Class 2.2.2 (except for the LiRA, Calibrated Score attacks)\,\textgreater\,Class 3.1.3\,\textgreater\,Class 3.1.2\,\textgreater\,Class 1.1.2, 
and for another particular ES50 on ImageNet could result in the following ranking: Class 2.1 (except for the Label-only and Top1\_Threshold attacks)\,\textgreater\,Class 2.2.1\,\textgreater\,Class 1.1.1 (except for the NN attack)\,\textgreater\,Class 2.2.2 (except for the PPV, Calibrated Score attacks)\,\textgreater\,Class 3.1.3\,\textgreater\,Class 3.1.2\,\textgreater\,Class 3.1.1\,\textgreater\,Class 1.1.2. 
The reason is that the Shapley Values\cite{duddu2021shapr} attack mainly set meaningful thresholds for SHAPr scores to decide whether a data record is susceptible to MI attacks, and when only the CV4 changes, the Shapley Values\cite{duddu2021shapr} attack will select the data samples that have high shapley values to infer their membership, which will affect evaluation results.

We observe that the ranking results of a particular MI attack (i.e. BlindMI-Diff-w\cite{hui2021practical} (Class 3.1.1 see Figure \ref{figs:Classification Tree_MI attacks})) remain unchanged across almost all the ESs in Group A (e.g., ES49: C100\_U+4.325+0.085+20\%, ES50:C100\_U + 4.325+0.085+40\% and ES51:C100\_U + 4.325+0.085+45\%). We also observe that the ranking results of the same MI attack remain unchanged across almost all the ESs in Group B (e.g., ES55:C100\_U+
4.325+0.157+45\%, ES56:C100\_
U+4.325+0.157+49\%). Note that Group A and Group B use the same tuning variable such as the CV4. However, the ranking results are different between the two groups. This indicates that the MI attack is probably affected by the differences between the values of the three non-tuning factors in the two groups. We look into the differences and find that the differences between the values of the three non-tuning factors in the two groups are the CV3 (e.g., 0.085 and 0.157) and BlindMI-Diff-w\cite{hui2021practical} mainly utilized
the differences between two datasets before and after a sample was moved. Therefore, the BlindMI-Diff-w attacks are most affected by the CV3.

We also observe that the ranking results of the Shapley Values\cite{duddu2021shapr} attack (Class 2.2.1 see Figure \ref{figs:Classification Tree_MI attacks}) remain unchanged across almost all the ESs in Group A (e.g., ES31:C100\_U+2.893+0.085+45\%, ES33:C100\_U +2.893 
+ 0.119+45\% and 
ES35:C100\_U+2.893+0.157+45\%) and Group B (e.g., ES34:C100\_
U+2.893+0.119+49\%,
ES36:C100\_U+2.893+0.157+49\%). Note that Group A and Group B use the same tuning variable such as the CV3. However, the ranking results are different between the two groups (e.g., the second and the third). We look into the differences and find that the differences between the values of the three non-tuning factors in the two groups are the CV4 (e.g., 45\% and 49\%) and Shapley Values attacks set meaningful thresholds for SHAPr
scores to decide distinguish members, and when only CV4 changes, the Shapley Values attack will select the data samples that have high shapley values to infer their membership. Therefore, the Shapley Values attacks are most affected by the CV4.


\section{Related Work}
\label{sec RELATED WORK}

\noindent \textbf{Membership Inference Attacks.} \textit{Homer et al.}~\cite{homer2008resolving} first proposed an MI attack on biological data. 
\textit{Shokri et al.}~\cite{shokri2017membership} proposed the first black-box MI attack against ML. 
Huge literature followed these works 
to different scenarios (e.g., location data~\cite{pyrgelis2020measuring}, 
language models~\cite{carlini2021extracting}, sentence embeddings~\cite{song2020information}, speech recognition models~\cite{shah2021evaluating}, federated learning~\cite{jere2020taxonomy},  transfer learning~\cite{zou2020privacy}, generative models~\cite{chen2020gan}, white box access~\cite{nasr2019comprehensive,sablayrolles2019white,leino2020stolen}).

\noindent \textbf{Categories of Membership Inference Attacks.} 
There are main three categories 
\textbf{1) Binary classifier-based MI attacks}, which  utilize the output predictions of shadow models to train a binary classifier to launch the MI attacks~\cite{shokri2017membership,salem2019ml}. \textbf{2) Evaluation metric-based MI attacks}, which used the defined evaluation metrics to distinguish members and nonmembers~\cite{hui2021practical, yeom2018privacy,salem2019ml, li2021membership,choquette2021label,long2020pragmatic}. \textbf{3) Differential Comparisons-based MI attacks (BlindMI-Diff)}, which 
mainly utilized the differences between two datasets. 

\noindent \textbf{Defenses against MI attacks.} Multiple defenses~\cite{abadi2016deep,nasr2018machine,shokri2017membership,jia2019memguard,salem2019ml,srivastava2014dropout,shejwalkar2021membership,zheng2021resisting} have been proposed to mitigate MI attacks. 

\section{Conclusion}
\label{sec Conclusion}

In this paper, 
we seek to develop a comprehensive benchmark for comparing 
different MI attacks. The primary design goal of our benchmark, called \textbf{MIBench}, is 
to meet {\em all} of above-mentioned 4 requirements (see Section \ref{sebsec:Introduction}). Our benchmark consists not only the evaluation metrics, but also the evaluation scenarios. And we design the evaluation scenarios 
from {\bf four perspectives}: the distance distribution of data samples in the target dataset, the distance between data samples of the target dataset, the differential distance between two datasets (i.e., the target dataset and a generated dataset with only nonmembers), and the ratio of the samples that are made no inferences by an MI attack. The evaluation metrics consist of ten typical evaluation metrics. We have identified three principles for the proposed ``comparing different
MI attacks'' methodology, and we have designed and implemented 
the MIBench benchmark with 84 evaluation scenarios for each dataset. In total, we have used our benchmark to fairly and systematically 
compare 15 state-of-the-art MI attack algorithms across 588 evaluation scenarios, and these evaluation scenarios cover 7 widely used datasets and 7 representative types of models.

\bibliographystyle{IEEEtran}
\bibliography{ref}{}

\appendix

\section*{A. Datasets Description}
\label{subsec:Datasets Description}

\noindent \textbf {CIFAR100.} CIFAR100 is a widely used benchmark dataset to image classification, which consists of 60,000 images in 100 classes. We randomly select two disjoint sets of 10,000 images 
as the target models and the shadow models' training datasets, respectively.

\noindent \textbf {CIFAR10.} CIFAR10 is a widely used dataset to evaluate the image classification, which consists of 60,000 images in 10 classes. We randomly select two disjoint sets of 10,000 images 
as the target models and the shadow models' training datasets, respectively. 

\noindent  \textbf {CH$\_$MNIST.} CH\_MNIST is a benchmark dataset of histological images used to evaluate human colorectal cancer, consisting of 5,000 histological images in 8 classes of tissues. We follow the same image processing methods and classification tasks as prior BlindMI-DIFF\cite{hui2021practical} to resize all images to 64$\times$64. We randomly select two disjoint sets of 2,500 images 
as the target models and the shadow models' training datasets, respectively.
   
\noindent \textbf {ImageNet.} Tiny-imagenet is a widely used benchmark dataset to image classification, which is a subset of the ImageNet dataset and consists of 100,000 images in 200 classes. We randomly select two disjoint sets of 10,000 images 
as the target models and the shadow models' training datasets, respectively. 

\noindent \textbf {Location30.} Location30 
contains location “check-in” records of 
individuals. We obtain a 
pre-processed
dataset from 
~\cite{shokri2017membership} 
which contains 5,010 data samples with with 446 binary features corresponding to whether an individual has visited a particular location. 
All data samples are clustered into 30
classes representing different geosocial types. The classifi-
cation task is to predict the geosocial type based on the 466
binary features. Following \emph{Jia et al.}~\cite{jia2019memguard}, we use 1,000 data
samples to train a target model.

\noindent \textbf {Purchase100.} Purchase100 
contains shopping records
of 
different individuals. We obtain a 
pre-processed dataset from 
\cite{shokri2017membership}
containing 197,324 data samples with 600 binary features corresponding to a specific product. 
All data samples are clustered into 100 classes representing different purchase styles. The classification task is to predict the purchase
style based on the 600 binary features. We follow \emph{Nasr et
al.} \cite{nasr2019comprehensive,nasr2018machine} to use 10\% data samples (19,732) to train a target model.

\noindent \textbf {Texas100.} Texas100 consists of Texas Department of State Health Services’ information about patients discharged from public hospitals.
Each data
record contains information about the injury, diagnosis, the
procedures the patient underwent and some demographic details.
We obtain 
preprocessed dataset from 
\cite{shokri2017membership} which contains 100 classes of patient’s procedures consisting 67,330 data
samples with 6,170 binary features.
The classification task is to predict the patient’s main procedure based on the patient’s information. 
Following 
\cite{nasr2019comprehensive,nasr2018machine,jia2019memguard}, we use 10,000 data samples to train a target model.

\begin{figure*}
\vspace{0.1pt}
\centering
\subfigure[CIFAR100]{
\begin{minipage}{0.2\textwidth}
\label{Fig.sub.5.1}
\includegraphics[width=1.2\textwidth]{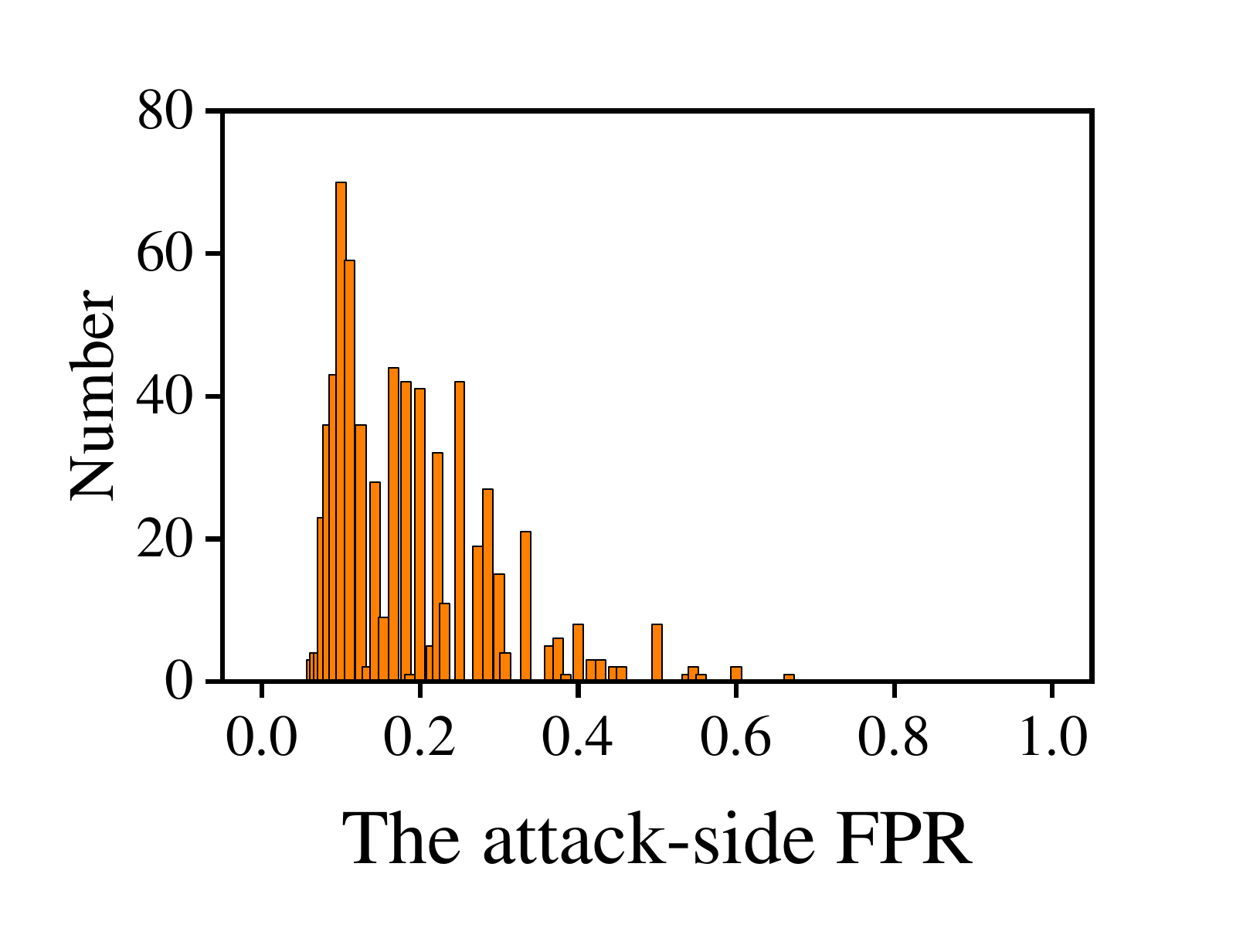}
\end{minipage}}
\subfigure[CIFAR10]{
\begin{minipage}{0.2\textwidth}
\label{Fig.sub.5.2}
\includegraphics[width=1.2\textwidth]{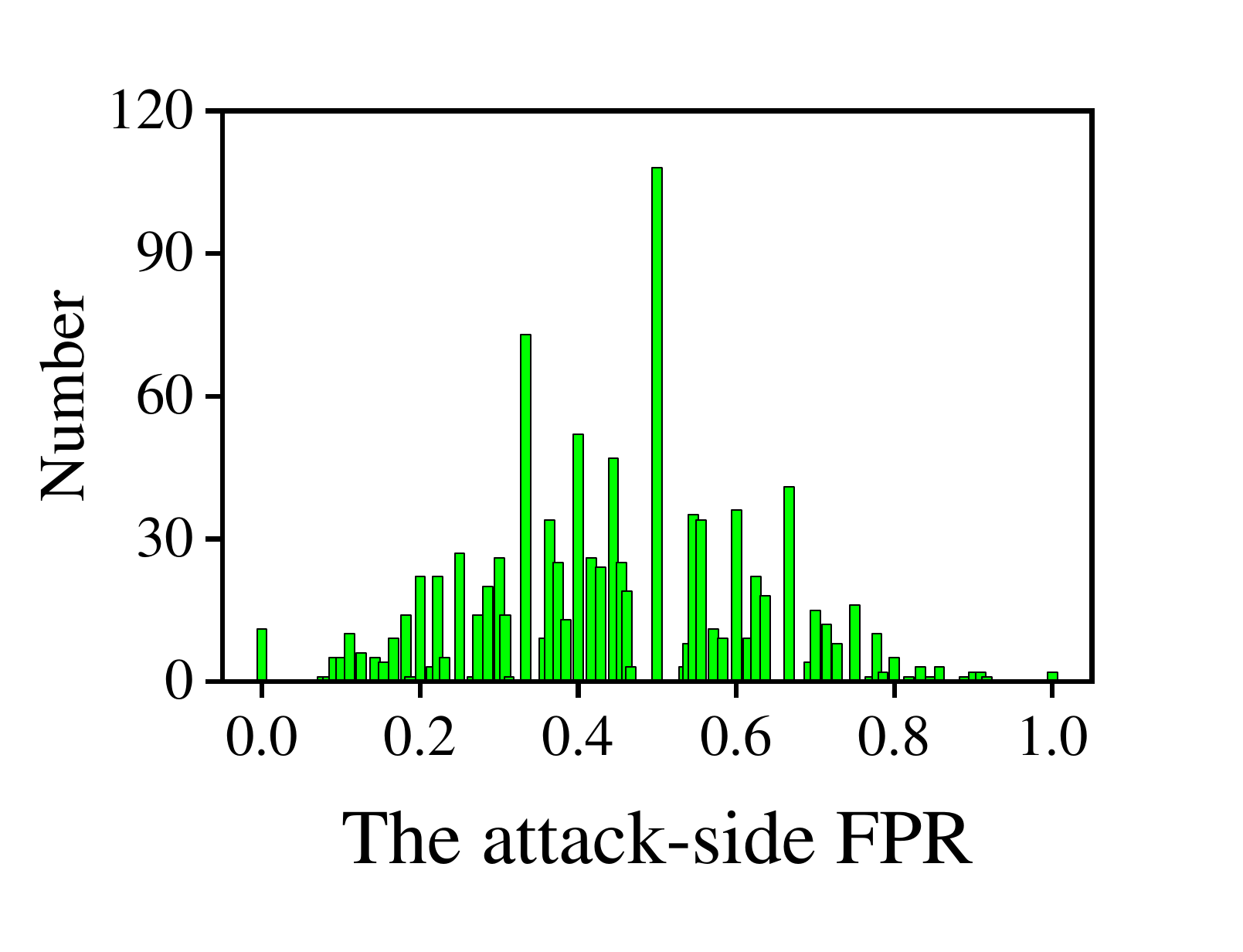}
\end{minipage}}
\subfigure[CH\_MNIST]{
\begin{minipage}{0.2\textwidth}
\label{Fig.sub.5.3}
\includegraphics[width=1.2\textwidth]{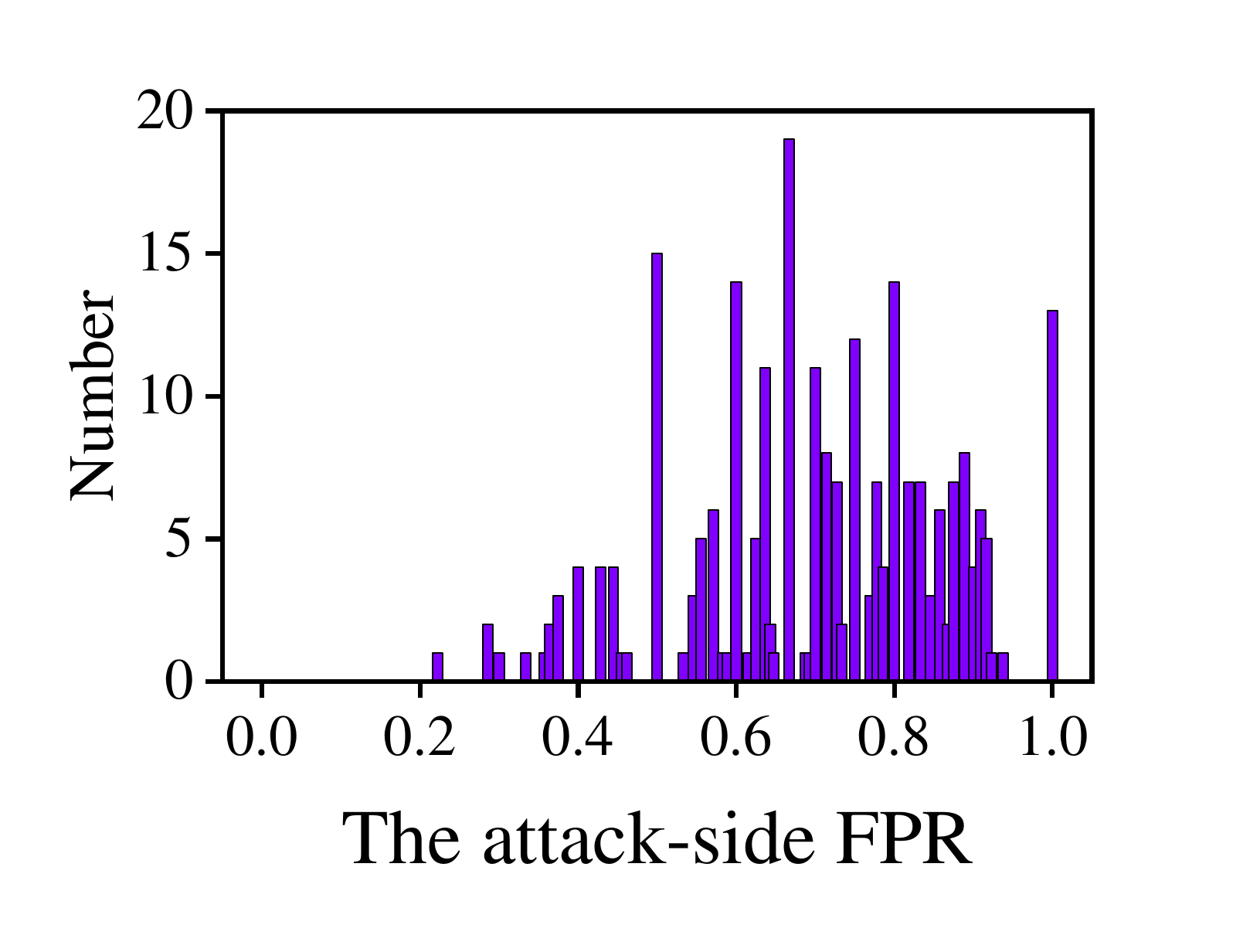}
\end{minipage}}
\subfigure[ImageNet]{
\begin{minipage}{0.2\textwidth}
\label{Fig.sub.5.4}
\includegraphics[width=1.2\textwidth]{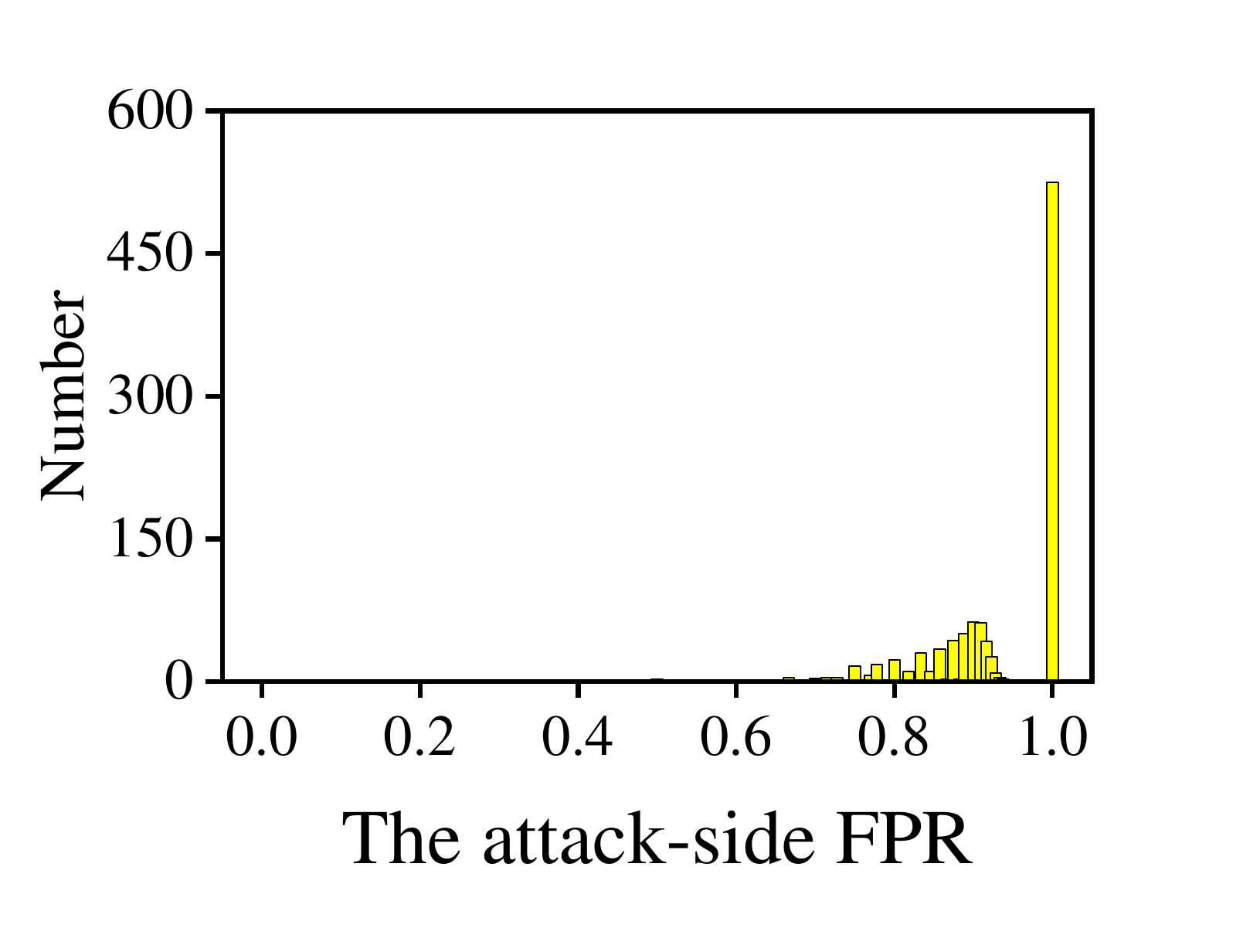}
\end{minipage}}
\caption{\label{figs:fig3}The distribution of the 
FPR of the BlindMI-Diff-w/\cite{hui2021practical}.} 

\vspace{0.1pt}
\end{figure*}

\begin{figure*}
\vspace{0.1pt}
\centering
\subfigure[CIFAR100]{
\begin{minipage}{0.2\textwidth}
\label{Fig.sub.1}
\includegraphics[width=1.2\textwidth]{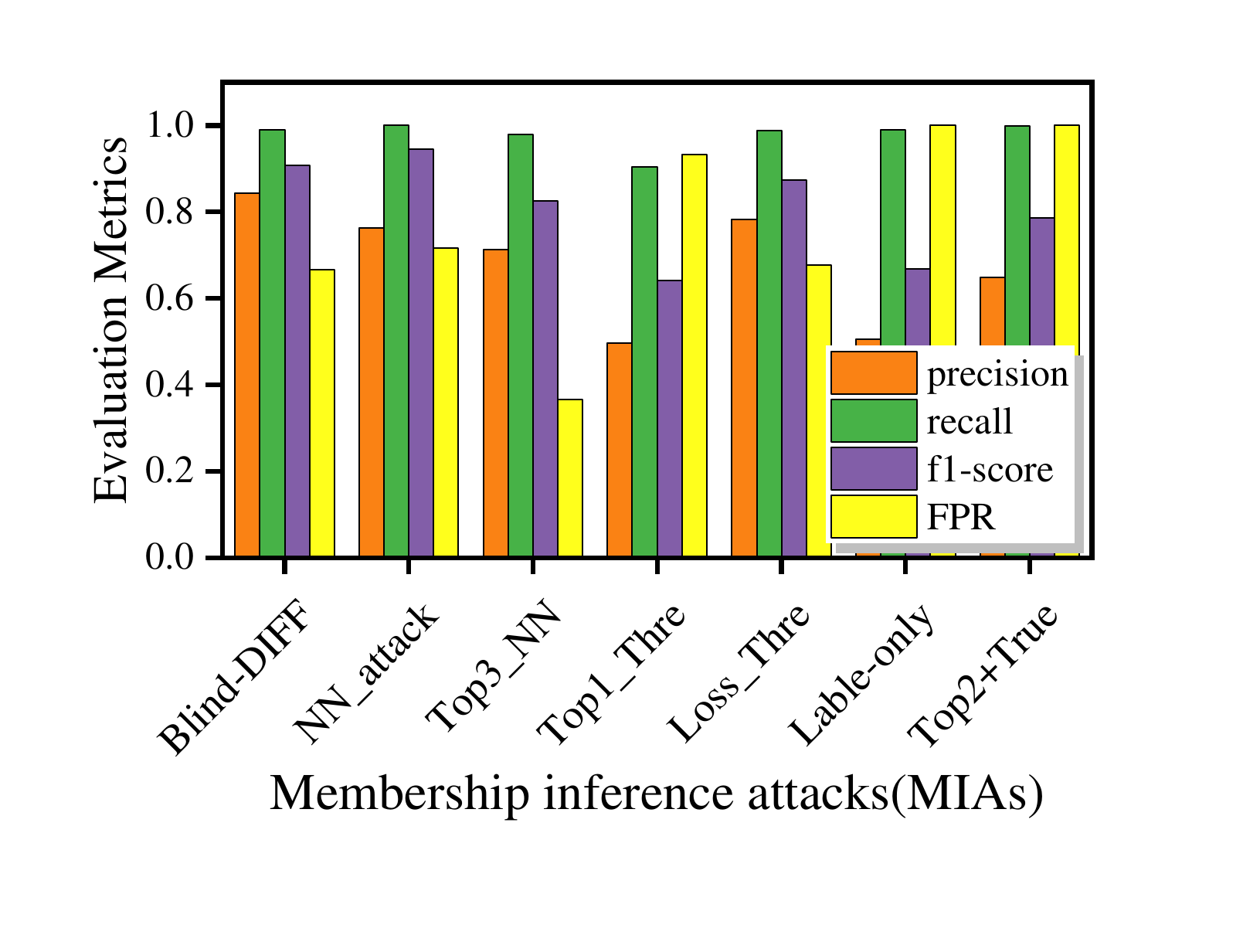}
\end{minipage}}
\subfigure[CIFAR10]{
\begin{minipage}{0.2\textwidth}
\label{Fig.sub.2}
\includegraphics[width=1.2\textwidth]{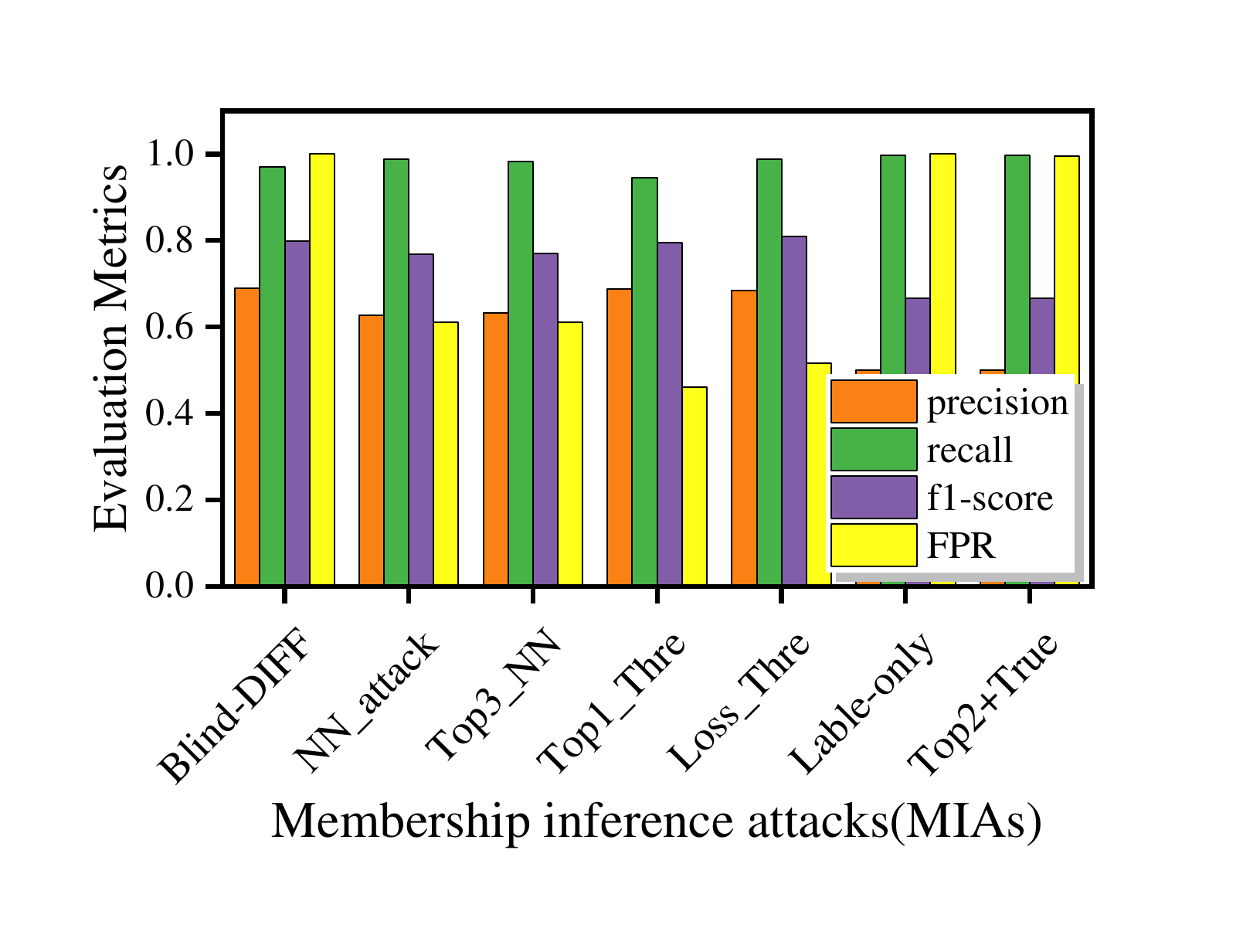}
\end{minipage}}
\subfigure[CH\_MNIST]{
\begin{minipage}{0.2\textwidth}
\label{Fig.sub.3}
\includegraphics[width=1.2\textwidth]{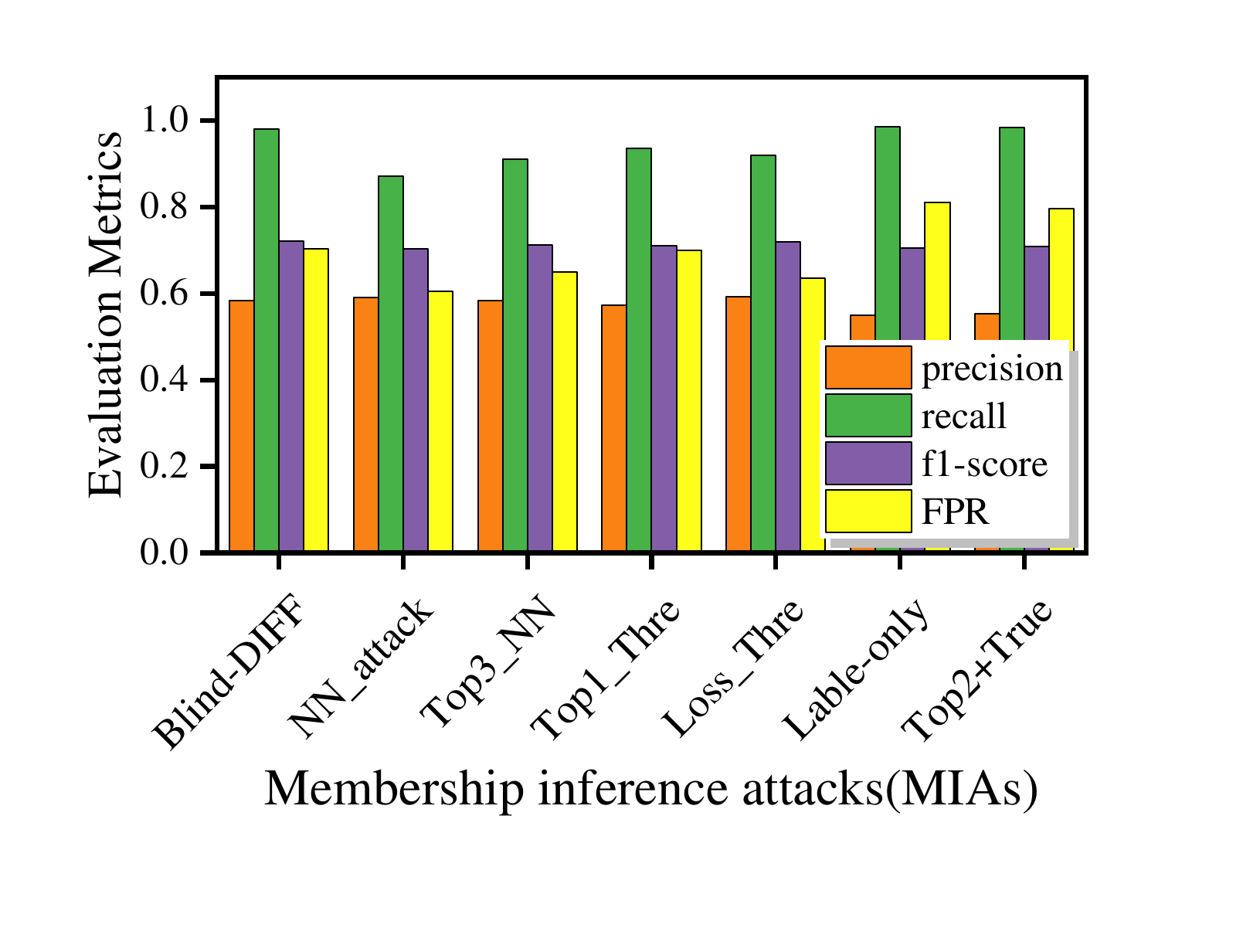}
\end{minipage}}
\subfigure[ImageNet]{
\begin{minipage}{0.2\textwidth}
\label{Fig.sub.4}
\includegraphics[width=1.2\textwidth]{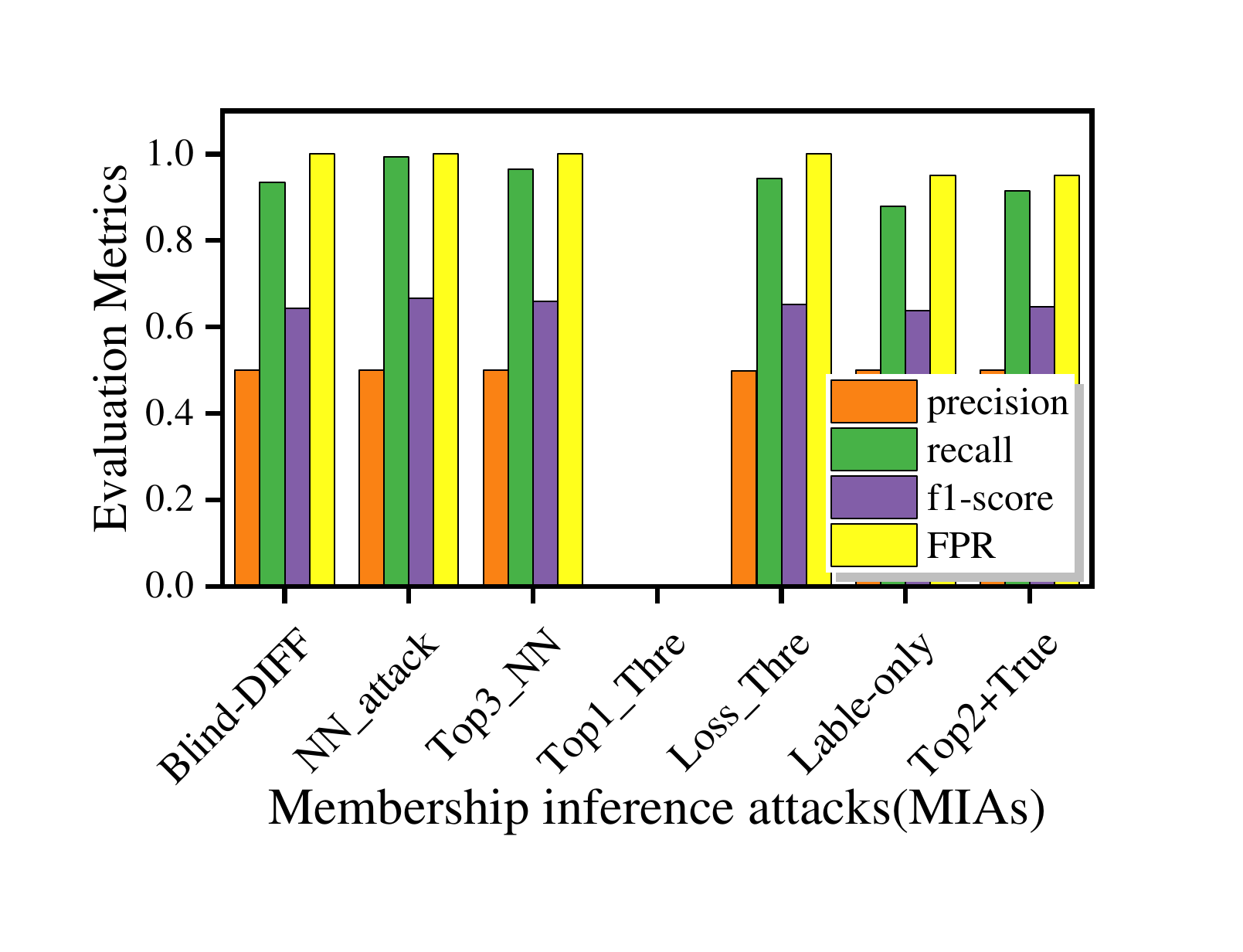}
\end{minipage}}
\caption{\label{figs:fig4}The evaluation of the existing MI attacks (e.g., \emph{the attacker-side precision, recall, f1-score and FPR}).}.
\vspace{0.1pt}
\end{figure*}

\begin{figure*}
\vspace{0.1pt}
\centering
\subfigure[CIFAR100]{
\begin{minipage}{0.2\textwidth}
\label{Fig.sub.1}
\includegraphics[width=1.2\textwidth]{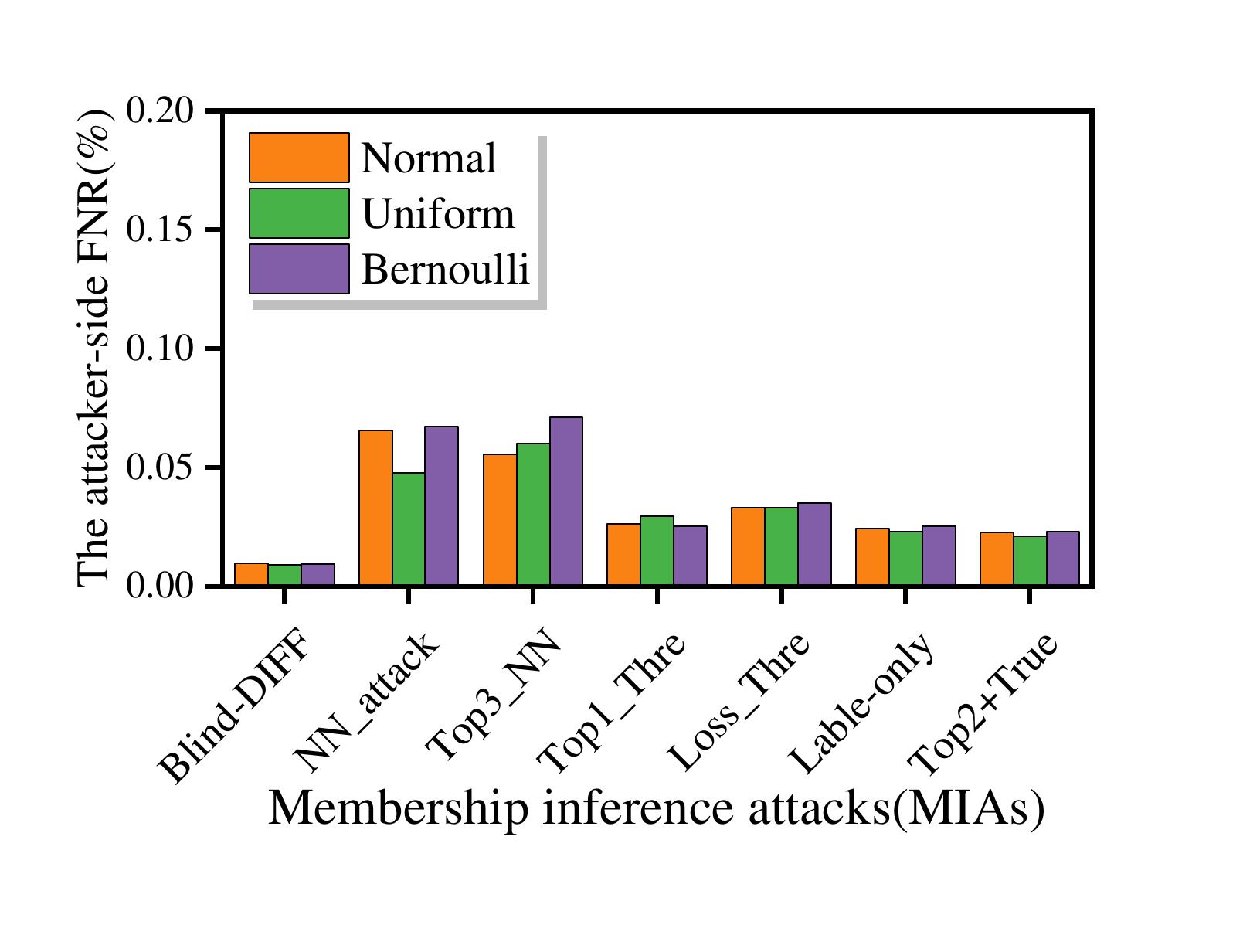}
\end{minipage}}
\subfigure[CIFAR10]{
\begin{minipage}{0.2\textwidth}
\label{Fig.sub.2}
\includegraphics[width=1.2\textwidth]{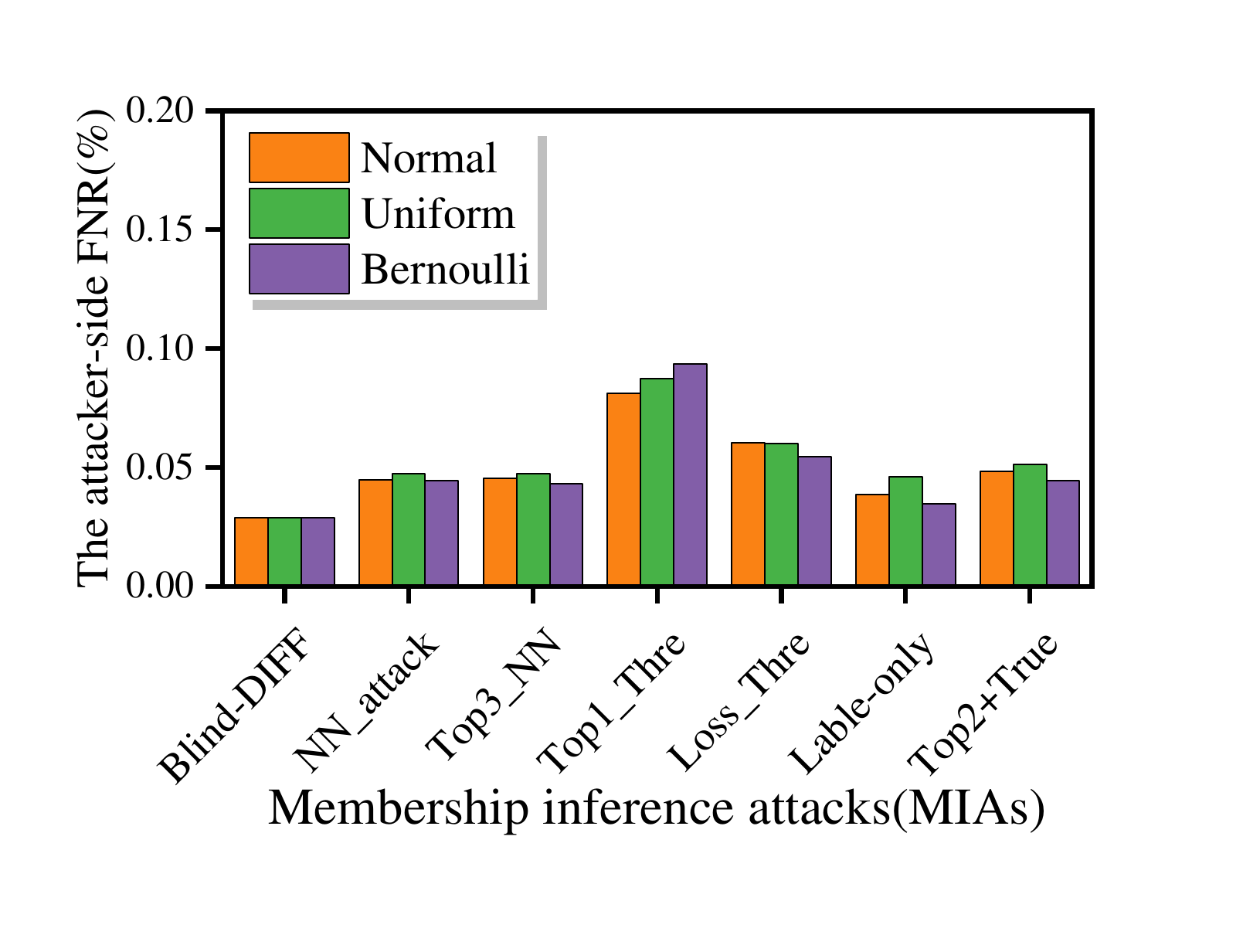}
\end{minipage}}
\subfigure[CH\_MNIST]{
\begin{minipage}{0.2\textwidth}
\label{Fig.sub.3}
\includegraphics[width=1.2\textwidth]{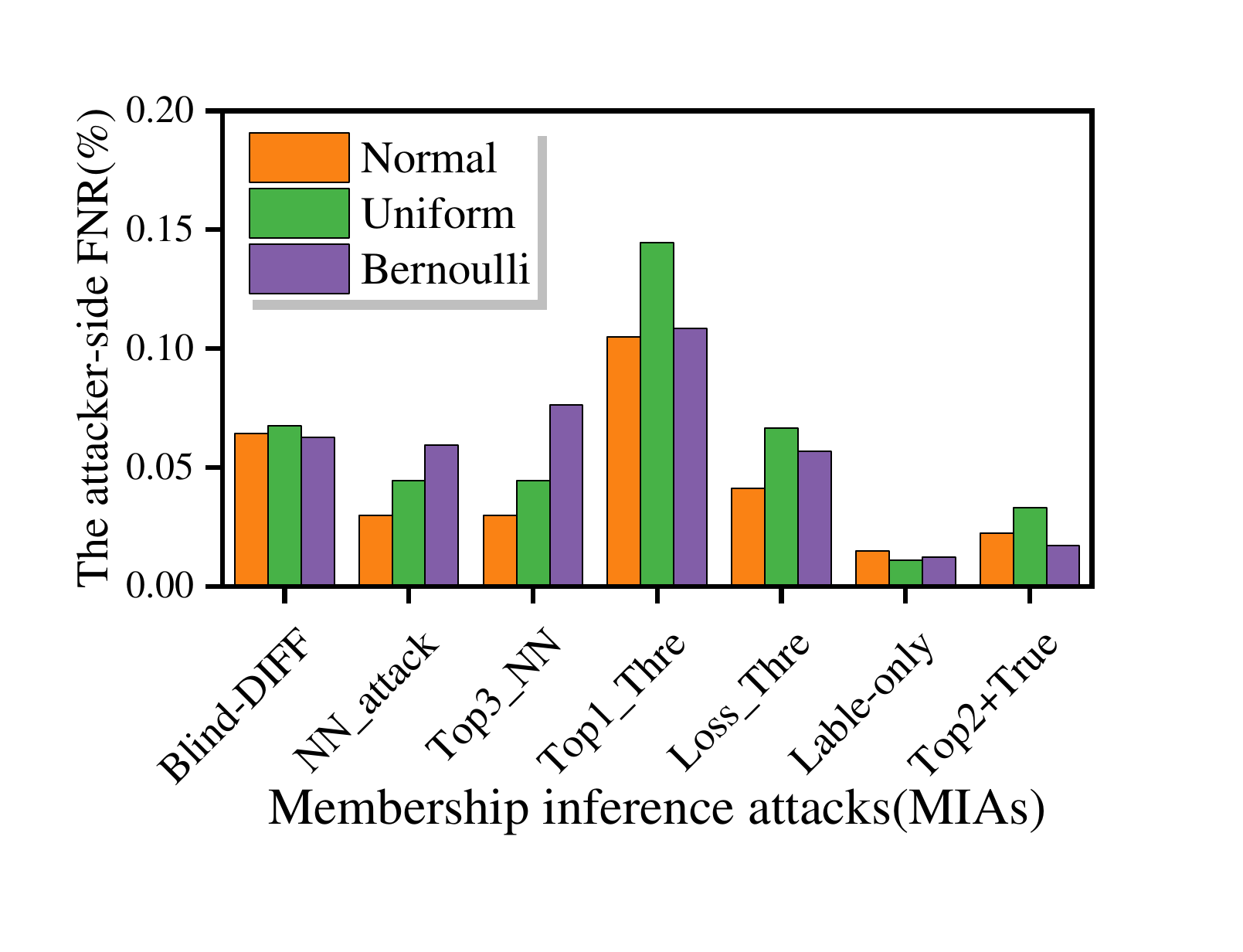}
\end{minipage}}
\subfigure[ImageNet]{
\begin{minipage}{0.2\textwidth}
\label{Fig.sub.4}
\includegraphics[width=1.2\textwidth]{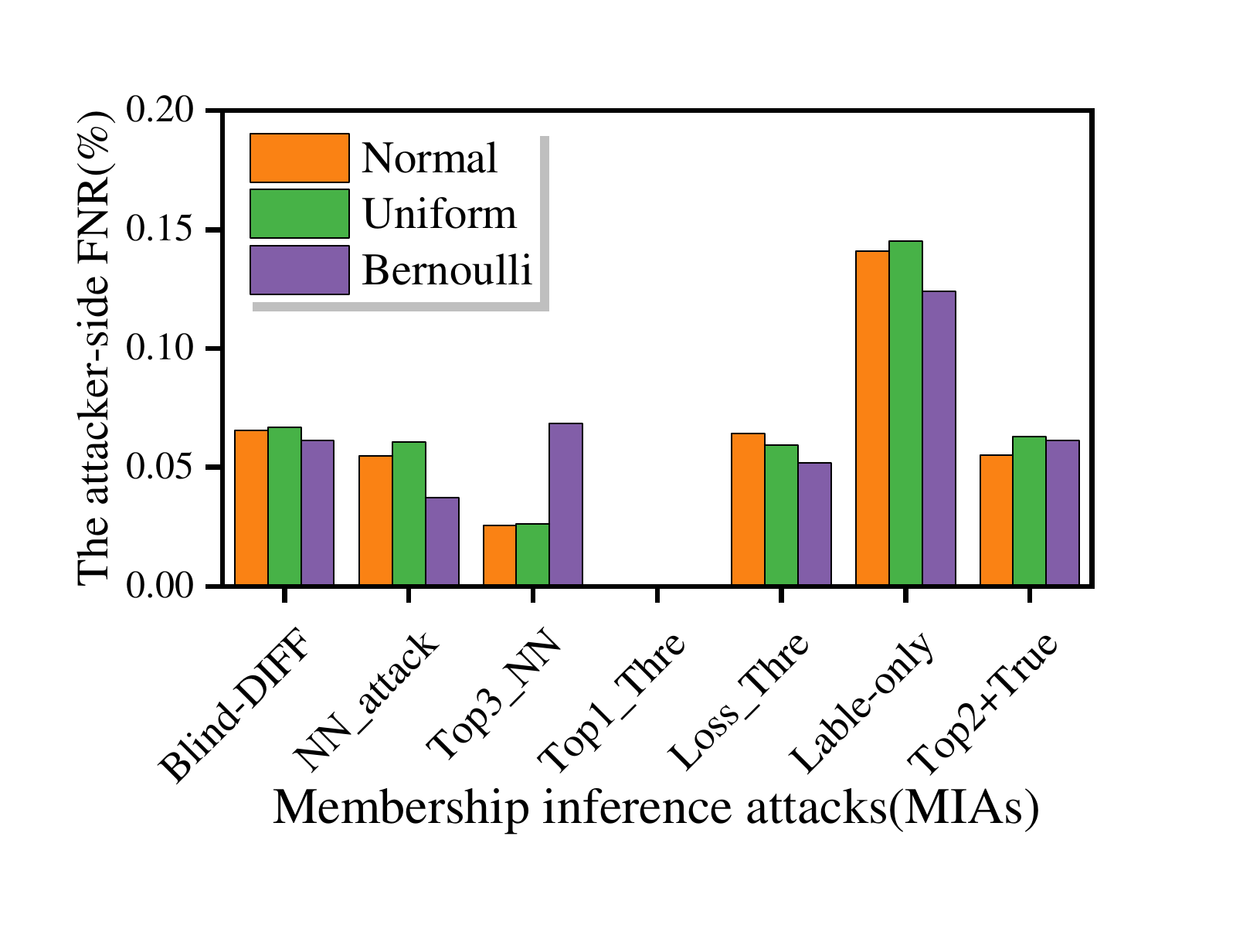}
\end{minipage}}
\caption{\label{figs:fig7}The attacker-side FNR of the distance distribution of the target datasets obeying normal, uniform and bernoulli distributions.}
\vspace{0.1pt}
\end{figure*}

\begin{figure*}
\vspace{0.1pt}
\centering
\subfigure[CIFAR100]{
\begin{minipage}{0.2\textwidth}
\label{Fig.sub.1}
\includegraphics[width=1.2\textwidth]{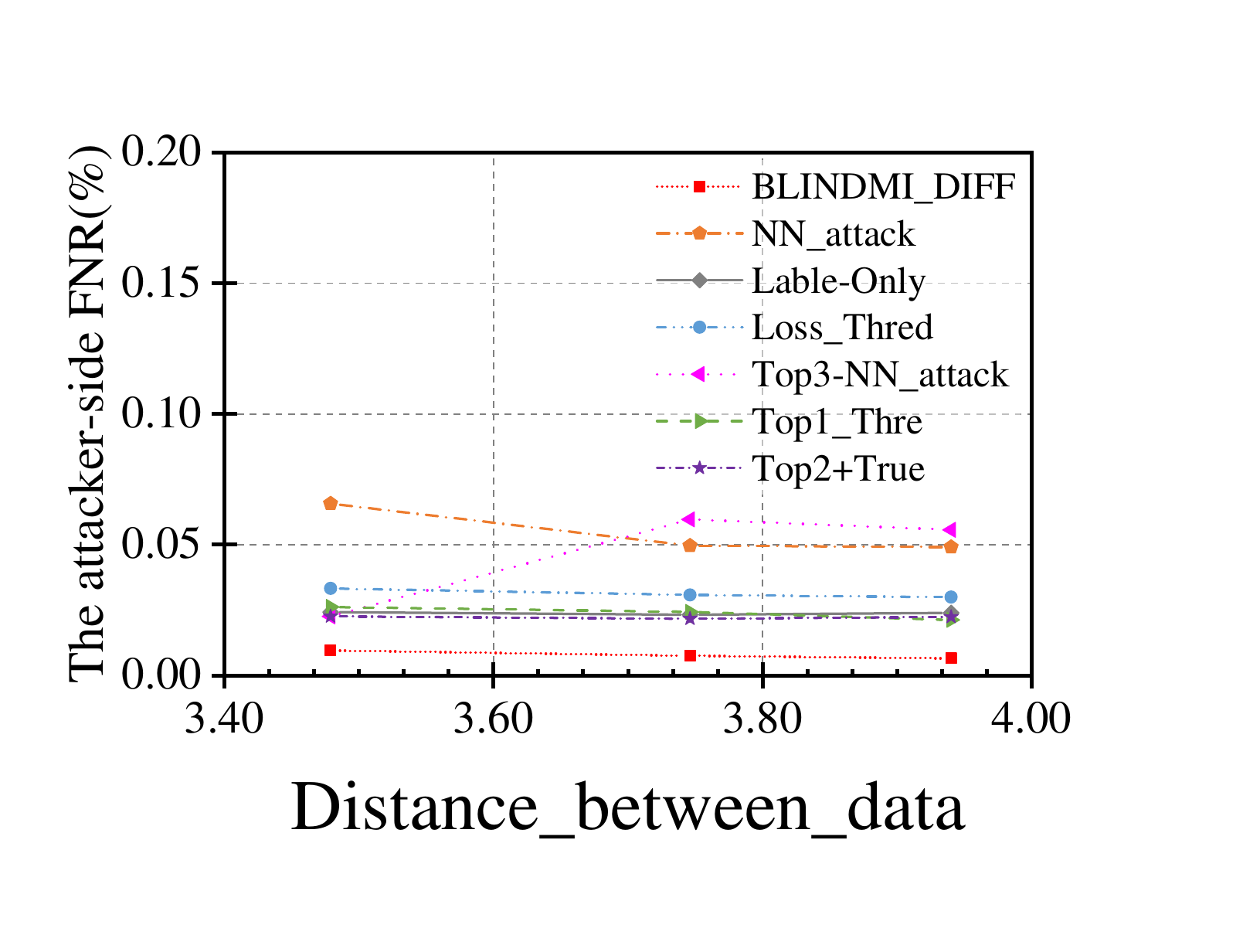}
\end{minipage}}
\subfigure[CIFAR10]{
\begin{minipage}{0.2\textwidth}
\label{Fig.sub.2}
\includegraphics[width=1.2\textwidth]{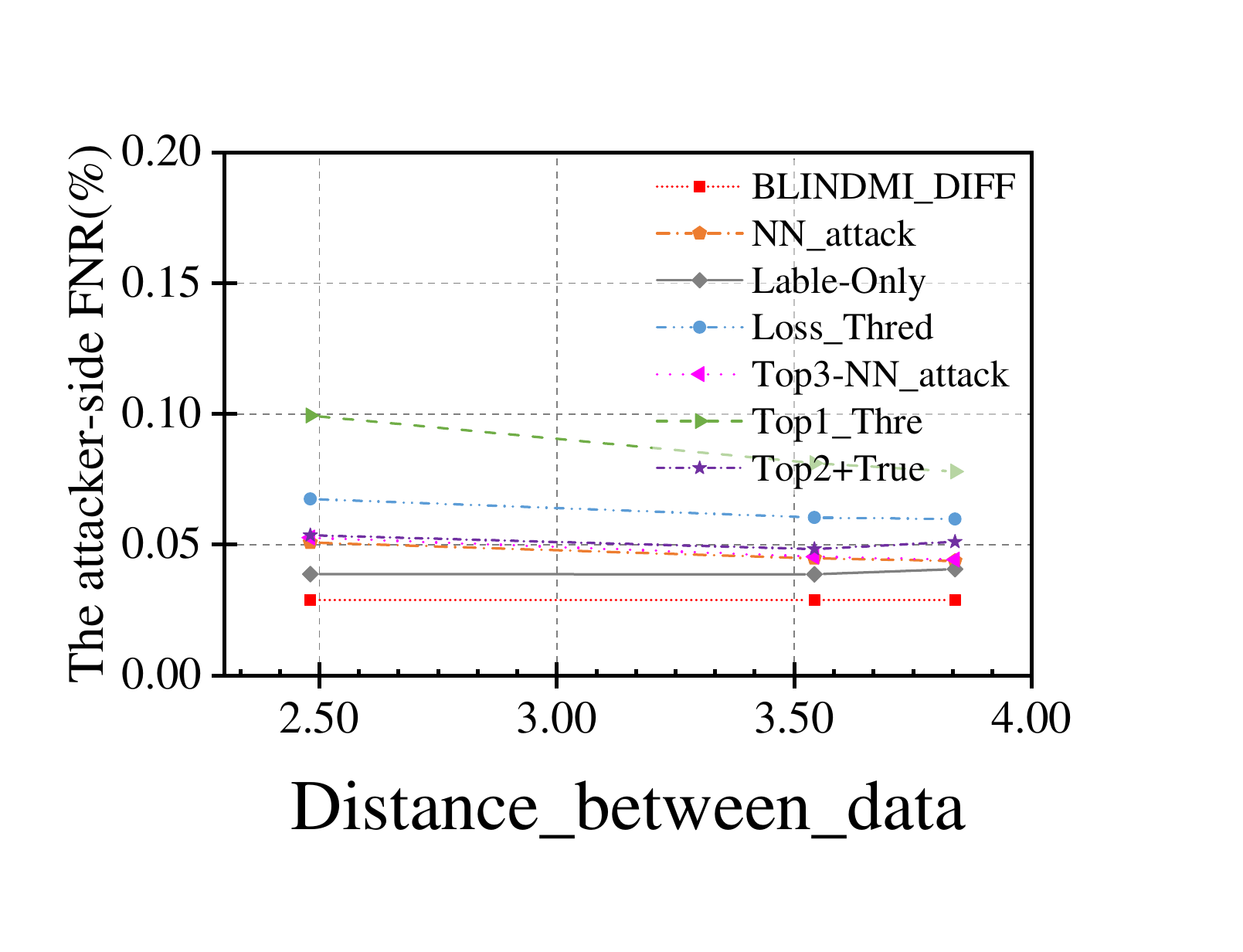}
\end{minipage}}
\subfigure[CH\_MNIST]{
\begin{minipage}{0.2\textwidth}
\label{Fig.sub.3}
\includegraphics[width=1.2\textwidth]{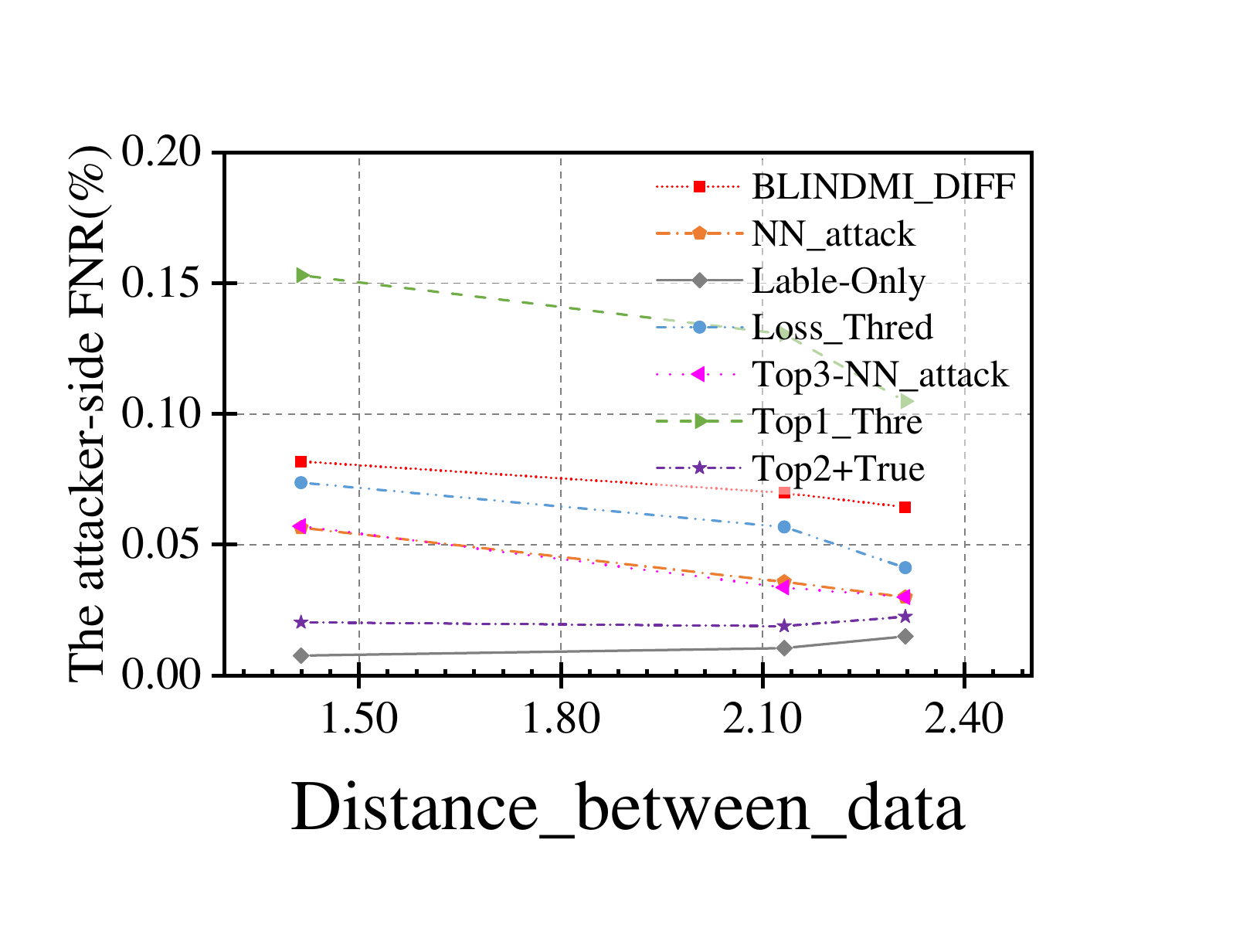}
\end{minipage}}
\subfigure[ImageNet]{
\begin{minipage}{0.2\textwidth}
\label{Fig.sub.4}
\includegraphics[width=1.2\textwidth]{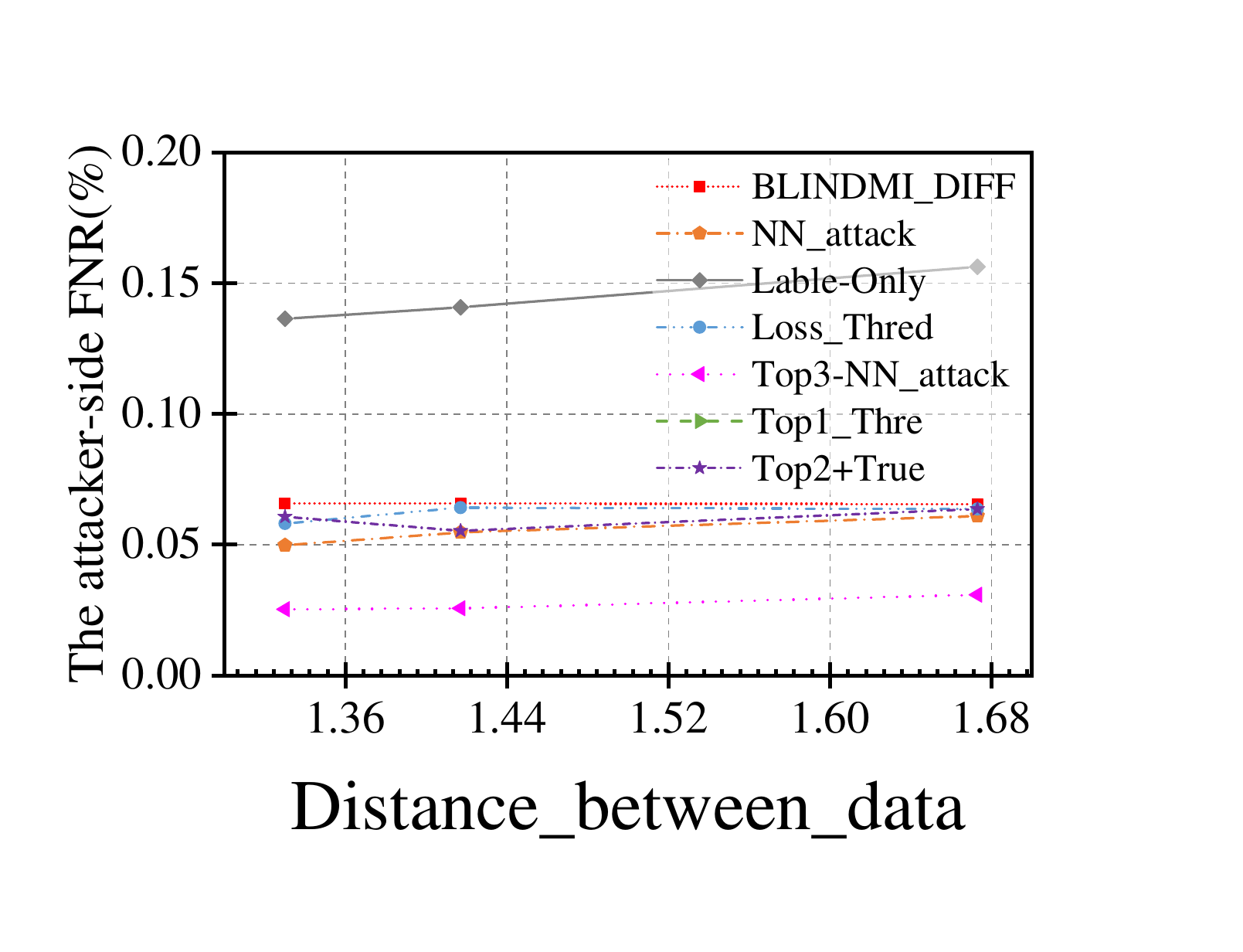}
\end{minipage}}
\caption{\label{figs:fig5}The effect of the distance between
data samples of the target dataset on the attacker-side FNR.}
\vspace{0.1pt}
\end{figure*}

\begin{table*}[htp]\tiny
    \caption{The effect of the distance between data samples of the target dataset (DisData) on the attacker-side Membership Advantage.}
    \centering
    \setlength{\tabcolsep}{1.8mm}{
    \begin{tabular}{ccccccccccc}
        \toprule[1.1pt]
                 \multirow{2}{*}{\textbf{Dataset}} &
                 \multirow{2}{*}{\textbf{DisData}} &
                 \multicolumn{9}{c}{\textbf{\textbf{Attacker-side Membership Advantage (MA)}}}\\ 
            \cmidrule(r){3-11} & & \textbf{BlindMI-Diff} & \textbf{NN\_attack} & \textbf{Label-only} & \textbf{Loss-Thres} & \textbf{Top3-NN} & \textbf{Top2+True}& \textbf{PPV} & \textbf{Calibrated Score} & \textbf{Distillation-based Thre}\\
    
         \midrule[1.1pt]  
        
    CIFAR100 & \textbf{3.823}
    & \textbf{61.63\%} & \textbf{56.13\%} & \textbf{75.38\%} & \textbf{73.37\%} & \textbf{60.25\%} & \textbf{69.25\%} &  \textbf{1.83\%} & \textbf{35.60\%} & \textbf{25.86\%} \\ 
        
        CIFAR10 &  2.573 
        & 53.34\% & 38.50\% & 33.50\% & 50.50\% & 38.38\% & 46.13\% &  -0.07\% & 33.61\% & 24.56\% \\  
        
         CH\_MNIST & 1.315 
         & 27.72\% & 26.60\% & 17.60\%  & 20.13\% & 26.24\% & 18.84\% &  -0.92\% & 22.04\% & 23.45\% \\ 
         
         ImageNet & 1.138 
         & 0.47\% & 0.15\% & 0.56\% &  0.05\% & 0.08\% & 0.09\% &  -1.08\% & -0.07\% & 23.12\% \\
         


         
         \bottomrule[1.1pt]
    \end{tabular}}
    \label{tabs:tab9}
     \vspace{0.1pt}
\end{table*}

\begin{figure*}
\vspace{0.1pt}
\centering
\subfigure[CIFAR100]{
\begin{minipage}{0.2\textwidth}
\label{Fig.sub.1}
\includegraphics[width=1.2\textwidth]{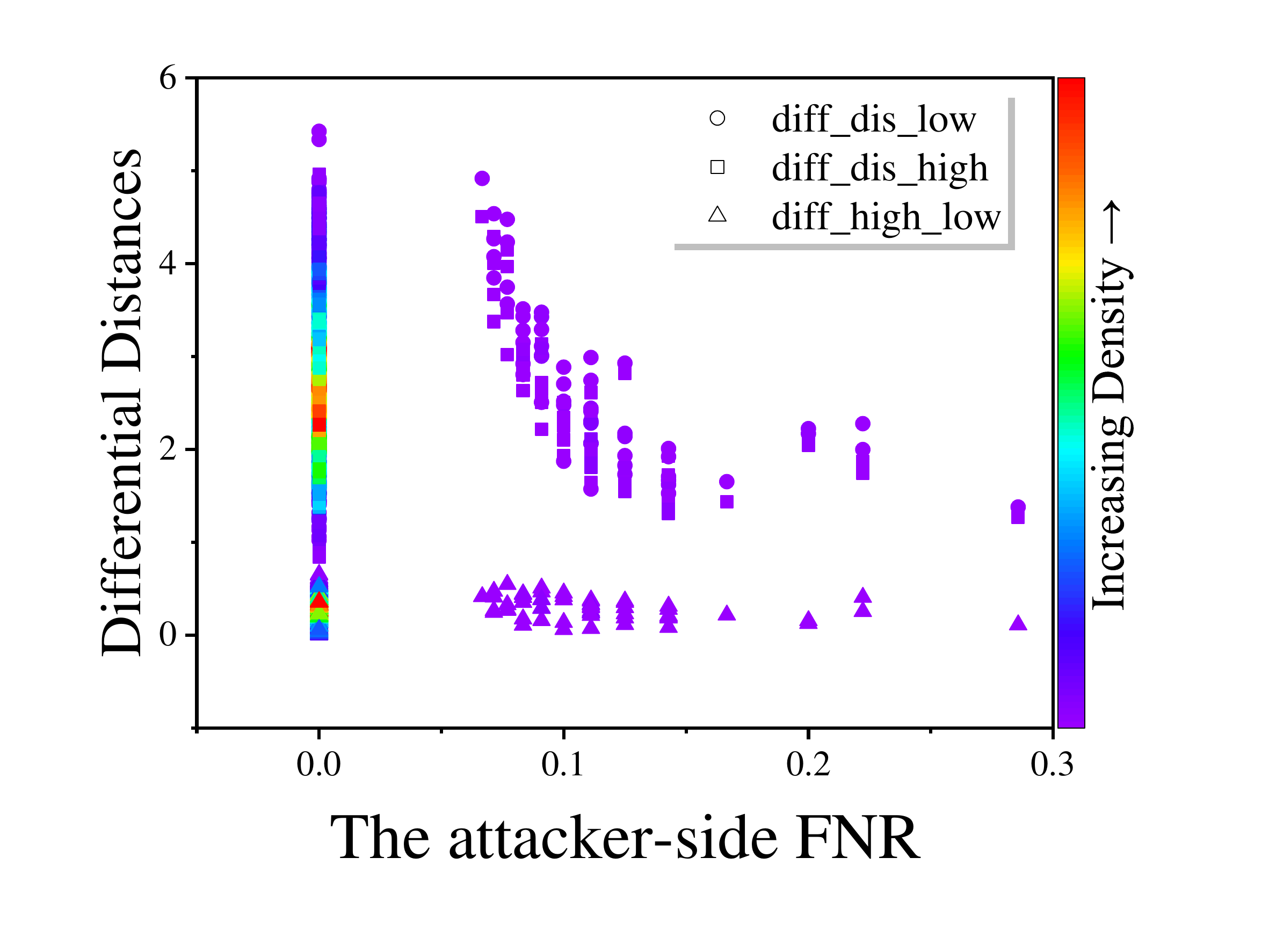}
\end{minipage}}
\subfigure[CIFAR10]{
\begin{minipage}{0.2\textwidth}
\label{Fig.sub.2}
\includegraphics[width=1.2\textwidth]{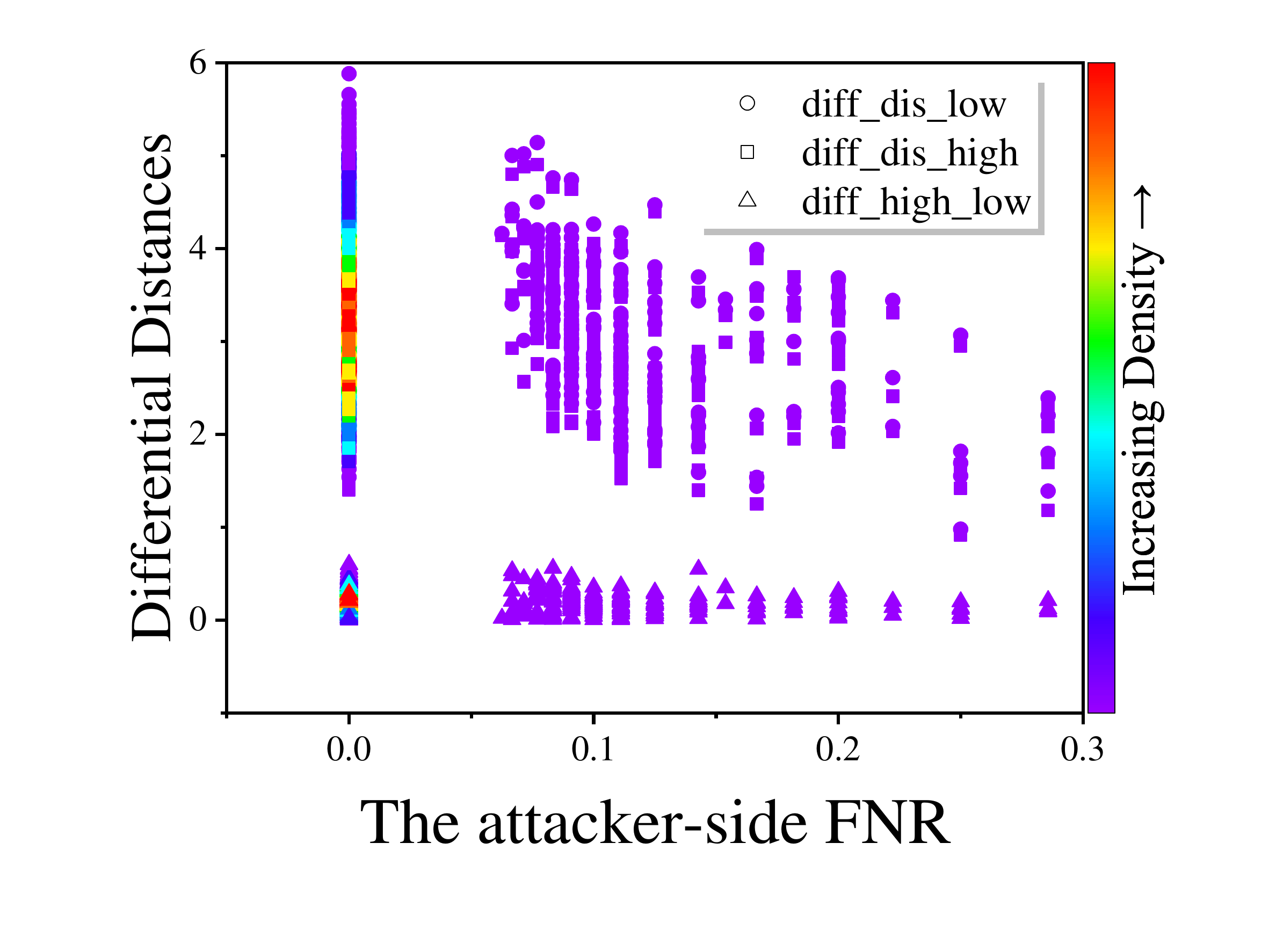}
\end{minipage}}
\subfigure[CH\_MNIST]{
\begin{minipage}{0.2\textwidth}
\label{Fig.sub.3}
\includegraphics[width=1.2\textwidth]{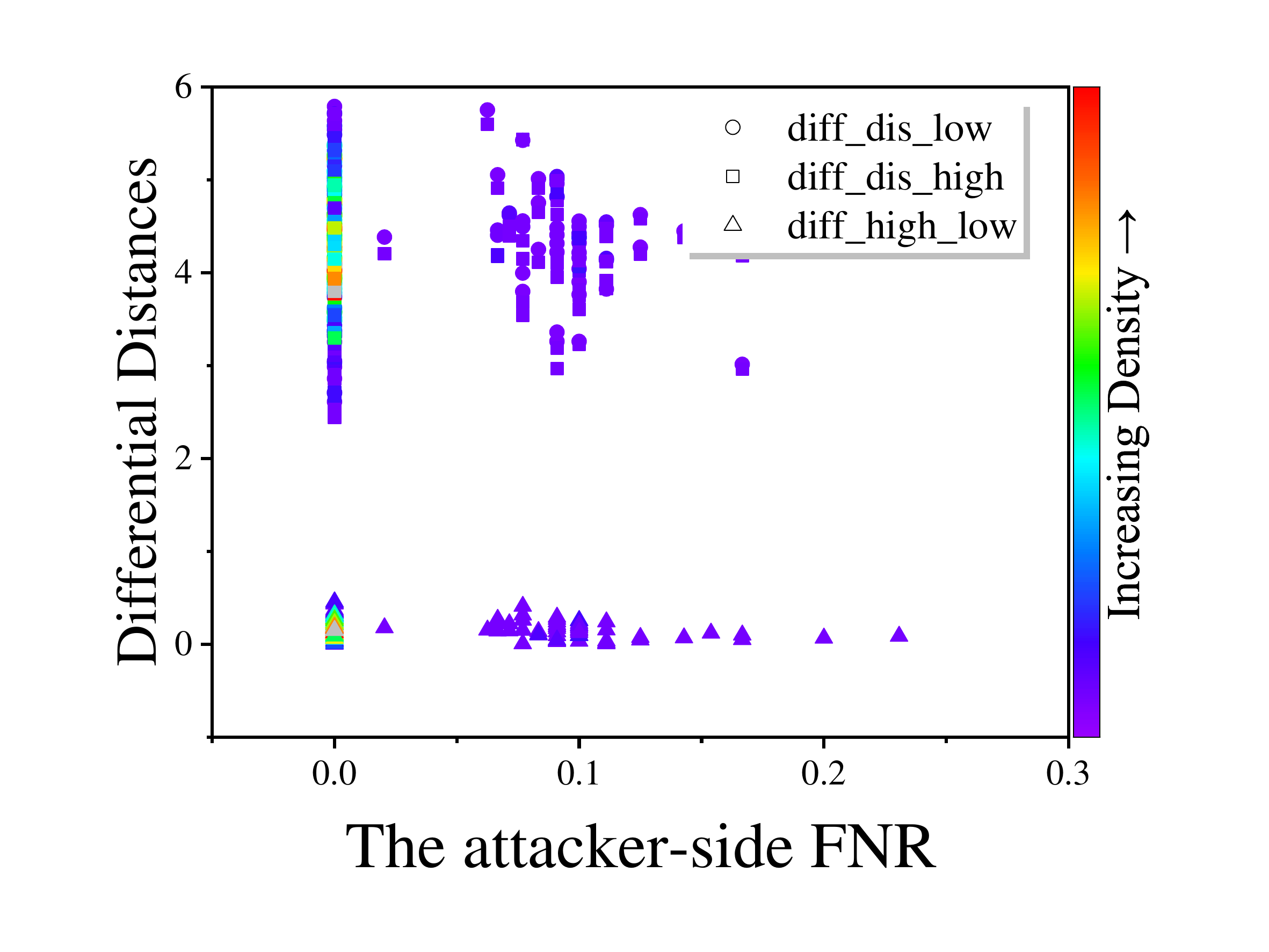}
\end{minipage}}
\subfigure[ImageNet]{
\begin{minipage}{0.2\textwidth}
\label{Fig.sub.4}
\includegraphics[width=1.2\textwidth]{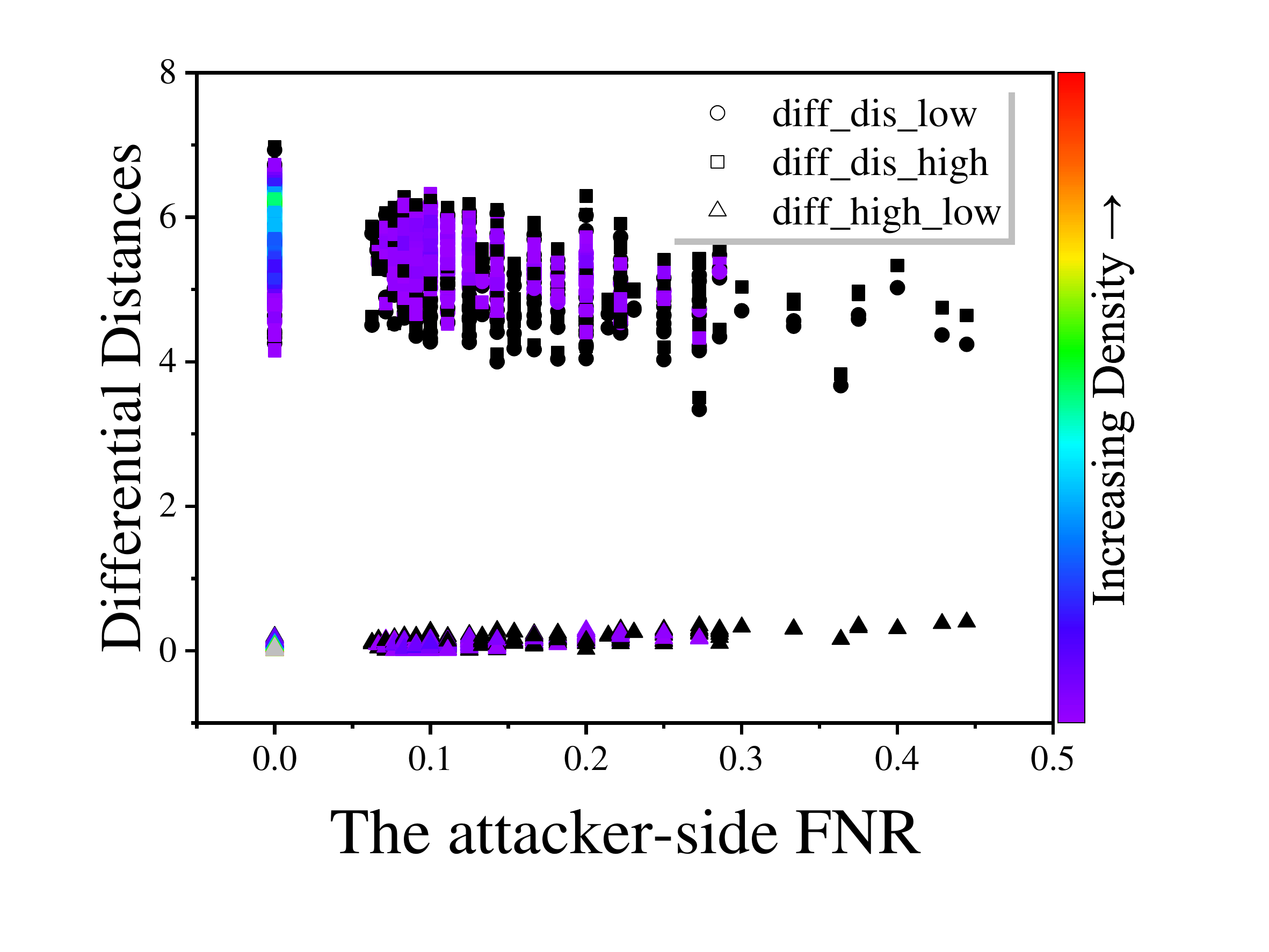}
\end{minipage}}
\caption{\label{figs:fig6}The effect of the differential distance between two datasets on attacker-side FNR.}
\vspace{0.1pt}
\end{figure*}

\begin{table}[htp]\tiny
    \caption{The average distance between data samples (DisBD) and the differential distance between two datasets (DiffDis) change on the MA (C\_MA).}
    \centering
    \begin{tabular}{c|ccc}
        \toprule[1.1pt]
            \textbf{Dataset} &  \textbf{C\_MA} &
           \textbf{average DisBD} & \textbf{average DiffDis}
            \\
         \midrule[1.1pt]
        \multirow{5}{*}{CIFAR100} & 
        10\% & 0.143 & \textbf{0.107} \\
        & 20\% & 0.230 & \textbf{0.099} \\
        & 30\% & 0.349 & \textbf{0.138} \\
        & 40\% & 0.460 & \textbf{0.143} \\
        & 50\% & 0.630 & \textbf{0.257} \\ \hline
        
        \multirow{7}{*}{CIFAR10} & 
        10\% & 0.113 &  \textbf{0.082} \\
        & 20\%  & 0.179 & \textbf{0.115} \\
        & 30\% & 0.299 &  \textbf{0.182} \\
        & 40\% & 0.440 &  \textbf{0.238} \\
        & 50\% & 0.544 &  \textbf{0.320} \\
        & 60\% & 0.659 &  \textbf{0.389} \\
        & 70\% & 0.793 &  \textbf{0.545} \\ \hline
        
        \multirow{6}{*}{CH\_MNIST} & 
        10\% & 0.140 &  \textbf{0.091} \\
        & 20\% & 0.249 &  \textbf{0.158} \\
        & 30\% & 0.335 &  \textbf{0.237} \\
        & 40\% & 0.476 &  \textbf{0.328} \\
        & 50\% & 0.638 &  \textbf{0.408} \\
        & 60\% & 0.843 &  \textbf{0.547} \\ \hline
        
        \multirow{3}{*}{ImageNet} & 
        10\% & 0.268 &  \textbf{0.225} \\
        & 20\% & 0.580 &  \textbf{0.288} \\
        & 30\% & 0.805&  \textbf{0.379} \\

        \bottomrule[1.1pt]
    \end{tabular}
   \label{tabs:The average distance change on the Attacker-side Membership Advantage.}
    \vspace{0.01pt}
\end{table}

\begin{table}[htp]\tiny
    \caption{The Selected Distances Between Data Samples (Dis\_Between\_Data) and Differential Distances Between Two Datasets of the Target Dataset.}
    \centering
    \begin{tabular}{c|cc}
        \toprule[1.1pt]
            \textbf{Dataset} &  \textbf{Dis\_Between\_Data} &
           \textbf{Differential Distances} 
            \\
         \midrule[1.1pt]
        \multirow{3}{*}{CIFAR100} & 
        2.893 & 0.085 \\ 
        & 3.813 & 0.119 \\ 
        & 4.325 & 0.157 \\ \hline

        \multirow{3}{*}{CIFAR10} & 
        1.908 & 0.155 \\ 
        & 2.501 & 0.213 \\ 
        & 3.472 & 0.291 \\ \hline

        \multirow{3}{*}{CH\_MNIST} & 
        0.954 & 0.083 \\ 
        & 1.355 & 0.108 \\ 
        & 1.720 & 0.133 \\ \hline

        \multirow{3}{*}{ImageNet} & 
         0.934 & 0.046 \\ 
        & 1.130 & 0.080 \\ 
        & 1.388 & 0.145 \\ \hline

        \multirow{3}{*}{Location30} & 
        0.570 & 0.041 \\ 
        & 0.724 & 0.076 \\ 
        & 0.801 & 0.094 \\ \hline

        \multirow{3}{*}{Purchase100} & 
        0.550 & 0.087 \\ 
        & 0.625 & 0.110 \\ 
        & 0.729 & 0.156 \\ \hline
        
        \multirow{3}{*}{Texas1000} & 
        0.530 & 0.038 \\  
        & 0.641 & 0.073 \\  
        & 0.734 & 0.107 \\ 

        \bottomrule[1.1pt]
    \end{tabular}
   \label{tabs:The Selected Distances}
    \vspace{0.01pt}
\end{table}

\section*{B. Original and Constructed Distance Distributions of Data Samples in the Target Dataset}
\label{subsec:Original and Constructed Distance Distributions of Data Samples in the Target Dataset}

\begin{figure*}[]
\vspace{0.1pt}
\begin{center}
\subfigure[CIFAR100\_N\_Original]{
\begin{minipage}{0.20\textwidth}
\label{CIFAR100_Normal_V4}
\includegraphics[height=0.6\textwidth,width=1.1\textwidth]{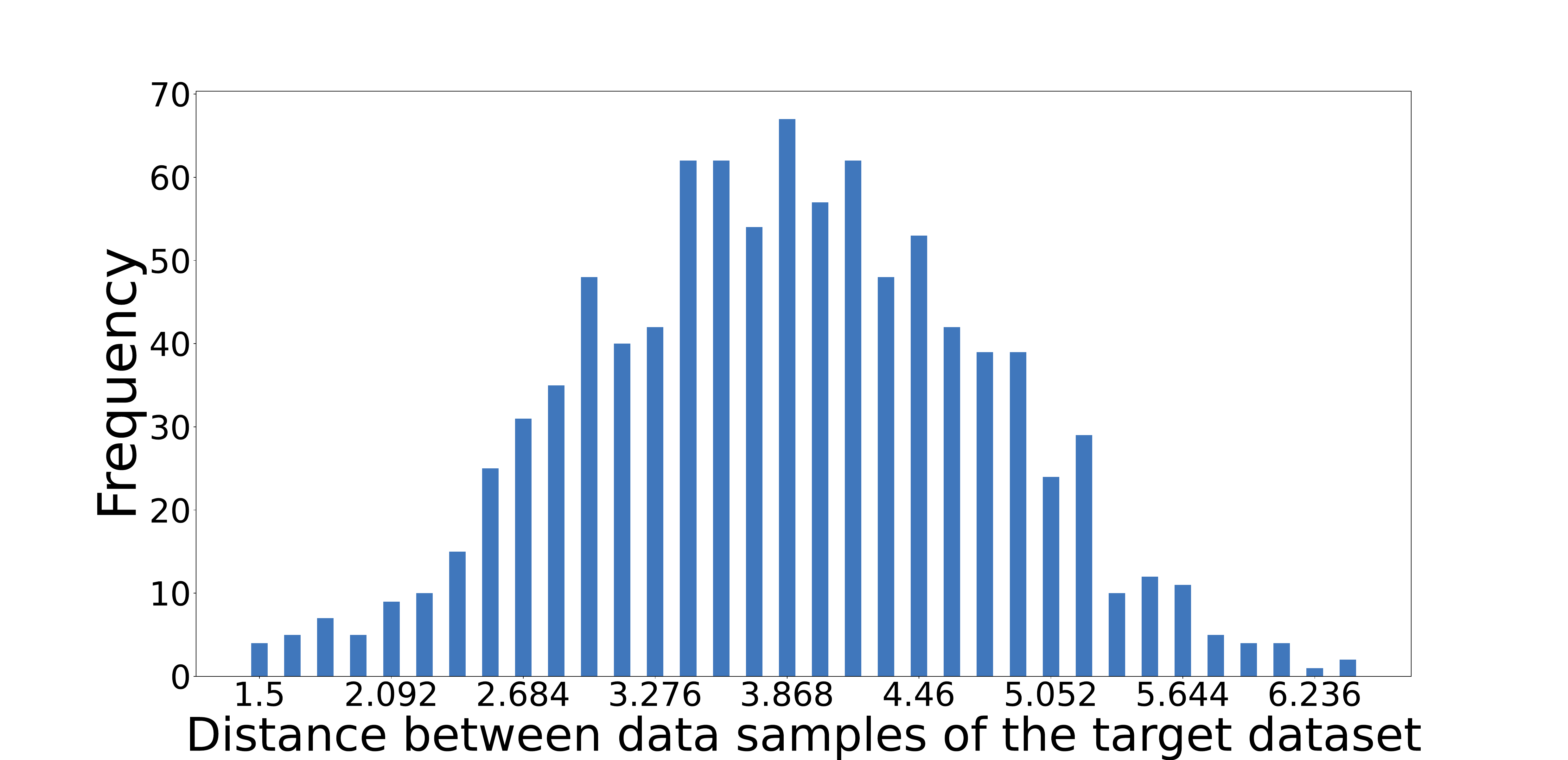}
\end{minipage}}
\subfigure[CIFAR10\_N\_Original]{
\begin{minipage}{0.20\textwidth}
\label{Fig.sub.1}
\includegraphics[height=0.6\textwidth,width=1.1\textwidth]{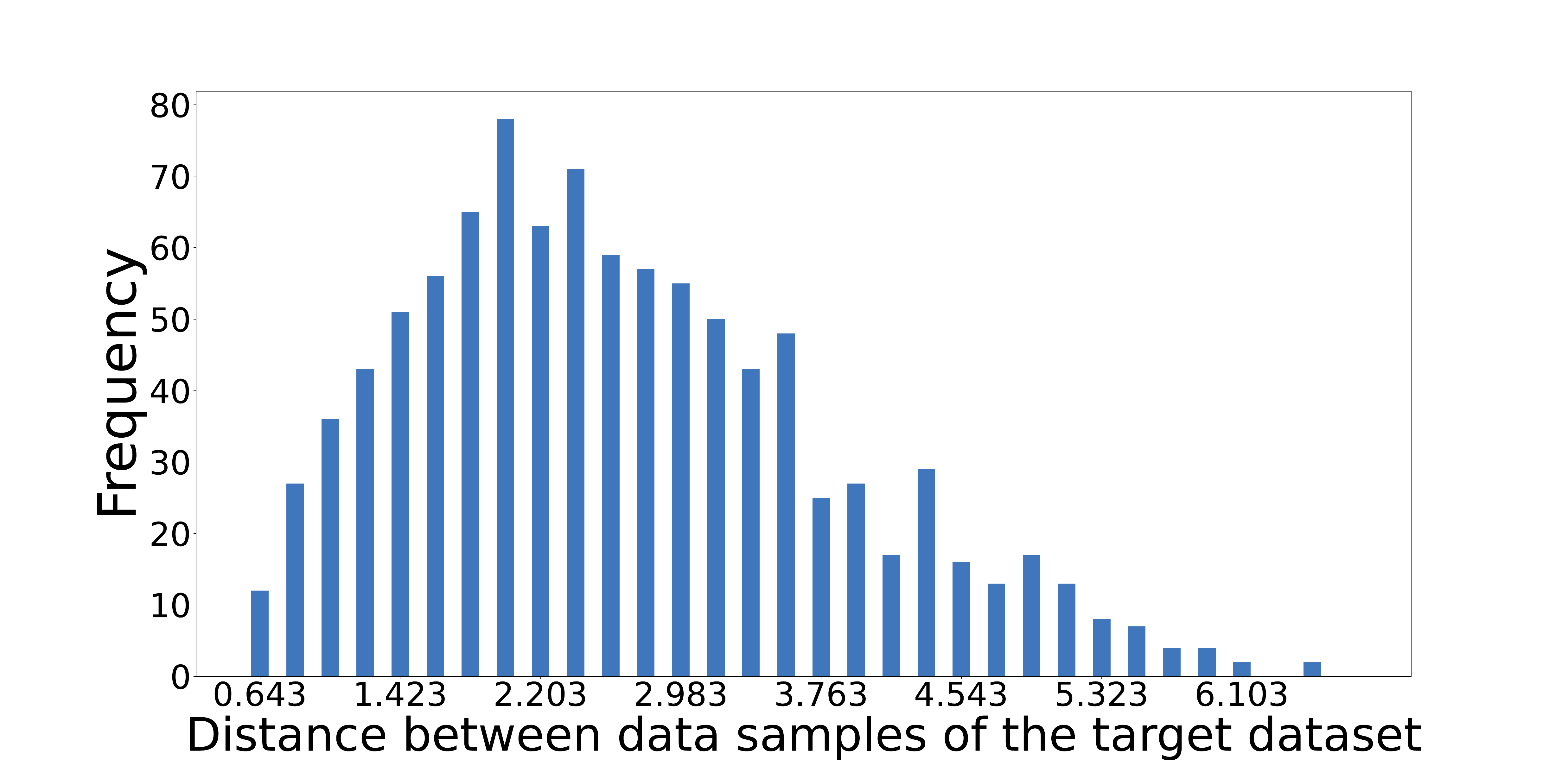}
\end{minipage}}
\subfigure[CH\_MNIST\_N\_Original]{
\begin{minipage}{0.20\textwidth}
\label{Fig.sub.1}
\includegraphics[height=0.6\textwidth,width=1.1\textwidth]{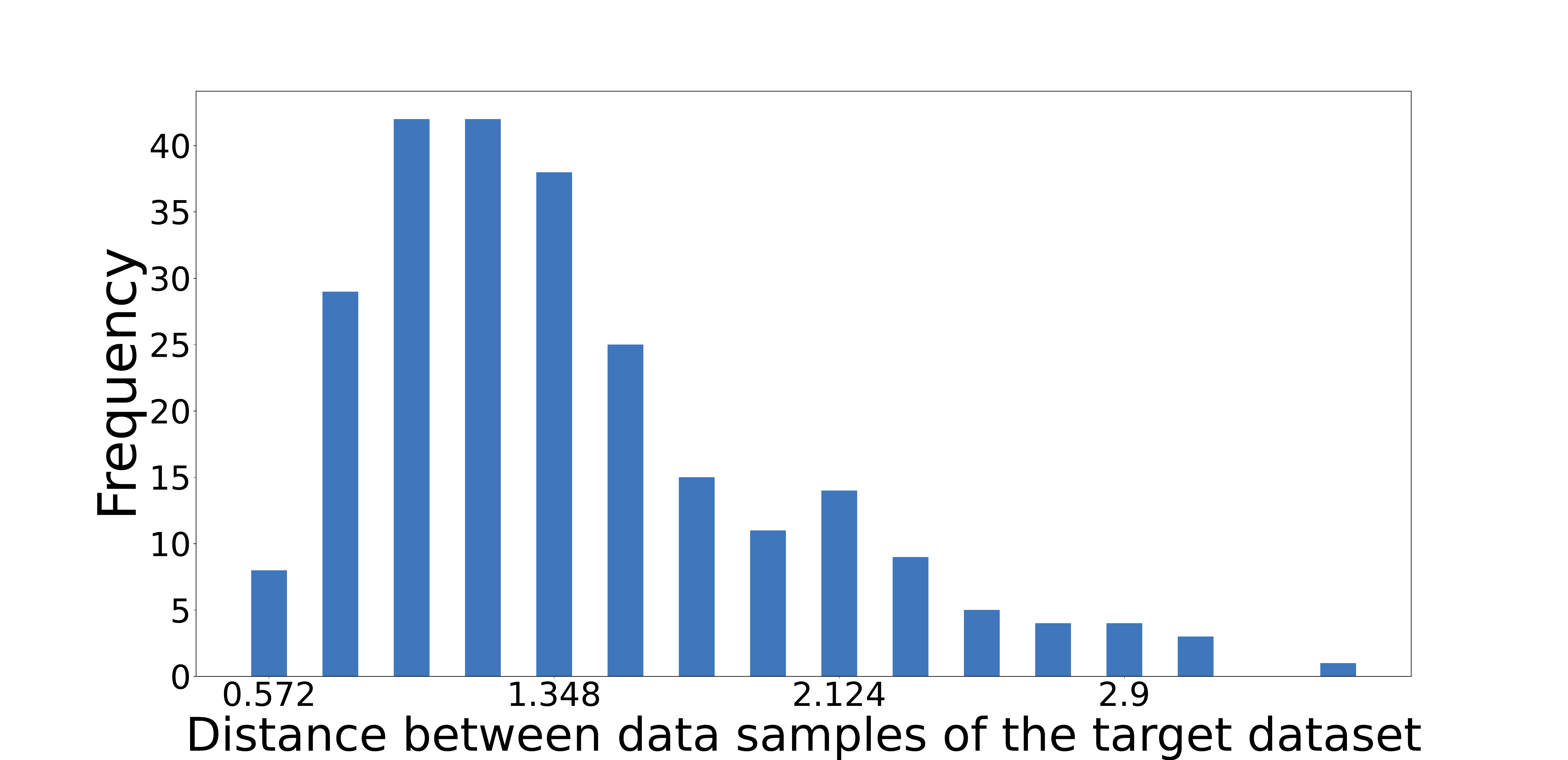}
\end{minipage}}
\subfigure[ImageNet\_N\_Original]{
\begin{minipage}{0.20\textwidth}
\label{Fig.sub.1}
\includegraphics[height=0.6\textwidth,width=1.1\textwidth]{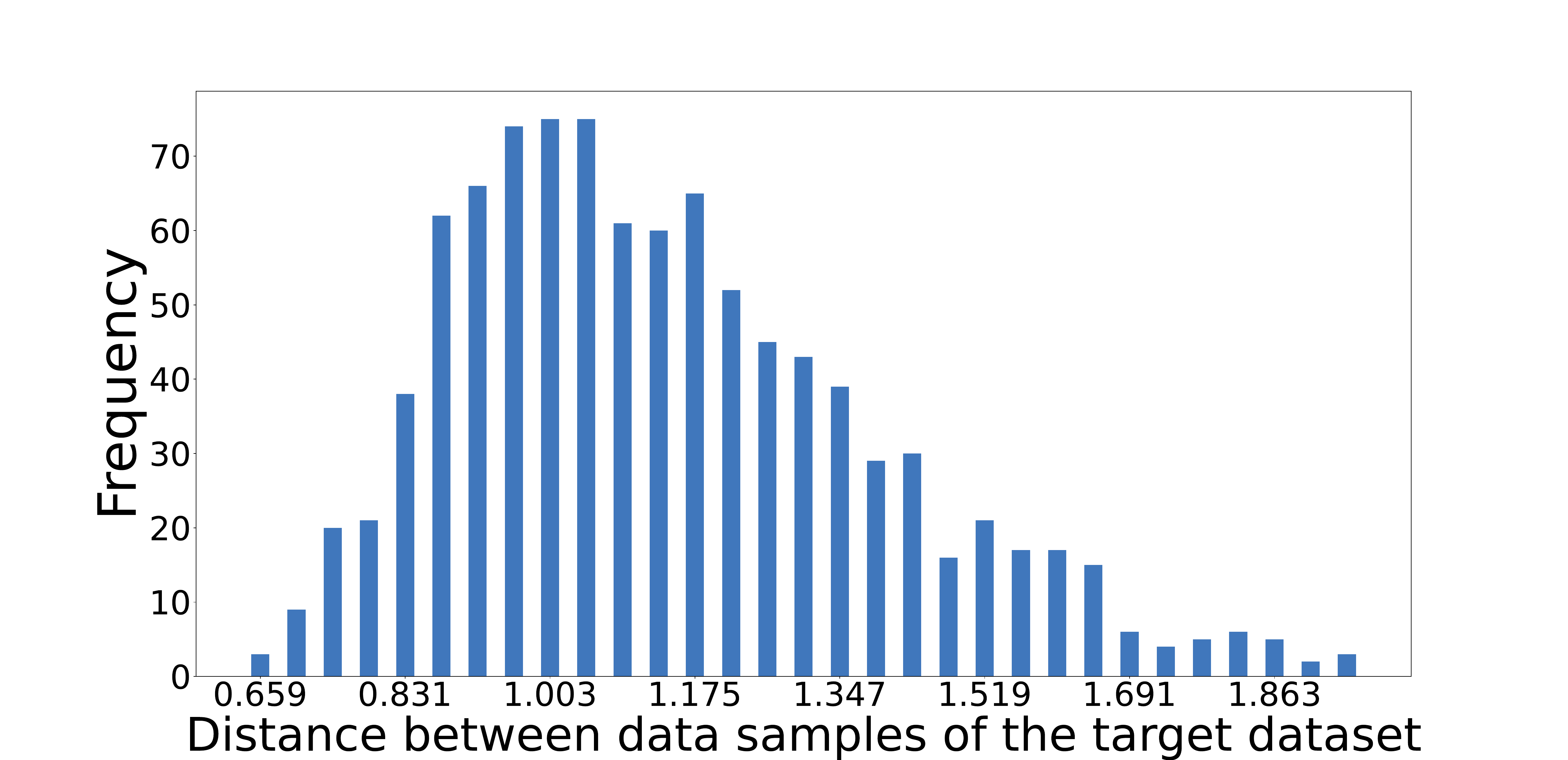}
\end{minipage}}

\subfigure[Location30\_N\_Original]{
\begin{minipage}{0.20\textwidth}
\label{Fig.sub.1}
\includegraphics[height=0.6\textwidth,width=1.1\textwidth]{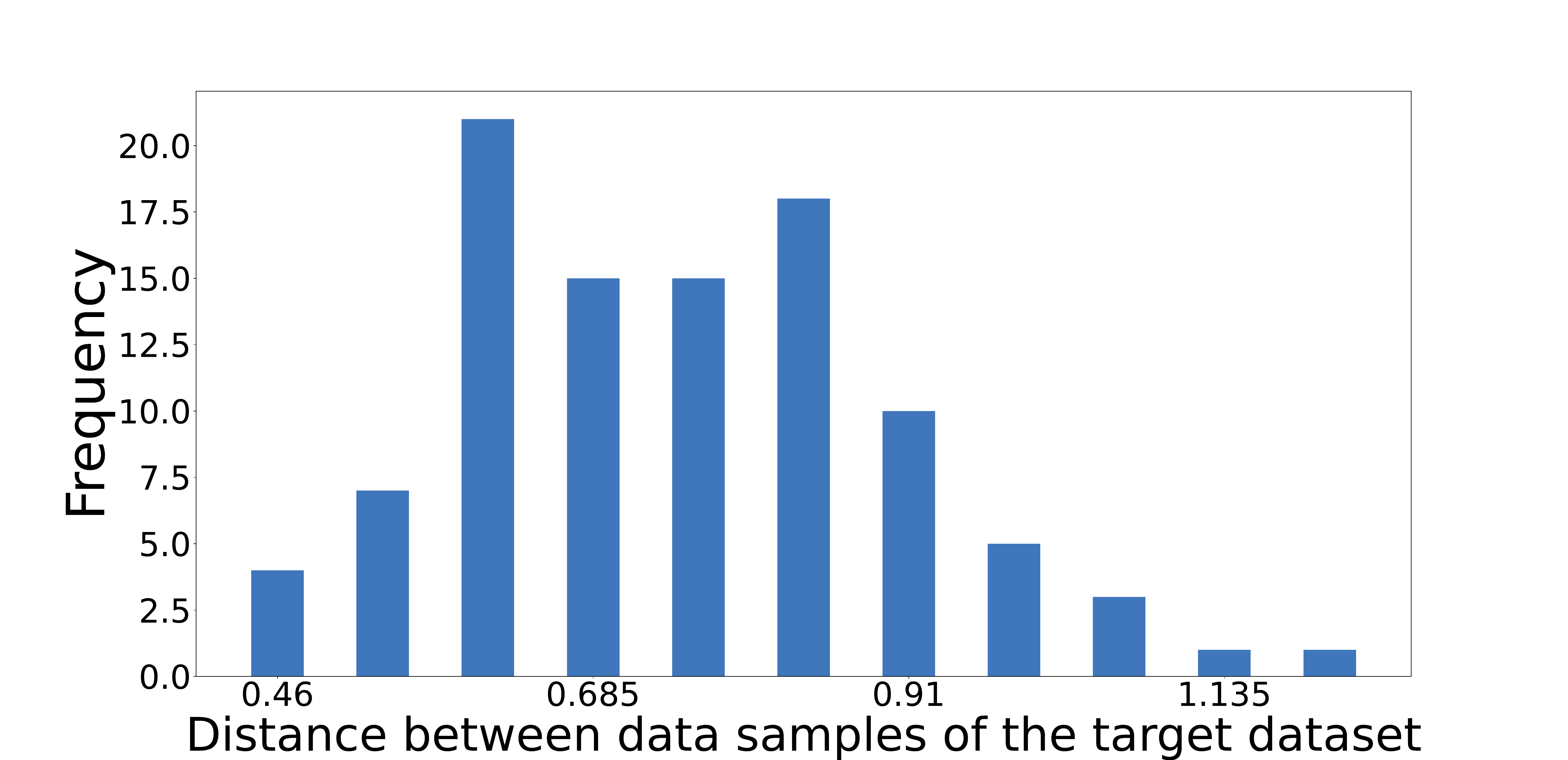}
\end{minipage}}
\subfigure[Purchase100\_N\_Original]{
\begin{minipage}{0.20\textwidth}
\label{Fig.sub.1}
\includegraphics[height=0.6\textwidth,width=1.1\textwidth]{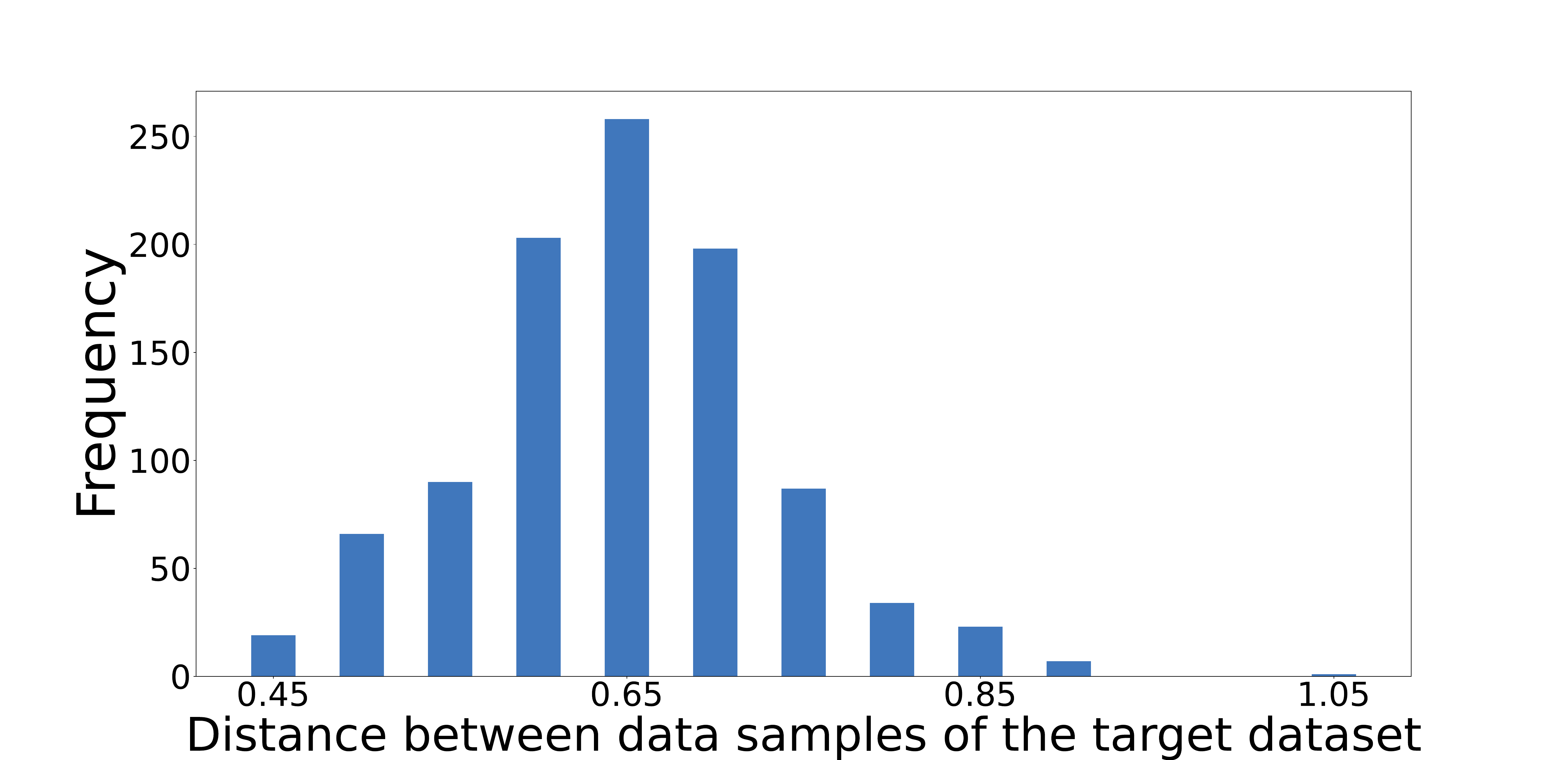}
\end{minipage}}
\subfigure[Texas100\_N\_Original]{
\begin{minipage}{0.20\textwidth}
\label{Fig.sub.1}
\includegraphics[height=0.6\textwidth,width=1.1\textwidth]{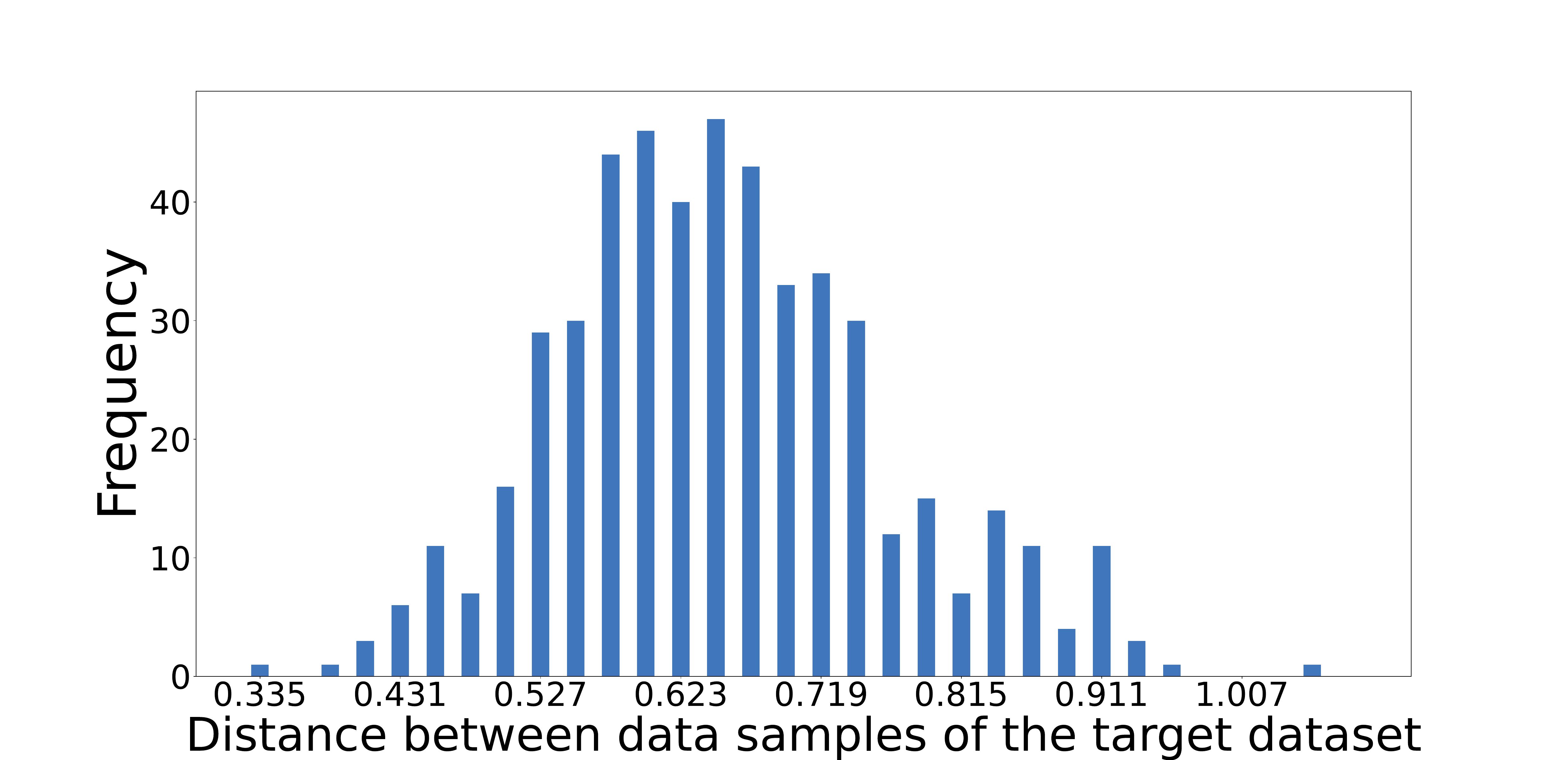}
\end{minipage}}
\subfigure[CIFAR100\_U\_Original]{
\begin{minipage}{0.20\textwidth}
\label{CIFAR100_Uniform_V4}
\includegraphics[height=0.6\textwidth,width=1.1\textwidth]{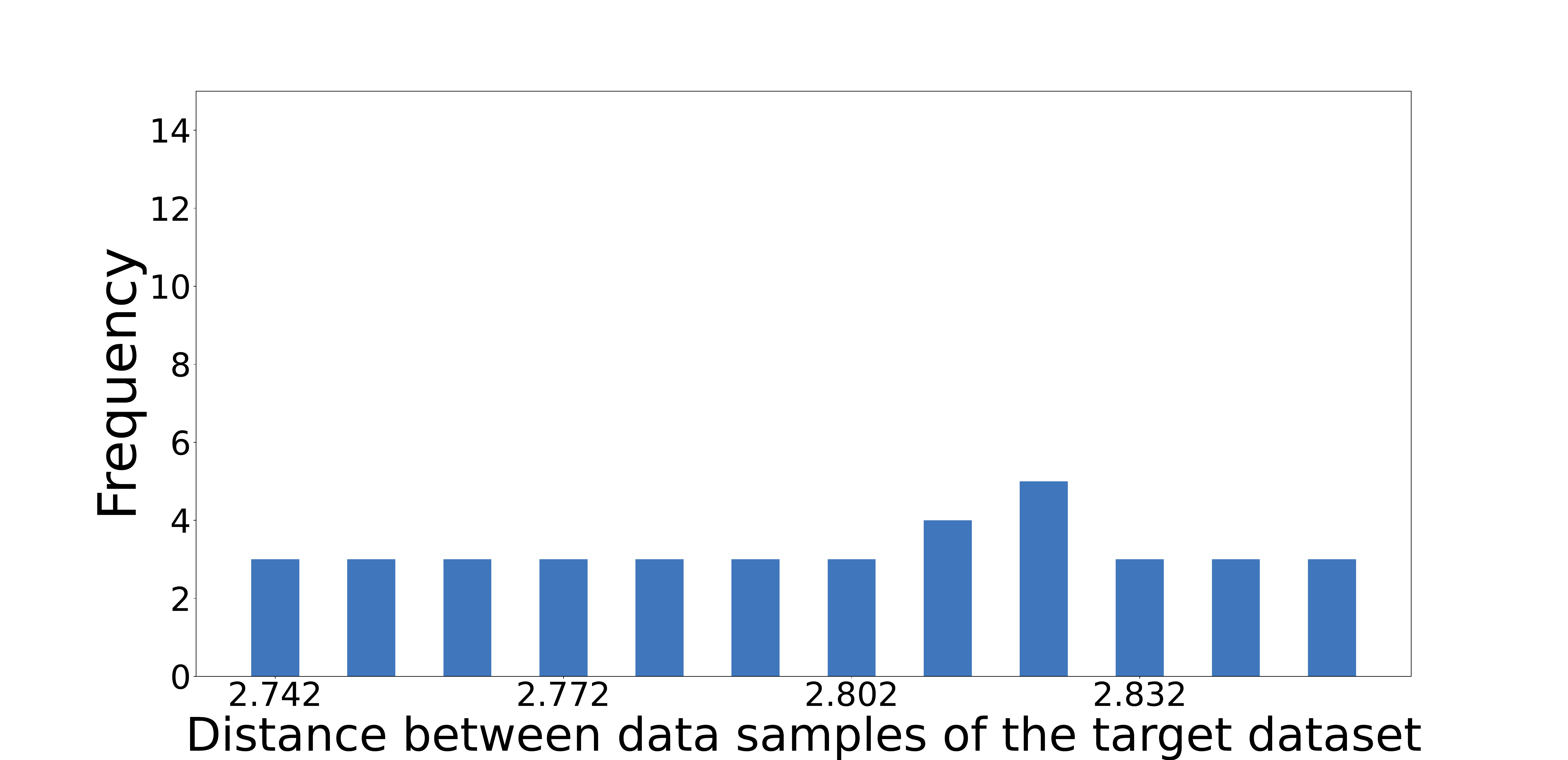}
\end{minipage}}

\subfigure[CIFAR10\_U\_Original]{
\begin{minipage}{0.20\textwidth}
\label{Fig.sub.1}
\includegraphics[height=0.6\textwidth,width=1.1\textwidth]{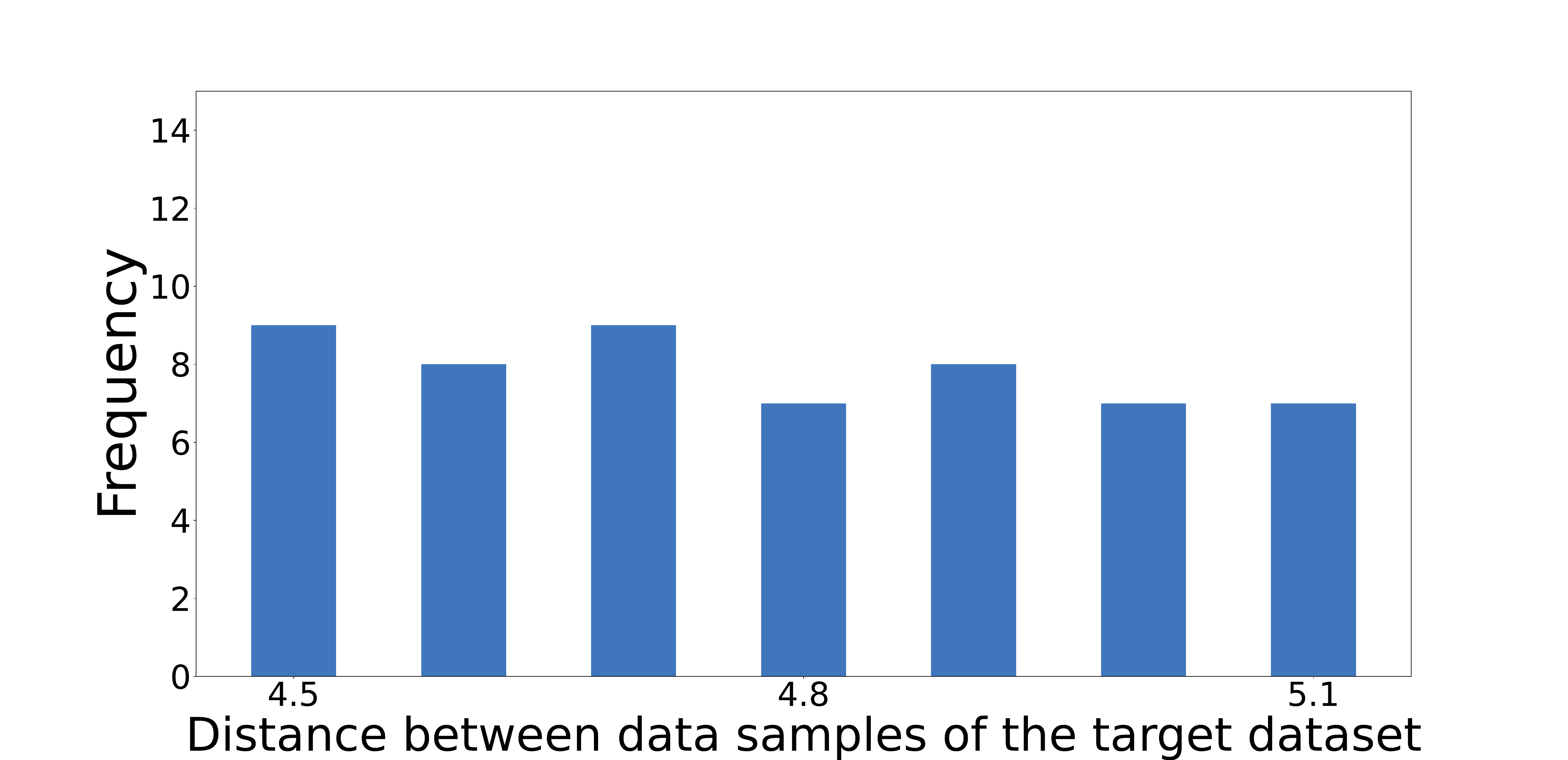}
\end{minipage}}
\subfigure[CH\_MNIST\_U\_Original]{
\begin{minipage}{0.20\textwidth}
\label{Fig.sub.1}
\includegraphics[height=0.6\textwidth,width=1.1\textwidth]{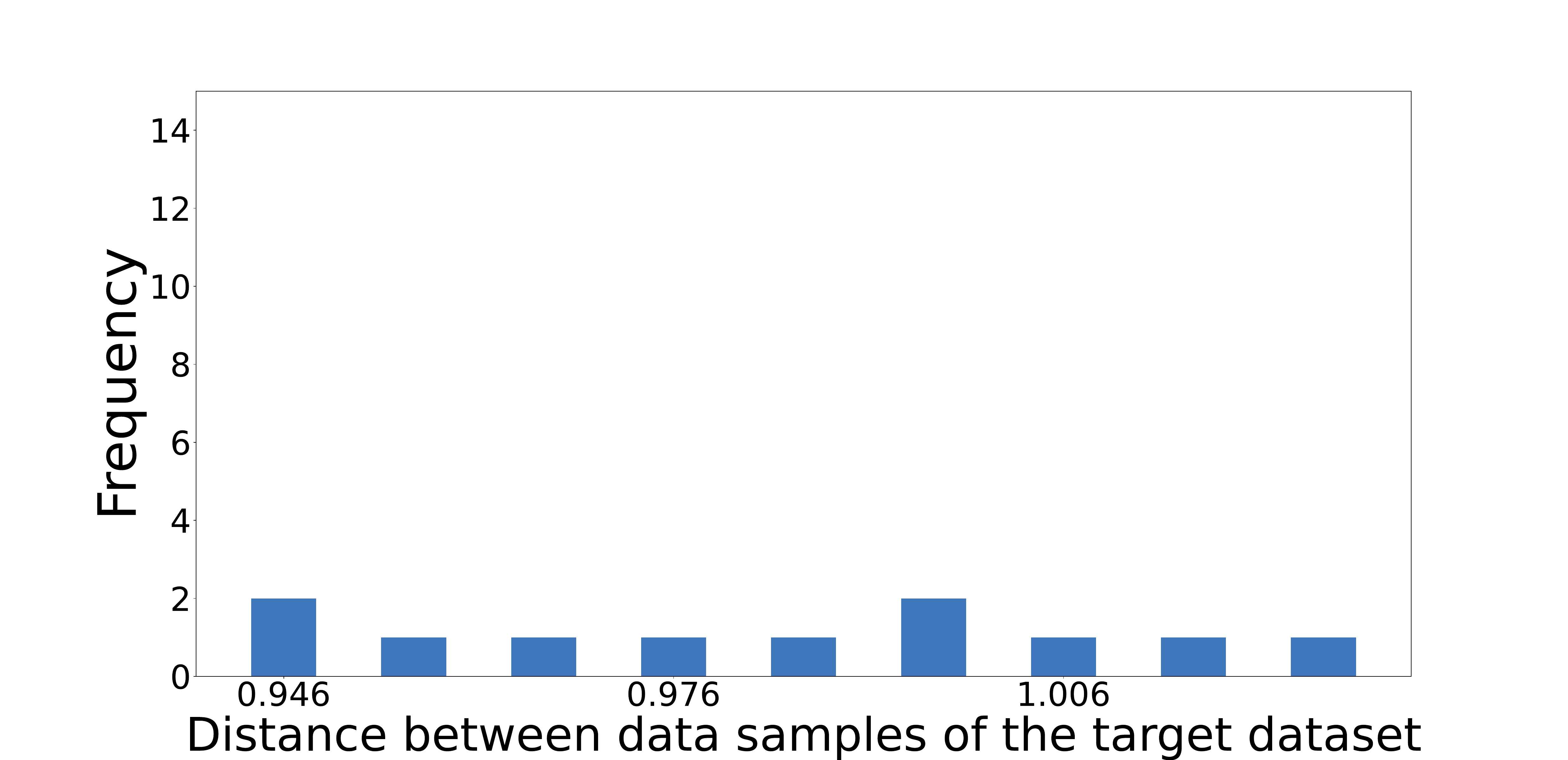}
\end{minipage}}
\subfigure[ImageNet\_U\_Original]{
\begin{minipage}{0.20\textwidth}
\label{Fig.sub.1}
\includegraphics[height=0.6\textwidth,width=1.1\textwidth]{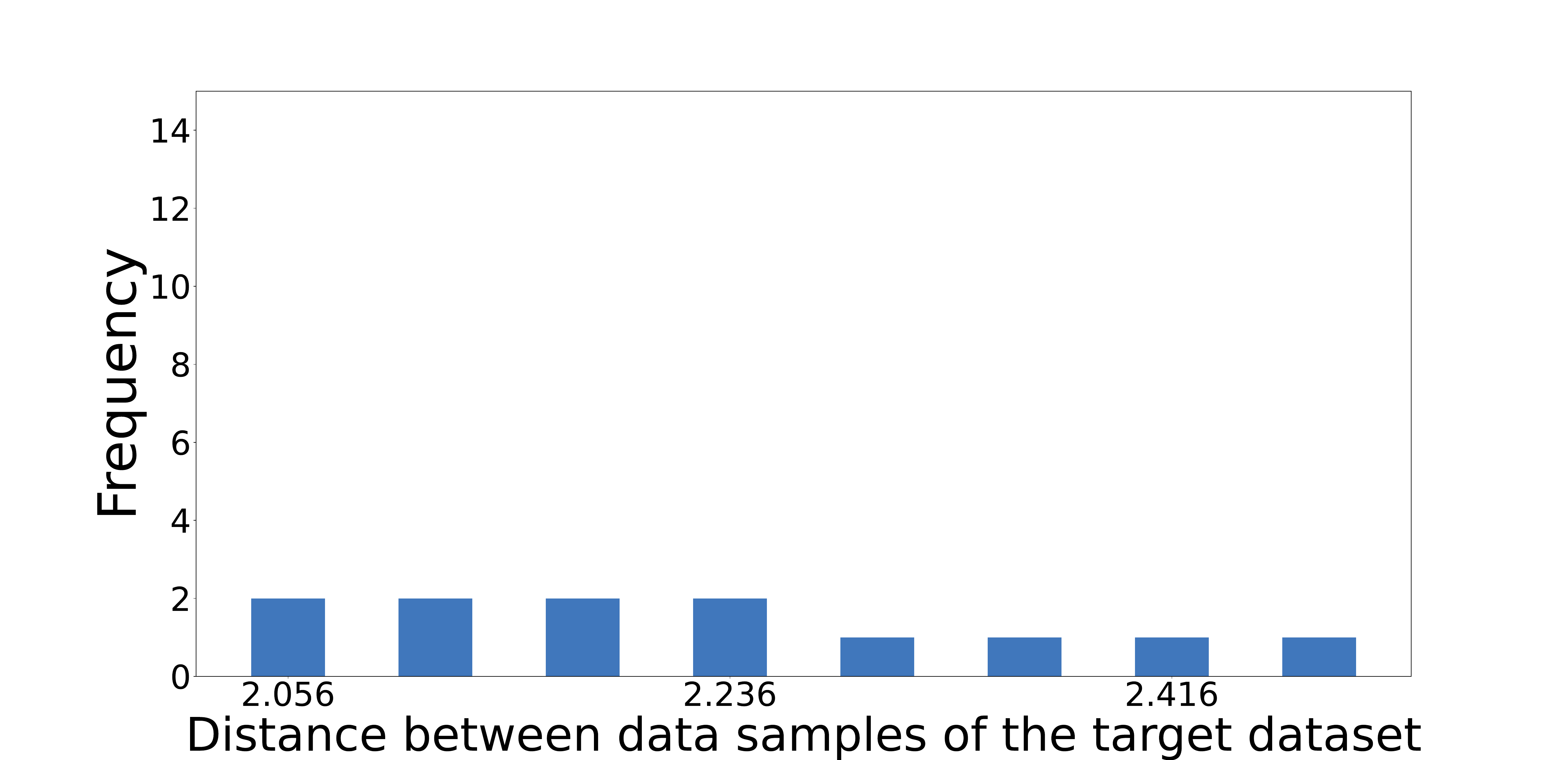}
\end{minipage}}
\subfigure[Location30\_U\_Original]{
\begin{minipage}{0.20\textwidth}
\label{Fig.sub.1}
\includegraphics[height=0.6\textwidth,width=1.1\textwidth]{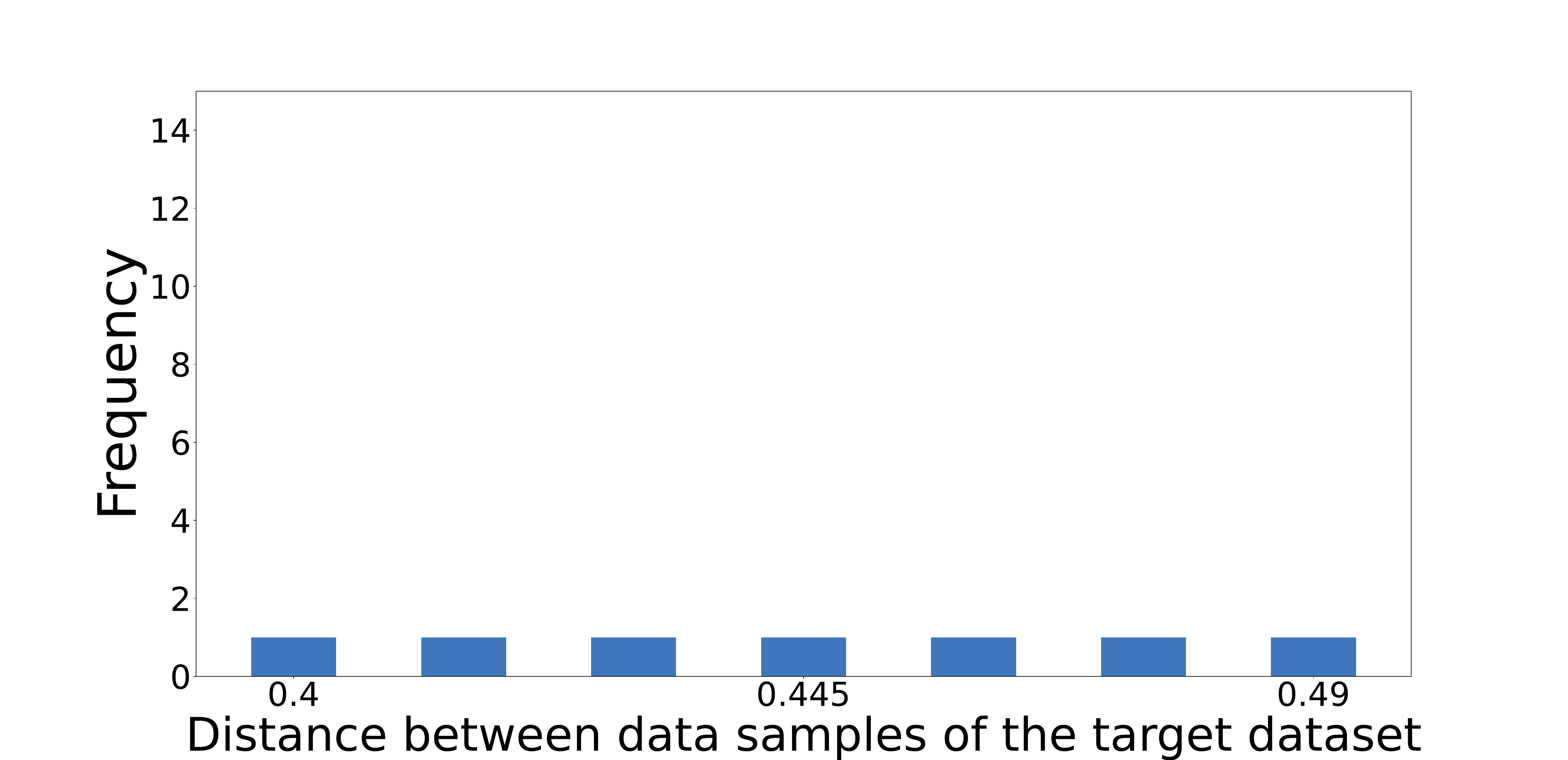}
\end{minipage}}

\subfigure[Purchase100\_U\_Original]{
\begin{minipage}{0.20\textwidth}
\label{Fig.sub.1}
\includegraphics[height=0.6\textwidth,width=1.1\textwidth]{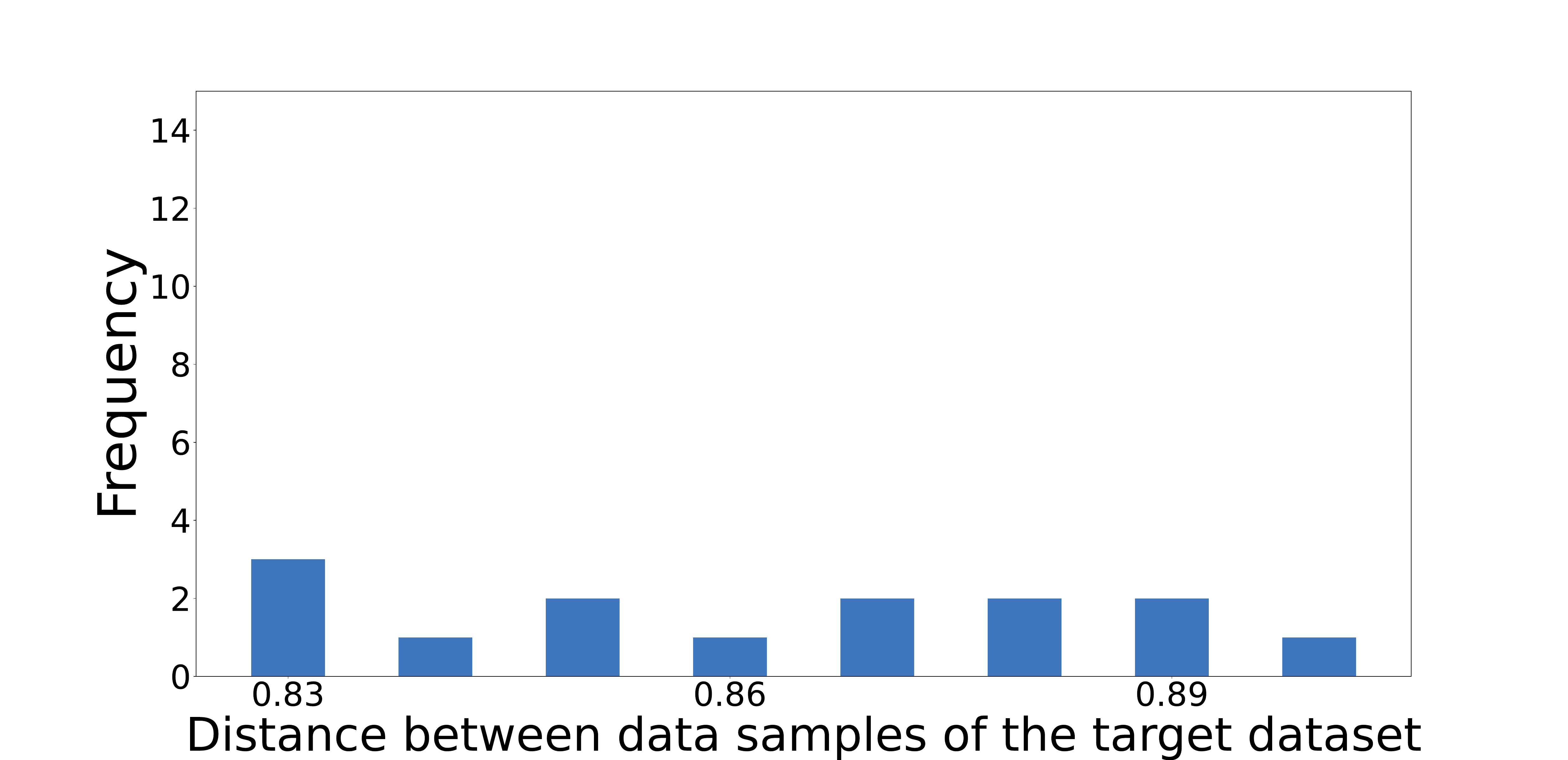}
\end{minipage}}
\subfigure[Texas100\_U\_Original]{
\begin{minipage}{0.20\textwidth}
\label{Fig.sub.1}
\includegraphics[height=0.6\textwidth,width=1.1\textwidth]{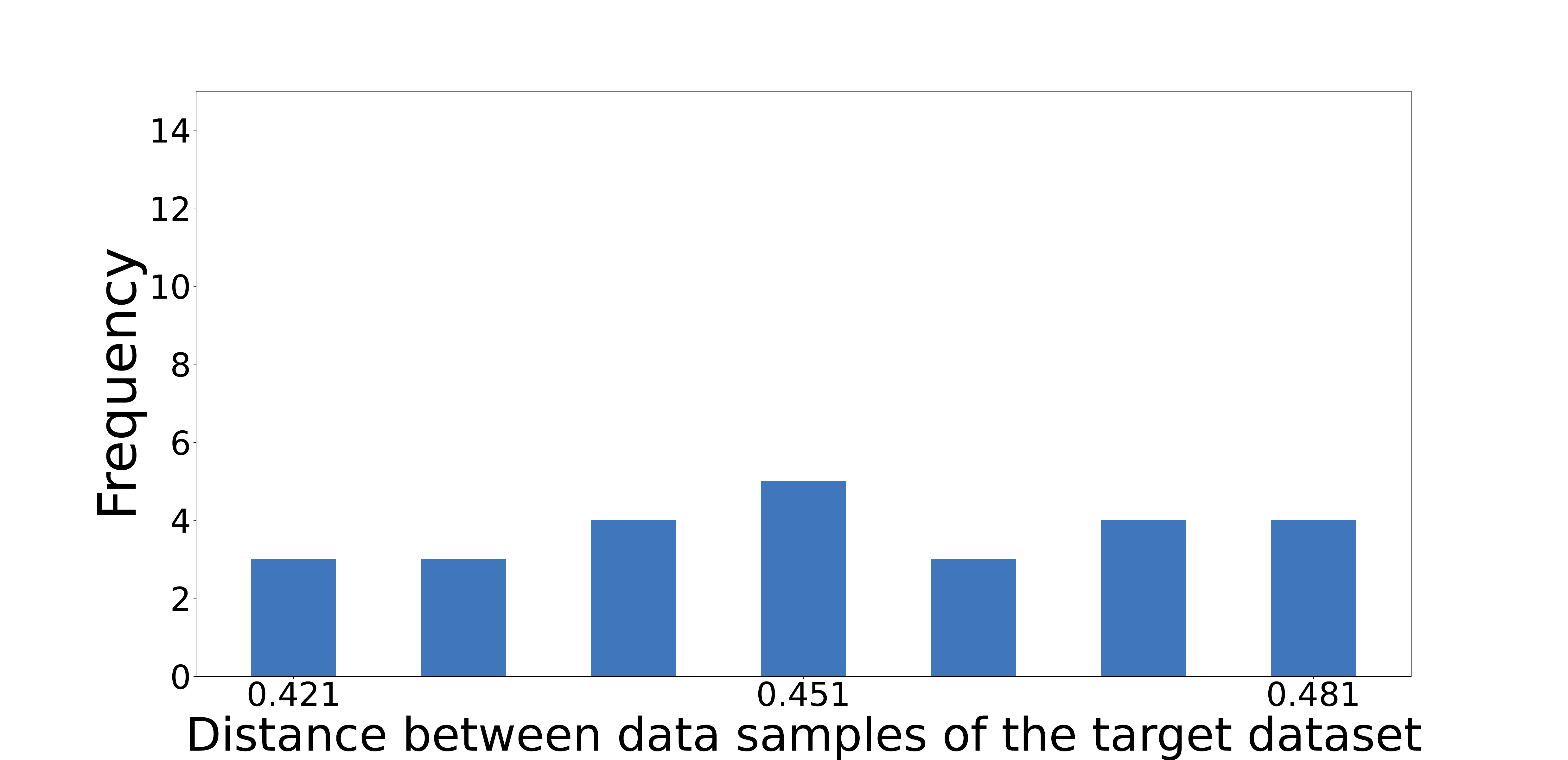}
\end{minipage}}

\end{center}
\caption{\label{figs:Fig33 Original} The Original Distance Distribution of Data Samples in the Target Dataset. Inside (a)-(g) are the original distance distributions of data samples in the target CIFAR100, CIFAR10, CH\_MNIST, ImageNet, Location30, Purchase100 and Texas100 datasets approximately obey normal distributions, and (h)-(n) are the original distance distributions of data samples in the target seven datasets approximately obey uniform distributions.}
\vspace{0.1pt}
\end{figure*}

\begin{figure*}[]
\vspace{0.1pt}
\begin{center}
\subfigure[C100\_N\_3.813\_0.085\_20\%]{
\begin{minipage}{0.20\textwidth}
\label{CIFAR100_N_3.813_d1_20}
\includegraphics[height=0.6\textwidth,width=1.1\textwidth]{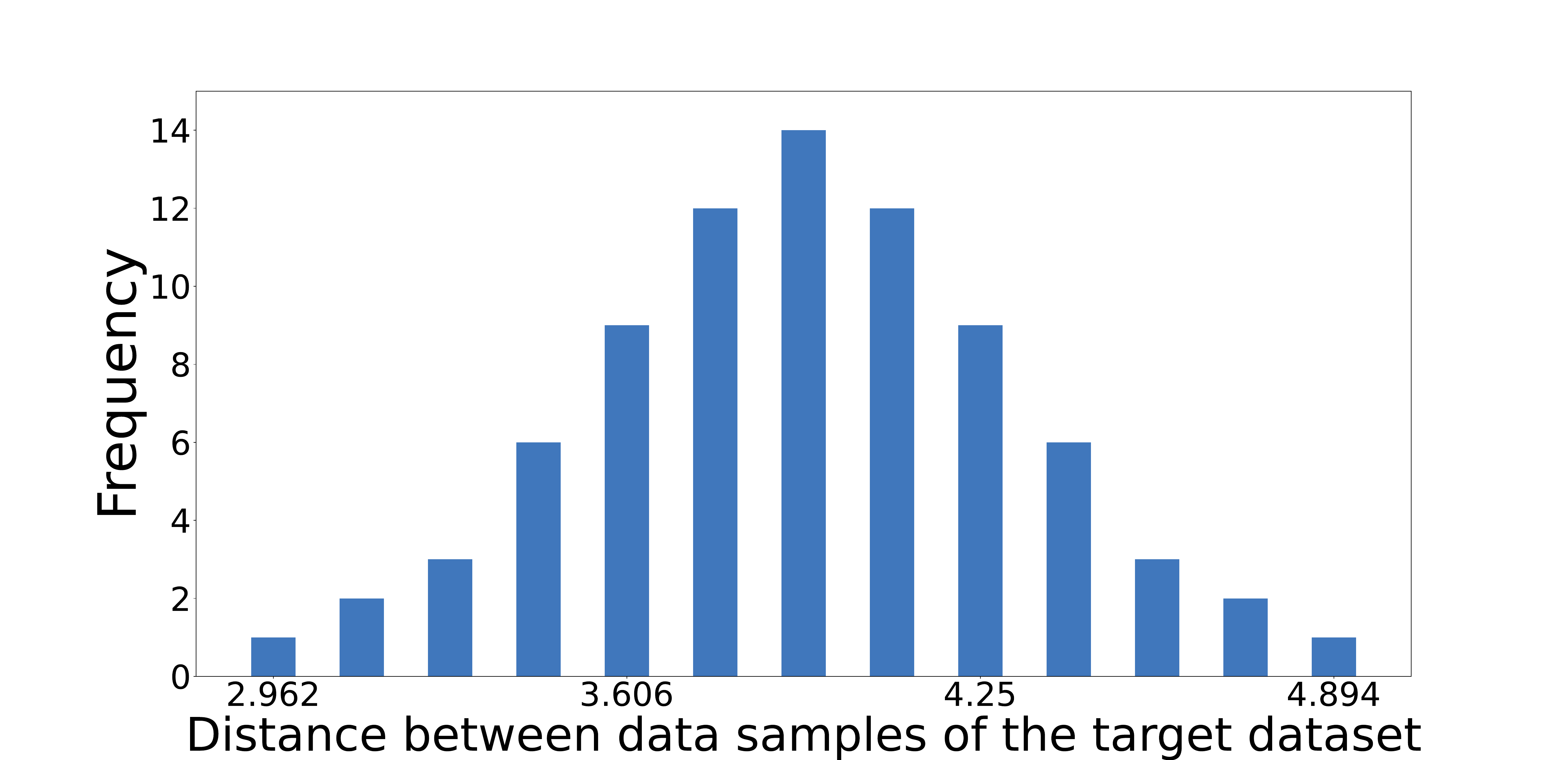}
\end{minipage}}
\subfigure[C10\_N\_2.501\_0.155\_20\%]{
\begin{minipage}{0.20\textwidth}
\label{Fig.sub.1}
\includegraphics[height=0.6\textwidth,width=1.1\textwidth]{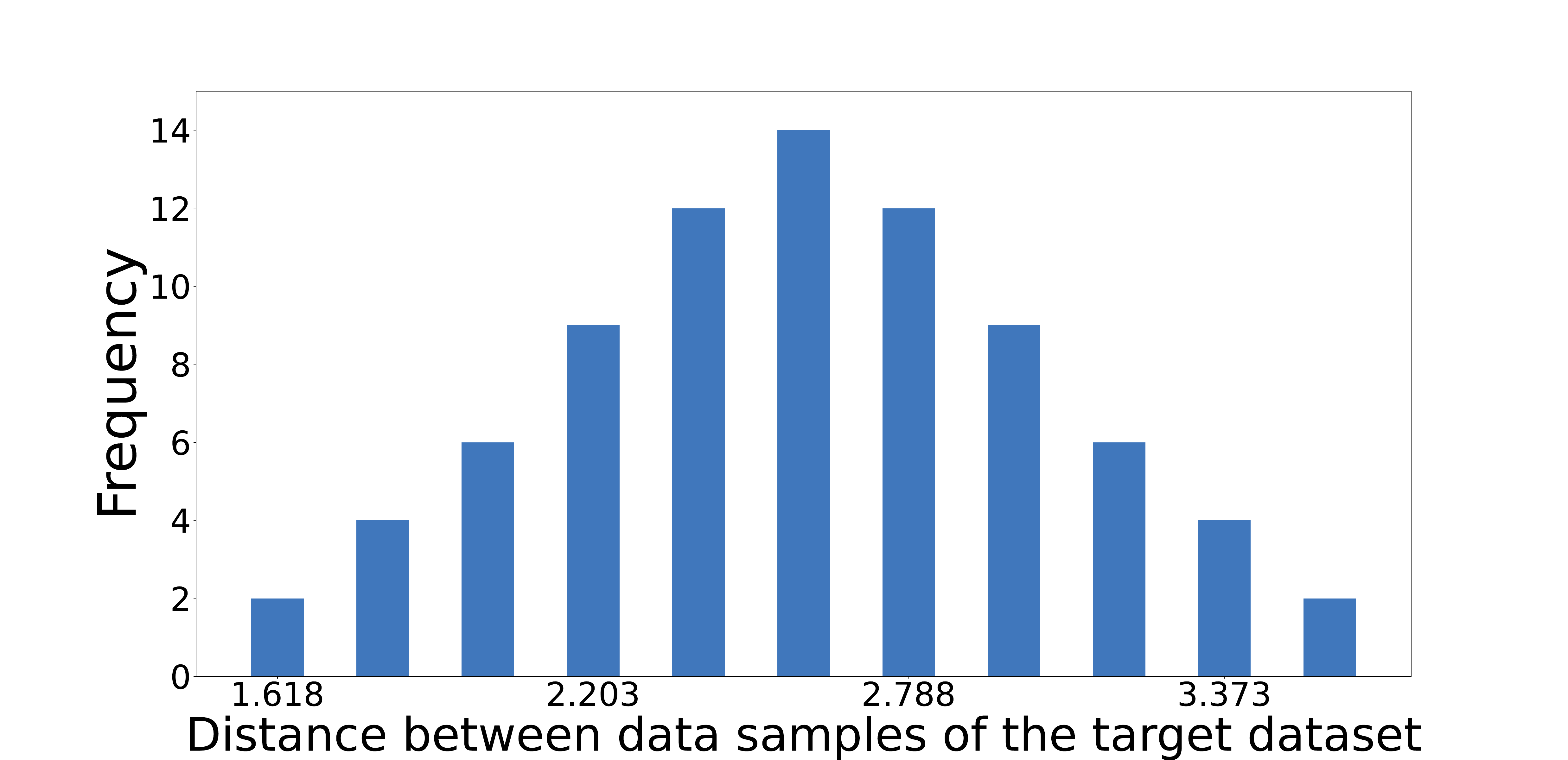}
\end{minipage}}
\subfigure[CHMN\_N\_1.72\_0.083\_20\%]{
\begin{minipage}{0.20\textwidth}
\label{Fig.sub.1}
\includegraphics[height=0.6\textwidth,width=1.1\textwidth]{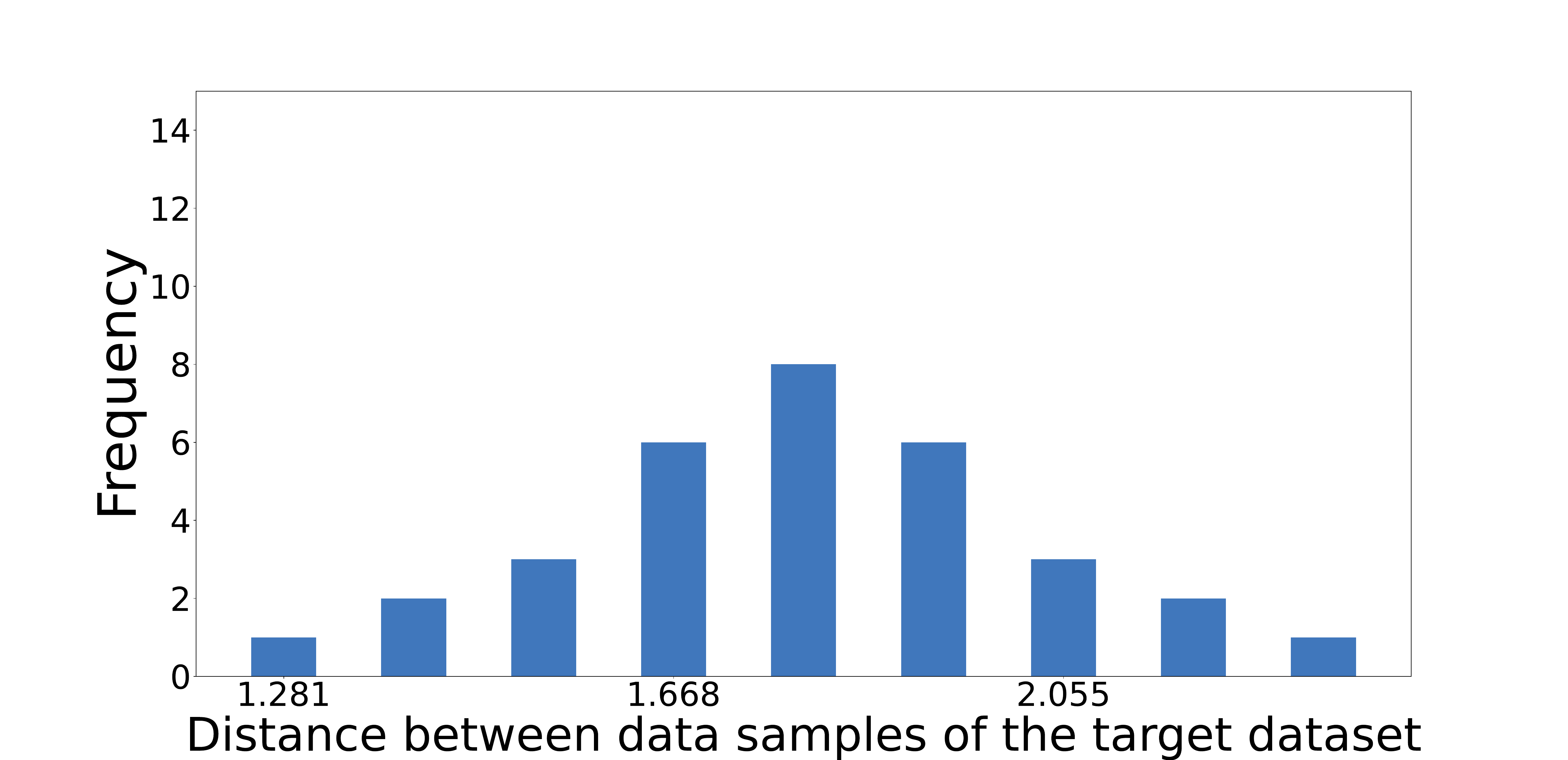}
\end{minipage}}
\subfigure[ImageNet\_N\_1.13\_0.046\_20\%]{
\begin{minipage}{0.20\textwidth}
\label{Fig.sub.1}
\includegraphics[height=0.6\textwidth,width=1.1\textwidth]{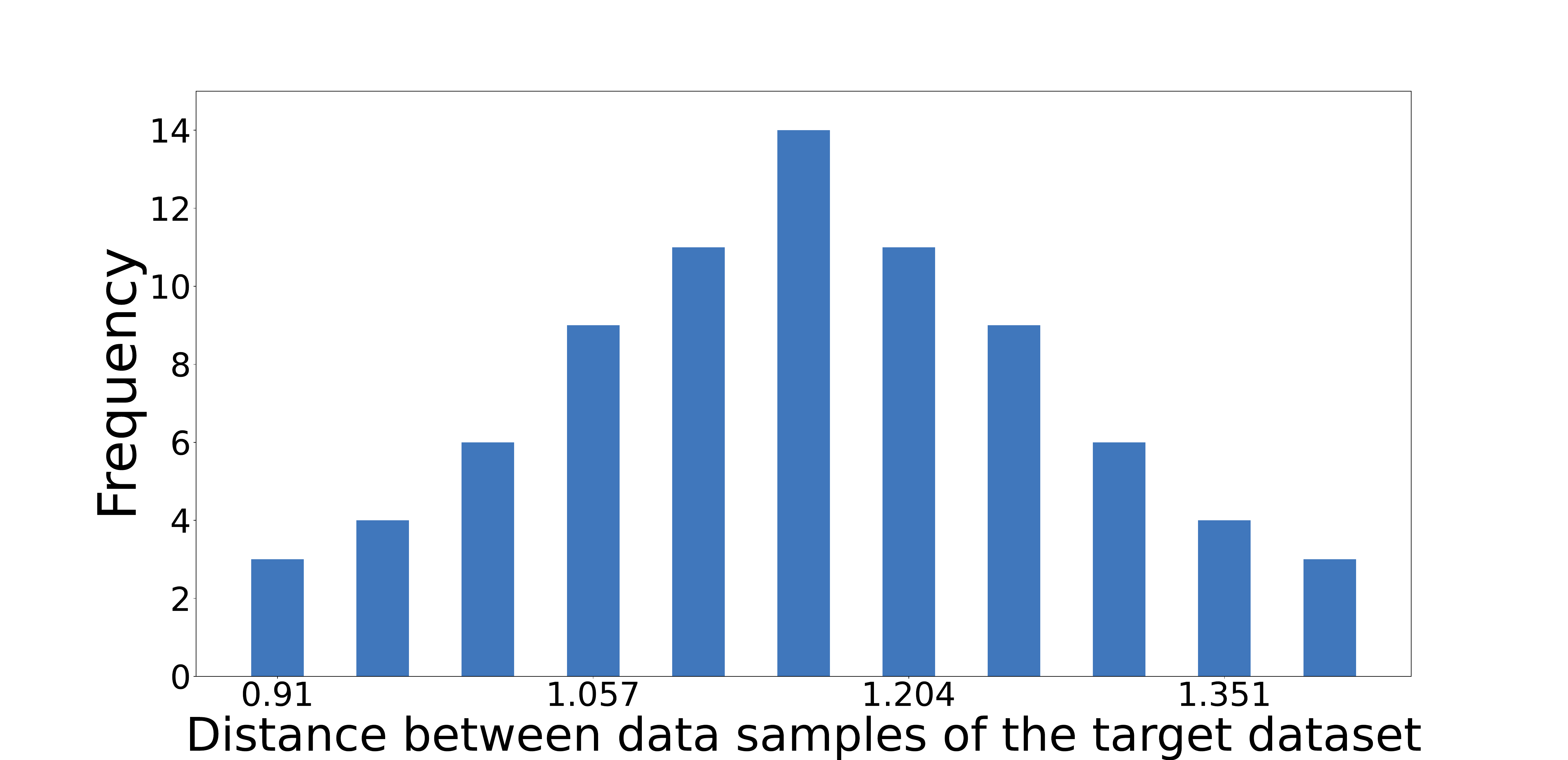}
\end{minipage}}

\subfigure[L30\_N\_0.724\_0.076\_4\%]{
\begin{minipage}{0.20\textwidth}
\label{Fig.sub.1}
\includegraphics[height=0.6\textwidth,width=1.1\textwidth]{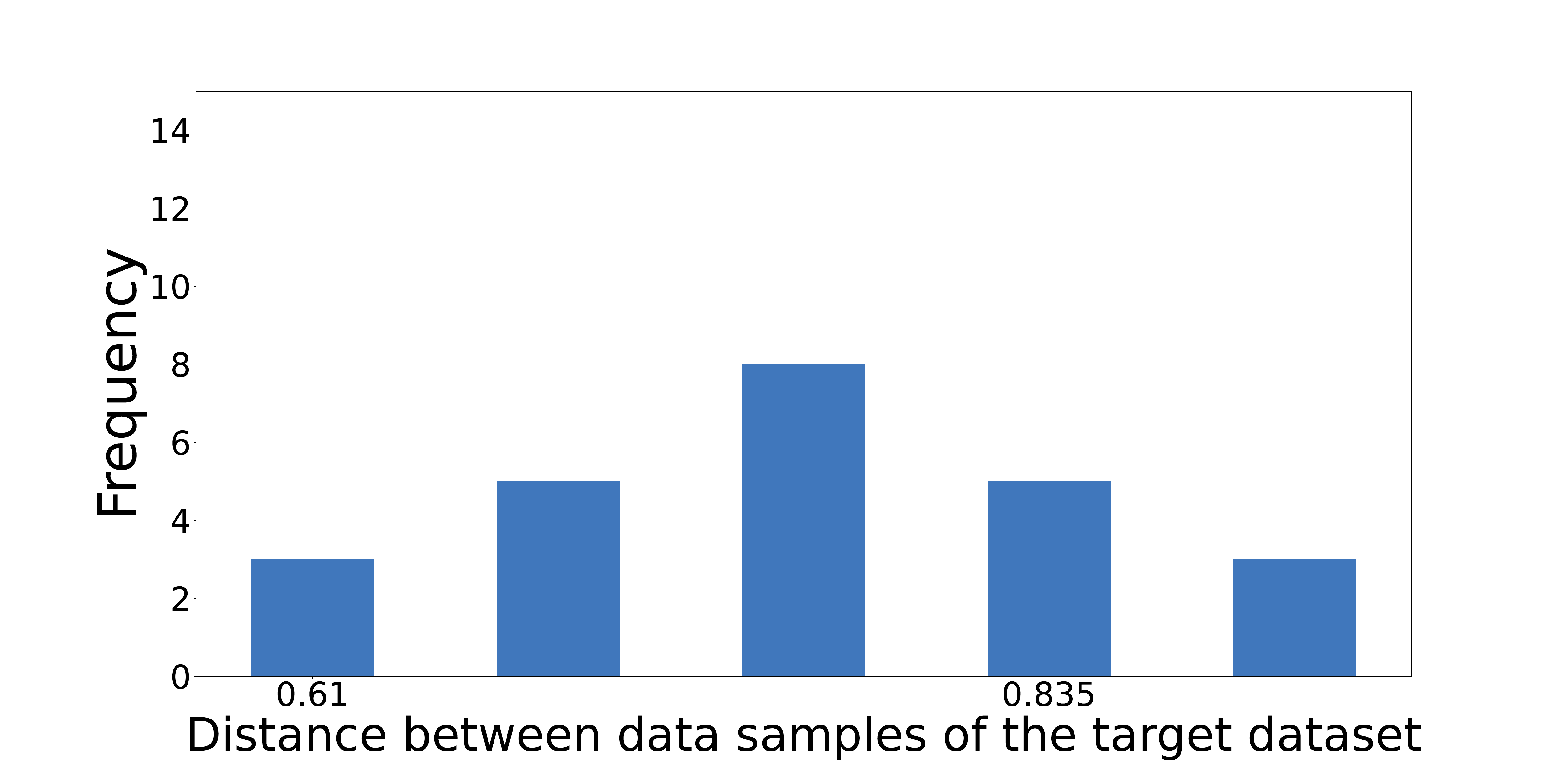}
\end{minipage}}
\subfigure[P100\_N\_0.625\_0.110\_2\%]{
\begin{minipage}{0.20\textwidth}
\label{Fig.sub.1}
\includegraphics[height=0.6\textwidth,width=1.1\textwidth]{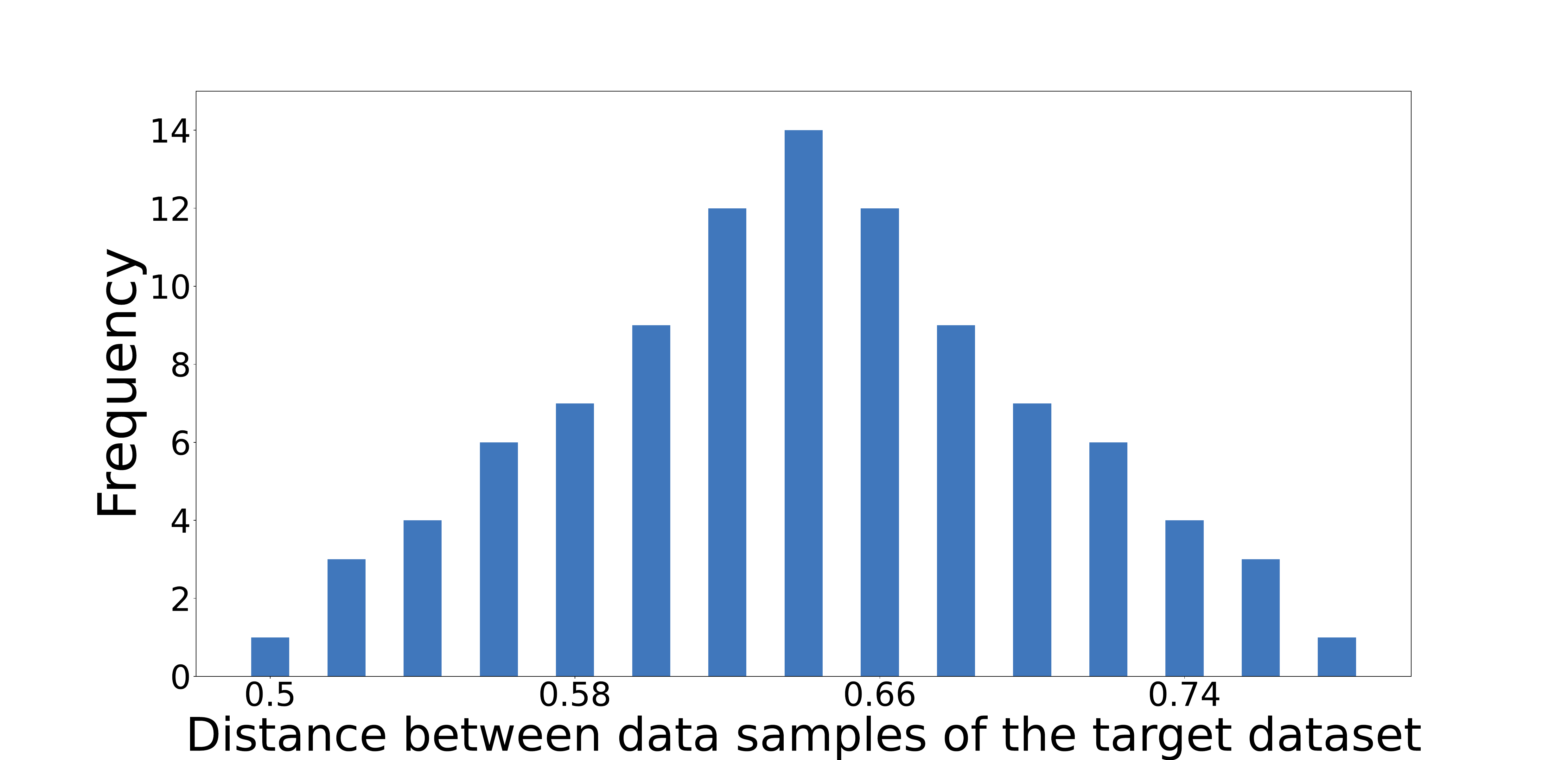}
\end{minipage}}
\subfigure[T100\_N\_0.641\_0.073\_2\%]{
\begin{minipage}{0.20\textwidth}
\label{Fig.sub.1}
\includegraphics[height=0.6\textwidth,width=1.1\textwidth]{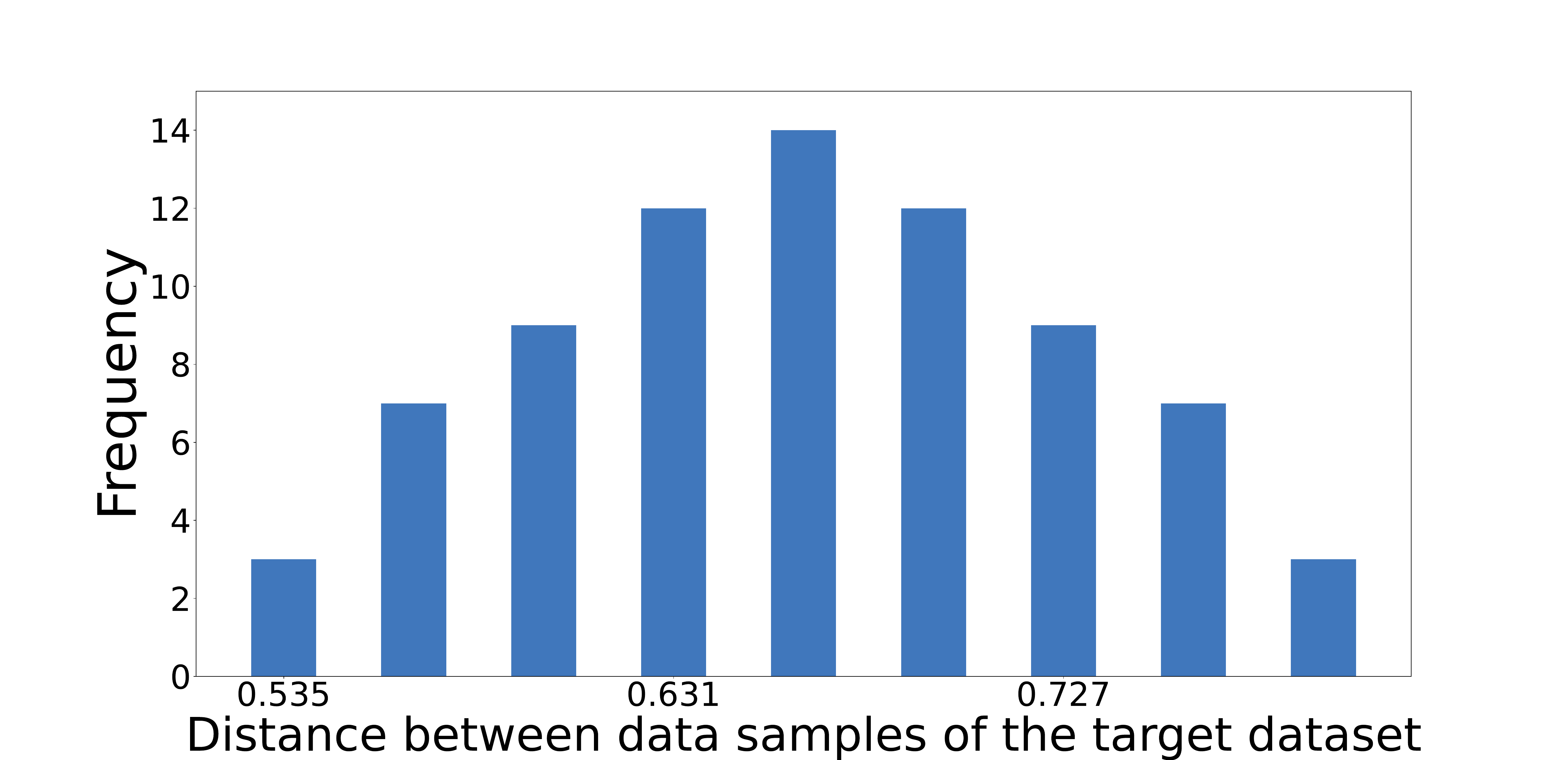}
\end{minipage}}
\subfigure[C100\_U\_3.813\_0.085\_20\%]{
\begin{minipage}{0.20\textwidth}
\label{CIFAR100_Uniform_V4}
\includegraphics[height=0.6\textwidth,width=1.1\textwidth]{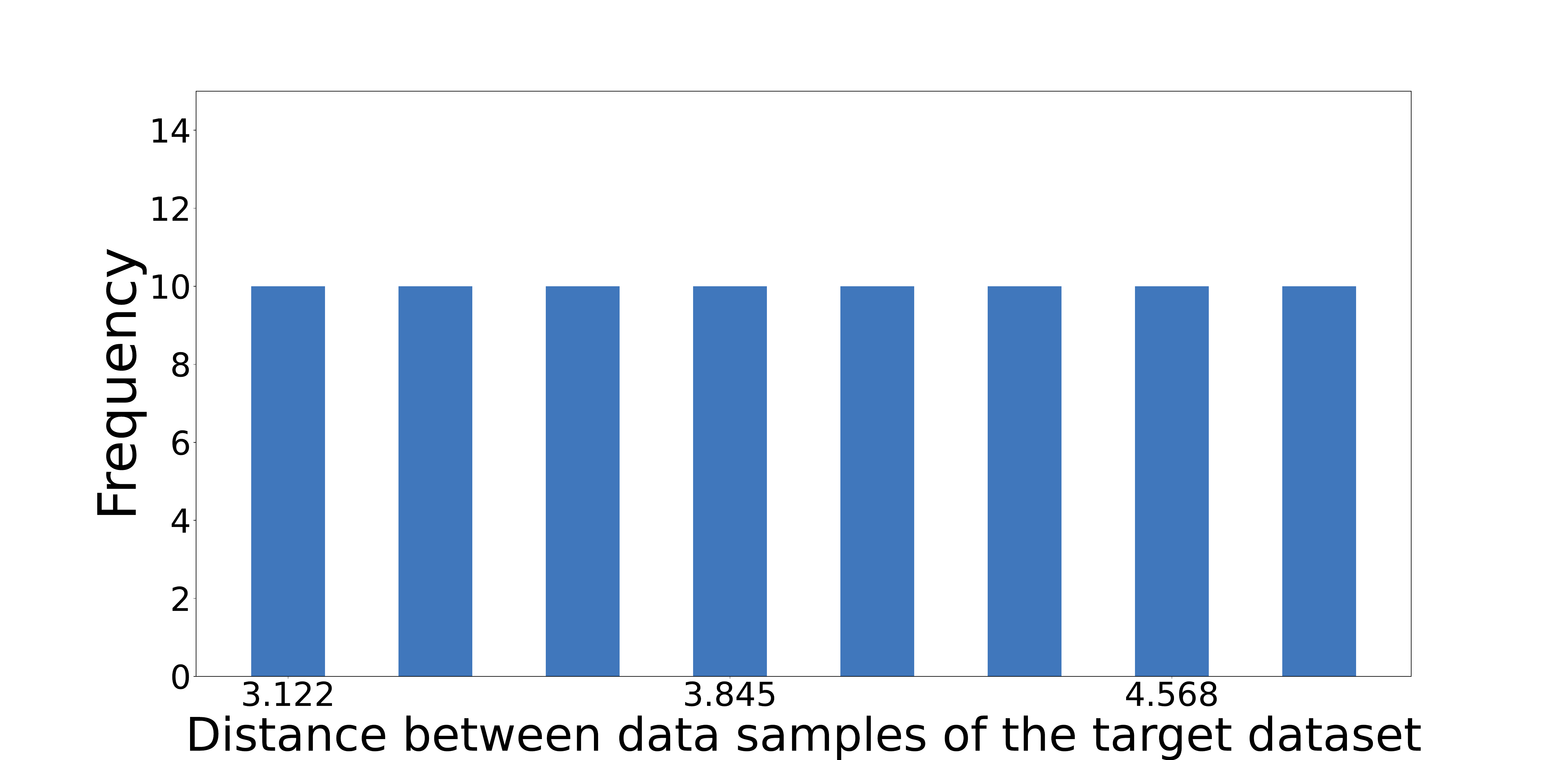}
\end{minipage}}

\subfigure[C10\_U\_2.501\_0.155\_20\%]{
\begin{minipage}{0.20\textwidth}
\label{Fig.sub.1}
\includegraphics[height=0.6\textwidth,width=1.1\textwidth]{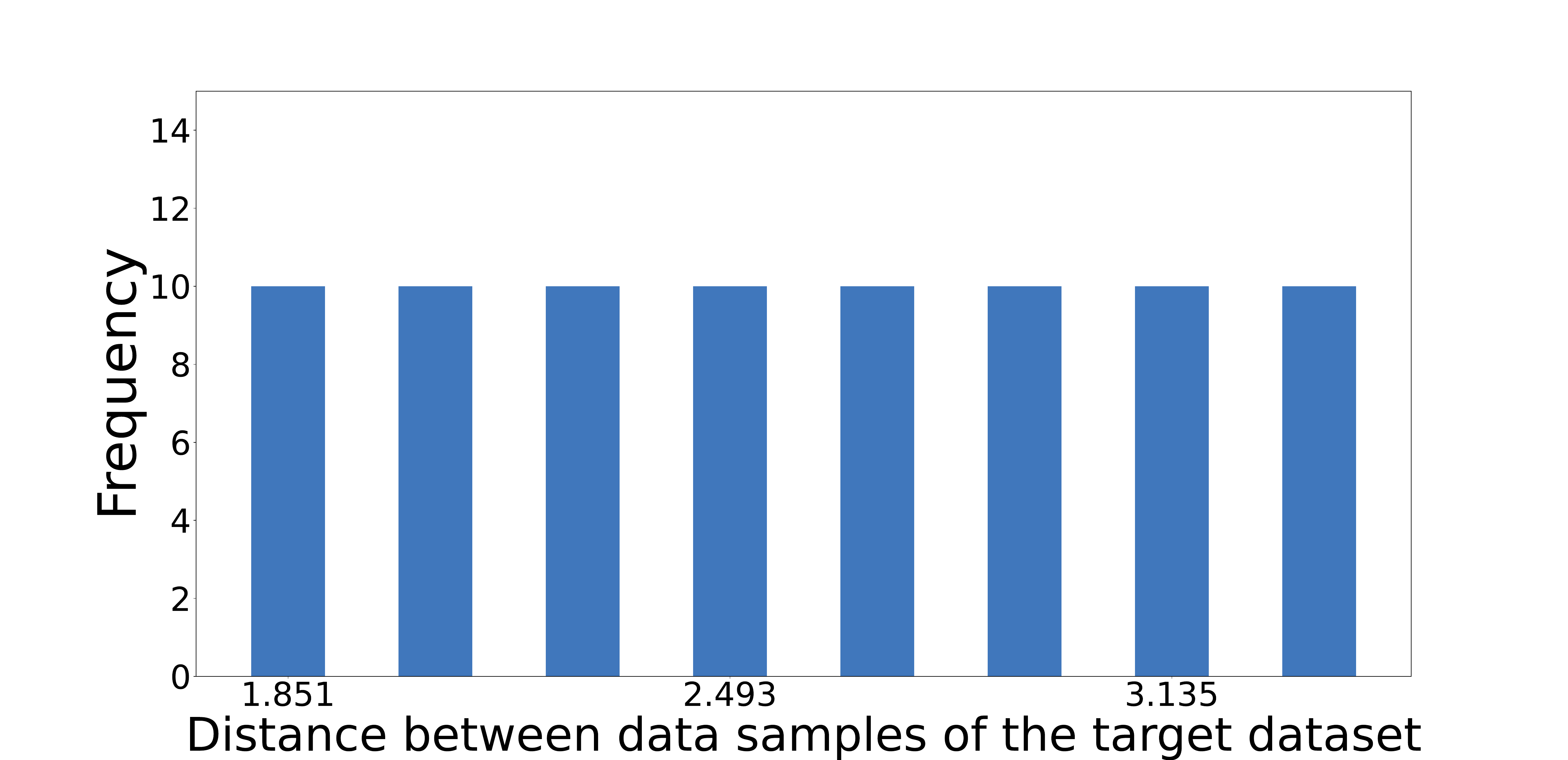}
\end{minipage}}
\subfigure[CHMN\_U\_1.72\_0.083\_20\%]{
\begin{minipage}{0.20\textwidth}
\label{Fig.sub.1}
\includegraphics[height=0.6\textwidth,width=1.1\textwidth]{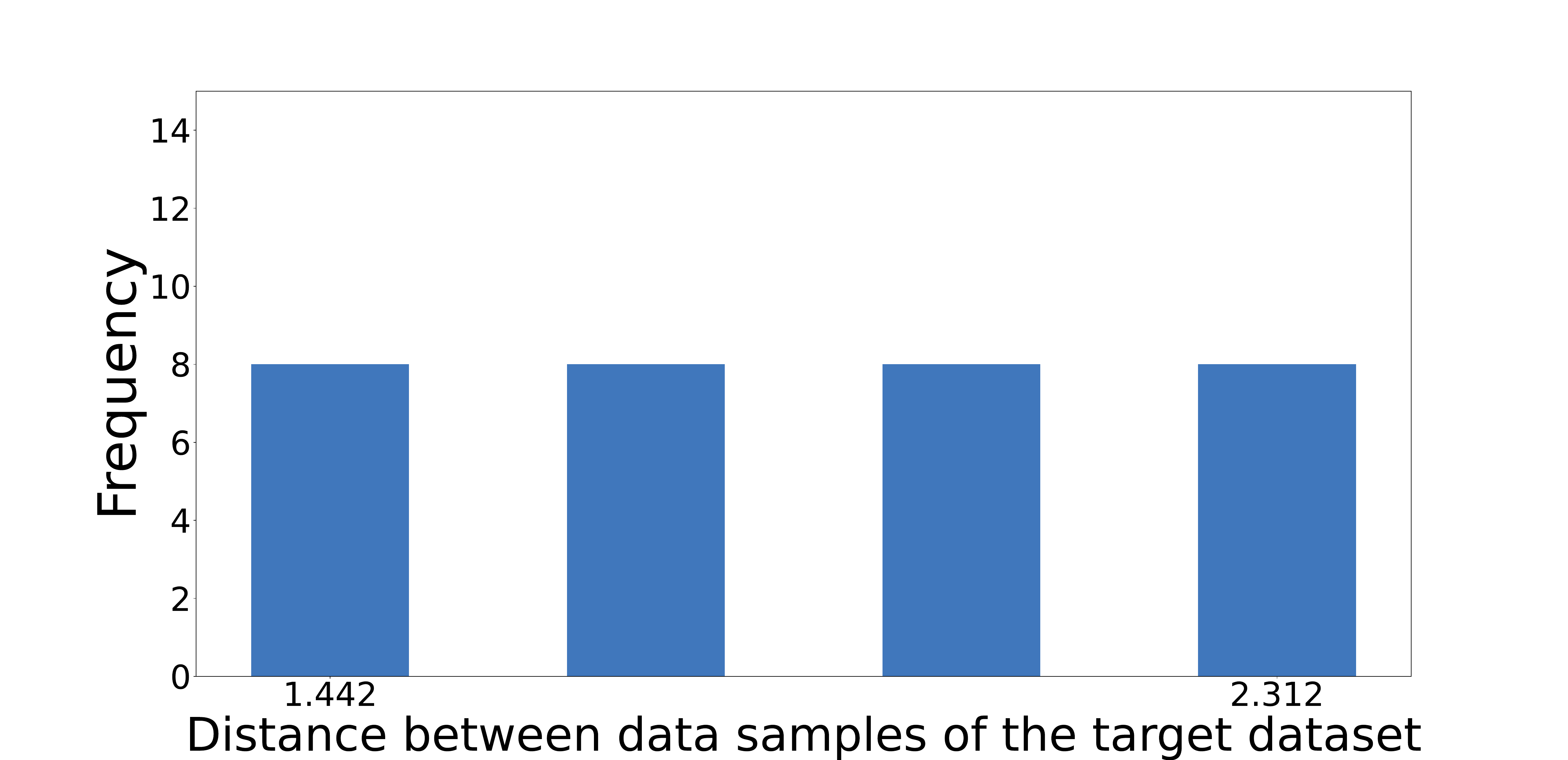}
\end{minipage}}
\subfigure[ImNet\_U\_1.13\_0.046\_20\%]{
\begin{minipage}{0.20\textwidth}
\label{Fig.sub.1}
\includegraphics[height=0.6\textwidth,width=1.1\textwidth]{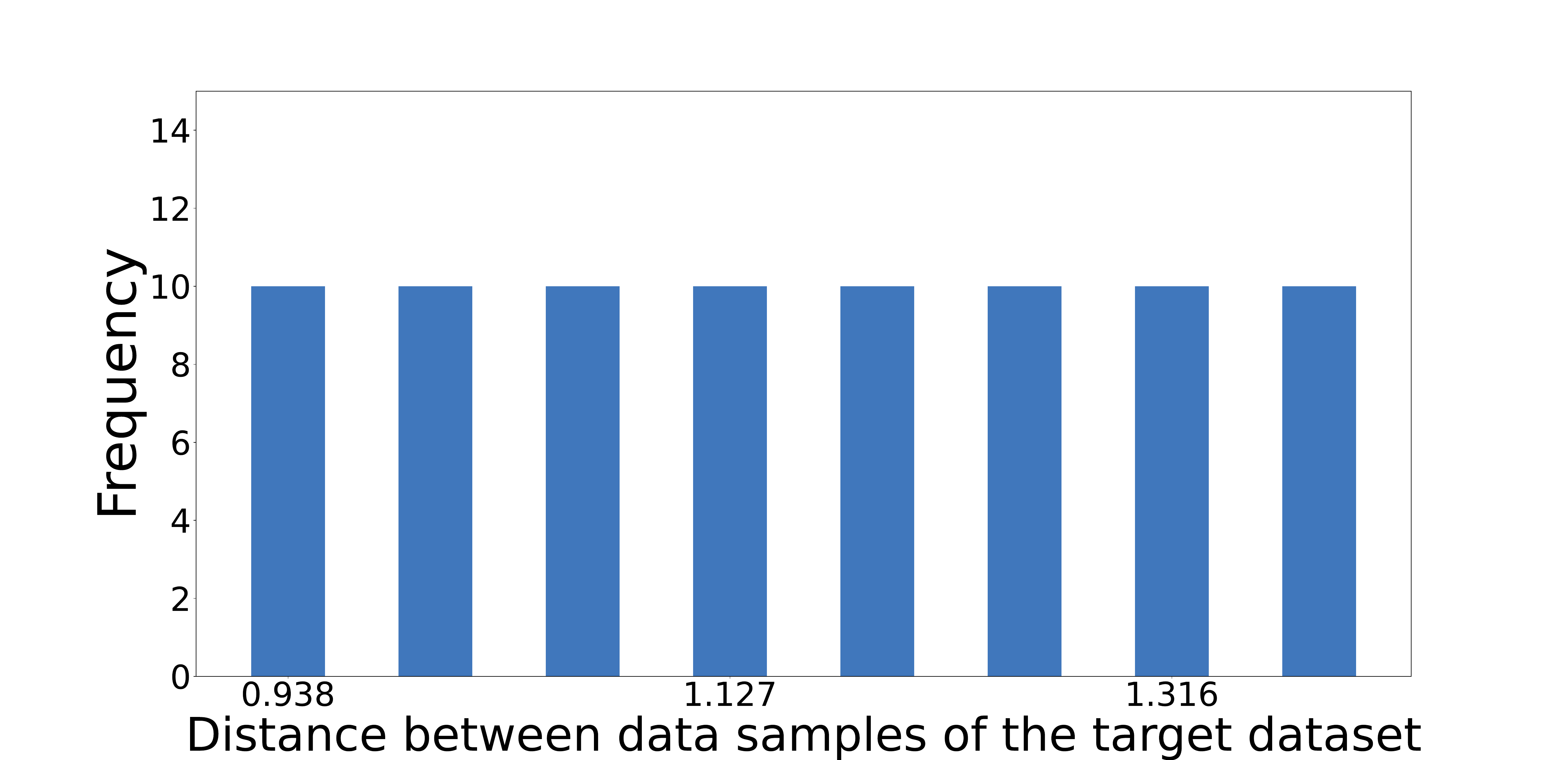}
\end{minipage}}
\subfigure[L30\_U\_0.724\_0.041\_4\%]{
\begin{minipage}{0.20\textwidth}
\label{Fig.sub.1}
\includegraphics[height=0.6\textwidth,width=1.1\textwidth]{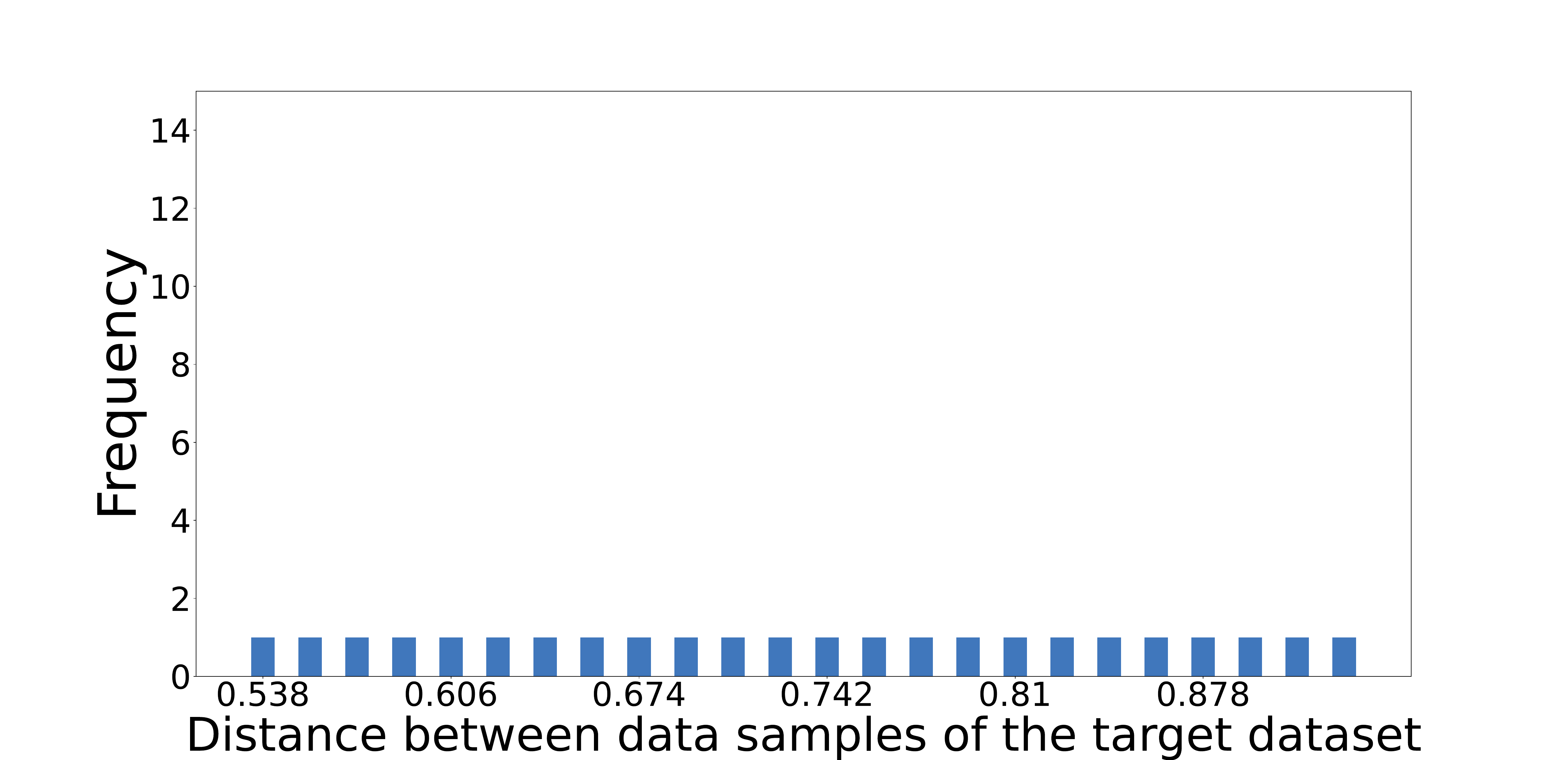}
\end{minipage}}

\subfigure[P100\_U\_0.625\_0.110\_2\%]{
\begin{minipage}{0.20\textwidth}
\label{Fig.sub.1}
\includegraphics[height=0.6\textwidth,width=1.1\textwidth]{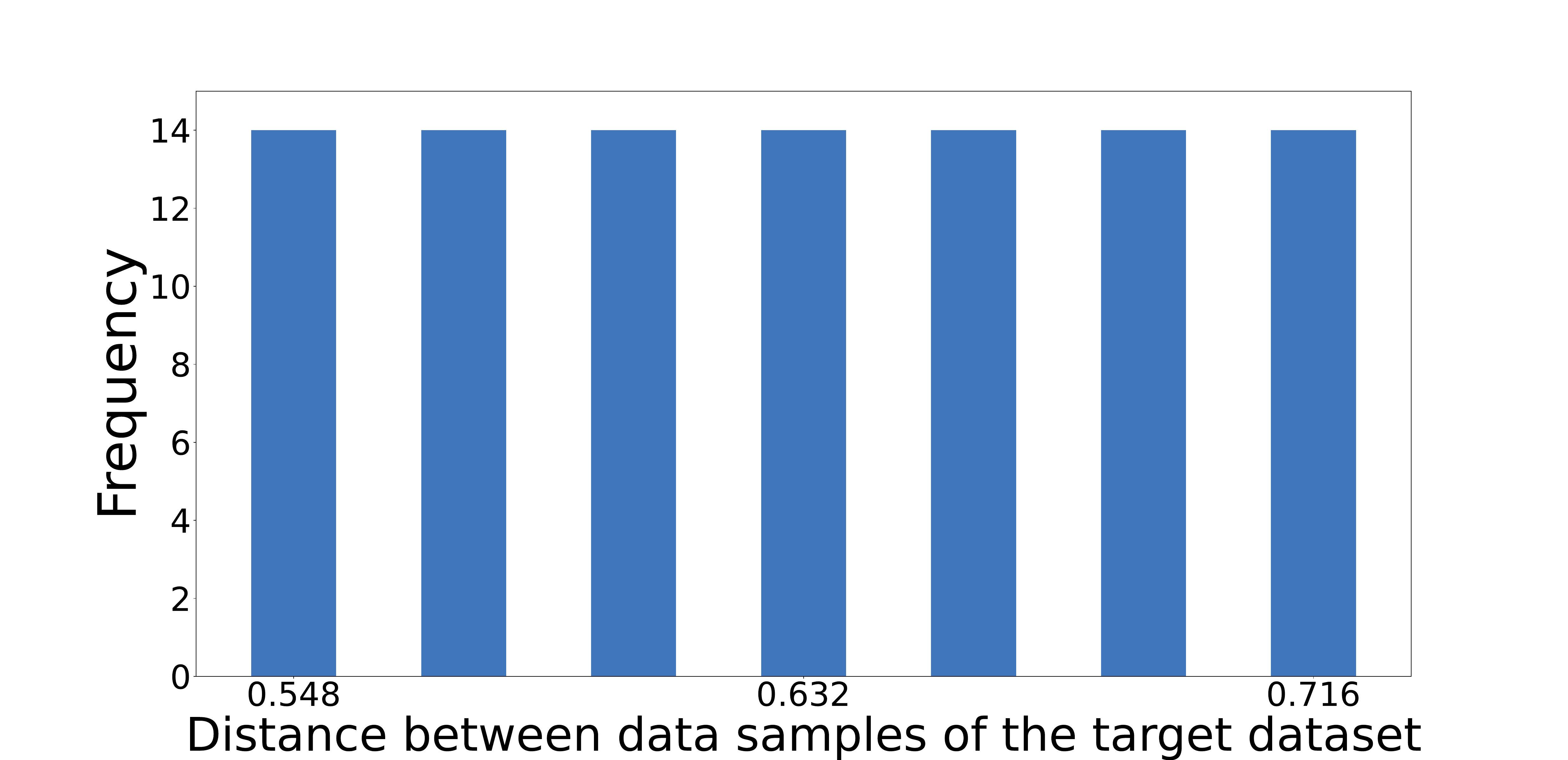}
\end{minipage}}
\subfigure[T100\_U\_0.641\_0.073\_4\%]{
\begin{minipage}{0.20\textwidth}
\label{Fig.sub.1}
\includegraphics[height=0.6\textwidth,width=1.1\textwidth]{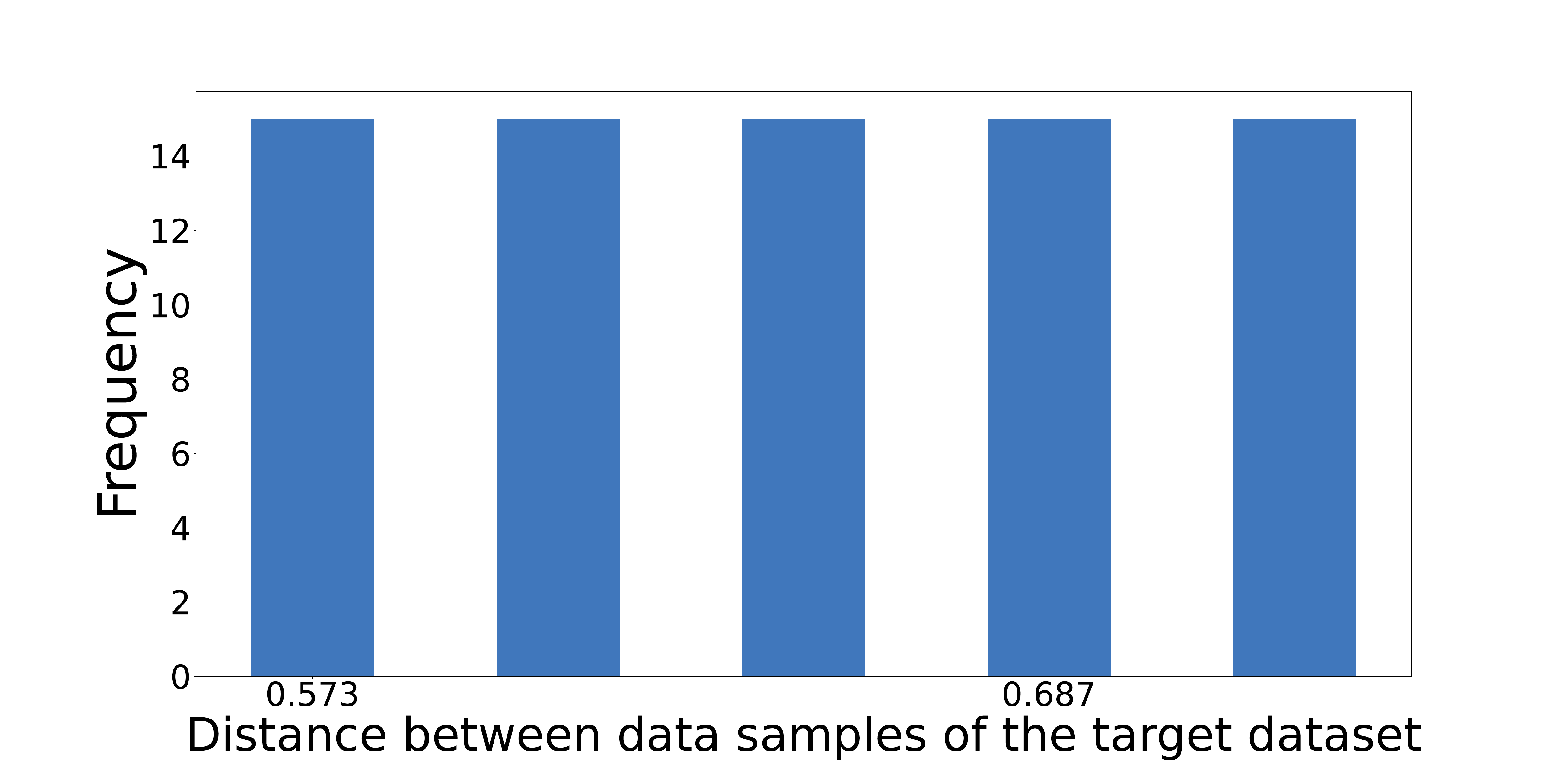}
\end{minipage}}

\end{center}
\caption{\label{figs:Fig3333} The Constructed Distance Distribution of Data Samples in the Target Dataset obeys Normal and Uniform Distribution, respectively. Inside (a)-(g) are the constructed distance distributions of data samples in the target CIFAR100 (C100), CIFAR10 (C10), CH\_MNIST (CHMN), ImageNet, Location30 (L30), Purchase100 (P100) and Texas100 (T100) datasets obey normal distributions, and (h)-(n) are the constructed distance distributions of data samples in the target seven datasets obey uniform distributions.}
\vspace{0.1pt}
\end{figure*}

Figure \ref{figs:Fig33 Original} and Figure \ref{figs:Fig3333} show the original and constructed distance distribution of data samples in the target target seven datasets, respectively.

\section*{C. The Selected Distances of the Target Dataset}
\label{subsec:The Selected Distances of the Target Dataset}

Table~\ref{tabs:The Selected Distances} shows the selected distances between data samples (Dis\_Between\_Data) and differential distances between two datasets of the target dataset in our experiments.

\begin{table}[htp]\tiny
    \caption{The Parameter (\emph{p}) and thresholds of the Bernoulli Distribution of Data Samples in the Target Dataset.}
    \centering
    \begin{tabular}{c|ccccccc}
        \toprule[1.1pt]
            \multirow{2}{*}{\textbf{Parameter (p)}} & \multicolumn{7}{c}{Thresholds of Bernoulli Distribution in Different Datasets} \\

             \cmidrule(r){2-8} & 
            \textbf{C\_100} & \textbf{C\_10} & \textbf{CH} & \textbf{Image} & \textbf{L\_30} & \textbf{P\_100} & \textbf{T\_100} 
            \\
         \midrule[1.1pt]
       0.1  & 2.623 & 1.147 & 0.729 & 0.835 & 0.520 & 0.504 & 0.512 \\ \hline

       0.2  & 2.97 & 1.529 & 0.837 & 0.906 & 0.574 & 0.554 & 0.551 \\ \hline

       0.3  & 3.306 & 1.84 & 0.935 & 0.964 & 0.608 & 0.566 & 0.578 \\ \hline

       0.4  & 3.535 & 2.097 & 1.081 & 1.021 & 0.660 & 0.608 & 0.605 \\ \hline

       0.5  & 3.778 & 2.39 & 1.186 & 1.081 & 0.705 & 0.620 & 0.630 \\ \hline

       0.6 & 4.013 & 2.707 & 1.30 & 1.153 & 0.750 & 0.635 & 0.661 \\ \hline

       0.7 & 4.259 & 3.064 & 1.45 & 1.227 & 0.784 & 0.675 & 0.692 \\ \hline

       0.8 & 4.555 & 3.513 & 1.765 & 1.323 & 0.845 & 0.688 & 0.729 \\ \hline

        0.9 & 4.913 & 4.285 & 2.19 & 1.491 & 0.920 & 0.741 & 0.813\\ 

        \bottomrule[1.1pt]
    \end{tabular}
   \label{tabs:The Parameter and thresholds}
    \vspace{0.01pt}
\end{table}

\begin{figure*}[]
\vspace{0.1pt}
\begin{center}
\subfigure[C100\_N\_2.893\_0.085\_20\%]{
\begin{minipage}{0.20\textwidth}
\label{ROC_CIFAR100_N_2.893_d1_20}
\includegraphics[height=0.7\textwidth,width=1.05\textwidth]{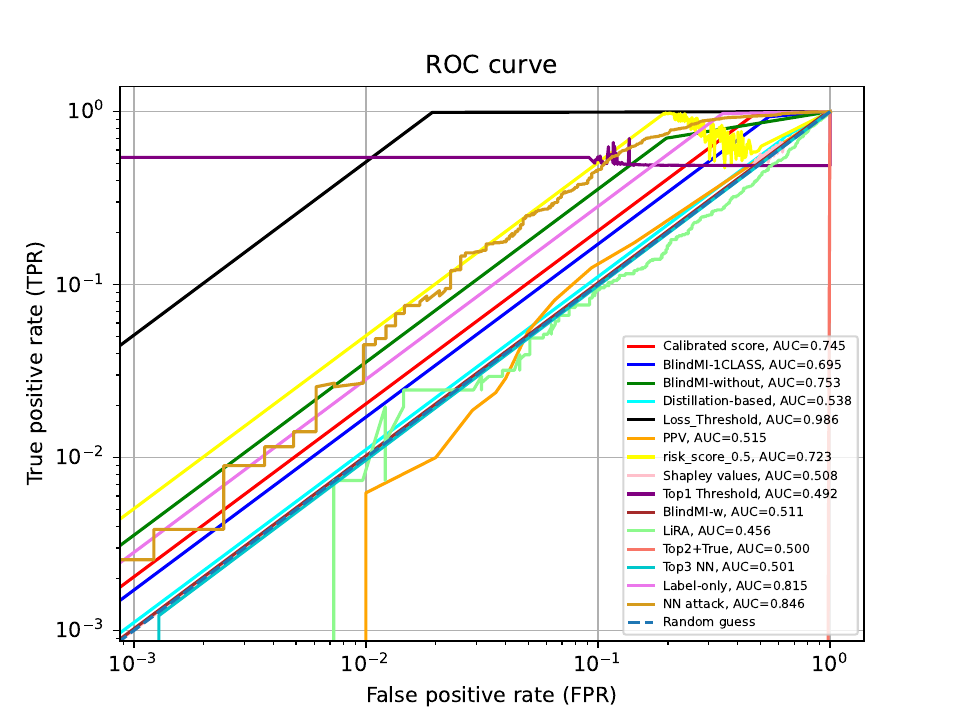}
\end{minipage}}
\subfigure[C100\_U\_2.893\_0.085\_20\%]{
\begin{minipage}{0.20\textwidth}
\label{ROC_CIFAR100_U_2.893_d1_20}
\includegraphics[height=0.7\textwidth,width=1.05\textwidth]{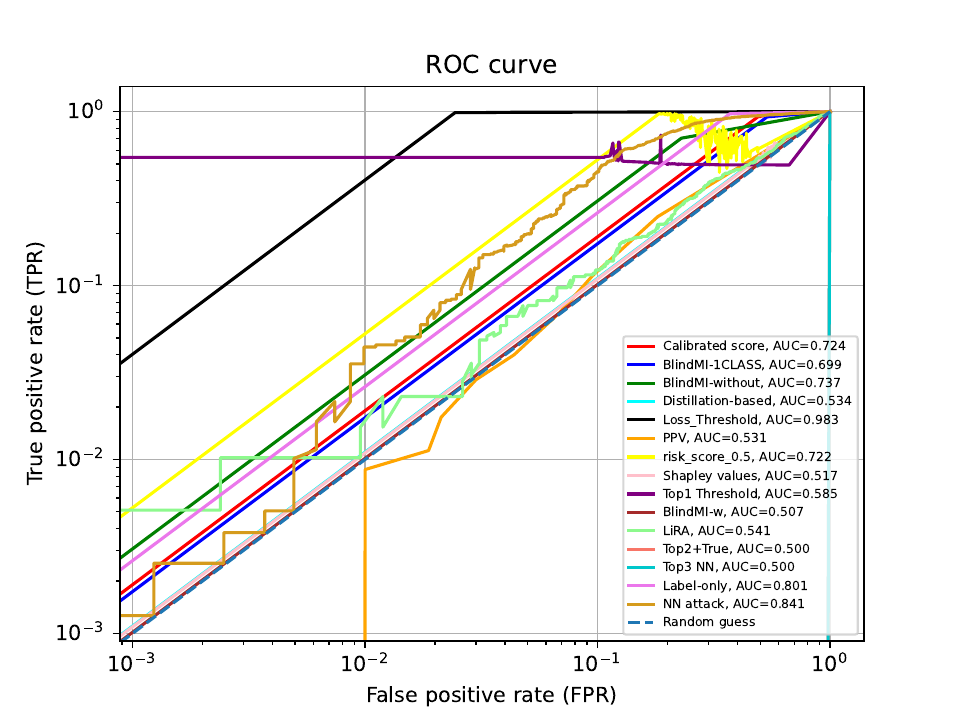}
\end{minipage}}
\subfigure[C100\_B\_2.893\_0.085\_20\%]{
\begin{minipage}{0.20\textwidth}
\label{ROC_CIFAR100_B_2.893_d1_20}
\includegraphics[height=0.7\textwidth,width=1.05\textwidth]{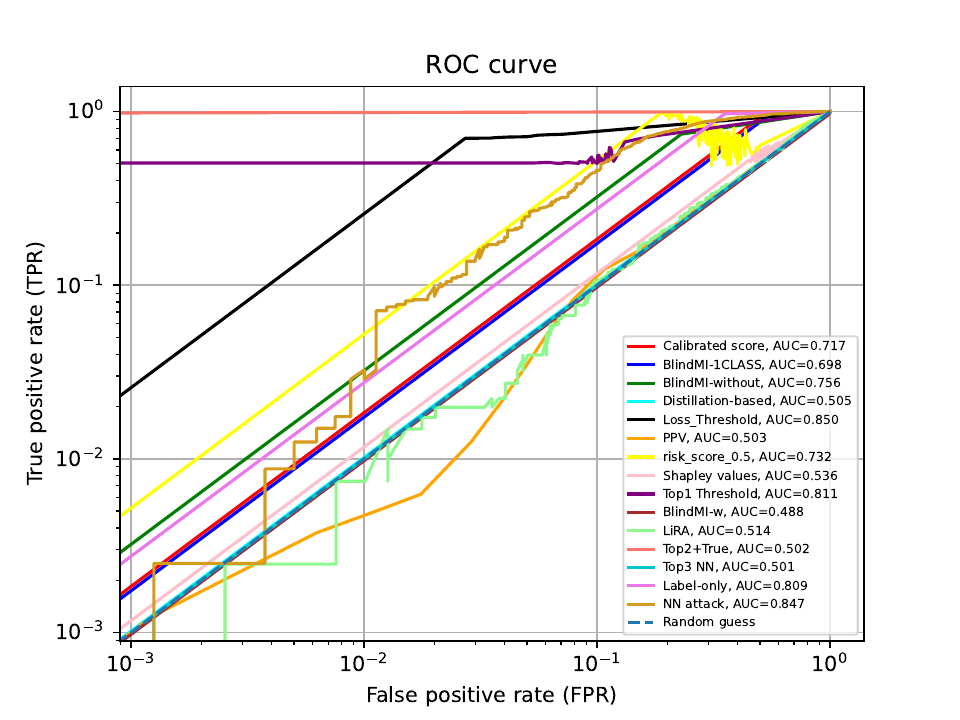}
\end{minipage}}

\end{center}
\caption{\label{figs:ROC} Comparing the true positive rate vs. false positive rate
of prior 15 state-of-the-art membership inference attacks on CIFAR100 by the AUC computed as the Area Under
the Curve of attack ROC in logarithmic scale. Inside (a)-(c) are the ROCs in logarithmic scale of the constructed distance distributions of data samples in the target CIFAR100 (C100) dataset obey normal, uniform and bernoulli distributions, respectively.}
\vspace{0.1pt}
\end{figure*}

\end{document}